\pdfoutput=1   
\documentclass[11pt,twoside,titlepage]{book}
\usepackage{fancyhdr}
\usepackage{subfigure} 
\fancypagestyle{plain}{%
\fancyhf{} 
\fancyfoot[R]{\textsf{SiD Letter of Intent\hspace{0.5cm}\thepage}}}
\usepackage{rotate}
\usepackage[figuresright]{rotating}
\usepackage{url}
\usepackage{graphicx}   
\usepackage[usenames]{color}
\usepackage[sectionbib]{chapterbib}
\usepackage{hyperref}  
\def\Ecm{\ifmmode {\,\mathrm{ E_{cm}}}\else
                   \textrm{E_{cm}}\fi}%
\def\TeV{\ifmmode {\,\mathrm{ Te\kern -0.05em V}}\else
                   \textrm{Te\kern -0.1em V}\fi}%
\def\GeV{\ifmmode {\,\mathrm{ Ge\kern -0.05em V}}\else
                   \textrm{Ge\kern -0.1em V}\fi}%
\def\MeV{\ifmmode {\,\mathrm{ Me\kern -0.05em V}}\else
                   \textrm{Me\kern -0.1em V}\fi}%

\def\lsim{\raise0.3ex\hbox{$\;<$\kern-0.75em\raise-1.1ex\hbox{$\sim\;$}}}
\def\gsim{\raise0.3ex\hbox{$\;>$\kern-0.75em\raise-1.1ex\hbox{$\sim\;$}}}

\begin{document}

\setcounter{chapter}{0}
\setcounter{secnumdepth}{3}
\setcounter{tocdepth}{2}
\pagenumbering{roman}

\begin{titlepage}
\vspace{3cm} {\Huge\begin{center} SiD Letter of Intent\end{center}}
\vspace{3cm} {\Large\begin{center} 31 March 2009\end{center}}

\begin{figure}[htbp]
\centerline{\includegraphics[width=6.0in]{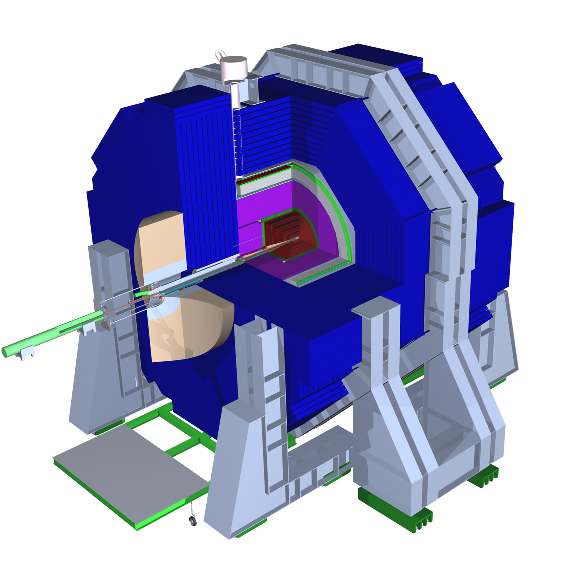}}
\label{fig:SiDdet}
\end{figure}

\newpage
\begin{center}
Editors:

H. Aihara, P. Burrows, M. Oreglia
\vspace{0.4in}

Cosigners:


E.~L.~Berger, V.~Guarino, J.~Repond, H.~Weerts, L.~Xia, J.~Zhang

\vspace{-.1in} {\it Argonne National Laboratory, USA} \vspace{0.1in}

Q. Zhang

\vspace{-.15in} {\it Argonne National Laboratory and IHEP, Beijing,
PR China} \vspace{0.1in}

A.~Srivastava

\vspace{-.15in} {\it Birla Institute for Technology and Science,
Pilani, India} \vspace{0.1in}

J.~M.~Butler

\vspace{-.15in} {\it Boston University, USA} \vspace{0.1in}

J.~Goldstein, J.~Velthuis

\vspace{-.15in} {\it Bristol University, UK} \vspace{0.1in}

V.~Radeka

\vspace{-.15in} {\it Brookhaven National Laboratory, USA}
\vspace{0.1in}

R.-Y.~Zhu

\vspace{-.15in} {\it California Institute of Technology, USA}
\vspace{0.1in}

P.~Lutz

\vspace{-.15in} {\it CEA-Saclay (IRFU), France} \vspace{0.1in}

A.~de~Roeck, K.~Elsener, A.~Gaddi, H.~Gerwig C.~Grefe, W.~Klempt,
L.~Linssen, D.~Schlatter, P.~Speckmayer

\vspace{-.15in} {\it CERN, Switzerland} \vspace{0.1in}

J.~Thom

\vspace{-.15in} {\it Cornell University, LNS, USA} \vspace{0.1in}

%

J.~Yang

\vspace{-.15in} {\it Ewha Womans University, South Korea}
\vspace{0.1in} \vspace{0.5in} 

D.~C.~Christian, S.~Cihangir, W.~E.~Cooper, M.~Demarteau,
H.~E.~Fisk, L.~A.~Garren, K.~Krempetz, R.~K.~Kutschke, R.~Lipton,
A.~Para, R.~Tschirhart, H.~Wenzel, R.~Yarema

\vspace{-.15in} {\it Fermi National Accelerator Laboratory, USA}
\vspace{0.1in}

M.~Grunewald

\vspace{-.15in} {\it Ghent University, Belgium} \vspace{0.1in}

A.~Pankov

\vspace{-.15in} {\it Gomel State Technical University, Belarus}
\vspace{0.1in}

T.~Dutta

\vspace{-.15in} {\it GSI, Germany} \vspace{0.1in}

P.~D.~Dauncey

\vspace{-.15in} {\it Imperial College London, UK} \vspace{0.1in}

J.P. Balbuena, C. Fleta, M. Lozano, M. Ull\'a{n}

\vspace{-.15in} {\it Insitute of Microelectronics of Barcelona
IMB-CNM (CSIC), Spain} \vspace{0.1in}

G.~B.~Christian

\vspace{-.15in} {\it Institute of Nuclear Research (ATOMKI),
Hungary} \vspace{0.1in}

A.~Faus-Golfe, J.~Fuster, C.~Lacasta, C.~Marin\~{n}as, M.~Vos

\vspace{-.15in} {\it Instituto de Fisica Corpuscular, IFIC
CSIC-Univ. Valencia, Spain} \vspace{0.1in}

J.~Duarte,  M.~Fernandez, J.~Gonzalez , R.~Jaramillo, A. Lopez-Virto,
C.~Martinez-Rivero, D.~Moya, A.~Ruiz-Jimeno,  I.~Vila

\vspace{-.15in} {\it Instituto de Fisica de Cantabria (IFCA),
CSIC-Univ. Cantabria, Spain} \vspace{0.1in}

C.~Colledani, A.~Dorokhov, C.~Hu-Guo, M.~Winter

\vspace{-.15in} {\it IPHC-IN2P3/CNRS, Strasbourg, France}
\vspace{0.1in}

G.~Moortgat-Pick

\vspace{-.15in} {\it IPPP, Durham, UK} \vspace{0.1in}

D.~V.~Onoprienko

\vspace{-.15in} {\it Kansas State University, USA}
 \vspace{0.1in}

G.~N.~Kim, H.~Park

\vspace{-.15in} {\it Kyungpook National University, South Korea}
\vspace{0.1in} \vspace{0.5in} 

C.~Adloff, J.~Blaha , J.-J.~Blaising, S.~Cap, M.~Chefdeville,
C.~Drancourt, A.~Espargiliare, R.~Gaglione, N.~Geffroy,
J.~Jacquemier, Y.~Karyotakis, J.~Prast, G.~Vouters

\vspace{-.15in} {\it Laboratoire d'Annecy-le-vieux de Physique des
Particules, Universite de Savoie, CNRS/IN2P3, France} \vspace{0.1in}

J.~Gronberg, S.~Walston, D.~Wright

\vspace{-.15in} {\it Lawrence Livermore National Laboratory, USA}
\vspace{0.1in}

L.~Sawyer

\vspace{-.15in} {\it Louisiana Tech University, USA} \vspace{0.1in}

M.~Laloum

\vspace{-.15in} {\it LPNHE, Faculte des Sciences de Paris VI et VII,
France} \vspace{0.1in}

C.~Ciobanu, J.~Chauveau, A.~Savoy-Navarro

\vspace{-.15in} {\it LPNHE, Universite Pierre et Marie Curie, Paris,
France} \vspace{0.1in}

L.~Andricek, H.-G.~Moser

\vspace{-.15in} {\it Max-Planck-Institute for Physics, Munich,
Germany} \vspace{0.1in}

R.~F.~Cowan, P.~Fisher, R.~K.~Yamamoto

\vspace{-.15in} {\it MIT, LNS, USA} \vspace{0.1in}

C.~J.~Kenney

\vspace{-.15in} {\it Molecular Biology Consortium, USA}
\vspace{0.1in}

E.~E.~Boos, M. Merkin

\vspace{-.15in} {\it Moscow State University, Russia} \vspace{0.1in}

S.~Chen

\vspace{-.15in} {\it Nanjing University, China} \vspace{0.1in}

D.~Chakraborty,
 A.~Dyshkant,
 D.~Hedin,
 V.~Zutshi

\vspace{-.15in} {\it Northern Illinois University, USA}
\vspace{0.1in}

V.~Galkin, N.~D'Ascenzo, D.~Ossetski, V.~Saveliev

\vspace{-.15in} {\it Obninsk State Technical University for Nuclear
Power Engineering, Russia} \vspace{0.1in}

F.~Kapusta

\vspace{-.15in} {\it LPNHE, Univ. VII, Paris, France} \vspace{0.1in}

R.~De~Masi

\vspace{-.15in} {\it PHC-IN2P3/CNRS, Strasbourg, France}
\vspace{0.1in}

V.~Vrba

\vspace{-.15in} {\it Institute of Physics, Prague, Czech Republic}
\vspace{0.1in}

C.~Lu, K.~T.~McDonald, A.~J.~S.~Smith

\vspace{-.15in} {\it Princeton University, USA} \vspace{0.1in}

D.~Bortoletto

\vspace{-.15in} {\it Purdue University, USA} \vspace{0.1in}

R.~Coath,
 J.~Crooks,
 C.~Damerell,
 M.~Gibson,
 A.~Nichols,
 M.~Stanitzki,
 J.~Strube,
 R.~Turchetta,
 M.~Tyndel,
 M.~Weber,
 S.~Worm,
 Z.~Zhang

\vspace{-.15in} {\it Rutherford Appleton Laboratory, UK}
\vspace{0.1in}

T.~L.~Barklow,
 A.~Belymam,
 M.~Breidenbach,
 R.~Cassell,
 W.~Craddock,
 C.~Deaconu,
 A.~Dragone,
 N.~A.~Graf,
 G.~Haller,
 R.~Herbst,
 J.~L.~Hewett,
 J.~A.~Jaros,
 A.~S.~Johnson,
 P.~C.~Kim,
 D.~B.~MacFarlane,
 T.~Markiewicz,
 T.~Maruyama,
 J.~McCormick,
 K.~Moffeit,
 H.~A.~Neal,
 T.~K.~Nelson,
 M.~Oriunno,
 R.~Partridge,
 M.~E.~Peskin,
 T.~G.~Rizzo,
 P.~Rowson,
 D.~Su,
 M.~Woods

\vspace{-.15in} {\it SLAC National Accelerator Laboratory, USA}
\vspace{0.1in}

S.~Chakrabarti,

\vspace{-.15in} {\it SUNY, Stony Brook, USA} \vspace{0.1in}

A.~Dieguez,
 Ll.~Garrido

\vspace{-.15in} {\it University of Barcelona, Spain}
 \vspace{0.1in}

J.~Kaminski

\vspace{-.15in} {\it University of Bonn, Germany} \vspace{0.1in}

J.~S.~Conway, M.~Chertok, J.~Gunion, B.~Holbrook, R.~L.~Lander,
S.~M.~Tripathi

\vspace{-.15in} {\it University of California, Davis, USA}
\vspace{0.1in}

V.~Fadeyev,
 B.~A.~Schumm

\vspace{-.15in} {\it University of California, Santa Cruz, USA}
\vspace{0.1in}

M.~Oreglia

\vspace{-.15in} {\it University of Chicago, USA} \vspace{0.1in}

J.~Gill, U.~Nauenberg, G.~Oleinik, S.~R.~Wagner

\vspace{-.15in} {\it University of Colorado, USA} \vspace{0.1in}

K.~Ranjan, R.~Shivpuri

\vspace{-.15in} {\it University of Delhi, India} \vspace{0.1in}

G.~S.~Varner

\vspace{-.15in} {\it University of Hawaii, USA} \vspace{0.1in}

R.~Orava

\vspace{-.15in} {\it University of Helsinki, Finland}
 \vspace{0.1in}

R.~Van~Kooten

\vspace{-.15in} {\it University of Indiana, USA} \vspace{0.1in}

B.~Bilki, M.~Charles\thanks{Now at Oxford}, T.~J.~Kim, U.~Mallik,
E.~Norbeck, Y.~Onel

\vspace{-.15in} {\it University of Iowa, USA} \vspace{0.1in}

B.~P.~Brau, S.~Willocq

\vspace{-.15in} {\it University of Massechusetts, Amherst, USA}
\vspace{0.1in}

G.~N.~Taylor

\vspace{-.15in} {\it University of Melbourne, Australia}
\vspace{0.1in}

K.~Riles, H.-J.~Yang

\vspace{-.15in} {\it University of Michigan, USA} \vspace{0.1in}

R.~Kriske

\vspace{-.15in} {\it University of Minnesota, USA} \vspace{0.1in}

L.~Cremaldi, R.~Rahmat

\vspace{-.15in} {\it University of Mississippi, USA} \vspace{0.1in}

G.~Lastovicka-Medin

\vspace{-.15in} {\it University of Montenegro, Montenegro}
\vspace{0.1in}

S.~Seidel

\vspace{-.15in} {\it University of New Mexico, USA} \vspace{0.1in}

M.~D.~Hildreth, M.~Wayne

\vspace{-.15in} {\it University of Notre Dame, USA} \vspace{0.1in}

J.~E.~Brau, R.~Frey, N.~Sinev, D.~M.~Strom, E.~Torrence

\vspace{-.15in} {\it University of Oregon, USA} \vspace{0.1in}

Y.~Banda, P.~N.~Burrows, E.~Devetak, B.~Foster, T.~Lastovicka,
Y.-M.~Li, A.~Nomerotski

\vspace{-.15in} {\it University of Oxford, UK}
\vspace{0.1in} \vspace{0.5in} 

J.~Riera-Babures, X.~Vilasis-Cardona

\vspace{-.15in} {\it Universitat Ramon Llull, Spain} \vspace{0.1in}

S.~Manly

\vspace{-.15in} {\it University of Rochester, USA} \vspace{0.1in}

B.~Adeva, C.~Iglesias~Escudero, P.~Vazquez~Regueiro,
J.~J.~Saborido~Silva, A.~Gallas~Torreira

\vspace{-.15in} {\it University of Santiago de Compostela (IGFAE),
Spain} \vspace{0.1in}

D.~Gao, W.~Jie, Y.~Junfeng, C.~Li, S.~Liu, Y.~Liu, Y.~Sun, Q.~Wang,
J.~Yi, W.~Yonggang, Z.~Zhao

\vspace{-.15in} {\it University of Science and Technology of China}
\vspace{0.1in}

K.~De, A.~Farbin, S.~Park, J.~Smith, A.~P.~White, J.~Yu

\vspace{-.15in} {\it University of Texas at Arlington, USA}
 \vspace{0.1in}

X.~C.~Lou

\vspace{-.15in} {\it University of Texas at Dallas, USA}
 \vspace{0.1in}

T.~Abe, H.~Aihara, M.~Iwasaki

\vspace{-.15in} {\it University of Tokyo, Japan} \vspace{0.1in}

H.~J.~Lubatti

\vspace{-.15in} {\it University of Washington, Seattle, USA}
\vspace{0.1in}

H.~R.~Band, F.~Feyzi, R.~Prepost

\vspace{-.15in} {\it University of Wisconsin, Madison, USA}
 \vspace{0.1in}

P.~E.~Karchin, C.~Milstene

\vspace{-.15in} {\it Wayne State University, USA} \vspace{0.1in}

C.~Baltay, S.~Dhawan

\vspace{-.15in} {\it Yale University, USA} \vspace{0.1in}

Y.-J.~Kwon

\vspace{-.15in} {\it Yonsei University, Japan} \vspace{0.1in}

\end{center}
\end{titlepage}

\fancypagestyle{plain} {
    \fancyhead{}
    \fancyfoot{}
}   
\pagestyle{empty} \tableofcontents
\newpage
\pagestyle{fancyplain}
\pagenumbering{arabic}

\chapter{Introduction}
\label{chap:introduction}

This document presents the current status of SiD's effort to develop
an optimized design for an experiment at the International Linear
Collider. It presents detailed discussions of each of SiD's various
subsystems, an overview of the full GEANT4 description of SiD, the
status of newly developed tracking and calorimeter reconstruction
algorithms, studies of subsystem performance based on these tools,
results of physics benchmarking analyses, an estimate of the cost of
the detector, and an assessment of the detector R\&D needed to
provide the technical basis for an optimised SiD.

\section{The ILC Physics Menu} \label{sec:ILCphys}
The Silicon Detector (SiD) has
been designed to address questions of fundamental importance to
progress in particle physics:
\begin{itemize}
\item What is the mechanism responsible for electroweak symmetry breaking and the generation of mass?
\item How do the forces unify?
\item Does the structure of space-time at small distances show evidence of extra dimensions?
\item What are the connections between the fundamental particles and forces and cosmology?
\end{itemize}

These questions are addressed through precision measurements by SiD at the
International Linear Collider (ILC) of the following:
\begin{itemize}
\item Higgs boson properties;
\item Gauge boson scattering;
\item Effects resulting from the existence of extra dimensions;
\item Supersymmetric particles; and
\item Top quark properties.
\end{itemize}

The Higgs boson studies will measure in detail the properties of the
Higgs boson in order to determine its consistency with Standard
Model expectations, and the nature of any deviations. These
measurements will include the mass, width, spin, branching ratios
and couplings, and the Higgs self-coupling. The estimated precision
in these measurements is sufficient to
discriminate between competing theories for electroweak symmetry
breaking. With the decay-independent detection of Higgstrahlung
events by Z tagging, sensitivity to a wide range of models is
possible. Such measurements would establish the role of the Higgs
boson in electroweak symmetry breaking, and the generation of mass.
Should Nature choose a Higgs-less scenario, it could be addressed by
studying the coupling of gauge bosons.
If additional Higgs bosons beyond that of the Standard Model exist, 
they can be detected and studied with SiD, even if they have very 
large masses.
For example, in the MSSM, the additional four Higgs bosons can be
detected if they are within the kinematic reach of the ILC. 

If
supersymmetric particles are produced at the ILC, 
their masses and couplings will be measured by SiD, allowing studies of the evolution of the strength of the 
fundamental forces and experimentally 
testing their unification at very high energies.
SiD has a reach up to $\sim $3 TeV in
Higgs-less strong coupling scenarios. Such models include those
with strongly-interacting W and Z and some with dimensions. 
A universe with extra dimensions of a scale within
reach of the ILC can be probed, with high-precision sensitivity to the separate
parameters of scale and number of dimensions. By observing masses
and widths of excited graviton states, the length scale and the
curvature in an additional fifth dimension of space-time can be
determined.
SiD plans to study the top quark, including the precise
measurement of its mass and Yukawa coupling. Finally. a number
of connections to issues of interest in cosmology would be realized
through many of the measurements described above. These focus on two
fundamental issues, namely the mechanism behind the present day
baryon asymmetry and the nature of the cold dark matter

\section{Detector Overview}

\begin{figure}[htbp]
\centering
\includegraphics[width=5in]{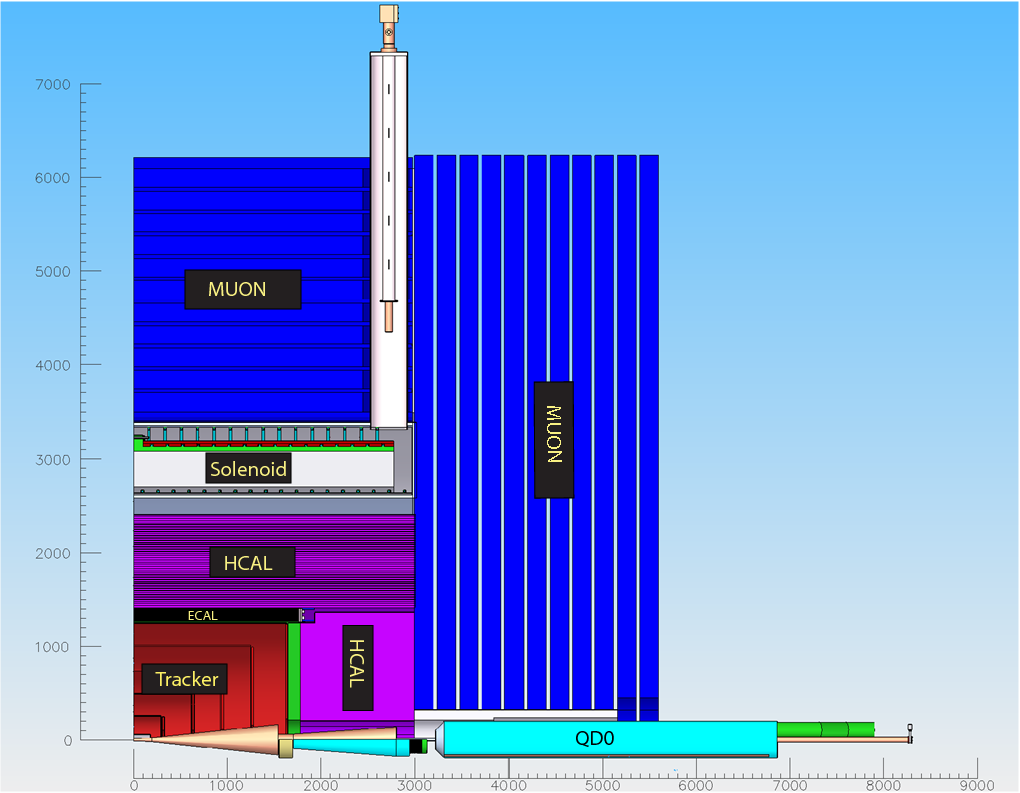}
\caption{Illustration of a quadrant of SiD (dimensions in mm).}
\label{fig:quadrant}
\end{figure}

SiD is a linear collider detector concept designed to make precision measurements,
and at the same time be sensitive to a wide range of possible
new phenomena at the ILC. A plan view of one
quadrant of the detector is shown in Figure~\ref{fig:quadrant}.
SiD is based on silicon
tracking, silicon-tungsten electromagnetic calorimetry, highly
segmented hadronic calorimetry, and a powerful silicon pixel vertex
detector. SiD also incorporates a high field solenoid, iron flux
return, and a muon identification system. Particle Flow Analysis (PFA)
(see below) is an important consideration for the basic philosophy
and layout of the detector.

The choice of silicon
detectors for tracking and vertexing ensures that SiD is robust
with respect to beam backgrounds or beam loss, provides superior
charged particle momentum resolution, and eliminates out of time
tracks and backgrounds. The main tracking detector and calorimeters
are ``live'' only during a single bunch crossing, so beam-related
backgrounds and low mass backgrounds from 2-photon processes will be reduced to
the minimum possible levels. The SiD design has been cost-conscious
from the beginning, and the present global design represents a
careful balance between cost and physics performance. The SiD
proponents are convinced that two detectors are required
scientifically, technically, and sociologically to exploit fully the ILC
physics
potential. Therefore SiD is engineered to make the push pull process
efficient and to minimize the time required for alignment and
calibration after a move.

\begin{table}[htbp]
\begin{center}
\begin{tabular}{|l|l|l|l|p{83pt}|}
\hline SiD BARREL& Technology& Inner radius& Outer radius& Z max \\
\hline Vertex detector& Pixel& 1.4& 6.0& $\pm \quad 6.25$ \\
\hline Tracker& Silicon strips& 21.7& 122.1& $\pm \quad 152.2$ \\
\hline EM calorimeter& Silicon-W& 126.5& 140.9& $\pm \quad 176.5$ \\
\hline Hadron calorimeter& RPCs& 141.7& 249.3& $\pm \quad 301.8$ \\
\hline Solenoid& 5 Tesla & 259.1& 339.2& $\pm \quad 298.3$ \\
\hline Flux return& RPCs& 340.2 & 604.2& $\pm \quad 303.3$ \\
\hline
\end{tabular}

%
\begin{tabular}{|p{107pt}|l|l|l|l|}
\hline SiD FORWARD& Technology& Inner Z& Outer Z& Outer radius \\
\hline Vertex detector& Pixel& 7.3& 83.4& 16.6 \\
\hline Tracker& Silicon strips& 77.0& 164.3& 125.5 \\
\hline EM calorimeter& Silicon-W& 165.7& 180.0& 125.0 \\
\hline Hadron calorimeter& RPCs& 180.5& 302.8& 140.2 \\
\hline Flux return& RPCs& 303.3& 567.3& 604.2 \\
\hline LumCal& Silicon-W& 158.0& 173.0& 19.0 \\
\hline BeamCal& Silicon-W& 295.0& 320.0& 14.5 \\
\hline
\end{tabular}
\caption[Key parameters of the baseline SiD design.]{Key parameters of the baseline SiD design. (All dimension 
are given in
cm.)\label{SiDparms}}
\end{center}
\end{table}
%
%

The key parameters of the SiD design starting point are listed in
Table~\ref{SiDparms}.
The innermost tracking sub-system is the Vertex Detector
(VXD), which comprises 5 cylinders and 4 sets of endcaps, composed of pixilated
sensors closely surrounding the beampipe. The impact parameter resolution will
surpass $\sigma _{r\phi
}=\sigma _{rz}$= 5 $\oplus $ 10/(p sin$^{3/2} \theta )$ [$\mu $m].
SiD has chosen a 5 T solenoidal field in part to control the
e$^+$e$^-$ pair background, and the cylinder and disk geometry is
chosen to minimize scattering and ensure high performance in the
forward direction. The VXD sensor technology is not yet chosen
because the relatively high luminosity per train at the ILC makes
integration through the train undesirable, and optimal technologies
for separating the train into small temporal segments, preferably
bunches, have not been determined. This is not a problem, since this
choice will have almost no effect on the rest of the SiD design, and
the VXD can be built and installed after many of the main detector components are
complete.

SiD has chosen Si strip technology, arrayed in 5 cylinders and 4
endcaps for precision tracking and momentum measurement. Particular
attention has been given to fabricating the endcaps with minimal
material to enhance forward tracking. The sensors are single sided
Si, approximately 15 cm square, with a pitch of 50 $\mu $m. With an
outer cylinder radius of 1.25 m and a 5 T field, the charged track
momentum resolution will be better than $\sigma $(1/p$_{T})$ = 5 x
10$^{-5 }$ (GeV/$c$)$^{-1}$ for high momentum tracks.
The endcaps utilize two sensors bonded together
for small angle stereo measurements.
Highly efficient track finding has been demonstrated in simulation
using the integrated tracking system.

SiD calorimetry is optimized for jet energy measurement, and is
based on a Particle Flow strategy, in which charged particle momenta
are measured in the tracker; neutrals are measured in the
calorimeter, and then the charged and neutral components are added.
The challenge is identifying the energy which charged particles
deposit in the calorimeters, and discriminating it from the energy
photons and neutral hadrons deposit, so it can be removed. This
requires highly segmented readout, both transversely and
longitudinally, and in the ECAL puts a premium on minimizing the
spread of electromagnetic showers.
SiD calorimetry begins with an
exceptionally dense, highly pixilated Silicon -- Tungsten
electromagnetic section. The ECAL has alternating layers of W and
silicon pixel detectors; there are 20 layers of 2.5 mm tungsten
followed by 10 layers of 5 mm tungsten. The silicon detector layers
are only 1.25 mm thick. This results in a Moliere radius for the
thin section of 13.5 mm.
The sensor is divided into 1024 hexagonal pixels,
forming
an imaging calorimeter with a track resolution of $\sim $1 mm. The
ECAL has a total of 26 X$_{0}$. The same technology is used in
the endcaps.
Silicon detector technology is used in the  vertex detector, tracker, and ECAL
because it is robust against machine backgrounds, only sensitive to  backgrounds in a single bunch crossing, 
and has high precision.

The Hadronic Calorimeter (HCAL) is made from 4.5 $\lambda $ of
Stainless Steel, divided into 40 layers of steel and detector. The
baseline detectors are RPCs with 1 cm square pixels,
inserted into 8 mm gaps between the steel layers. The same
technology is used for the endcaps.
The jet energy resolution has been studied with Pandora PFA and also
with SiD PFA (see online appendices for details); it is $\Delta E/E$
= 4\%. Further improvement is expected as the algorithm matures.
Having such superb jet energy resolution will allow, for the first
time, the invariant masses of W's, Z's, and tops to be reconstructed
with resolutions approaching the natural widths of these particles.

The calorimetric coverage is completed in the forward direction by a
LumCal and a BeamCal. The LumCal overlaps the endcap ECAL, and is
designed for a $\Delta $L/L precision of $1 \times 10^{-3}$. The
Lumcal is Si-W, with the pixilation designed to optimize the
luminosity measurement precision. The BeamCal is the smallest angle
calorimeter and is mounted to the inboard side of QD0. The BeamCal
sensor technology may be diamond or low resistivity Si. Both
calorimeters are designed for a 14 mrad crossing angle and an
anti-DID correction winding on the main solenoid (see Section~\ref{sec:magnet}).

The SiD 5 T superconducting solenoid is based on the CMS design,
but has 6 layers of conductor. The stored energy is $\sim $1.6 GJ.
The critical cold mass parameters, such as stored energy/Kg, are
similar to CMS. The CMS conductor is the baseline choice, but SiD is
developing an advanced conductor that would eliminate the e-beam
welding of structural alloy and be easier to wind. SiD will carry
all the solenoid utilities (power supply, quench protection, etc)
except for the He liquefier, which will be connected by a vacuum
insulated flex line.

The flux is returned with an iron structure, configured as a barrel with movable endcaps.
The present design limits field leakage to $<$100~G at
1~m. The flux return is 11 layers of 20 cm iron. The flux return
also is the absorber for the muon identifier and is an important
component of SiD self shielding. The barrel is composed of full
length modules to help keep the structure stable during push pull
and to enable full length muon detectors. The endcaps support the final-focus
QD0 magnets, with provision for transverse
alignment of the quads and vibration isolation. A platform fixed to
the barrel supports the 2K cryogenic system for the QD0s. SiD is
designed for rapid push pull and rapid recovery after such motion.
Precision alignment will utilise a geodetic network of frequency
scanning interferometers.

\section{Polarimeters and Energy Spectrometers} \label{sec:POLE}

The ILC physics program demands precise polarization and beam energy
measurements~\cite{role-pol}, and SiD will have major
responsibilities for these. The ILC baseline configuration is
defined in the RDR~\cite{rdr} and specifies many key features of the
machine that impact these measurements and the related physics
program.  These features are summarized in
Reference~\cite{pole-technote} and include:
\begin{itemize}
\item{a highly polarized electron beam with $P>80\%$;}
\item{a polarized positron beam with an initial polarization of $\sim30-45\%$;}
\item{spin rotator systems for both electron and positron beams to achieve longitudinal polarization at the collider IP;}
\item{polarimeters and energy spectrometers upstream and downstream of the IP for both beams.  The polarimeters utilize Compton 
scattering and aim to measure the polarization with an accuracy  of $\Delta P/P = 0.25\%$.  The energy spectrometers aim to 
achieve an accuracy of 100-200 parts per million (ppm);}
\item{capability to rapidly flip the electron helicity at the injector using the source laser.  (The possibility of fast positron 
helicity flipping is not included in the baseline configuration, but a scheme for fast positron helicity flipping has been 
proposed).}
\end{itemize}

The baseline ILC provides beam energies in the range 100-250 GeV;
precise polarization and energy measurements are required for this
full range.  The baseline also provides for detector calibration at
the Z-pole with 45.6 GeV beam energies, but does not require
accurate polarimetry or energy spectrometer measurements at the
Z-pole. As discussed in References~\cite{pole-technote}
and~\cite{Aurand}, accurate polarization and energy measurements at
the Z-pole would allow important cross-checks and calibration of the
polarimeters and energy spectrometers. SiD endorses making
polarimetry and energy measurement checks and calibrations at the Z.

The polarimeter and energy spectrometer systems need to be a joint
effort of the ILC Beam Delivery System (BDS) team and the Detector collaborations as
summarized in Reference~\cite{pole-technote}.  SiD intends to take
significant responsibility for the design, development, operation
and performance of these systems. SiD participants are already
active in making significant contributions to their design and
development.  Data from the polarimeters and spectrometers must be
delivered to the SiD DAQ in real time so as to be logged and permit fast
online analysis. Fast online analysis results must also be provided
to the ILC controls system for beam tuning and diagnostics.  Details
for integrating the polarimeter and energy spectrometer data with
the SiD DAQ remain to be worked out, but SiD experts will assume
responsibility for integrating the polarimeter and spectrometer data
streams with the DAQ.

\section{SiD Detector Optimization} \label{sec:Opt}

It is the design of the calorimeters, more than other subsystems,
which determines the global parameter choices for the whole
detector. The calorimeters are costly, and their performance and
costs depend critically on how far from the interaction point they
are placed and how thick they must be to contain most of the energy
of jets, and therefore how big and costly the detector must be. The
problem of optimizing the global detector design simplifies, crudely,
to optimizing the parameters of the
calorimeters and solenoid. Bigger is more performant, but
also more costly. A balance must be struck.

Optimizing SiD's global parameters required three tools. First, a
parametric cost model was developed which could calculate the total
cost based on the choice of global parameters, assumptions about
what materials and detectors comprised each detector element, and
per unit cost assumptions appropriate for the materials and
detectors chosen.

Secondly, algorithms to estimate the jet energy resolution of the detector or
the momentum resolution and impact parameter resolution of the tracker were
needed to evaluate the subsystem performance as a function of the global
parameters. The key parameters include the outer radius of the tracker, $R$,
the tracker's length, $z$, the magnetic field, $B$, and the thickness of the
hadronic calorimeter absorber $\lambda$. SiD utilized its own algorithms to
estimate the tracking and vertexing performance, and employed
parameterizations of the PFA performance derived 
from 
Pandora PFA studies~\cite{Thomson:2008zz,Thomson:2007xb}, to
estimate the jet energy resolution as a function of the key
parameters.

\begin{figure}[htbp]
\centerline{\includegraphics[width=4.00in,height=4.0in]{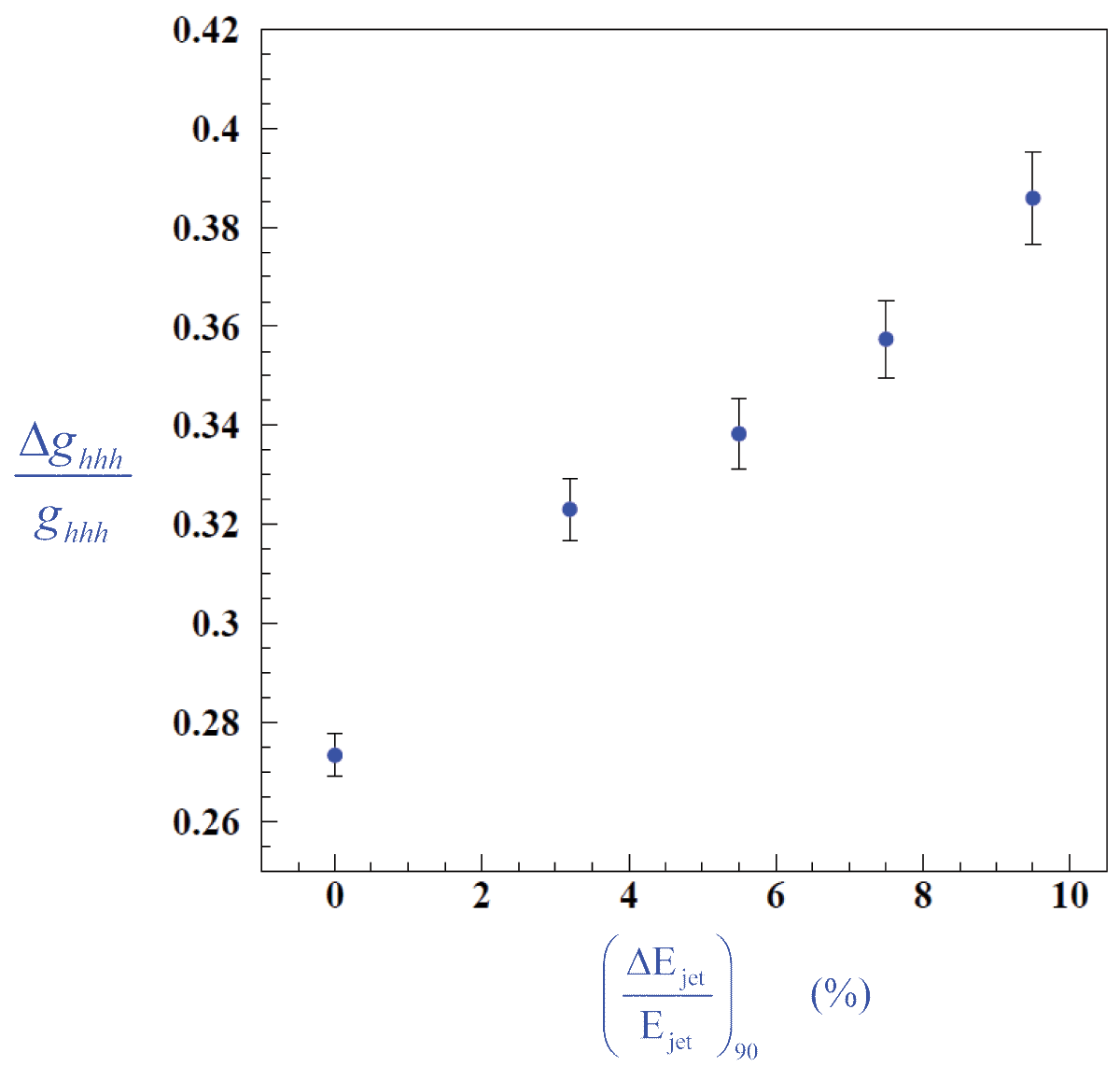}}
\caption{Fractional error in the triple Higgs coupling, g$_{hhh}$,
as a function of the jet energy resolution, $\Delta E/E$ [{\%}],
from a fast Monte Carlo study of the measurement of the cross
section for e$^{+}$e$^{-} \to ZHH \to qqqqbb$ at 500 GeV with an
integrated luminosity of 2000 fb$^{-1}$.} \label{fig:opt1}
\end{figure}

Thirdly, the ultimate physics performance
of SiD was studied as a function of the jet
energy resolution. Fast Monte Carlo
studies of these and other important physics measurements were conducted for
a full range of possible jet energy resolutions, and the resulting
measurement accuracy determined as a function of input jet energy
resolution. Figure~\ref{fig:opt1} shows an example.

These studies confirmed the
results of earlier studies that showed slow improvements in physics
measurement capability as jet energy resolution is improved. For
several possible measurements, improving the jet energy resolution
(i.e. $\Delta E/E$) by a factor of two, from 60{\%}/$\surd E$ to
30{\%}/$\surd E$, reduced the error in the measurement by about
30{\%}, equivalent to roughly 60{\%} more luminosity. This is a
significant improvement, but not enough to make night and day
differences in measurement capability.

\begin{figure}[htbp]
\centerline{
\includegraphics[height=2.5in]{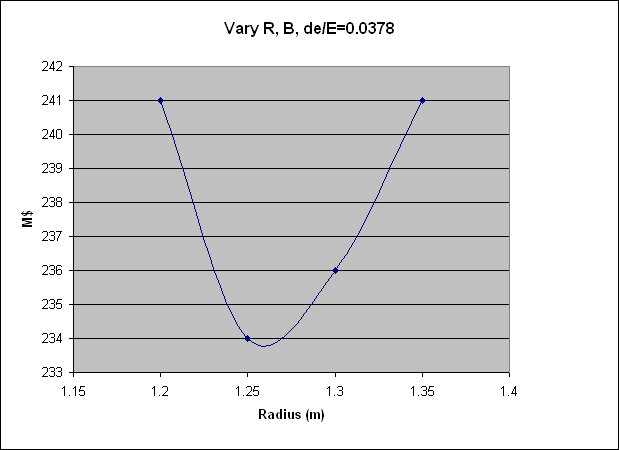}\hspace*{3mm}
\includegraphics[height=2.5in]{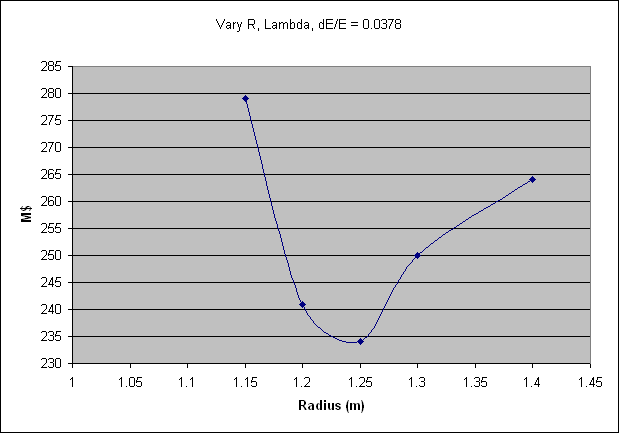} }
\vspace*{0mm} \caption{SiD cost vs radius. In the plot at left, the
radius and B field are varied in such a way that a fixed value of
jet energy resolution, $\Delta E/E$ = 0.0378, is maintained. At
right, the radius and HCal depth, $\lambda $, are varied. In both
cases, Thomson's parameterization of the jet energy resolution for
180 GeV jets is used, along with the SiD costing model. The cost
optimal point in this case is near the SiD baseline, with $R$=1.25
m, $B$=5 T, and $\lambda $=4.5.} \label{fig:opt2}
\end{figure}

Given these tools, it is possible to choose
what set of global parameters minimize the cost of the detector for a desired jet
energy resolution.
Figure~\ref{fig:opt2} shows an example.
This study, and a similar study based on jet energy resolutions for 100
GeV jets, lead to the choices of $R$ = 1.25m, $B$ = 5T, and $\lambda $ =
5 interaction lengths.

This exercise can of course be repeated for a full range of jet
energy resolutions, yielding an optimal cost (and selection of $R$,$B$,
$\lambda )$ for each jet energy resolution.

\begin{figure}[htbp]
\centerline{\includegraphics[width=5.00in,height=3.0in]{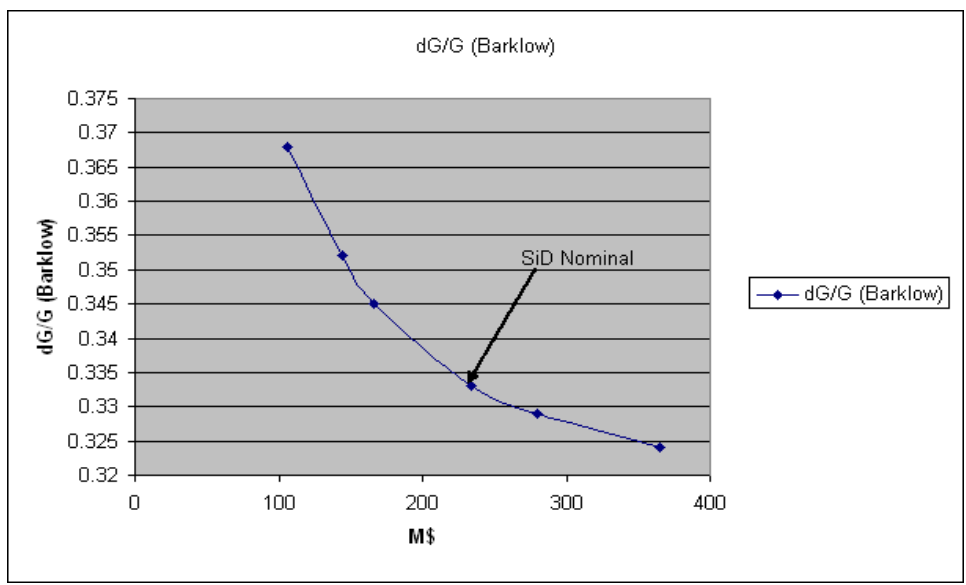}}
\caption{Fractional error in the measurement of the triple Higgs
coupling in SiD, as a function of the cost of the detector. The SiD
baseline is the point at
{\$}238M (ILC units),
slightly beyond the ``knee''. Similar curves
can be generated for other physics measurements.} \label{fig:opt3}
\end{figure}

The final step is to use the cost vs. jet energy resolution and the
physics performance vs. jet energy resolution inputs, to see how
errors in a measured physics quantity change as the cost of the
detector is varied. An example is shown in Figure~\ref{fig:opt3}.
What performance is good enough? Obviously, the detector performance
must be adequate for making measurements to the accuracy motivated
by the physics. Once that is satisfied, it is desirable to pay as
little for such performance as possible. Ideally, the chosen design
should sit near the ``knee'' of the performance vs cost plot.

The parameters selected for the SiD baseline place SiD's cost
somewhat beyond the cost vs performance ``knee''; this fact reflects
the existence of additional performance constraints and some
conservatism. The optimization process outlined above has been based
solely on jet energy resolution. Of course other detector
characteristics must be optimized as well, especially track momentum
resolution and impact parameter resolution. The choice of $B$ = 5T
does improve the jet energy resolution compared to choices of lower
$B$, and it helps to optimize the cost by making the detector
compact, but it also provides superb track momentum resolution.
Furthermore, since it allows the smallest possible beam pipe radius
by constraining the orbits of beamstrahlung-produced e$^{+}$e$^{-}$
pairs, it optimizes vertex detector resolution as well. While less
expensive detectors, with poorer jet energy resolution, may still be
adequate for ILC jet physics, their tracker radii are typically
below the chosen value of 1.25 m. Such trackers would not deliver
the desired tracking performance, so more compact designs have not
been chosen. Technical uncertainties restrict the maximum B field to
5T and the maximum R to 1.25 m in the case where B=5T.

The optimization process will be iterated in tandem with ongoing consideration of
improvements to the detector design.

\section{ILC Environmental Concerns}

The ILC beams comprise trains consisting of 2820
bunches, separated by 308ns; the train repetition rate is 5 Hz.
Consequently, the bunch trains are about one millisecond long,
separated by intervals of 199 milliseconds.
The resulting machine backgrounds can be classified as being either directly due to the beam collisions or 
non-collision-related. Events contributing to the first category are:
\begin{itemize}
\item disrupted primary beam exiting the IP
\item beamstrahlung photons
\item $e^+e^-$-pairs from beam-beam interactions
\item radiative Bhabha events
\item hadrons or muons from $\gamma \gamma$ interactions.
\end{itemize}
The second category is populated with events from:
\begin{itemize}
\item direct beam losses
\item beam-gas interactions
\item collimator edge scattering
\item synchrotron radiation
\item neutron back-shine from the beam dump
\item extraction line losses
\end{itemize}

Each of these sources has been studied.
GUINEAPIG~\cite{Schulte:1997tk} was used to simulate the beam-beam
interaction and to generate pairs, radiative Bhabhas, disrupted
beams and beamstrahlung photons. The ILC nominal 500 GeV beam
parameters were used for the simulation. Table~\ref{tab:env1}
summarizes the number of particles per bunch and their average
energy. $\gamma \gamma $ interactions were included without a
transverse momentum cut so as to provide sensitivity to the entire
cross sections for these processes. The $\gamma \gamma $ hadronic
cross section was approximated in the scheme of Peskin and
Barklow~\cite{Chen:1993ye}.

\begin{table}[htbp]
\begin{center}
\caption{Background sources for the nominal ILC 500 GeV beam
parameters.}
\begin{tabular}{|c|c|c|}
\hline
Source & {\#}particles/bunch & $<$E$>$ (GeV) \\
\hline \hline
Disrupted primary beam & $2 \times 10^{10}$ & 244 \\
\hline
Beamstrahlung photons & $2.5 \times 10^{10}$ & 4.4 \\
\hline
e$^+$e$^-$ pairs from beam-beam interactions & 75K & 2.5 \\
\hline
Radiative Bhabhas & 320K & 195 \\
\hline
$\gamma \gamma  \quad \to $ hadrons/muons & 0.5 events/1.3 events & - \\
\hline
\end{tabular}
\label{tab:env1}
\end{center}
\end{table}

The backgrounds impose a number of constraints on the detector
technologies and on the readout electronics.
The e$^+$e$^-$ pairs are produced at the IP, interact with the beam pipe, inner
vertex detector layers, mask and beamline magnets, and produce a
large number of secondary e$^+$e$^-$, photons and neutrons which in
turn contribute background in the vertex detector and in the Si
tracker and calorimeter at large radius.
Furthermore, the two-photon processes yield about 200 hadronic events per bunch train.
High energy interactions comprise only one
event in about every ten bunch trains. Therefore, the pile up of the
two-photon events could significantly confuse detection of the
principal signal of interest unless the detector can cleanly select
single bunch crossings, which SiD is designed to do.

\begin{figure}[htbp]
\centerline{\includegraphics[width=6.0in]{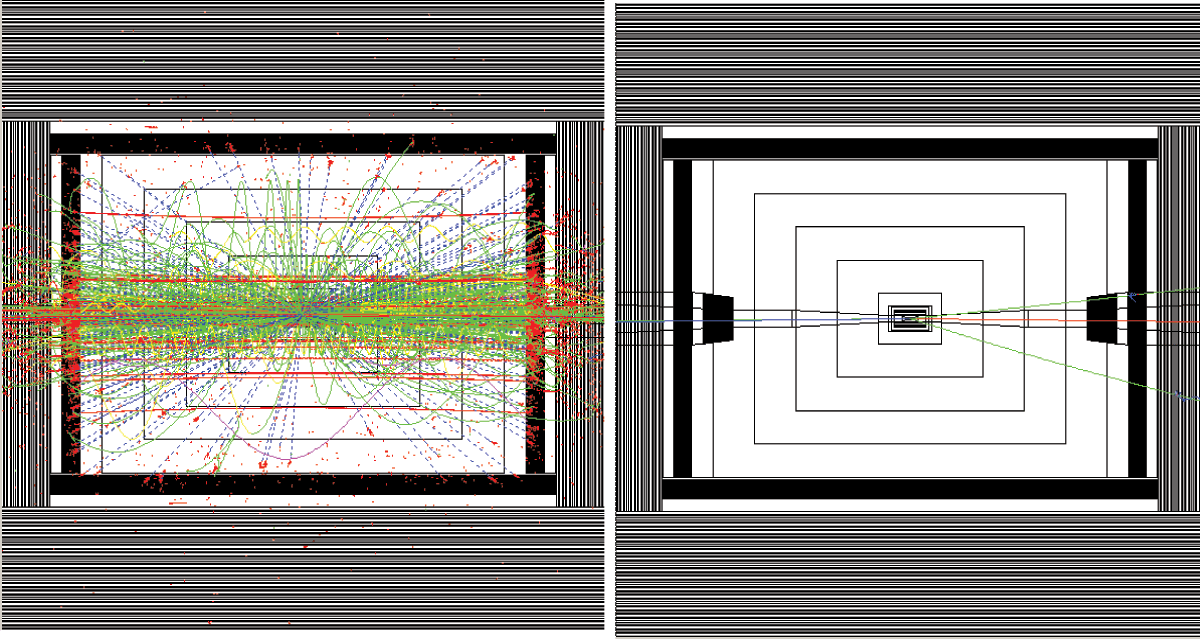}}
\caption{Physics backgrounds from $\gamma \gamma $ produced
e$^{+}$e$^{-}$ pairs, muon pairs, and hadronic events integrated
over 150 bunch crossings (left) and a single bunch crossing
(right).} \label{fig:Env1}
\end{figure}

The silicon microstrip and pixel detectors used in
the vertexing, tracking, and electromagnetic calorimetry can be made
sensitive to ionization which is deposited within just 1 $\mu $s of
the interaction time, and the resultant hits can be uniquely
associated with a single bunch crossing. A channel--by-channel
buffer, which is 4 deep in current designs, will store hits over the
course of the entire bunch train, and record each responsible bunch
crossing. Consequently, SiD is sensitive only to the physics and
backgrounds of a single bunch crossing, in contrast to detectors
with longer inherent livetimes, which can be sensitive to $\sim $150
bunch crossings or more. Figure~\ref{fig:Env1} contrasts the $\gamma
\gamma $ backgrounds when the detector livetime is short or long.
Pattern recognition benefits tremendously from this cleanest
possible environment, and physics ambiguities are minimized.
Channel-by-channel deadtime inefficiencies are minimal, and the
event buffering insures that essentially no physics data is lost
because of noise hits.

A further complication comes from lessons learned at the first
linear collider, the SLC. There, bunch-to-bunch variations in the
beam parameters were large, and hard to predict, model, and control.
Individual bunches with anomalous backgrounds were problematic to
operation of the SLD. Significant precautions are being
taken at ILC to deal with this, but experience suggests the need for
robust detectors. SiD's reliance on silicon sensors for vertexing,
tracking, and electromagnetic calorimetry promises the needed
robustness.

\section{SiD Organization} \label{sec:origins}

\begin{figure}[!htb]
\centerline{\includegraphics[width=6.50in]{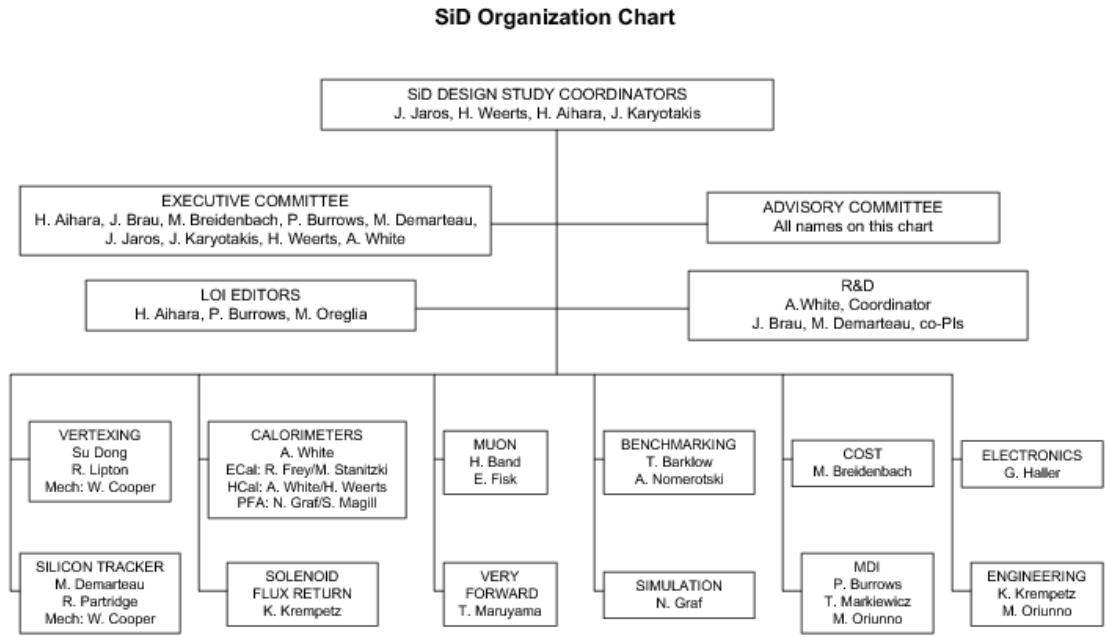}}
\caption{Organization chart for the Silicon Detector Design Study.}
\label{fig:orgchart}
\end{figure}

 The SiD Design Study was initiated in 2004
at the Victoria LCWS meeting, and introduced in subsequent regional
meetings at Durham and Taipei. The organization, shown in its
present form in Figure~\ref{fig:orgchart}, was put in place during
this period. Working groups were created, leaders recruited, and
plans debated on how to proceed toward the first major SiD Design
Study Meeting, which was held in conjunction with the ALCPG Snowmass
Meeting in 2005.  That gathering marked the first opportunity for
the Design Study participants to meet and interact for a protracted
period, and it was the first meeting where the design had been
captured completely in GEANT4, and made available for detailed
study.

Since then the organization has expanded considerably. A series of
Workshops has been conducted at Fermilab, SLAC, RAL, and Colorado,
in addition to informational meetings conducted at the regional
linear collider workshops at Vienna, Valencia, Beijing, Sendai, and
Warsaw. The computing infrastructure developed for the ALCPG,
org.lcsim, has been adopted and expanded with new reconstruction
codes, detector descriptions, and a repository of simulated data for
detailed study. The Design Study has assembled several major
documents leading up to the production of this Letter of Intent. In
2006, SiD produced a Detector Outline Document, followed the next
year with major contributions to the Detector Concept Report. SiD
also produced reports in 2007 to the Tracking, Calorimetry, and
Vertex Detector Reviews which were organized by the World Wide
Study.

Since its inception, the various SiD Working Groups have conducted
regular meetings to monitor progress and organize activities. These
meetings and the other Design Study activities are summarized on the
SiD website: {\verb$ http://silicondetector.org/display/SiD/home $}.
The meetings are of course open to all interested participants. From
time to time, SiD has found it useful to organize SiD Collaboration
Phone Meetings to discuss important issues when face-to-face
meetings are inconvenient. These meetings are conducted on Webex.

The SiD Executive Board is ultimately responsible for making design
and policy decisions for SiD, but it actively seeks input from the
SiD Advisory Board, those attending SiD workshops, and the SiD
Collaboration proper. Major decisions are reviewed by all SiD
collaborators, and all voices are heard, before actions are taken.

Besides the linear collider detector R\&D support enjoyed by some
SiD collaborators, SiD has received  significant support from the US
national laboratories ANL, FNAL, and SLAC, as well as Bristol,
Imperial College, Oxford University, and the Rutherford Appleton
Laboratory in the UK, and LAPP-Annecy in France. At the request of
the US funding agencies, SiD (along with the other concepts) has
recently taken on responsibility for organizing and presenting
linear collider detector R\&D proposals from US universities, and is
optimistic that this support for linear collider detector R\&D, well
aligned with the needs of the concept, can grow.

\bibliographystyle{unsrt}



\chapter{Subsystems}\label{chap:subsystems}

\section{Vertex and Tracking System}
\label{sec:vxdtrk}
\subsection{Introduction}

The tracking system of the SiD detector uses a barrel-disk
layout. Five cylindrical layers of pixel detectors surround the
interaction point, complemented by four disks of pixel sensors
on either end.  These inner layers are followed by a set of
five barrels of silicon strip sensors in the central region,
capped by four nested disks of silicon strip sensors at both ends.
To provide uniform hit coverage, three disks with pixel sensors
are provided in the transition region between the inner and outer
disks in both the forward and backward region.

Within the SiD detector concept the tracking system is
regarded as an integrated tracking system. Although individual detector
components can be identified in the vertexing and tracking system,
the overall design is driven by the combined performance
of the pixel detector at small radius, the outer strip detector
at large radius and the electromagnetic calorimeter for the
identification of minimum ionizing track stubs.
The physics at the ILC requires good track reconstruction
and particle identification for a wide array of topologies.
The main elements for the pattern recognition are the highly
pixellated vertex detector and the low occupancy outer strip
detector.

Early track finding studies relied on identifying tracks in
the vertex detector, where pattern recognition is simplified
by the fact that precise three-dimensional information is available
for each hit.  Tracks found in the vertex detector are then
propagated into the outer tracker, picking up additional hits.
While good performance was achieved using this approach,
an important class of events, notably highly boosted b-quarks,
will decay at radii that do not allow for pattern recognition
in the vertex detector alone.  To provide additional flexibility,
a more general tracking algorithm has been developed that can
seed tracks using any three layers in the tracker, either from
the outer tracker, the vertex detector or a combination of both.
Tracks produced by the decay products of long-lived particles,
however, can leave too few hits in the tracker to be reconstructed
using only hits in the tracking volume.
Obvious examples are long-lived particles such as
$K_s^0$'s and $\Lambda$'s.
The detector should also be capable of detecting
new physics signatures that would include long-lived
exotic particles like those predicted by some gauge-mediated
supersymmetry breaking scenarios. There are also issues of
reconstructing kinked tracks produced by particles that lose
a substantial portion of their energy in the tracker,
as well as reconstructing backscatters from the calorimeter.
To capture the tracks from these event topologies a
calorimeter-assisted tracking algorithm has been employed.
This algorithm uses
the electromagnetic calorimeter to provide seeds for pattern
recognition in the tracker.
The very fine segmentation of the EM calorimeter allows
for detection of traces left by minimum ionizing particles.
These can be used to determine the track entry point, direction,
and sometimes curvature with a precision sufficient for
extrapolating the track back into the tracker.
This set of complementary algorithms provides for very robust
pattern recognition and track finding and it is the performance
of this integrated tracking system that determines the overall
physics reach of the detector. In this section the design and
performance of the overall tracking system will be described.
More details can be found in the companion documentation.

\subsection{Vertex Detector Design}

\begin{figure}[ht]
\centerline{\includegraphics[width=5in]{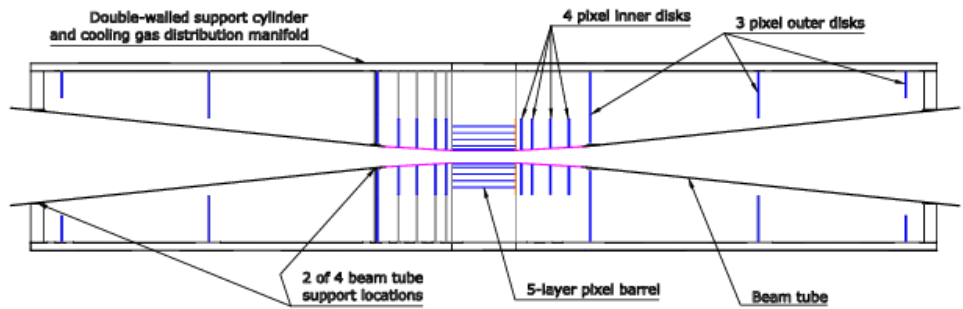}}
\caption{R-z view of the vertex detector. The right hand side has
been
         drawn without the support structures.}
\label{fig:vxd}
\end{figure}

The vertex detector integrates with the outer tracker and remainder of
the detector to provide significantly extended physics reach through
superb vertex reconstruction -- primary, secondary and tertiary.
To date, all vertex detectors at collider experiments are silicon based,
and the vertex detector for the SiD concept is no exception.
The vertex detector consists of a central barrel section with
five silicon pixel layers and forward
and backward disk regions, each with four silicon pixel disks.
Three silicon pixel disks at larger $\mid z\mid$ provide uniform
coverage for the transition region between the vertex detector
and the outer tracker. Barrel layers and disks are arranged to
provide good hermeticity for $\cos\vartheta \leq 0.984$ and to
guarantee good pattern recognition capability for charged tracking
and excellent impact parameter resolution over the whole solid angle.
A side-view of the vertex detector is shown in Fig.~\ref{fig:vxd}.
For clarity, the silicon support structures have not been drawn
in the right hand side of this figure.

Vertex detectors are generally plagued by a mismatch in thermal
expansion coefficients between the silicon and its support structures.
Moreover, these supports in general add to the material budget
in a region of physics phase space where it is least desired.
To partially address those considerations, an `all-silicon' structure
is proposed for the vertex detector barrel. In this context,
`all-silicon' means that sensors of each barrel layer are joined
along their edges with adhesive to approximate an arc of a
circular cylinder, and no other structural materials are present
in that limited region. Thermal distortions are reduced by
limiting material to that of the sensors themselves and adhesive,
which has a low elastic modulus relative to silicon.
Longitudinal deflection of a layer is controlled by the
cylindrical shape, thereby minimizing additional material.
The quasi-cylindrical shape of a layer is maintained by annular,
flat rings at each end. In turn, the end rings are joined to one
another and connected to an outer support cylinder via web-like
support disks. Though various possibilities are still under
consideration for the end ring and support disk material,
we would like a material, such as carbon fiber laminate or
silicon-based foam, which has a longer radiation length than
that of silicon. Because other materials than silicon are used
for external mechanical connections, the term
\lq\lq all-silicon\rq\rq\ is placed in quotes.

\begin{figure}[tp]
\begin{center}
\begin{tabular}{cc}
\includegraphics[width=2.8in]{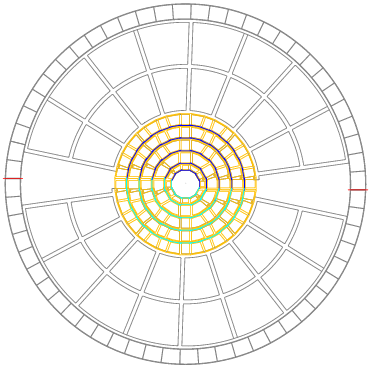}  &
\includegraphics[width=2.8in]{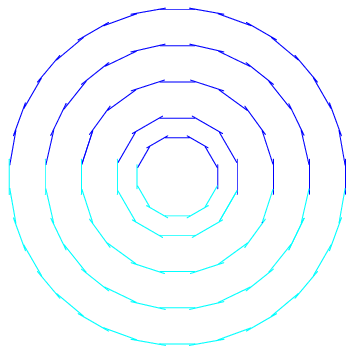} \\
\end{tabular}
\end{center}
\caption{Barrel end view of the vertex detector (left) and layer arrangement
of the silicon sensors only (right). }
\label{fig:vxdend}
\end{figure}

At this time, many sensor options remain under investigation and
we have yet to choose a specific sensor technology.
The design presented here
makes the assumption, independent of sensor technology,
that fabrication and assembly of the detector occur at room
temperature and that the sensors are operated at a temperature
$>$ -10~$^0$C. Because sensors are used as a structural element
and other material has been minimized, our design favors relatively
thick sensors. The sensor thickness has been taken to be 75~$\mu$m.
Sensor cut width is 8.68 mm in the innermost layer and 12.58 mm
in all other layers. The cut sensor length for all layers is 125 mm.

To allow assembly about the beam pipe and later servicing,
all barrels, disks, and support elements of the vertex detector
are split about approximately the horizontal plane into
top and bottom sub-assemblies. Once mated, the two sub-assemblies
are supported from the beam pipe and stiffen the portion of the
beam pipe passing through them.
Fig.~\ref{fig:vxdend} is an end view of the barrel region,
showing the five silicon barrel layers and their spoked support disk.
The outer rings indicate the double-walled carbon fiber support
tube. Since the silicon is very thin on the scale of this drawing,
the layer arrangement of the individual sensors is shown in
the right drawing in Fig.~\ref{fig:vxdend} for clarity.

The five layers are arranged at radii ranging from 14
to 60 mm. The vertex detector also has four disk layer
sensors which are attached to carbon fiber support disks at
z positions ranging from about 72 to 172 mm.
The innermost disk covers radii from 14 mm out
to 71 mm; the outermost, from 20 mm to 71 mm. Forward tracking
continues beyond the vertex detector proper with three additional
small pixel disks, extending in $z$ from about 207 to 834 mm.
Their inner radii range from 29 to about 117~mm, and
their outer radius is about 166 mm.

The beam pipe through the central portion of the vertex
detector has been taken to be all-beryllium. Within the barrel
region of the vertex detector, the beryllium beam pipe has been
taken to be a straight cylinder with inner radius of 1.2 cm and a
wall thickness of 0.04 cm. At $z  = \pm 6.25$~cm, a transition is
made to a conical beam pipe with a wall thickness of 0.07 cm. The
half angle of the cone is ~3.266$^o$. Transitions from beryllium to
stainless steel are made beyond the tracking volume, at
approximately $z = \pm 20.5$~cm. The initial stainless steel wall
thickness is 0.107~cm; it increases to 0.15~cm at approximately $z =
\pm 120$~cm. The half angle of the stainless steel cone is
~5.329$^\circ$. The inner profile of the beam pipe is dictated by the
need to avoid the envelope of beam-strahlung produced
$e^+e^-$-pairs.

To prevent bending of the small-radius portion of the beam pipe and
ensure good stability of vertex detector position,
the outer vertex detector support cylinder is coupled to the
beam pipe at four longitudinal locations: $\pm$ 21.4 and $\pm$ 88.2 cm.
Inner and outer support cylinder walls are 0.26 mm thick.
They are made from four plies of high modulus
carbon fiber, resin pre-preg. Wall separation is 15 mm.
We propose to deliver cooling air via the vertex detector outer
support cylinder. To allow that, the two walls of the cylinder
would be separated by radially-oriented ribs running the full
cylinder length. Calculations assumed ribs at
60 azimuths. Openings, each approximately 12.2 mm x 15 mm,
at 18 z-locations in the inner cylinder wall distribute
flow to the various disk locations and to the barrel.

\begin{figure}[ht]
\centerline{\includegraphics[width=5in]{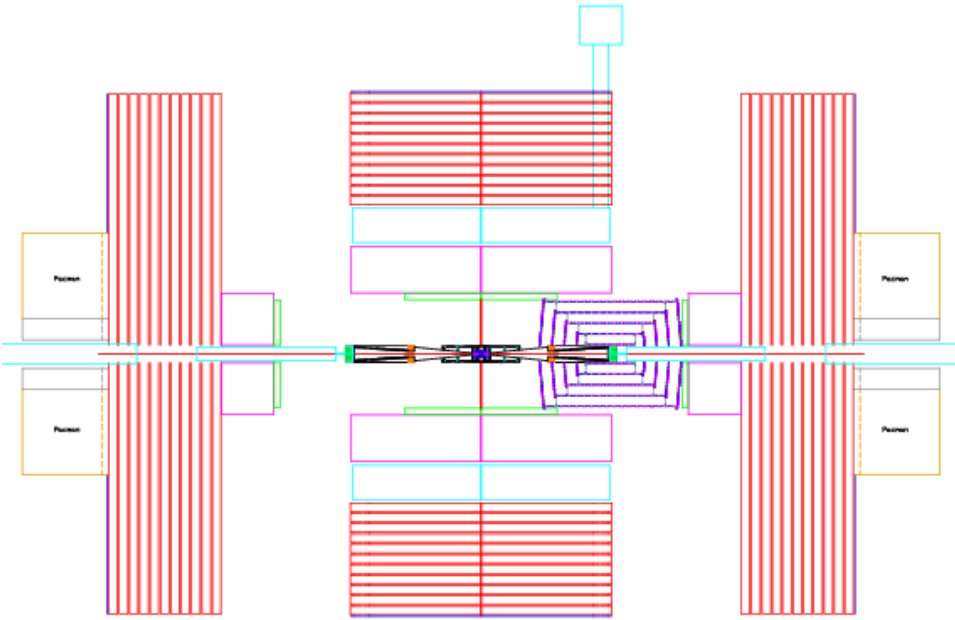}}
\caption{Tracker in the open position for servicing of the vertex
         detector. }
\label{fig:trkservice}
\end{figure}

During silicon servicing, the vertex detector and beam pipe
remain fixed while the outer silicon tracker rolls longitudinally
(see Fig.~\ref{fig:trkservice}).
To allow that motion, to enable placement of the outer silicon
tracker elements at the smallest
possible radius, and to leave space for any additional thermal
insulation which might be needed, the outer radius of the vertex
detector, including its support structures, has been
limited to 18.5 cm. To maximize physics potential, the inner
radius of vertex detector elements has been chosen to be as small
as practical consistent with beam-related
backgrounds and the beam pipe profile. In the barrel region,
the minimum radius to a sensor surface is 1.4 cm,
governed by the beam backgrounds as discussed earlier.

\begin{figure}[ht]
\begin{center}
\begin{tabular}{cc}
\includegraphics[width=2.8in]
           {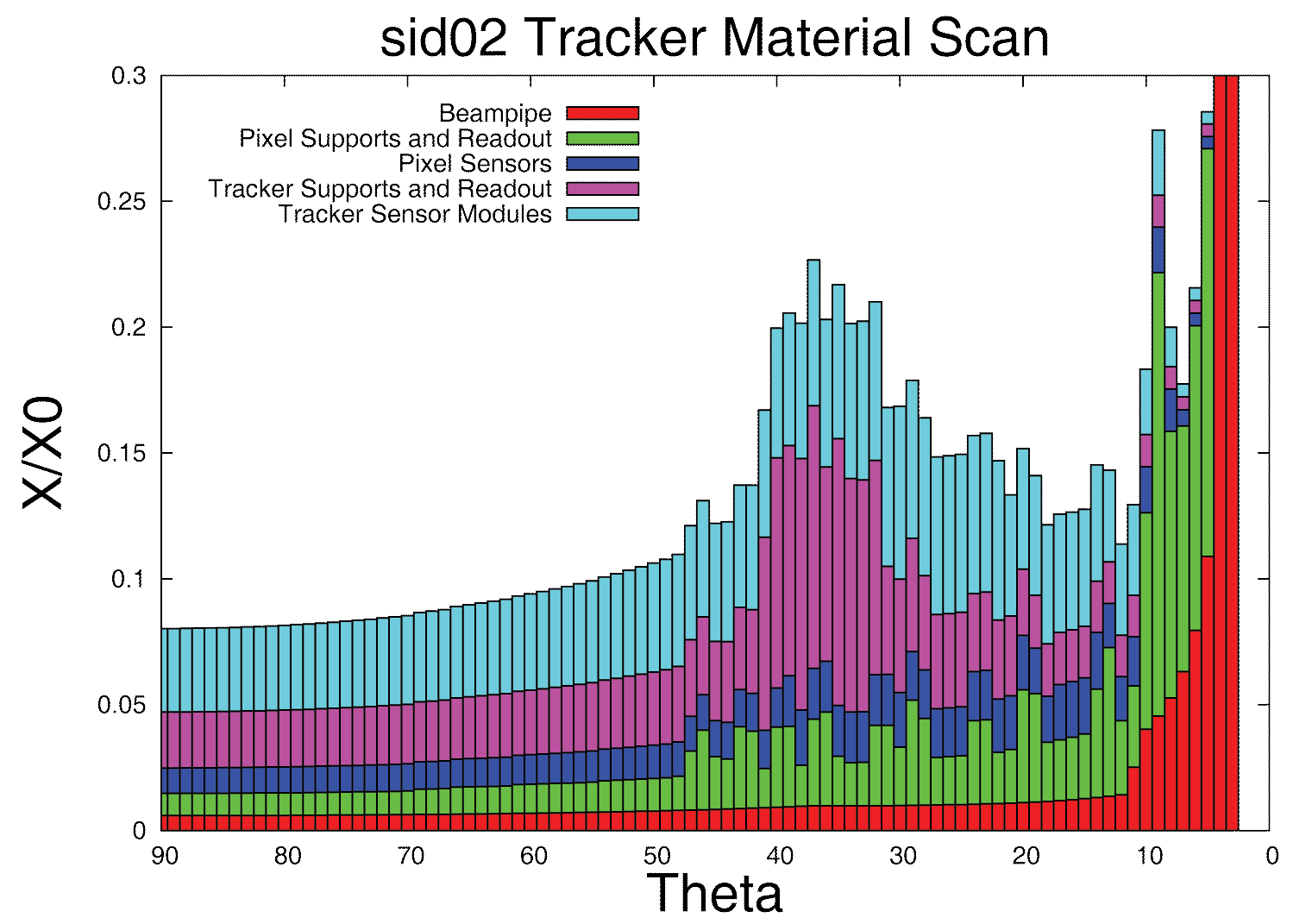}
           &
\includegraphics[width=2.8in]
           {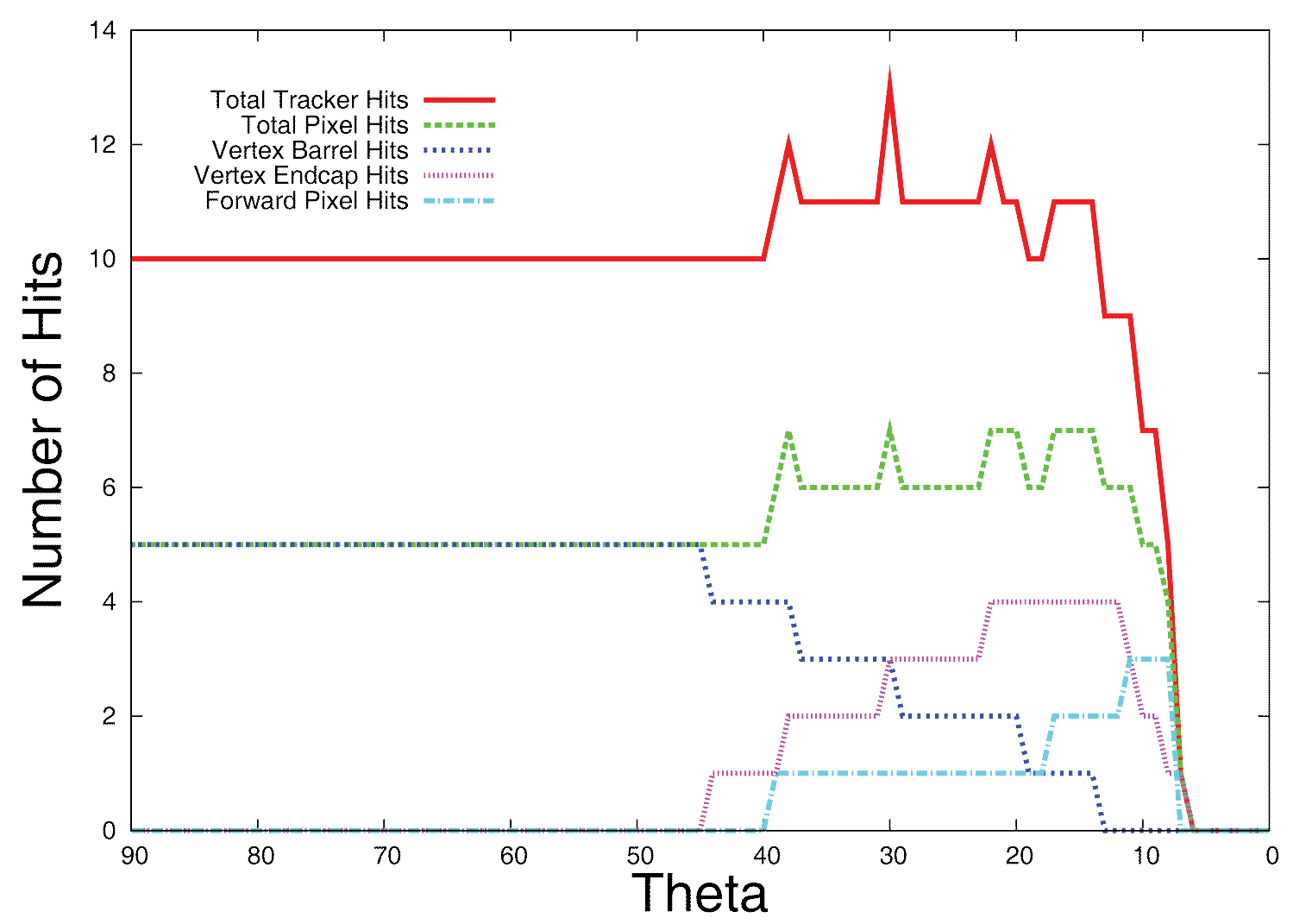} \\
\end{tabular}
\end{center}
\caption{Material budget of the tracking system (left) and
         number of hit layers in the tracking system as a function
         of polar angle (right). }
\label{fig:x0}
\end{figure}

The number of radiation lengths represented by vertex detector
structures, averaged over $\varphi$ in most cases, and the
vertex detector hit pattern are shown in Fig.~\ref{fig:x0}.
Shown are the contributions of the individual subdetectors to the
total number of hits on a particle track with infinite momentum.
The irregular features correspond to the transition regions.
Overall there are more than five pixel hits on a particle trajectory
down to angles of about $10^{\circ}$.

The irregular features of the readout and service contributions
to the material budget are due to discrete elements at the end
of the sensors.
Most of the readout material is beyond the first few layers of
the vertex detector, so that their influence on the impact parameter
resolution is limited.  The fact that the amount of material in
these elements is comparable to that of the sensors or mechanical
supports calls for close attention to the design of low mass power
delivery and signal transmission components.
If the readout and service material can indeed
meet what is in the current model, the material balance would
be more favorable for a
considerable portion of the endcap region compared to
the $1/\sin\vartheta$ growth for a long barrel
geometry. With this material balance, the benefit of the endcap
geometry in spatial resolution with a better track entrance angle
and smaller radial alignment effect, is a meaningful advantage.
Table~\ref{tab:vxd} summarizes the main parameters of the vertex
detector.

\begin{table}[t]
\begin{center}
\begin{tabular}{||c|c|c|c||}
\hline\hline
    Barrel   & R     &  Length    &  Number of \\
    Region   & (mm)  &  (mm)      &  sensors in $\varphi$ \\
\hline
    Layer 1  & 14    &  125   &  12 \\
    Layer 2  & 21    &  125   &  12 \\
    Layer 3  & 34    &  125   &  20 \\
    Layer 4  & 47    &  125   &  28 \\
    Layer 5  & 60    &  125   &  36 \\
\hline\hline
    Disk     & R$_{inner}$ & R$_{outer}$  &  z$_{center}$   \\
\hline
    Disk 1   & 15    &  75    &  76  \\
    Disk 2   & 16    &  75    &  95  \\
    Disk 3   & 18    &  75    &  125 \\
    Disk 4   & 21    &  75    &  180 \\
\hline\hline
    Forward Disk & R$_{inner}$ & R$_{outer}$  &  z$_{center}$   \\
\hline
    Disk 1   & 28    & 166    &  211 \\
    Disk 2   & 76    & 166    &  543 \\
    Disk 3   & 118   & 166    &  834 \\
\hline\hline
\end{tabular}
\caption[]{Parameters of the vertex detector. Units are mm. }
\label{tab:vxd}
\end{center}
\end{table}

\subsection{Tracker Design}

\begin{figure}[h]
\centerline{\includegraphics[width=5in]{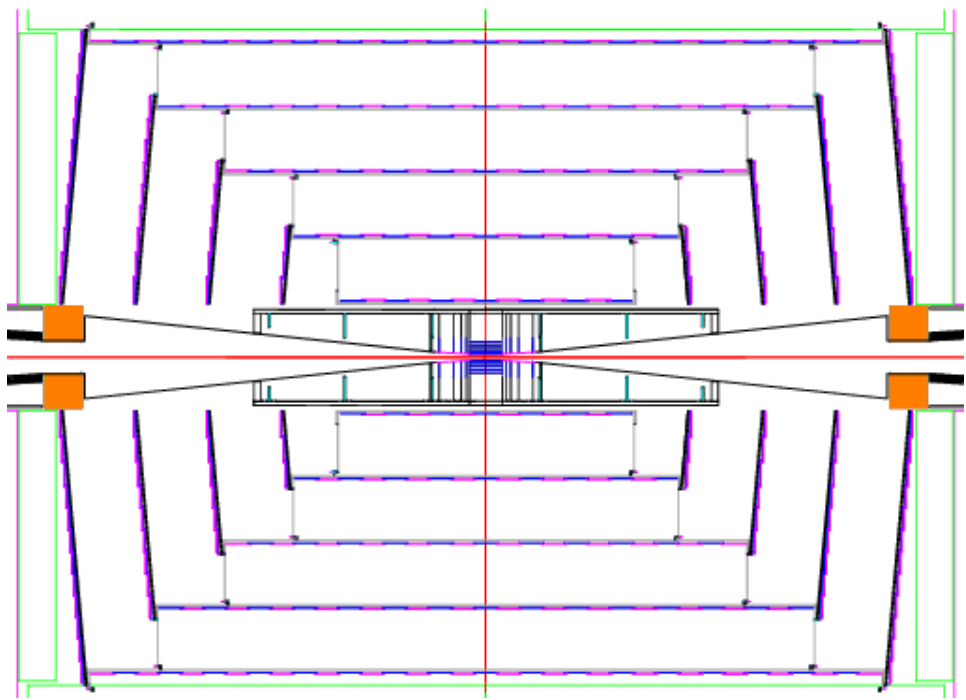}}
\caption{R-z view of the whole tracking system. } \label{fig:vxdtrk}
\end{figure}

The ILC experiments demand tracking systems unlike any previously envisioned.
In addition to efficient and robust track-finding, the momentum resolution
required to enable precision physics at ILC energies must improve
significantly upon that of previous trackers.
The design must minimize material in front of the calorimeter
that might endanger particle-flow jet reconstruction.
Even with the largest feasible magnetic field, the tracking volume
is quite large so that tracker components must
be relatively inexpensive and easily mass-produced.
Finally, the tracker must be
robust against beam-related accidents and aging.
These requirements have led
to the choice of silicon microstrip detectors for the tracker.
The outer silicon tracker design consists of five nested barrels in the
central region and four cones in each of the end regions.
The support material of disks follows a conical surface with an
angle of 5-degrees with respect to the normal to the beamline.
Sensors on the disk modules are normal to the beam line.
The barrel supports are continuous cylinders formed from a
sandwich of pre-impregnated carbon fiber composite around a Rohacell
core. The support cones are also double-walled carbon fiber structures
around a Rohacell core.
Each support cone is supported off a barrel.
Spoked annular rings support the ends of each barrel cylinder from
the inner surface of the next barrel out.
It is expected that openings will be cut in the support structures to
reduce material, once module mounting locations are known.
These openings not only reduce the number of radiation lengths,
but also reduce the weight to be supported.
Openings may also be needed for an optical alignment system.
It is envisioned that the electronics and power
cables that supply entire segments of the detector are mounted on
these spoked rings. The dimensions of the barrels and cones are
given in Table~\ref{tab:trk}.
Fig.~\ref{fig:vxdtrk} shows an elevation view of the tracking
system.

Because of the very low occupancies in the outer barrel, the nominal
design for the outer tracker employs only axial readout in the
barrel region. In the baseline design, the barrels
are covered with silicon modules. Modules are comprised of a
carbon fiber composite frame with rohacell/epoxy cross bracing
and have one single-sided silicon sensor bonded
to the outer surface. Sensors are obtained from one single
6-inch wafer and are approximately 10~cm $\times$ 10~cm.
This size sets the longitudinal readout segmentation of the barrel
detectors. The sensors are 300$\mu$m thick with a readout pitch of
50~$\mu$m and intermediate strips.
Full coverage is obtained by ensuring small overlap both longitudinally and
azimuthally. Azimuthal overlap is obtained by slightly tilting the sensors.
The angle by which the sensor is tilted partially compensates for the
Lorentz angle
of the collected charge in the 5T field of the solenoid.
Longitudinal overlap is obtained by placing alternate sensors at
slightly different radii.

Modules are attached to the
cylinder using a PEEK (Poly Ether Ether Ketone) mounting clip. The readout chips and cables
are mounted directly to the outer surface of the silicon sensors.
The cables supply power and control to the readout chip from
electronics located at the ends of the barrel.

\begin{figure}[h]
\centerline{\includegraphics[width=4in]{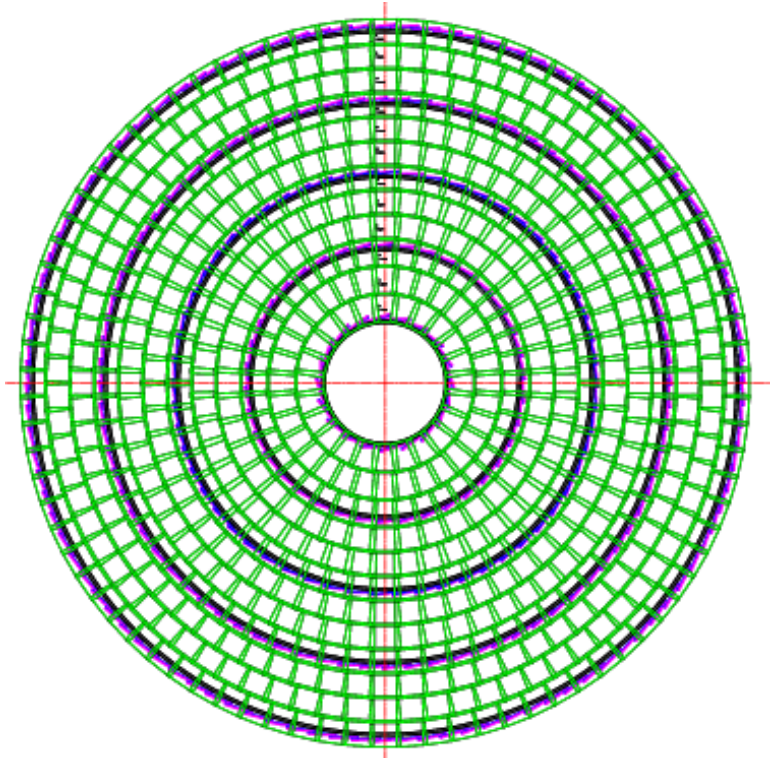}}
\caption{R$\varphi$ projection view of the tracker barrels and
         disks. }
\label{fig:trk_rphi}
\end{figure}

Fig.~\ref{fig:trk_rphi} shows an R$\varphi$-view of the
barrel region. The outermost disk is projected onto the barrel layout
in this figure. For pattern recognition in the disks, small angle
stereo will provide 3d-space points.
The current design has two single-sided wedge detectors back-to-back,
with strips at $\pm 6^{\circ}$ with respect to the long axis of the
wedge for a stereo angle of $12^{\circ}$. Please note that in
Fig.~\ref{fig:x0} the hits from a pair of sensors, corresponding
to one 3d-space point, is represented as one hit.
Two types of sensors
are needed to tile the disks, one type at the inner radii and a
second sensor type to populate the area at the outer radii.
Also in the forward region sensors will be 300$\mu$m thick with
intermediate strips.
The conical support disk design (\lq\lq lampshade design\rq\rq )
provides an elegant way to implement module overlaps, eliminating
any dead areas, and allows for easy module mounting and cable routing.

The inner radius of the outer tracker is set by forward,
beam-monitoring calorimetry and beamline elements, over which the
tracker is intended to slide.  Once the tracker inner radius is set,
the outer radius of vertex detector structures follows.
During servicing, the vertex detector and beam pipe remain
fixed while the outer silicon tracker rolls longitudinally,
as shown in Figure~\ref{fig:trkservice}.
To allow that motion, no element
from the outer tracker can be at a radius smaller than the radius
of the vertex detector outer support cylinder.
To allow for good acceptance and pattern recognition, the small
angle region is covered by three small silicon disks at each end
with radius below 20~cm, which has been described in the section
on the mechanical layout of the vertex detector.
Figure~\ref{fig:x0} shows the cumulative amount of material as a function
of polar angle as modeled in the Monte Carlo. The lowest curve
shows the contribution from the beampipe followed by the contribution
of the support and readout structures for the pixel detector.
The material corresponding to
the various readout elements has conservatively been assumed
to be uniformly distributed in the tracker
volume. The next curve indicates the material due to the active
vertex detector elements.
The outer two curves give the amount of material of the tracker supports
and readouts and the silicon modules, respectively.
It should be noted that the material corresponding to the silicon modules
includes the module supports in addition to the silicon.

Overall a material budget of
about 0.8\% $X_0$ per layer is achieved for the outer tracker.
Table~\ref{tab:trk} lists some of the parameters of the tracker for
the current design. There are 8130 modules in the barrel region and
2848 modules for the end regions combined.

\begin{table}[h]
\begin{center}
\begin{tabular}{||c|c|c|c|c||}
\hline\hline
    Barrel   & $\langle R\rangle$
             & Length of sensor
             & Number of
             & Number of  \\
    Region   & (cm)
             & coverage (cm)
             & modules in $\varphi$
             & modules in $z$       \\
\hline
    Barrel 1  & 21.95    & 111.6   & 20  & 13   \\
    Barrel 2  & 46.95    & 147.3   & 38  & 17   \\
    Barrel 3  & 71.95    & 200.1   & 58  & 23   \\
    Barrel 4  & 96.95    & 251.8   & 80  & 29   \\
    Barrel 5  & 121.95   & 304.5   & 102 & 35   \\
\hline\hline
    Disk     & $z_{inner}$
             & R$_{inner}$
             & R$_{outer}$
             & Number of        \\
    Region   & (cm)
             & (cm)
             & (cm)
             & modules per end  \\
\hline
    Disk 1   & 78.89     & 20.89  & 49.80   &  96    \\
    Disk 2   & 107.50    & 20.89  & 75.14   &  238   \\
    Disk 3   & 135.55    & 20.89  & 100.31  &  428   \\
    Disk 4   & 164.09    & 20.89  & 125.36  &  662   \\
\hline\hline
\hline\hline
\end{tabular}
\caption[]{Parameters of the tracking detector. }
\label{tab:trk}
\end{center}
\end{table}

The central component of the readout architecture proposed for the
SiD tracker is the 1024-channel KPiX readout chip.
The KPiX chip is designed for pulsed power,
minimizing the input current between bunch trains.
This reduces the power consumption to 20~mW on average for a 1024-channel
KPiX chip, or 40~mW for a single-sided module
allowing for gas cooling of the tracker.
The chip has four time-stamped analog buffers per channel that store
signals from the
detector until the inter-train period for digitization and readout.
As a result, the only digital activities on KPiX during the bunch train
are a synchronous LVDS clock and individual comparators firing when a
channel crosses the readout thresholds. This low-noise mode of operation
during the bunch train allows KPiX to be mounted directly to the sensor
without inducing large RF pickup on the strips.

\subsection{Simulation Infrastructure}

The vertex detector and tracker designs have been incorporated in
the compact $xml$ detector description that drives our simulation
studies. This detector description serves as an input both to {\it slic.org},
the GEANT4-based detector simulation used by SiD, as well as the
event reconstruction software.

\begin{figure}[ht]
\centerline{\includegraphics[width=5in]
           {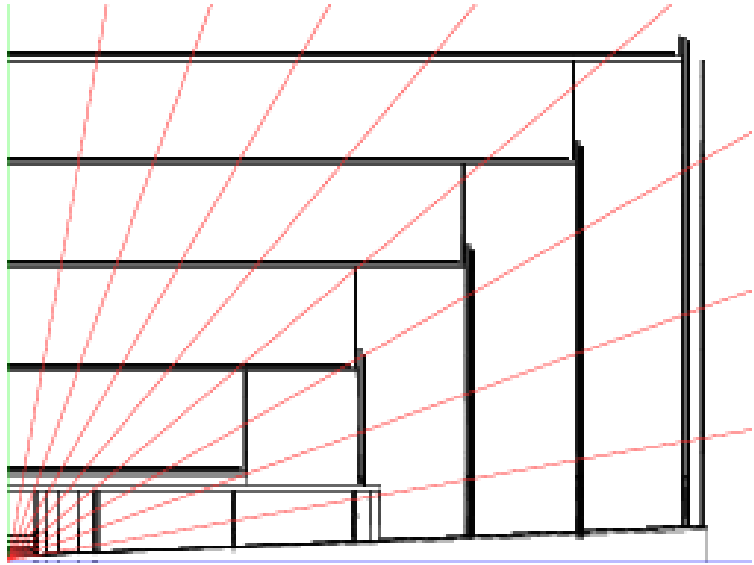}}
\caption{R-z view of the simplified tracking system as implemented
in SiD02. } \label{fig:sid02Tracker}
\end{figure}

The current detector description includes both the active sensing
elements, as well as our estimates of the dead material required to
provide mechanical support, beam tube, readout electronics, and
required services (including power and cooling). For the tracking
studies reported here, the barrel sensors have been approximated by
thin cylinders, while the disk sensors have been approximated by
planar disks perpendicular to the beamline. The dead material is
modeled as a cylinder, planar disk, or cone as appropriate.
Fig.~\ref{fig:sid02Tracker} shows an x-y quarter view of the
tracking system as implemented in this simplified geometry.
This model has also been used to simulate
the detector response for the large number of events generated for
the physics benchmarking.

We have also developed a fully detailed tracker description that
closely matches the engineering designs. It incorporates each
individual sensor as a planar device with its mounting hardware. The
fully segmented tracker description provides a highly realistic
model of the tracker geometry, allowing us to study the effects of
sensor overlaps and gaps, sensor mis-alignment, support material and
more generally improve the precision of our detector modeling.
Having the complete geometry fully defined and configurable at
runtime (using a plain-text file), and immediately available to the
reconstruction software, provides enormous flexibility to the design
and optimization process.

We do not include the individual pixels and strips in the GEANT4
simulation, preferring to defer this to the reconstruction stage.
Instead, we store the full Monte Carlo information about the
particle's interaction with the sensitive sensor material (e.g.
track ID, deposited energy, time, position). Hit digitization is the
process of transforming these GEANT4 energy deposits into strip and
pixel charge depositions, and then clustering these charge
depositions to form tracker hits. This allows us to study different
readout configurations without the overhead of having to rerun the
full simulation.

\subsection{Track Reconstruction}

%
The standard pattern recognition algorithm developed by SiD is
designed to efficiently find tracks using pixel and strip hits
in the tracker.  The pattern recognition algorithm treats the
tracker as a single integrated device that is \lq\lq agnostic\rq\rq\
as to the origin of the hits (pixel or strip, barrel or endcap).
This approach provides uniform pattern recognition throughout
the tracking volume, effortlessly handling transitions between
the different parts of the detector.  Typically, 6-7 hits are
sufficient for finding a track, which allows the standard
pattern recognition algorithm to efficiently track particles
originating near the interaction point with $p_T>200$ MeV.

Since pattern recognition is of utmost importance in a sparse
hit environment, additional track finding algorithms are explored.
Of particular note is the calorimeter assisted track finder, which
uses the tracking capability of the electromagnetic calorimeter to
associate hits in the outer tracker with calorimeter
\lq\lq track stubs\rq\rq .  Calorimeter assisted tracking is
particularly well suited to reconstructing tracks that
originate outside the vertex detector, as often occurs in
$K_s$ and $\Lambda$ decays.
Both the standard pattern recognition and calorimeter assisted
tracking algorithm are described below.

\subsubsection{Standard Pattern Recognition Algorithm}

The standard pattern recognition algorithm is explicitly designed
for the task of optimizing the design of an all-silicon tracker.
Variations in tracker geometry and layout can be easily studied with
no change to the software.
The algorithm bases all its decisions on a global $\chi^2$.
A high level of user control over the
tracking ``strategies'' is available if desired, but more typically
a ``strategy builder'' tool is used to automate the process of
developing an optimized set of strategies for a given detector
configuration.

The first step in track finding is to convert the digitized hits
into a common hit format. This format encapsulates all the
information needed by the standard pattern recognition algorithm,
while insulating the track finding from differences and changes
in the digitization algorithms.

Three types of hits are supported: pixel hits that have two measured
coordinates, axial strip hits that have one measured coordinate and
one bounded coordinate, and stereo hits formed from a pair of strip hits.
The pixel and stereo hits may be associated with either barrel or
disk geometries, while the axial strip hits have the bounded coordinate
parallel to the beam axis and are intrinsically associated with barrel
geometries.  One further limitation is placed on stereo hits:
the planes of the two strip sensors must be parallel to each other.

Track finding is controlled by a set of strategies.  A strategy
consists of the list of detector layers to be used, the role
of each layer (seed, confirm, or extend), kinematic constraints
($p_T$, impact parameters), requirements on the number of hits,
and the $\chi^2$ cut.
Multiple strategies can be processed by the track finding
algorithm, and the resulting tracks are the collection of all distinct
tracks found.  The set of strategies is contained in an $xml$ file
that is easily understood and can be viewed/edited
with a text editor.

The track finding algorithm is exhaustive in the sense that all
combinations of hits that could potentially lead to a successful
track fit are considered.
The algorithm proceeds in four steps:

\begin{enumerate}
\item The first step is to form a 3-hit track seed candidate
      by taking all 3-hit combinations possible among the 3 seed
      layers.  A helix fit is performed on the seed candidate, and
      those seeds that fail the $\chi^2$ cut are eliminated.
\item The second step tries to ``confirm'' the seed by adding
      additional hit(s) from the confirm layer(s).
      A helix fit is performed on the new seeds and those that
      fail the $\chi^2$ cut are eliminated.  Typically, it is found
      that good performance is achieved with one confirmation layer.
\item The third step seeks to ``extend'' a confirmed
      seed by trying to add additional hits from the extend layers.
      Each time a new hit is considered, a helix fit is performed
      and the hit is discarded if it fails the $\chi^2$ cut.
      If no hits in a given extend layer give a satisfactory helix fit,
      then the original track seed is kept and the next
      extend layer is tried.
\item Track seeds that meet the strategy's requirement on the minimum
      number of hits are merged to form a list of distinct tracks.
      Two track candidates are allowed to share a single hit, but if
      a track candidate shares more than one hit with another
      candidate, an arbitration scheme is used to select the better
      candidate.
      Precedence is given to the candidate with the greatest number
      of hits, while the candidate with smaller $\chi^2$ is selected
      when the number of hits is equal.
\end{enumerate}

Consistency checks and hit sorting algorithms are used to minimize
the number of helix fits performed,
substantially improving the performance of the algorithm.
Furthermore, a ``bad hit $\chi^2$'' cut is used to identify track
candidates with an outlier hit and allows preference to be given to
track candidates without an outlier hit.

A key component of the pattern recognition algorithm is a fast helix
fitter.  The helix fitter takes as input 3 or more tracker hits.
The hits can be any combination of pixel, axial strip, or stereo hits
in the barrel and/or endcap detectors. The fitter is the one place
in the tracking code that distinguishes between the various
types of hits. The fast fitter is used to estimate the helix
parameters and helix fit $\chi^2$.  First,
a circle fit to the x,y coordinates of all hits is performed
using the Karim\"aki algorithm to determine the helix parameters
$\omega$, $\phi_0$, and $d_0$.
If there are two or more pixel/stereo hits,
then a line fit in the $s-z$ plane
is used to determine the $z_0$ and $\tan\lambda$ helix parameters.
In order to provide full helix fits for the case where there are
fewer than two pixel/stereo hits,
a new fitting algorithm was developed.

While an axial strip does not measure the $z$ coordinate,
it has well defined bounds that impose the constraint
$z_{min} < z < z_{max}$.  These bounds lead
to each axial strip having an allowed band in the $z_0$-$\tan\lambda$
plane.  For two or more axial strips, the intersection of these
bands will produce a polygonal allowed region, the centroid of which
is taken to be the measured values for $z_0$ and $\tan\lambda$.
If there is no region of intersection, the hits are inconsistent
with being from a helix and the fit fails.
For the case of a single pixel/stereo hit,
the pixel/stereo hit is treated like a very short strip and the
above algorithm is used.

For all but the highest momentum particles, the multiple scattering
errors will exceed the intrinsic hit resolution.
Multiple scattering errors for both the active and dead
materials are estimated and included in the helix fit.
Correlations in the multiple scattering errors are ignored,
leading to an under-estimate of the helix parameter errors
by a factor of $\approx 1.5$.  For stereo hits, full account is
taken for the separation between the two stereo layers in the calculation
of both the hit position and hit covariance matrix.

The performance of the standard pattern recognition algorithm is
shown in the section on tracker performance and is also reflected
in the benchmarking studies.  Unless
otherwise noted, the tracking strategies require 1 confirmation hit,
a total of at least 7 hits (6 hits for barrel-only tracks),
$p_T>0.2$ GeV,
$xy$ distance of closest approach $d_0 < 10$~mm,
and distance of closest approach along the beam direction $z_0<10$~mm.

\subsubsection{Calorimeter-Assisted Tracking}
\label{sec:cal-ass}

The development of the calorimeter assisted track finding algorithm was
primarily motivated by the need to reconstruct non-prompt tracks and
long-lived particles in the SiD detector. As will be shown later,
the standard track finding algorithm achieves excellent efficiency
in reconstructing prompt tracks that originate close to the interaction
point. However, using the same algorithm for finding non-prompt tracks is
difficult because those tracks often do not produce enough hits in the
vertex detector to seed the track finder, and creating seeds from 3-hit
combinations in outer layers without limiting combinatorics by
constraining the track origin to the interaction region can
be problematic.

The calorimeter assisted tracking solves the problem by seeding the track
finder from traces left by charged particles in the electromagnetic
calorimeter - so called MIP stubs.
The standard SiD track finder is
run first, and the found tracks are then propagated through the calorimeter.
Clusters are created and attached to the tracks. After that, topological
clustering is applied to the remaining calorimeter hits. Those of the
created clusters that include hits in inner electromagnetic calorimeter
layers are analyzed for consistency with being produced by minimum ionizing
particles. Clusters that pass the test are converted into track seeds that
contain information about the track position and direction at the
calorimeter entry point. Depending on the MIP stub quality, the seed can
also contain a track curvature estimate. The seeds are propagated back into
the tracker, picking up hits that are not attached to any existing tracks.
The search window in each tracker layer is calculated based on the
trajectory parameters uncertainties, and the track parameters are
re-evaluated after attaching every new hit. If more than one hit can be
attached to the track in a certain layer, multiple track candidates are
created. Once all seeds have been processed, the newly created tracks are
rated based on quality, and duplicates are removed.
This algorithm is essential for
reconstructing all kinds of non-prompt tracks - $K_s^0$ and $\Lambda$ decay
products, new physics signatures that might include long-lived particles,
kinked tracks, and calorimeter backscatters. It also performs high purity,
topologically linked initial clustering in the calorimeter, and associates
clusters with tracks.

\subsubsection{Track Fitting}

The track fitting results presented here use the fast helix finder
described before. Two more precise approaches to track fitting have
been developed by SiD.
The first of these is based on a track fitting algorithm originally
used by SLD and used in previous studies of the performance of the
SiD tracking system.
It solves minimization equations for $\chi^2$, calculated from
the matrix of residual weights. This weight matrix takes into account
multiple  scattering  derived from the amount of material
passed by a track, correlations
between track deviations in the subsequent detector layers and independent
measurement errors due to the sensors spatial resolution.
Solving the matrix equation for 5 track parameters gives the parameter values
corresponding to the  best fit; the inverse  of the  weight matrix of the
parameters gives the covariance matrix of parameter errors.
Comparison of the expected parameter errors derived from the covariance
matrix with real track parameter residual distributions shows agreement
to within a few percent.
This fitter may be slow for a large number of layers, as it requires a
matrix inversion  with a matrix dimension equal to the number of layers.
For the SiD concept, however, it works fine as total number of layers
crossed by a track rarely exceeds  10.
A Kalman filter, which treats multiple scattering close to optimally,
is under development and updated results using that Kalman filter
will be presented at a later time.
The Kalman filter will be used as a final fitter to refit tracks
that are currently fitted by the simple fitter.

\subsection{Tracking Performance}

In this section the performance of the vertex and tracking detector
will be described, along with its associated track-parameter fitter.
The goal of these studies is to evaluate the overall performance of
the SiD tracking system with the most realistic simulation available.
The standard tracking algorithm was tuned for the benchmark processing
to find tracks having $p_T>0.2$~GeV that originate from near the
interaction region.  The strategies used generally required at least
7 hits to be associated with a track.  An additional strategy that
required 6 barrel hits in the vertex detector and first layer of the
outer tracker was put in place to provide low-$p_T$ coverage for central
tracks that may not pass through 7 different layers before curling back
around.  Additionally, for the benchmark processing the strategies placed
a 1~cm constraint on the x-y and z distances of closest approach.

We can break down tracking efficiency into two parts: (1) the fraction of
Monte Carlo charged particles that are in principle ``findable''\ given
the set of strategies used, and (2) the track reconstruction efficiency
for the findable tracks.  The starting point for the tracking efficiency
measurement is the set of long-lived charged particles identified as
being ``final state particles''\ by the event generator.  Final state
particles include short-lived particle decay products
(e.g. $\pi^\pm$ from $K_S\to\pi^+\pi^-$), but does not include long-lived
particle decay products (e.g. $\pi^\pm$ decay products) or secondaries
produced by GEANT as the result of interactions in the detector.

Among the final state particles in $e^+e^-\to t\bar t$ collisions at
$\sqrt{s}=500$ GeV, 6.5\% of the tracks are below
the 0.2 GeV $p_T$ cutoff in the standard tracking algorithm.  Of
the tracks satisfying the $p_T$ constraint, 9.1\% of the tracks have
fewer than 6 hits, which is the minimum number of hits for the
standard tracking algorithm.  Some of these tracks are from sources,
such as $K_s$ decay, that are potentially recoverable by the
calorimeter assisted tracking algorithm.  Findable tracks must
also satisfy the requirements for having seed and confirmation hits
and satisfy the 1 cm constraint on the x-y ($d_0$) and z ($z_0$) distances
of closest approach.  Taken together, 84.4\% of the final state
particles are findable by the standard tracking algorithm.  The
breakdown of these contributions to the findable track efficiency
can be found in Table~\ref{tab:eff}.

\begin{table}[t]
\begin{center}
\begin{tabular}{||c|c|c||}
\hline\hline
    Selection        & Selection Efficiency & Cumulative Efficiency\\
\hline
All Tracks           & -                 & 100\%   \\
$p_T \ge 0.2$ GeV    & $(93.54\pm 0.11)$\% & $(93.54\pm 0.11)$\% \\
$N_{hit} \ge 6$      & $(90.91\pm 0.13)$\% & $(85.04\pm 0.16)$\% \\
Seed Hits Present    & $(99.78\pm 0.02)$\% & $(84.85\pm 0.17)$\% \\
Confirm Hit Present  & $(99.95\pm 0.01)$\% & $(84.84\pm 0.17)$\% \\
$|d_0| \le 1$ cm     & $(99.80\pm 0.02)$\% & $(84.65\pm 0.17)$\% \\
$|z_0| \le 1$ cm     & $(99.69\pm 0.03)$\% & $(84.39\pm 0.17)$\% \\
Track Reconstruction & $(99.32\pm 0.04)$\% & $(83.81\pm 0.17)$\% \\
\hline\hline
\end{tabular}
\caption[]{
Fraction of findable tracks}
\label{tab:eff}
\end{center}
\end{table}

\begin{figure}[h]
\begin{center}
\begin{tabular}{cc}
\includegraphics[width=3in]{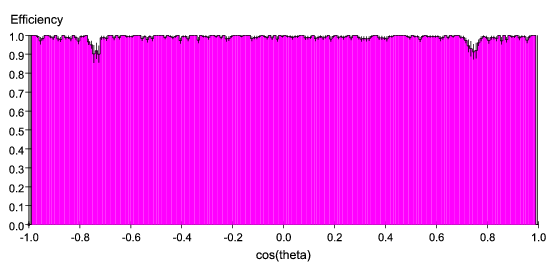}  &
\includegraphics[width=3in]{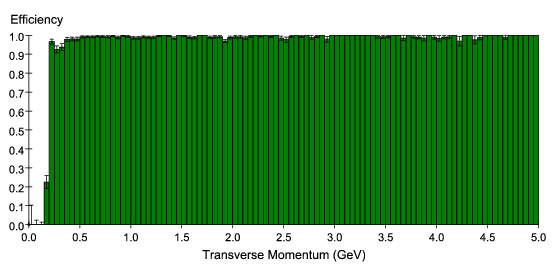}  \\
\end{tabular}
\end{center}
\caption{Track finding efficiency as a function of track
         $\cos\vartheta$ (left) and $p_T$ (right). }
\label{fig:trkeff}
\end{figure}

\begin{figure}[h]
\begin{center}
\begin{tabular}{cc}
\includegraphics[width=3in]
                {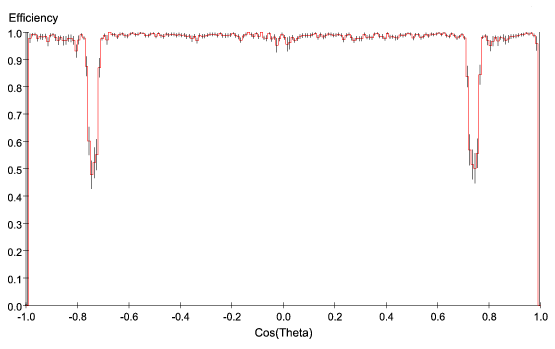}  &
\includegraphics[width=3in]
                {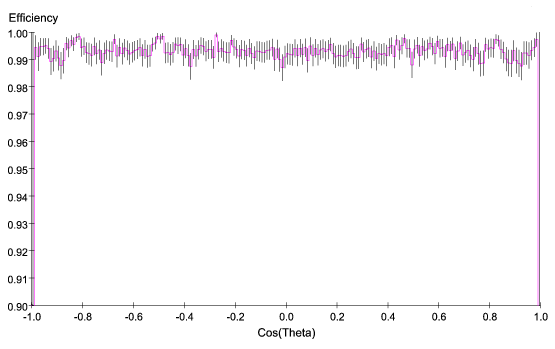}  \\
\end{tabular}
\end{center}
\caption{Track finding efficiency as a function of track
         $\cos\vartheta$ for tracks with $p_T < 500$~MeV (left) and
         $p_T > 500$~MeV (right). Please note the different vertical scales
         in the two figures. }
\label{fig:trkeff_ptbins}
\end{figure}

The track reconstruction efficiency measures the fraction of findable tracks
that are found by the standard tracking algorithm.  Using the sample described
above, the track reconstruction efficiency is found to be 99.3\%.
Fig.~\ref{fig:trkeff} shows the efficiency as a
function of $\cos\vartheta$ and $p_T$. The track finding efficiency
drops in the transition region between barrels and disks.
Fig.~\ref{fig:trkeff_ptbins} shows the efficiency as function of
$\cos\vartheta$ for two $p_T$ bins. The left plot is for tracks
with $p_T < 500 $~MeV and the right plot is for tracks with
$p_T > 500 $~MeV. All the inefficiency is due to low momentum
tracks in this transition region.
It is thought that the inefficiency is due to tracks just beyond
the pixel barrel acceptance that curl by more than 180 degrees
before they get to the seed layers that cover this acceptance region.
As such, this may be an artifact of the current tracking algorithm
and could be improved upon.

\begin{figure}[t]
\begin{center}
\includegraphics[width=4in]{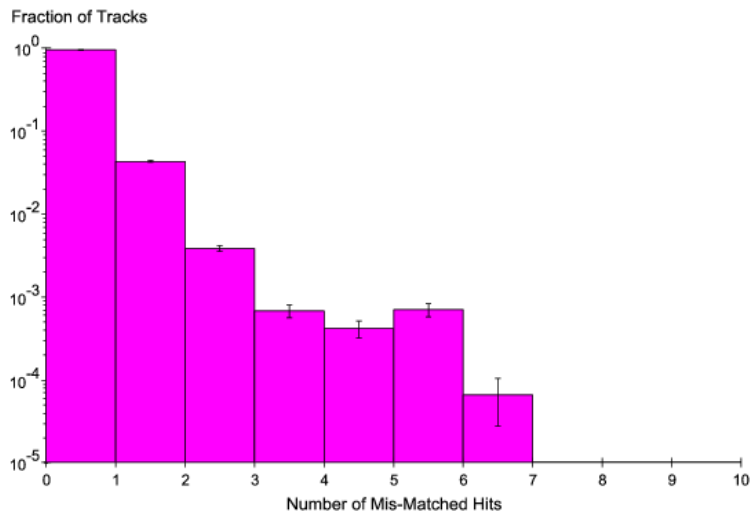}
\end{center}
\caption{Fraction of tracks versus the number of mis-assigned hits. }
\label{fig:trkmiss}
\end{figure}

Tracking algorithms must balance track finding efficiency against
the probability of finding ``fake tracks''\ that are not associated
with a Monte Carlo particle.  A key indicator for the number of fake
tracks is the number of mis-assigned hits on a track.  These hits
are generated by a different Monte Carlo particle than the one
with the preponderance of hits on the track.  More than 99\% of
tracks have at most one wrong hit on the track, as seen
from Fig.~\ref{fig:trkmiss}.
Fake tracks, where no single Monte Carlo particle is responsible
for the majority of hits, make up only 0.07\% of the tracks found.

\begin{figure}[p]
\begin{center}
\begin{tabular}{c}
\includegraphics[width=5in]{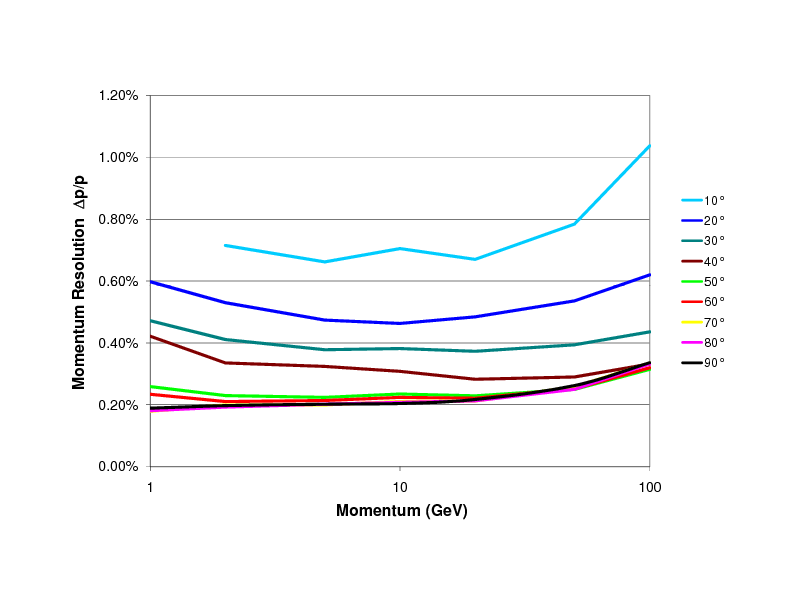} \\
\includegraphics[width=5in]{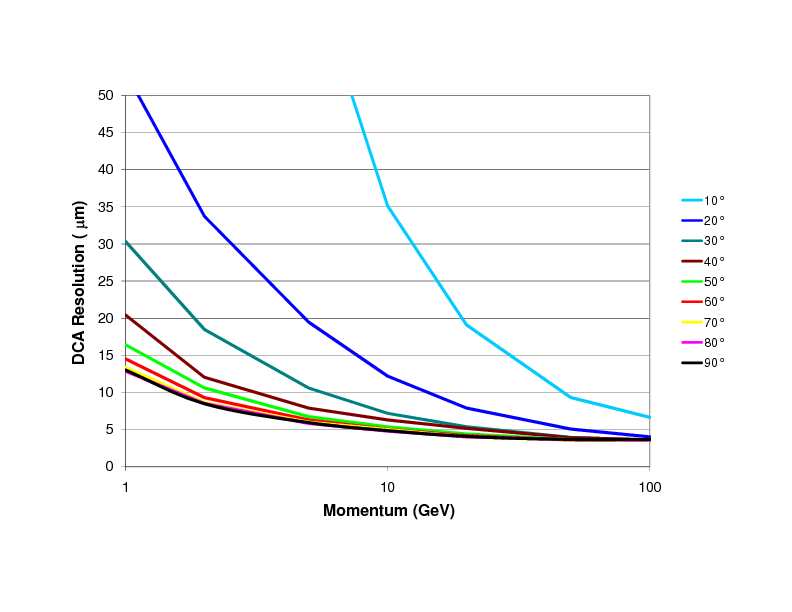}
\end{tabular}
\end{center}
\caption{Resolution in momentum (top) and $r-\varphi$ distance of
         closest approach, DCA (bottom), as function of
         track momentum at various angles. }
\label{fig:resolution}
\end{figure}

The momentum resolution of the tracker is shown in the top plot in
Fig.~\ref{fig:resolution}
as a function of momentum for various track angles.
The bottom figure shows the
impact parameter resolution for various track angles. An impact
parameter resolution of 4 $\mu$m is obtained in the high momentum limit.

\begin{figure}[p]
\begin{center}
\begin{tabular}{cc}
\includegraphics[width=3in]{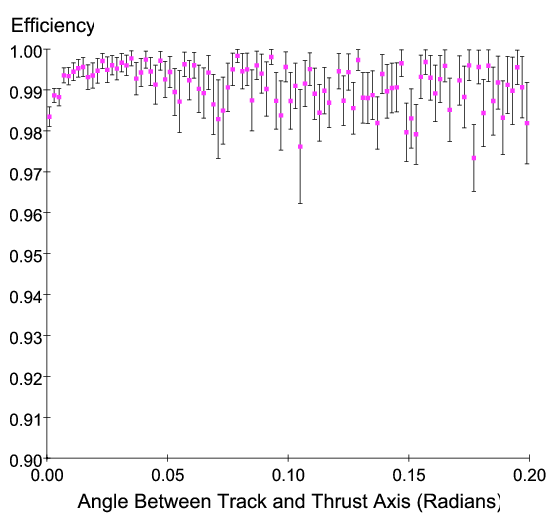}  &
\includegraphics[width=3in]{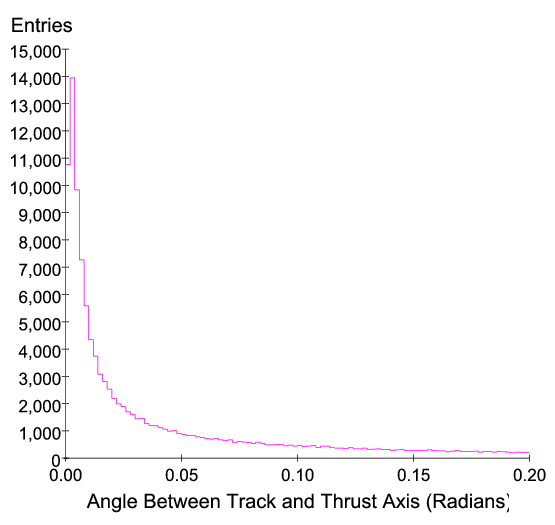}   \\
\quad & \quad \\
\quad & \quad \\
\quad & \quad \\
\end{tabular}
\begin{tabular}{c}
\includegraphics[width=4in]{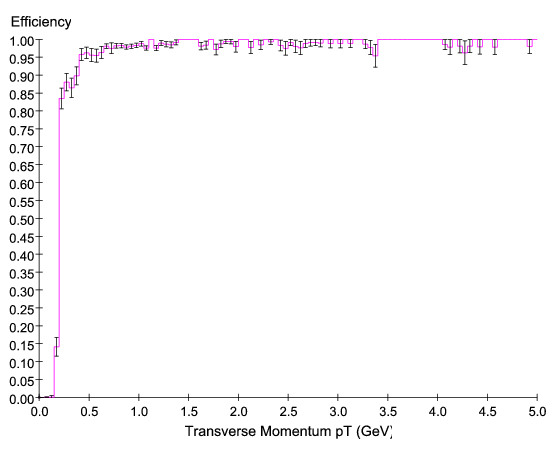}
\end{tabular}
\end{center}
\caption{Tracking efficiency as function of $\alpha$ (upper left)
         and distribution in $\alpha$ (upper right) for
         $e^+e^- \rightarrow q\overline{q}$ events at $\sqrt{s} = 1$~TeV.
         The bottom figure shows the tracking efficiency for
         $e^+e^- \rightarrow b\overline{b}$ events at $\sqrt{s} = 500$~GeV
         with the background from 10 bunch crossings overlaid. }
\label{fig:trkeff_occ}
\end{figure}

How the tracking performs in higher occupancy environments is
summarized in Fig.~\ref{fig:trkeff_occ}. Two studies have been
carried out.
First, the performance of the track finder has been studied
in the environment of dense jets. The plot in the upper left
corner in
Fig.~\ref{fig:trkeff_occ} shows the track finding efficiency as
function of the angle between the track
and the jet thrust axis for $e^+e^- \rightarrow q\overline{q}$ events at
$\sqrt{s} = 1$~TeV. The efficiency holds up rather well, dropping by
about 1\% for tracks within 2mrad of the jet core. The distribution
in $\alpha$, the angle of charged particles with respect to the jet
thrust axis, is shown in the upper right corner in
Fig.~\ref{fig:trkeff_occ}.
The bottom figure shows the track finding efficiency as a function
of track $p_T$ for $e^+e^- \rightarrow b\overline{b}$ events at
$\sqrt{s} = 500$~GeV with the backgrounds from 10 bunch crossings
overlaid.
In this study the effect of accumulating beam backgrounds
over 10 crossings has been mimicked by adding these hits to all
pixel devices in the detector. Hits in the silicon strip tracker
were added only for a single bunch crossing, in-time of course with
the physics event.
There is a small loss in efficiency at low $p_T$,
as anticipated. Also the fake track rate is higher, about 0.6\%.
Most of the fake tracks seem to be due to combinatorics.

\begin{figure}[t]
\begin{center}
\includegraphics[width=3.5in]{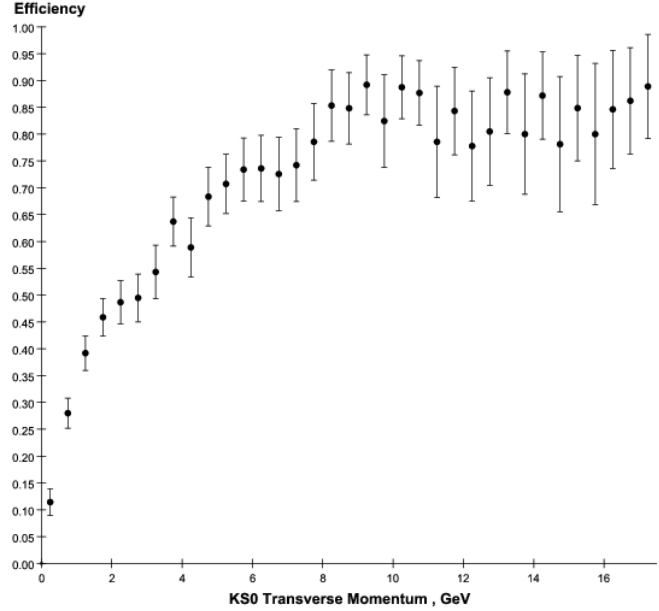}
\end{center}
\caption{$K_s^0$ finding efficiency as function of its transverse momentum. }
\label{fig:k0s}
\end{figure}

An often-voiced concern of a tracker with relatively few measurements on a
charged particle trajectory is the efficiency for the reconstruction of
long-lived particles. The SiD detector should be viewed as an integrated
detector where the overall performance derives from a combination
of all the subdetectors. As described in section~\ref{sec:cal-ass} a
calorimeter assisted track finding algorithm was developed to reconstruct
non-prompt tracks and long-lived particles.
Fig.~\ref{fig:k0s} shows the $K_s^0$ reconstruction efficiency for
$t \overline{t}$-events, obtained by running the standard tracking algorithm
followed by the calorimeter assisted tracking algorithm.
The efficiency is defined as the ratio of the number of successfully
reconstructed $K_s^0$'s to the total number of $K_s^0$'s that decayed into
a charged pion pair inside the third layer of the outer tracker.
The efficiency reaches 85\% for $K_s^0$'s with transverse momenta above
8~GeV. The result represents the current status of the software and
significant improvements, particularly for low momentum $K_s^0$, are
anticipated.



\begin{figure}[t]
\begin{center}
\begin{tabular}{cc}
\includegraphics[width=3in]
                {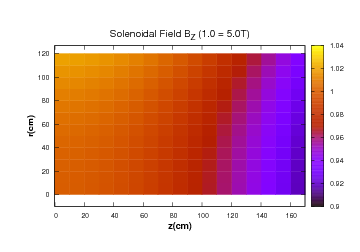}  &
\includegraphics[width=3in]
                {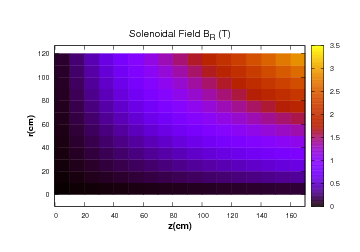}  \\
\end{tabular}
\end{center}
\caption{Map of the magnetic field components inside the tracking volume:
         $B_z$ in fractional deviation from the nominal magnetic field
         (left) and
         $B_R$ in absolute values (right) }
\label{fig:Bfield}
\end{figure}

All the simulations are performed using a uniform 5 Tesla magnetic
field with no radial component. ANSYS simulations of the solenoid
and the return flux show that the field is not uniform.
Fig.~\ref{fig:Bfield} shows the distribution inside the tracking
volume of the longitudinal (left) and radial (right) component of
the magnetic field. The SiD detector having a 5~T magnetic field is
defined as the field at $(R,z) = (0,0)$ being 5~T.
In Fig.~\ref{fig:Bfield} $B_z$ is given as fraction of the nominal
5~T field. In the center of the detector $B_z$ increases slightly
with increasing radius. There is a drop of about 6\% when reaching
the end of the tracker volume. The field component $B_R$, given in
absolute values in Fig.~\ref{fig:Bfield}, is not negligible.
The effect of a non-uniform magnetic field was studied by simulating
particle trajectories with the field map as shown in
Fig.~\ref{fig:Bfield} and reconstructing the trajectories assuming a
perfect uniform field configuration. Preliminary studies indicate
the effect on the pattern recognition to be minor. Studies are ongoing
to investigate how it affects the fitted track parameters and
whether the field uniformity must be improved.

\subsection{Tracker Alignment}

The unprecedented track momentum resolution contemplated for linear
collider detectors demands minimizing systematic uncertainties in
sub-detector relative alignments. At the same time, there is a
strong impetus to minimize the amount of material in the tracking
system, which might compromise its stability.
These two requirements put a premium on accurate alignment of
the various elements of the tracker. The short time
scales on which alignment could change (e.g., from beam-driven
temperature fluctuations) may preclude reliance on traditional
alignment schemes based on detected tracks, where it is assumed
the alignment drifts slowly, if at all, during the time required
to accumulate sufficient statistics.

The prospect of two ILC detectors swapping places in the beamline
only increases the importance of {\it in situ} alignment monitoring
that does not depend on tracks. It will be important to monitor
tracker distortions during the push-pull operations, not only for
later track reconstruction, but also to ensure that no
damage-inducing stresses are inadvertently applied to the tracker
components. Alignment systems that can also be used during
tracker assembly to monitor strains would also be useful.

A system that can monitor alignment drifts in real
time would be highly desirable in any precise tracker and probably
essential to an aggressive, low-material silicon tracker.
The trade-off one would make in the future between low material
budget and rigidity will depend critically upon what a feasible
alignment system permits. The SiD tracker is considering two
alignment methods, one based on Frequency Scanned Interferometry (FSI),
and one based on Infrared Transparent Silicon Sensors (IRSS).

The FSI system incorporates multiple interferometers fed by optical
fibers from the same laser sources, where the laser frequency is
scanned and fringes counted, to obtain a set of absolute lengths.
With a test apparatus the state of the art in precision
DC distance measurements over distance scales of a meter under
laboratory-controlled conditions has been reached and even extended.
Precisions better than 100 nm have been attained using
a single tunable laser when environmental conditions are carefully
controlled. Precisions under uncontrolled conditions
(e.g., air currents, temperature fluctuations) were, however,
an order of magnitude worse with the single laser measurements.

Hence a dual-laser FSI system is foreseen for the tracker,
that employs optical choppers to alternate the beams introduced
to the interferometer by the optical fibers. By using lasers
that scan over the same wavelength range but in opposite directions
during the same short time interval, major systematic
uncertainties can be eliminated. Bench tests have achieved a
precision of 200 nm under highly unfavorable conditions
using the dual-laser scanning technique.

It should be noted that complementary analysis techniques
of FSI data can be used either to minimize sensitivity to
vibrations in order to determine accurate mean shape
distortion or to maximize sensitivity to vibrations below
the Nyquist frequency of data sampling. The latter algorithm
could prove especially useful in commissioning in assessing
vibration effects, such as might arise from pulse powering
in a magnetic field.

The second method exploits the fact that silicon sensors have a
weak absorption of infrared (IR) light.
Consecutive layers of silicon sensors are traversed by IR laser
beams which play the role of infinite momentum tracks.
Then the same sophisticated alignment algorithms as employed for
track alignment with real particles can be applied to achieve
relative alignment between modules to better than a
few microns.
This method employs the tracking sensors themselves, with only a
minor modification to make them highly transparent to infrared light.
Only the aluminum metalization on the back of the sensor needs to
be swept away in a circular window with a diameter of few millimeters
to allow the IR beam to pass through.
Since IR light produces a measurable signal in the silicon bulk,
there is no need for any extra readout electronics.

A key parameter to understand the ultimate resolution of this method
is the transmittance of a silicon sensor and the diffraction of the
light. As a first approximation a silicon sensor is viewed as a stack
of perfectly homogeneous plano-parallel layers, each characterized by
its index of refraction and thickness.
The layers are, however, not continuous but
present local features, so that diffraction
phenomena will appear if the size of the
obstacle is comparable to the wavelength
used. For instance, the strips of the detector with
50~$\mu$m readout pitch, are good examples of an
optical diffraction grating for an incoming beam in the IR.
It has been determined that a key parameter that determines the
overall transmittance of a microstrip detector is the pitch to
strip width ratio, that is, the fraction of the sensor covered by aluminum.
The smaller the strip width, the more light is transmitted.
It was determined that good transmittance was achieved when the
strip width was set to 10\% of the pitch.
Tuning of sensor thickness was found to contribute up to 5\% over
the layout optimized value.
In bench tests, based on CMS strip detectors, a relative alignment
of a few microns has been achieved.




\section {Calorimeters}

\subsection{Introduction}
The SiD baseline design uses a Particle Flow Algorithm-based
approach to Calorimetry. PFAs have been applied to existing
detectors, such as CDF and ZEUS and have resulted in significant
improvements of the jet energy resolution compared to methods based
on calorimetric measurement alone. However, these detectors were not
designed with the application of PFAs in mind. The SiD baseline
design on the other hand considers a PFA approach necessary to reach
the goal of obtaining jet energy resolutions of the order of about
$\Delta E/E = 3-4 {\%}$ or better. SiD is therefore optimized
assuming the PFA approach and the major challenge imposed on the
calorimeter by the application of PFAs is the association of energy
deposits with either charged or neutral particles impinging on the
calorimeter. This results in several requirements on the calorimeter
design:

\begin{itemize}
\item{}
To minimize the lateral shower size of electromagnetic clusters the
Moli\`{e}re radius of the ECAL must be minimized. This promotes
efficient separation of electrons and charged hadron tracks.
\item{}
Both ECAL and HCAL must have imaging capabilities which allow
assignment of energy cluster deposits to charged or neutral
particles. This implies that the readout of both calorimeters needs
to be finely segmented transversely and longitudinally.
\item{}
The calorimeters need to be inside the solenoid to be able to do
track to cluster association; otherwise, energy deposited in the
coil is lost and associating energy deposits in the calorimeter with
incident tracks becomes problematic.
\item{}
The inner radius of the ECAL should match the outer radius of the
tracking system, within the requirements of access and removal.
\item{}
The calorimeter needs to be extendable to small angles to ensure
hermeticity, and be deep enough to contain hadronic showers.
\end{itemize}

Following is a short description of the baseline designs and options
for the ECAL and the HCAL, as defined in December 2008. More details
are available in the appendices and references. After that,
the performance of the calorimeter system  is discussed together
with the implications of running at 1 TeV center-of-mass energy.
Finally this leads into the discussion of an alternative
approach to calorimetry for SiD based on dual readout.


\subsection{Electromagnetic Calorimeter} \label{sec:ECAL}

The PFA approach of SiD is the main driver for the calorimetry
design. For the ECAL, this implies that electromagnetic showers be
confined to small volumes in order to avoid overlaps. Effective
shower pattern recognition is possible if the segmentation of
readout elements is small compared to the shower size.
\begin{figure}[htbp] 
\centering
\includegraphics[width=0.26\textwidth]{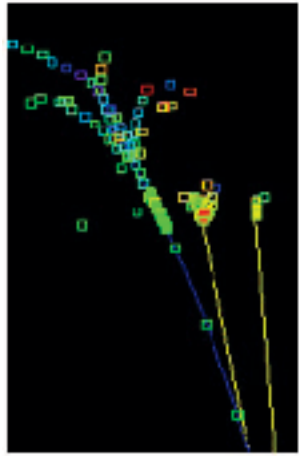}
\caption{Simulated $\rho^+\rightarrow \pi^+\pi^\circ$ decay in the
SiD detector.} \label{fig:ECAL_rho}
\end{figure}
This level of transverse segmentation then also facilitates
separating electromagnetic showers from the tracks of hadrons which
haven't interacted or muons. The longitudinal segmentation is chosen
not only to achieve the required electromagnetic energy resolution,
but also to provide discrimination between electromagnetic showers
and hadronic interactions which may occur in the ECAL. Finally,
there should be a sufficient number of longitudinal readout layers
to provide charged particle tracking in the ECAL. This is important
not only for the PFA algorithms, but also to provide
calorimeter-assisted track pattern recognition (see Section 2.1
Vertex and Tracking System). An important benefit of an ECAL which
meets these requirements is that it is an {\em imaging} calorimeter.
This is illustrated in Figure~\ref{fig:ECAL_rho}, which shows a
simulated $\rho^+\rightarrow \pi^+\pi^\circ$ decay in SiD. The
photons are clearly distinguished from each other and from the
charged track in the ECAL.

The ECAL described in this section according to the qualitative description
above is expected to have capabilites including:
\begin{itemize}
\item Precise measurement of beam-energy electrons and positrons (and photons) from
(radiative) Bhabha scattering. This is sensitive to contact terms and the
angular distribution provides important information on electroweak couplings, e.g.
in interference terms between $Z$, $\gamma$, and a new $Z^\prime$.
The Bhabha acollinearity distribution provides a key input for the measurement of the
differential luminosity spectrum\cite{ref:Boogert_2002jr}, which is needed for precision mass determinations
coming from cross-section measurements at threshold.
\item Identification of electrons from semileptonic decays, Dalitz decays, and photon conversions.
\item EM energy resolution of 17\%/$\sqrt{E}$ , which is more than adequate to contribute negligibly
to the overall jet energy resolution.
\item PFA reconstruction of photons in jets with high (95\%) efficiency and photon vertexing. The impact parameter
resolution for photons of $\sim 1$ cm would be important for
identifying decays where photons are the only visible decay
products, such as predicted from some gauge-mediated SUSY-breaking
models.
\item PFA tracking of charged particles in jets and ECAL-assisted tracking (especially of $V_0$'s)
\item $\pi^\circ$ reconstruction in jets and $\tau$ decays, to improve the jet energy
resolution\cite{ref:ECAL_intro_pi0GW} and to discriminate different
$\tau$ final states.
\end{itemize}

\subsubsection{Global ECAL design}
A sampling ECAL provides the required energy resolution for ILC physics, as discussed above.
Because of its small radiation length ($X_0$) and Moli\`{e}re radius, as well as its mechanical
suitability, we have chosen tungsten as absorber/radiator. Due to practical considerations
for ease of production of large plates and machining, the tungsten will be a (non-magnetic)
alloy. This currently chosen alloy includes 93\% W which has $X_0$ of $3.9$ mm
and Moli\`{e}re radius $9.7$ mm. An additional benefit of tungsten is that it has a relatively
large interaction length, which helps to minimize confusion between electromagnetic and
hadron showers in the ECAL.
\begin{figure}[htbp] 
\centerline{\includegraphics[width=0.6\textwidth]{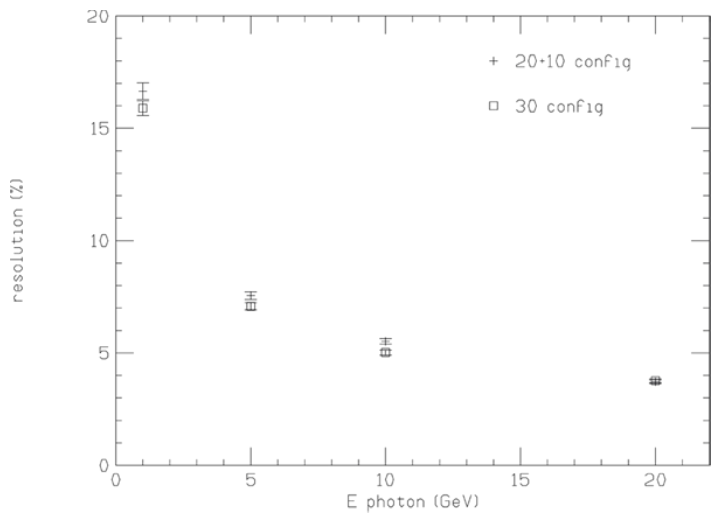}}
\caption[Energy resolution for two longitudinal
configurations.]{Energy resolution at low energy for two
longitudinal configurations. The ``20+10'' configuration is our
chosen design.} \label{fig:ECAL_Eres}
\end{figure}

The longitudinal structure has 30 total layers.
The first 20 layers each have $2.5$ mm tungsten thickness
and $1.25$ mm readout gap. The last 10 layers each have $5$ mm tungsten thickness
plus the same $1.25$ mm readout gap. This configuration attempts to
compromise between cost, shower radius, sampling frequency, and shower containment.
The cost is roughly proportional to the silicon area, hence the
total number of layers. Finer sampling for the first half of the total depth
improves the energy resolution for showers of typical energy. The total depth is 26 $X_0$, providing
reasonable containment for high energy showers. Figure \ref{fig:ECAL_Eres}
shows that the energy resolution at low energy does not suffer
relative to a design with 30 layers all having the thinner radiator.
The resolution of the baseline design is well fit by the
function 17\%/$\sqrt E$. These simulations were made with EGS4, but have been
verified by GEANT4 \cite{ref:ECAL_intro_GEANT4} simulations.
\begin{table}[htb]
\begin{center}
\begin{tabular}{|l|c|}
\hline
Inner radius of ECAL barrel & $1.27$ m \\ \hline
Maximum z of barrel & $1.7$ m \\ \hline
Longitudinal profile & (20 layers $\times$ 0.64 $X_0$ ) $+$ (10 layers $\times$ 1.3 $X_0$ )\\ \hline
EM energy resolution & $17\%/\sqrt{E}$ \\ \hline
Readout gap & $1.25$ mm   \\ \hline
Effective Moli\`{e}re radius (${\cal R}_{eff}$) & $14$ mm \\ \hline
\end{tabular}
\caption{Nominal parameters of the silicon-tungsten ECAL for SiD.}\label{tab:ECAL_parameters}
\end{center}
\end{table}
For both technology options (described below) we have chosen a transverse segmentation much
smaller than the typical shower radius. As discussed above, the scale for this is set by the shower size,
which we wish to be as small as feasible.  Since showers will
spread in the material between tungsten layers, it is crucial to keep the
readout gaps as small as possible. We can scale the shower radii by a simple
factor to provide a figure of merit. In our case, this factor is
$(2.5+1.25)/2.5 = 1.5$ for the crucial first 20 layers. We can then define
the {\em effective} Moli\`{e}re radius, ${\cal R}_{eff}$, as the Moli\`{e}re radius of
the radiator multiplied by this factor. In our case, this is about $14$ mm. A
crucial driving force in our design has been to provide as small a ${\cal R}_{eff}$ as
feasible, along with a transverse segmentation of the readout which is well
below ${\cal R}_{eff}$. Table \ref{tab:ECAL_parameters} summarizes the basic ECAL parameters.
\begin{figure}[htbp]
\includegraphics[width=0.9\textwidth]{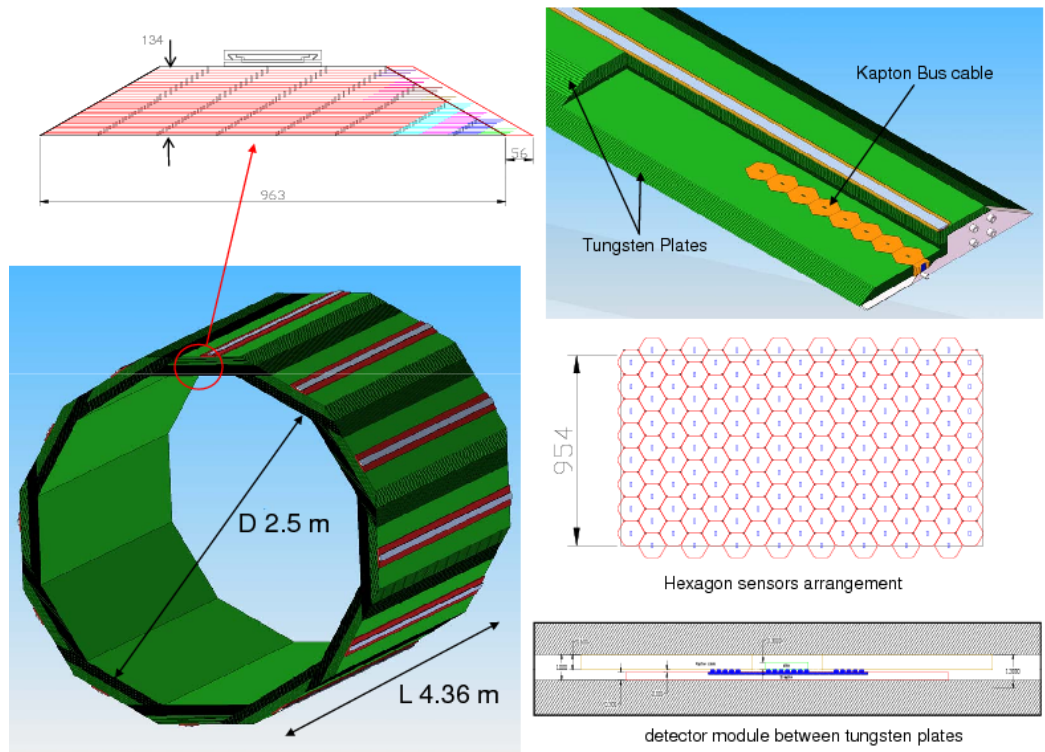}
\caption{Overall mechanical layout of the ECAL.}
\label{fig:ECAL_mechA}
\end{figure}

The mechanical design of the ECAL consists of 12 independent wedges,
fixed by rails on the inner radius of the HCAL. Each wedge is made of a stack of
tungsten plates with a readout gap of 1.25 mm. Tungsten plates are available
with a maximum size of 1 x 1 m,  therefore, the wedge assembly is done
interconnecting the plates by a screw-and-insert network, which transfers the
load from the bottom of the stack to the rail. Such a design allows also to
optimize the number of different silicon sensor sizes required to tile up each
tungsten plate surface. The mass of one wedge is about 5000 kg. Data
concentrator boxes are mounted on both sides of the wedge, reducing
locally the number of the services to be interfaced outside the detector and all
the services and cables run along the z-axis. Figure \ref{fig:ECAL_mechA} shows
the overall mechanical structure of the ECAL barrel, including  sensor layout
(for the baseline) and readout gap.\newline

\subsubsection{Technology choice}
The technologies under consideration both use silicon sensors. Silicon
will provide excellent transverse segmentation, while providing readout in a
thin gap, which is a crucial performance criterion, as discussed above. The
baseline design uses silicon sensors segmented into pixels of area 13
mm$^2$. The alternate option uses the CMOS-based Monolithic Active Pixel Sensor
(MAPS) technology with 50 x 50 $\mu$m pixels.
Both options are discussed below. An important feature of the SiD ECAL
mechanical design is that  it can accommodate either technology option without
change to the basic mechanical configuration.



\subsubsection{Baseline Design}
\label{sec:ECAL_pixels}
In the baseline design, the ECAL readout layers are tiled by large,
commercially feasible silicon sensors (presently from 15 cm wafers).

\begin{figure}[htb]  
\centerline{\includegraphics[width=0.45\textwidth]{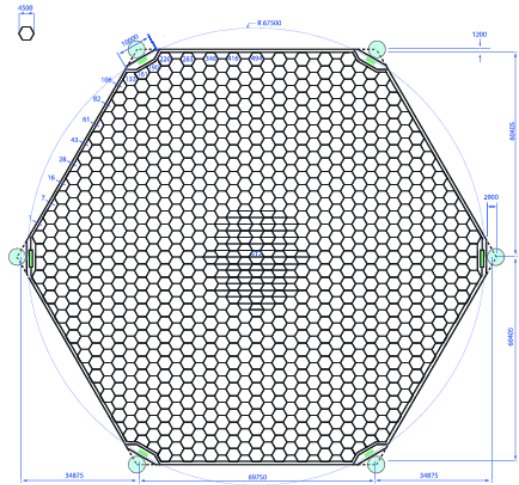}}
\caption[Drawing of a silicon sensor for the ECAL.]{Drawing of a
silicon sensor for the ECAL as delivered in 2008. The sensors are
segmented into 1024 13 mm$^2$ pixels.} \label{fig:ECAL_SiW_sensor}
\end{figure}

\begin{figure}[htbp]
\centerline{\includegraphics[height=2.0in]{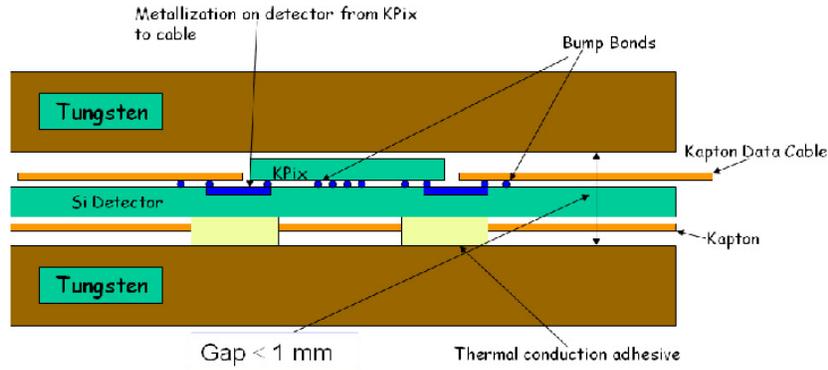}}
\caption[Readout gap in the vicinity of the KPiX readout chip.]{View
into a readout gap in the vicinity of the KPiX readout chip.
Representative bump bond connections are indicated by the small blue
circles. Traces (dark blue lines) connect the KPiX serial readout
stream, control signals, and power to the polyimide.}
\label{fig:ECAL_SiW_gap}
\end{figure}

The sensors are segmented into pixels which are individually read out with a wide dynamic range.
The complete electronics for the pixels is contained in a single chip (the KPiX ASIC)
which is bump bonded to the wafer. The low beam-crossing duty cycle allows a reduction of the
heat load by power pulsing, which permits passive thermal management within the ECAL modules.
The realization of this technology
has been the subject of an intensive, ongoing R\&D program.
The main parameters associated with the baseline technology choice are given in Table \ref{tab:ECAL_SiW_parameters}.
Some details of the design and R\&D results are given below. Further details
can be found in the references \cite{ref:Ray_Calor02} \cite{ref:WWS_review} \cite{ref:ANL_review}.
Figure \ref{fig:ECAL_SiW_sensor} shows a sensor with 1024 pixels. Not shown in the
drawing are the signal traces, part of the second layer metalization of the sensors, which connect
the pixels to a bump-bonding pad at the center of the sensor for input to the KPiX readout chip.
The pixels are DC-coupled to the KPiX, requiring only two metalization layers.
The pixels near the bump-bonding array at the center are split to reduce capacitance from the
large number of signal traces near the sensor center. The electronic noise due to the resistance
and capacitance of the traces has been minimized within the allowed trace parameters.
The cutouts at the corners of the sensor are to accommodate mechanical standoffs which support
the gaps between the tungsten layers.
Figure \ref{fig:ECAL_SiW_gap} depicts a cross-sectional view of the readout gap
in the vicinity of the center of the sensor. The silicon sensor is about
320 $\mu $m thick. The KPiX is bump-bonded to the silicon sensor at an array of 32 x 32
bump pads which are part of the second metalization layer from sensor fabrication.
Polyimid (kapton) flex cables connect near the center of the sensors. The cables bring
power and control signals into the KPiX chip and bring out the single digital output
line for the 1024 channels.

\begin{table}[htbp]
\begin{center}
\begin{tabular}{|l|c|c|}
\hline
             & Baseline  & MAPS option \\\hline
Pixel size               & 13 mm$^2$                                       & 50 x 50 $\mu$m\\ \hline
Pixels per silicon sensor& 1024                                            &  1,000,000\\ \hline
Channels per KPiX readout chip    & 1024                                   &      -      \\ \hline
Pixel dynamic range requirement   & $\sim 0.1$ to 2500 MIPs                & 1 MIP (digital) \\ \hline
Heat load requirement             & 20 mW per sensor                       & 20 mW per sensor \\ \hline
\end{tabular}
\caption{Parameters of both the baseline and MAPS options}\label{tab:ECAL_SiW_parameters}
\end{center}
\end{table}
Thermal management is a crucial feature of this design. The most power
hungry elements of the KPiX chip, particularly the analog front end, are
switched off for most of the interval between bunch trains.
Our requirement is to hold the average power dissipation
per wafer to less than 40 mW. This will allow the heat to be extracted
purely passively, providing a much simpler design, less subject to
destructive failure modes. The design of the KPiX chip in fact gives average
power less than 20 mW per wafer, which would not be possible without power pulsing.

\subsubsection{MAPS option}
\label{sec:ECAL_MAPS}
\begin{figure}[htbp]
\includegraphics[height=5.5cm]{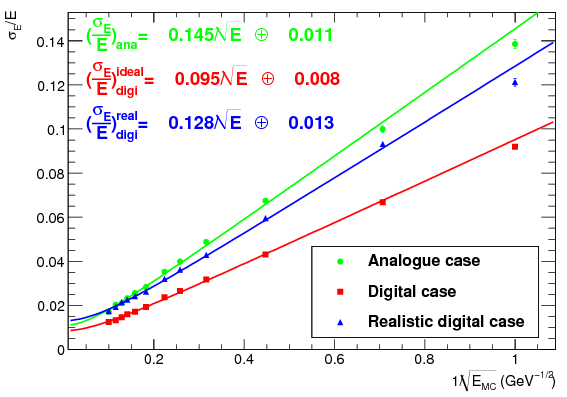}
\includegraphics[height=5.5cm]{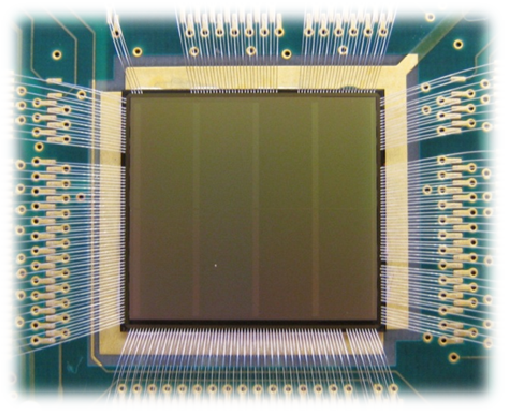}
\caption{Left: The energy resolution as a function of the incident
energy for single electrons for both analog and digital readout
using a GEANT4 simulation. The realistic digital cases includes
effects of saturation and charge sharing, leading to a degradation
of ~35\% \cite{ref:ECAL_MAPS_LCWS08}. Right: a picture of a bonded
TPAC sensor.} \label{fig:maps_digitalecal}
\end{figure}

The Monolithic Active Pixel Sensor option \cite{ref:ECAL_MAPS_basics} uses 50x50 $\mu$m silicon
pixels as readout material. The main difference here is the usage
of digital electromagnetic calorimetry where the ECAL is operated as a shower
particle counter.  Simulated performance \cite{ref:ECAL_MAPS_LCWS08} is illustrated in Figure
\ref{fig:maps_digitalecal}. These sensors could be manufactured in a commercial 180 nm mixed-mode
CMOS process using 300 mm wafers.
This is an industrial and widely available process, so pricing for these wafers should be competitive.

We have manufactured and tested two first-generation sensors for digital
electromagnetic calorimetry, TPAC 1.0 and 1.1 \cite{ref:ECAL_MAPS_details} which both consists of 168x168
pixels with the required size of 50x50 $\mu$m and TPAC 1.1 will be described in
a bit more detail. It uses a the pre-Shaper architecture and consists of a
charge preamplifier, a CR-RC shaper which generates a shaped signal pulse
 proportional to the amount of charge collected and a two-stage comparator which
triggers the hit-flag. The sensor supports single-bunch time stamping with up to
13 bits. Each pixel has a 6 bit trim to compensate for pedestal variations and
each pixel can be masked off individually. A bank of forty-two pixels shares nineteen memory buffers to
store the hits during the bunch train. The sensor also supports power-pulsing already and is able to
power off its front-end in the quiet time between bunch
trains. As the active sensor area is only about 20 $\mu$m thick, it allows
backthinning of the wafers down to 100 $\mu$m or less. Both sensors have been tested using sources and lasers\cite{ref:ECAL_MAPS_details}.
The MAPS option is designed to fit in the same mechanical structure as the baseline option and we
foresee a sensor size of 5 x 5 cm (baseline) for a final system. The main
parameters for the MAPS option are summarized in Table
\ref{tab:ECAL_SiW_parameters}.

\subsubsection{Calibration and Alignment}
\label{sec:ECAL_ALIGN}
Silicon detectors are inherently insensitive to gain variations with time and should not have significant inter-pixel
gain differences. Pixel to pixel gain differences in the electronic readout are calibrated by dedicated calibration
circuitry within the KPiX and TPAC chips. Perhaps the main calibration issue will be sensor to sensor gain differences.
These are not expected to be large, but we are investigating different options for this calibration. For mechanical alignment, the ECAL detector
sensor locations will be optically measured and related to external fiducials on the modules. The modules will be
located inside the HCAL using these fiducials to 1 mm or better, and the positions measured to approximately 100
micrometers.









\subsection{Hadronic Calorimeter}



\subsubsection{HCAL barrel geometry}

The PFA-based HCAL is a sandwich of absorber plates and instrumented
gaps with active detector elements. It is located inside the magnet
and surrounds the electromagnetic calorimeter, the latter being
fixed to it. The total absorber depth amounts to 4.5 $\lambda$, made
of stainless steel, divided into 40 layers, separated by 8mm gaps.
Thus the HCAL internal and external radii are respectively:
R$_{int}$=1419 mm and R$_{ext}$=2583 mm. The overall length is
6036mm long, centered on the interaction point.

\begin{figure}[htbp]
\centerline{\includegraphics[width=3in,height=3in]{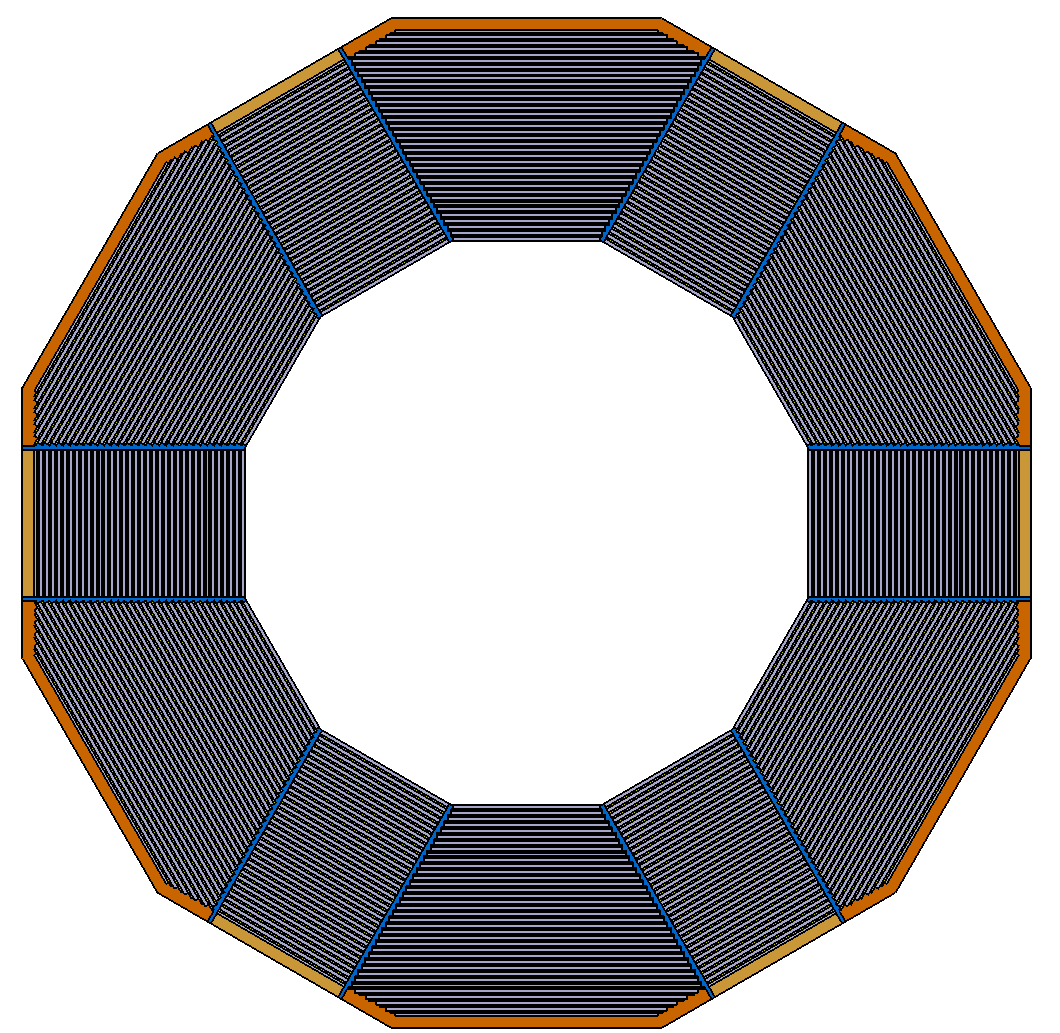}}
\caption{Cross-section of the HCAL barrel.} \label{fig:Hcal_bar_geo}
\end{figure}

 The HCAL is divided into twelve azimuthal modules. In order to
avoid cracks in the calorimeter, the module boundaries are not
projective with respect to the interaction point. Consequently, in
order to keep a symmetric shape two types of modules are used: 6
rectangles and 6 pseudo-trapezoids, as illustrated in
Fig.~\ref{fig:Hcal_bar_geo}. Each module covers the whole
longitudinal length. Chambers are inserted in the calorimeter along
the Z-direction from both ends and can eventually be removed without
taking out the absorber structure from the magnet. Special care of
the detector layout has to be taken into account to avoid a 90
degree crack.

The absorber plates are supported by several stringers fixed
radially on both sides of the modules. Stringers of two
consecutive modules are shifted in order to maximise the active
detector area. Although the space between two consecutive modules
is not instrumented, it is however filled by the absorber
material. The barrel will be fixed on the magnet at 3 and 9
o'clock or 5 and 7 o'clock.

\subsubsection{Forward HCAL} 
\begin{figure}[htbp]
\centerline{\includegraphics[width=2in]{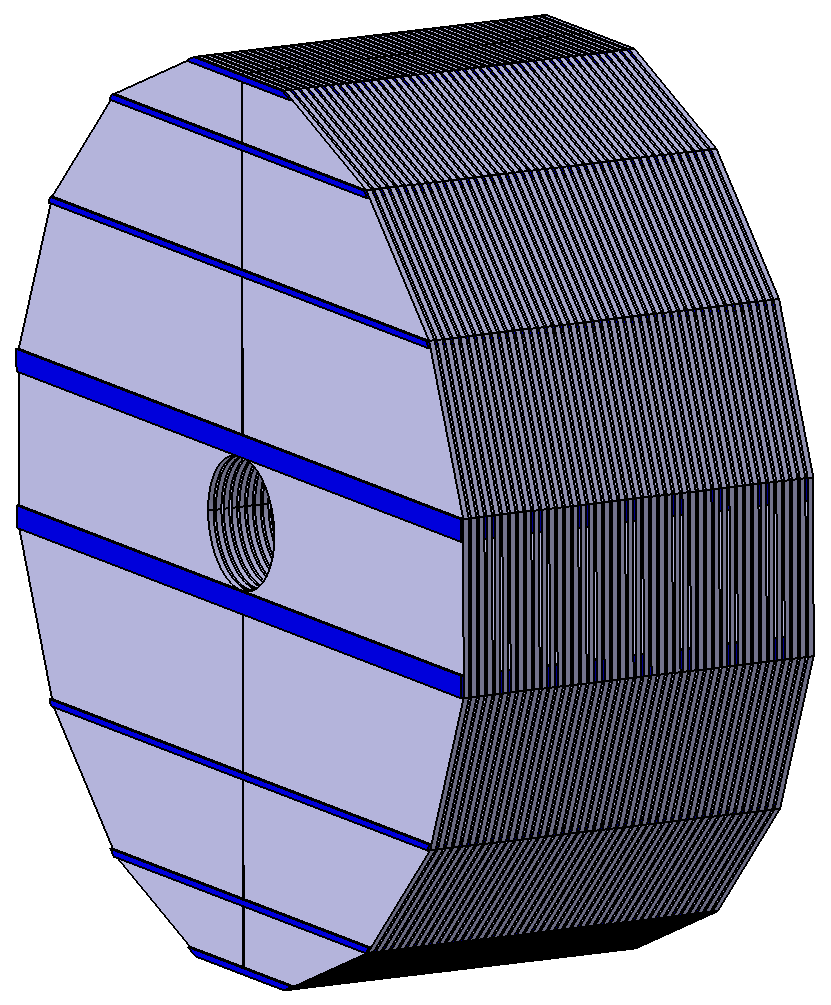}}
\caption{Face and top views of the Hcal forward.}
\label{fig:Hcal_fwd_geo}
\end{figure}

Each endcap forms a plug that is inserted into an end
of the barrel calorimeter. The layer structure of the end cap
calorimeters is the same as for the barrel.
Fig.~\ref{fig:Hcal_fwd_geo} shows a view of one endcap.

\subsubsection{Baseline Design (RPC)}
\label{sec:HCAL_RPC}



Resistive Plate Chambers (RPCs) are gaseous detectors primarily in
use for the large muon systems of colliding beam detectors. The
detectors feature a gas volume defined by two resistive plates,
typically Bakelite or glass. The outer surface of the plates is
coated with a layer of resistive paint to which a high voltage is
applied. Depending on the high voltage setting of the chamber,
charged particles crossing the gas gap initiate a streamer or an
avalanche. These in turn induce signals on the readout strips or
pads located on the outside of the plates.

In recent years RPCs have seen a wide range of applications. Using glass as
resistive plates, these detectors have shown excellent long-term stability
and reliability. The assembly of the chambers is straightforward and does
not require special skills or tools. The materials needed for their
construction are readily available and cheap.

{\em Current designs of the active layer}

\begin{figure}[htbp]
\centerline{\includegraphics[width=4.2in]{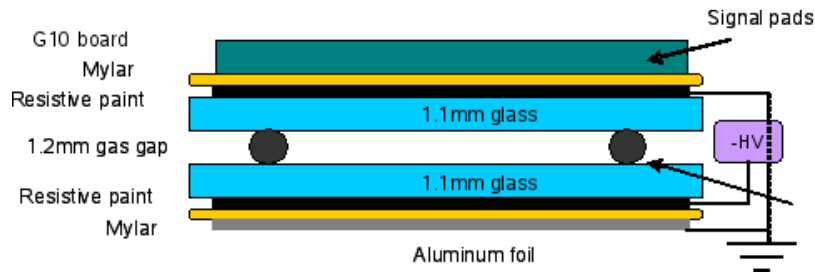}}
\caption[Schematic of the RPC design with two glass
plates.]{Schematic of the RPC design with two glass plates. Not to
scale.} \label{fig:HCAL_RPCA}
\end{figure}

\begin{figure}[htbp]
\centerline{\includegraphics[width=4.2in]{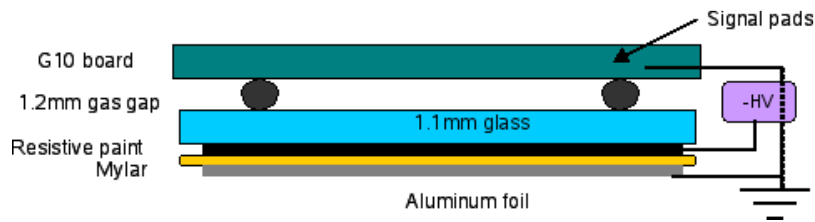}}
\caption[Schematic of the RPC design with one glass
plate.]{Schematic of the RPC design with one glass plate. Not to
scale.} \label{fig:HCAL_RPCB}
\end{figure}

Various chamber designs have been
investigated~\cite{ref:HCAL_RPC_designs}. Of these two are
considered particularly promising: a two-glass and a one-glass plate
design. Schematics of the two chamber designs are shown in
Figs.~\ref{fig:HCAL_RPCA} and ~\ref{fig:HCAL_RPCB}. The thickness of
the glass plates is 1.1 mm and the gas gap is maintained with
fishing lines with a diameter of 1.2 mm. The overall thickness of
the chambers, including layers of Mylar for high voltage protection,
is approximately 3.7 mm and 2.6 mm, respectively.

The chambers are operated in saturated avalanche mode with an
average high voltage setting around 6.3 kV. The gas mixes three
components: Freon R134A (94.5{\%}), isobutane (5.0{\%}) and
sulfur-hexafluoride (0.5{\%}).

The readout of the chambers consists of a pad board with 1$\times$1 cm$^{2}$ pads,
read out individually.
In such a calorimeter the energy of incident
particles is reconstructed as a function of the number of pads with signals
above threshold.


As part of the program of the CALICE
collaboration~\cite{ref:HCAL_RPC_calice}, the ANL group
~\cite{ref:HCAL_RPC_groups} built and tested about 20
chambers. The tests included extensive characterization of the
chambers with a 14-bit (analog) readout system.
The size of the signal charges, the MIP detection efficiency and the pad
multiplicity (as function of operating conditions) were measured with
cosmic rays~\cite{ref:HCAL_RPC_designs}.

\begin{figure}[htbp]
\centerline{\includegraphics[width=3.0in]{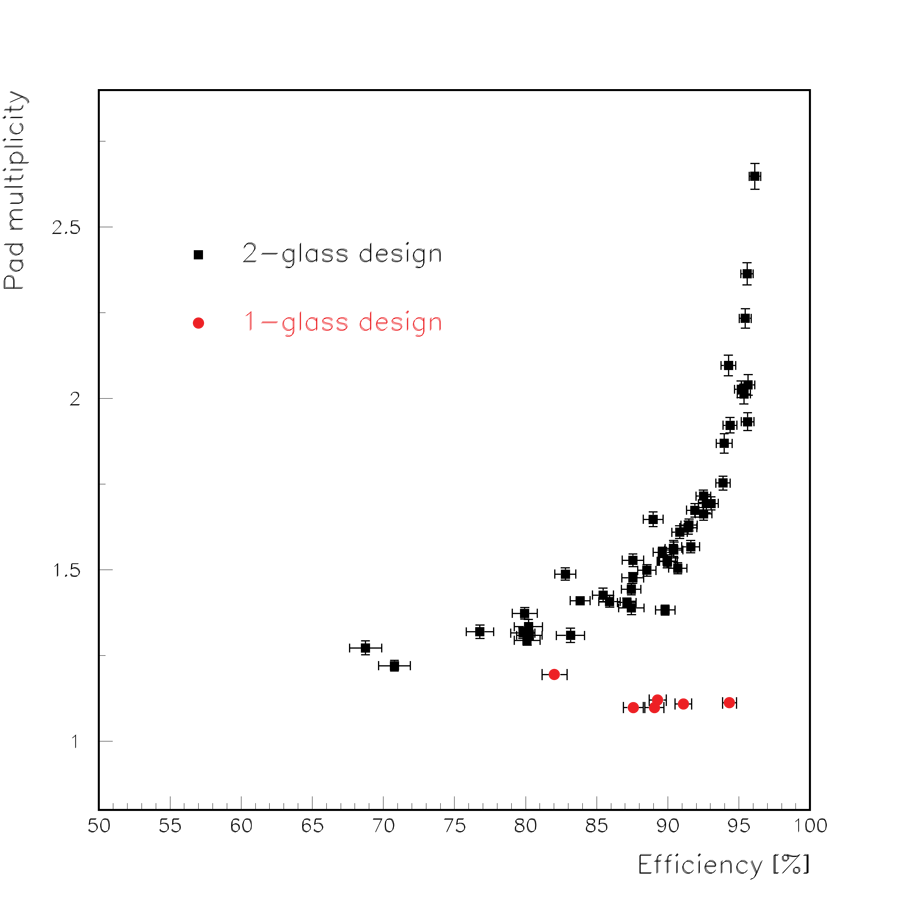}}
\caption[MIP detection efficiency]{Pad multiplicity versus MIP
detection efficiency for 2-glass RPC and 1-glass RPC. The results
were obtained with a broadband muon beam} \label{fig:HCAL_RPCC}
\end{figure}

A test calorimeter consisting of up to 10 chambers, each with an area of
$20\times 20$ cm$^{2}$, was assembled and tested with cosmic rays and in the
Fermilab test beam. The active media were interleaved with 20 mm thick Steel
plates. As an example, Fig.~\ref{fig:HCAL_RPCC}
shows the pad multiplicity versus MIP detection efficiency~\cite{ref:HCAL_RPC_muons}.
Note the constant pad multiplicity at 1.1 (shown as open black squares), independent
of efficiency, for the one-glass design.

\begin{figure}[htbp]
\centerline{\includegraphics[width=3in]{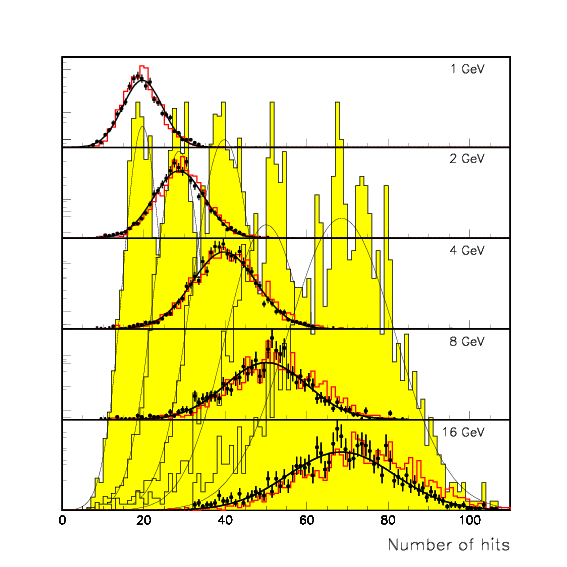}}
\caption[RPC stack response to positrons]{Response of a small size
prototype calorimeter to positrons with energies of 1, 2, 4, 8 and
16 GeV. The red histograms are the results of a GEANT4 based
simulation.} \label{fig:HCAL_RPCD}
\end{figure}

As an example of the results obtained with showering
particles, Fig.~\ref{fig:HCAL_RPCD} shows the response to
1, 2, 4, 8, and 16 GeV positrons.

Note that the response is highly non-linear, due to substantial
leakage out the back of the calorimeter (with a depth of 6.8 X$_0$)
and the high density of electromagnetic showers. The response is
adequately reproduced by Monte Carlo simulations based on GEANT4.
The results have been submitted for
publication~\cite{ref:HCAL_RPC_positrons}. Additional tests with
pions of various momenta were performed and showed the expected
results~\cite{ref:HCAL_RPC_pions}.

The stack was also exposed to 120 GeV protons at various beam intensities to
study the rate capability of RPCs. The results show no evidence of a dead
time in excess of 0.3 ms. For rates above 100 Hz/cm$^{2}$ the MIP
detection efficiency drops exponentially (with a time constant depending on
the beam intensity) until reaching a constant level.
The results have been submitted for publication \cite{ref:HCAL_RPC_rate}.

\begin{figure}[htbp]
\centerline{\includegraphics[width=4.4in]{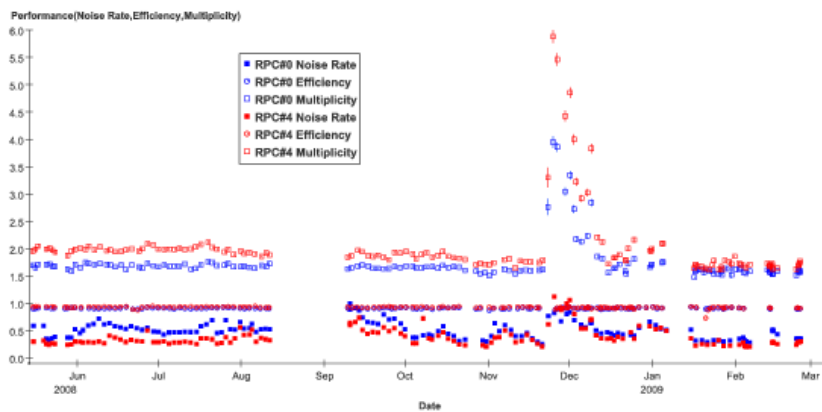}}
\caption[RPC performance as function of time]{Noise rate, MIP
detection efficiency and pad multiplicity as a function of time for
two 2-glass RPCs. Periods without measurements are due to safety
reviews and re-stacking of the chambers. The large increase in noise
rate and pad multiplicity seen in December 2008 are due to tests
with various gas flow rates.} \label{fig:HCAL_RPCE}
\end{figure}

The stack has now been operational for over 18 months. During this
time period, measurements of the noise rate, MIP detection
efficiency and pad multiplicity were performed using cosmic rays.
Fig.~\ref{fig:HCAL_RPCE} shows the results as a function of time. It
is seen that the noise rate and the pad multiplicity fluctuate over
time. This is due to changing environmental conditions (temperature
and barometric pressure). The efficiency of the chambers remains
constant. Within the time period of these studies there is no
evidence of long-term aging effects. Additional details on the
long-term studies can be obtained from~\cite{ref:HCAL_RPC_environ}.

The group is currently assembling a larger prototype calorimeter with 40
planes and approximately 400,000 readout channels. The completed calorimeter
will be tested in the Fermilab test beam in standalone mode and also
together with the CALICE Silicon-Tungsten electromagnetic calorimeter~\cite{ref:HCAL_RPC_calice}.
The tests will validate the concept of a digital hadron calorimeter with RPCs as
active medium and the proposed technical approach presented here. The
prototype calorimeter will provide precision measurements of hadronic
showers with unprecedented spatial resolution.
These measurements will be compared to predictions of various hadron shower models.


The analysis of the data from the small scale prototype is close to complete.
Two additional publications are planned: {\sl Hadron Showers in a Digital
Hadron Calorimeter} and {\sl Measurement of
the Environmental Dependence of the RPC Performance}.
The group is currently assembling a one meter cube prototype calorimeter.
Its completion is expected sometime during CY 2009, to be
followed by tests in the Fermilab test beam.
The data analysis will be carried through 2011.
The projected R{\&}D for the module/technical prototype will initiate in
2010, after the completion of the construction of the one meter cube prototype
calorimeter. This phase is expected to last 2 -- 3 years



\subsubsection{GEM Option}
\label{sec:HCAL_GEM}

\begin{figure}[htbp]
\centering
\includegraphics[width=0.4\textwidth]{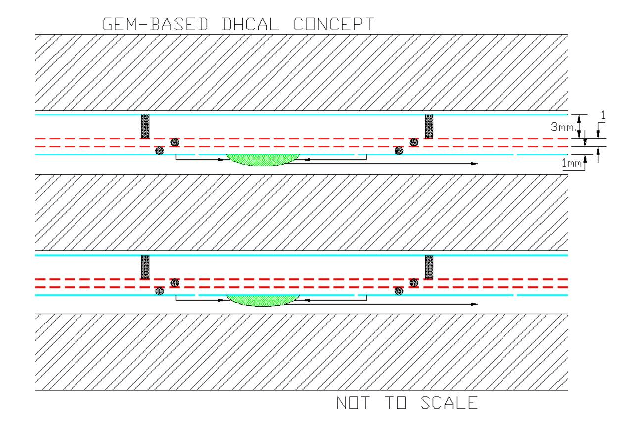}
\includegraphics[width=0.4\textwidth]{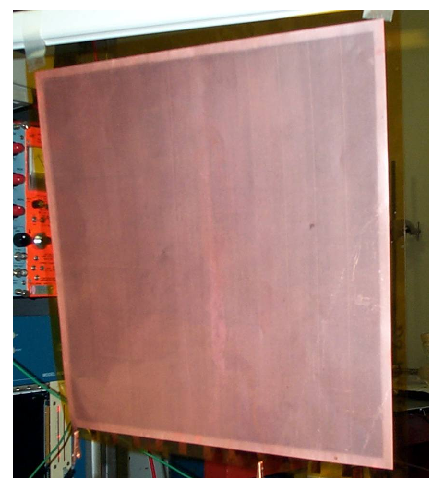}
\caption[Schematic layout of GEM-based DHCAL] {Left: GEM DHCAL
Concept.
 Right: A 30 cm x 30 cm foil from 3M Corporation.}
\label{fig:GEM_DHCAL_Schematic} \vspace*{-2mm}
\end{figure}

We have been developing a digital hadronic calorimeter
(DHCAL)\cite{ref:HCAL_GEM_DHCAL} using GEM
as the sensitive gap detector technology.  GEM can provide flexible
configurations which allow small anode pads for high granularity.  It is
robust and fast with only a few nano-second rise time, and has a short
recovery time which allows a higher rate capability than other detectors.
It operates at a relatively low voltage across the amplification layer,
and can provide high gain using a simple gas (ArCO2), which protects the detector
from long term degradation issues, and is stable.
The ionization signal from charged tracks passing through the drift section
of the active layer is amplified using a double GEM layer structure. The
amplified charge is collected at the anode layer with pads at zero volts.
The GEM design allows a high
degree of flexibility with, for instance, possibilities for microstrips
for precision tracking layer(s), variable pad sizes and shapes, and optional ganging
of pads for finer granularity of future readout, if allowed by cost
considerations and demanded by physics requirements.
Fig.~\ref{fig:GEM_DHCAL_Schematic} depicts
how the double GEM approach can be incorporated into a DHCAL scheme.

{\em Status of GEM Chambers development}

   A number of double GEM chambers have been built and tested with
cosmic rays sources, and test beam. Initial studies were conducted on
signal characteristics and gain from a small prototype GEM detector read
out using the QPA02 chip developed by Fermilab for Silicon Strip Detectors.
The gain of the chamber, with a 70\% Ar/20\% CO$_2$ gas mixture, was determined
to be of the order 3500, consistent
with measurements done by the CERN GDD group. The MIP efficiency was
measured to be 94.6\% for a 40 mV threshold, which agrees with a simulation
of chamber performance. The corresponding hit multiplicity for the same
threshold was measured to be 1.27, which will be beneficial for track
following and cluster definition in a final calorimeter system. A gas
mixture of 80\% Ar/20\% CO$_2$ was shown to work well and give an increase
in gain of a factor of three over the original mixture. A minimum MIP
signal size of 10 fC and an average size of 50 fC were observed from the
use of this new mixture. The prototype system has proved very stable in
operation over many months, even after deliberate disassembly and
rebuilding, returning always to the same measured characteristics.

Recently we have been using the SLAC KPiX chip to read out our chambers,
and will continue this using the expected 128 and eventual 1024 channel
KPiX versions. Our plans to build several 1m$\times$1m layers and expose them
to a testbeam in the CALICE 1 m$^3$  stack, are described in the appendices along
with more details of recent cosmic ray, source and test beam data.

\subsubsection{MICROMEGAS Option}
\label{sec:HCAL_MICROMEGAS}
A very interesting alternative for the DHCAL active medium is the
MICRO MEsh GAseous Structure (MICROMEGAS) \cite{micro1} based
on micro-pattern technology. Our prototypes consist
of a commercially available 20~$\mu$m thick mesh which separates
the drift gap (3~mm) from the amplification gap (128~$\mu$m). This
simple structure provides good gas gain uniformity over the whole
area and allows full efficiency for MIPs. The rate obtained is
not constrained, as is the case of glass RPCs.
Moreover, the tiny size of the electron avalanches results in fast
signals without physical crosstalk and leads to low hit
multiplicity. The chosen bulk technology, based on industrial PCB
processes, offers a robust detector with working voltages lower
than 500 V. MICROMEGAS, with anode pads as small as 1~cm$^{2}$, is
therefore a very appealing option for a DHCAL optimized for a
PFA.

\begin{figure}[h]
\centerline{
\includegraphics*[bb=14 4 759 548,height=1.6in]{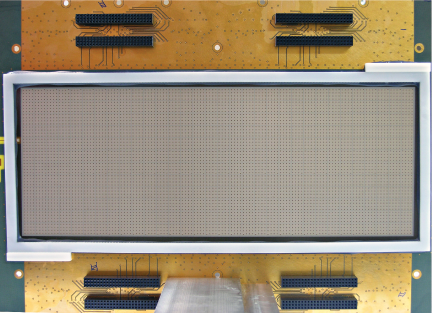}
\includegraphics*[bb=0 8 565 370,height=1.7in]{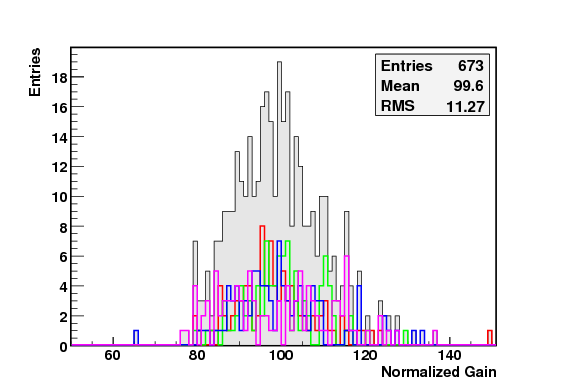}}
\caption{MICROMEGAS 12$\times$32 cm$^2$ analog readout prototype
with 1~cm$^2$ pads. MPV dispersion normalized to 100~\% for four
prototypes} \label{fig:Micro1}
\end{figure}

Different kinds of prototypes were developed, built and tested at
LAPP: one type with analog readout for characterization (see Fig.
\ref{fig:Micro1}) and two types with embedded digital ASICs. The
next step is a 1 m$^2$ assembly of six bulks with 24 ASICs each,
closed by two plates of 2 mm thick stainless steel which delimit a
6~mm active medium. Its construction is on-going and it should be
available for 2009 beam tests. This design is foreseen for large
quantity production in order to build a 1~m$^3$ DHCAL prototype.

Using an $^{55}$Fe X-ray source, the measured gain comes to 10$^{4}$
with an energy resolution down to 8.5~\% corresponding to 19.6~\% FWHM.
The gain and the resolution
were measured as a function of the drift field,
amplification field, gas flow and pressure variables.
Our prototypes were tested at CERN in 2008. With
200k muons, the Landau distribution obtained on each pad has a
Most Probable Value (MPV) around 45~fC. The chamber mapping was
performed in terms of pedestal mean and sigma, Landau distribution
MPV and sigma. The pedestal gaussian fits showed an average
standard deviation of 0.6~fC. The MPV relative variation is about
11~\% RMS (see Fig. \ref{fig:Micro1}), the efficiency 97~\% and the
multiplicity 1.08.
For the first time, a bulk was laid on a PCB with embedded electronics, the DIRAC chip
\cite{micro4}, and could be exposed to pions. With a threshold of
19~fC the beam profile was obtained. The next prototypes, with four HARDROC chips
\cite{micro3} are read out with the Detector InterFace board (DIF), designed at LAPP in the framework
of the DHCAL CALICE data acquisition system \cite{micro5}. More details of this options and future
plans may be found in the appendices.

\subsubsection{Scintillator with SiPM Option}
\label{sec:HCAL_SCINTILLATOR}

The CALICE Collaboration has been pursuing the design and prototyping of a
fine granularity scintillator-based hadron calorimeter. This option capitalizes
on the marriage of proven detection techniques with novel photodetector devices.
The main challenge for a scintillator-based calorimeter is the
architecture and cost of converting light, from a large number of channels, to electrical
signal. Studies demonstrate that small tiles (4-9 cm${^2}$) interfaced to Silicon Photomultipliers
(SiPMs)/Multi Pixel Photon Counter (MPPC) photodetectors \cite{ref:HCAL_scint_sipm},
\cite{ref:HCAL_scint_ham} offer
an elegant solution. SiPM/MPPCs are multi-pixel photo-diodes operating in the
limited Geiger mode. They have distinct advantage over conventional photomultipliers
due to their small size, low operating voltages and insensitivity to magnetic fields.
The {\it{in situ}} use of these photodetectors opens the door to integration
of the full readout chain to an extent that makes a high channel count scintillator calorimeter
entirely plausible. Also, in large quantities the devices are expected to cost a few dollars per
channel making the construction of a full-scale detector instrumented with these photo-diodes
financially feasible.

\begin{figure}
\begin{center}
\includegraphics*[scale=0.4]{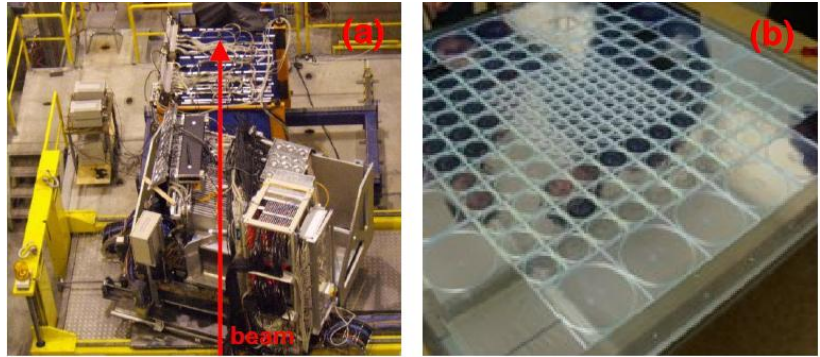}
\caption[CALICE test beam setup at CERN.]{CALICE test beam setup at
CERN (left) and an active layer of the scintillator-SiPM prototype
(right)} \label{fig:comb1}
\end{center}
\end{figure}

A scintillator-SiPM prototype of about 1 m$^{3}$ has been designed, constructed and exposed to
a test beam during the 2006-2008 period at CERN and Fermilab (see Fig. \ref{fig:comb1}).
The active layers of the HCAL prototype consist of 5 mm thick scintillator tiles sandwiched
between 2 cm thick steel absorber plates. Each tile comes with its own
1 mm diameter WLS fiber mated to a SiPM embedded in it.
Over six run periods millions of electron, pion and proton
events in the 2-180 GeV/c range were written to disk.
Ongoing analysis of the data collected, has gone a long way in establishing the scintillator-SiPM
option as a calorimeter technology, benchmarking hadron shower Monte Carlo's and testing
the particle-flow paradigm using hadrons from real data \cite{ref:HCAL_scint_analnotes}.

\begin{figure}
\begin{center}
\includegraphics*[scale=0.4]{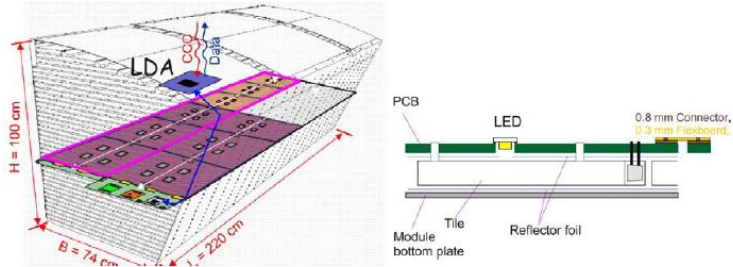}
\caption[Barrel wedge intrumented with IRL planes.]{Barrel wedge
intrumented with IRL planes (left) and conceptual design of an IRL
(right).} \label{fig:comb3}
\end{center}
\end{figure}

The focus of the current R\&D effort is the development of an Integrated Readout Layer (IRL).
In general for the IRL, it is proposed to have a printed circuit board (PCB) inside the detector
which will support the scintillator tiles, connect to the silicon photodetectors and carry
the necessary front-end electronics and signal/bias traces (see Fig. \ref{fig:comb3}).
This can however be achieved in
a number of ways and different groups are working on complementary approaches (e.g.
fiber vs. direct or fiberless coupling of SiPMs to the tiles) in a coordinated
fashion. It is planned that the R\&D being carried out on the scintillator-SiPM
hadron calorimeter option in America, Asia and Europe will be brought together for a next-generation
EUDET/CALICE 'technology'
prototype in 2010. This 'wedge' will be intrumented with IRL planes. Complementary versions of the
IRL planes being pursued by different groups in the collaboration will be tested and compared.

\subsection{Calorimeter Performance} \label{sec:Cal_performance}
The SiD detector is
designed to take advantage of the expected improvement in jet energy
resolution deriving from the use of a Particle Flow Algorithm (PFA).
The SiD PFA is described in the online appendices and was used in
the physics benchmarking reported in Chapter~\ref{chap:bench}. This
section first describes the necessary software calibrations, which
were then applied to check calorimeter/PFA performance. Three
results were compared: calorimeter only/no PFA, a perfect pattern
recognition (PPR) PFA, and the SiD PFA.

Full Geant4 simulations using SLIC~\cite{ref:slic} and the SiD
simplified detector geometry (SiD02) were performed on test samples.
The test samples were $qq$ events ($q = uds$) at fixed E$_{\rm cm}$,
and $ZZ \rightarrow qq \nu\nu$ at 500 GeV E$_{\rm cm}$. The $qq$
events have no beamstrahlung, no bremsstrahlung, and no gluon
radiation, so each event contained exactly 2 equal energy
back-to-back quark jets. The $ZZ$ events were generated with all
these effects, and allow mass reconstruction without jet finding
algorithms.

\begin{figure}[htbp]
\centering
\includegraphics[width=0.6\textwidth]{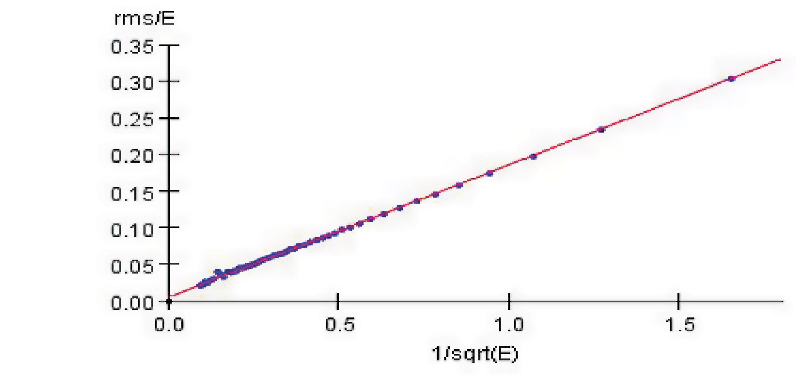}
\caption[Resolution plot for single photons from ttbar events.]
{Resolution plot for single photons from ttbar events. The line in
the resolution plot represents a fit to the data, 0.5\% +
18\%/$\sqrt{E_{Gen}}$.} \label{fig:calperf1}
\end{figure}

Software calibrations were performed separately for photons and hadrons using
the $ZZ$ test sample, to relate the simulated detector response to the
incident energy.
Details of the calibration procedure may be found in the appendices .
The calibrations were then applied to a sample of particles not used
in the calibration ($t\bar{t}$ events at 500 GeV E$_{\rm cm}$). For
photons a single calibration constant was determined for each of the
calorimeters, ECAL (barrel and endcap) and HCAL (barrel and endcap).
This was sufficient to give a maximum nonlinearity of less than
0.5\% and full rms of 0.5\% + 18\%/$\sqrt{\rm E}$  for single
photons in the $t\bar{t}$ sample. See Fig.~\ref{fig:calperf1}.

\begin{figure}[htbp]
\centering
\includegraphics[width=0.6\textwidth]{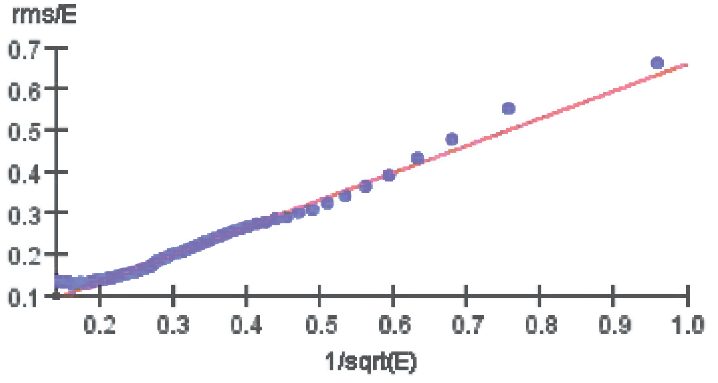}
\caption[Resolution plot for single neutral hadrons from ttbar
events.]{Resolution plot for single neutral hadrons from ttbar
events. The line in the resolution plot represents 65\%/EGen.}
\label{fig:calperf2}
\end{figure}

For neutral hadrons, one calibration constant was determined per
calorimeter, as was a polar angle correction for the hadron
calorimeters (barrel and endcap). After combining the deposits in
different calorimeters, an energy correction was applied. This
resulted in a maximum nonlinearity of less than 3\% and full rms of
65\%/$\sqrt{\rm E}$ for single neutral hadrons in the $t\bar{t}$
sample Fig.~\ref{fig:calperf2} .

The test sample events were then reconstructed using particle flow algorithms
(PFA's) with this calibration. Several PFA algorithms are being developed, and
the one yielding the best energy resolution for the test samples
 was chosen for the benchmark analyses.

The calorimeter only/no PFA reconstruction calculated a single
parameter for each calorimeter with no assumption as to the type of
particle depositing energy. For the digital hadron calorimeter, a
polar angle correction was applied. Using this calibration, event
energy was measured using only the calorimeters. In the PPR PFA
reconstruction, the Monte Carlo truth information was used to
associate calorimeter hits to the appropriate particles, showing the
potential of a PFA with the SiD detector.

\begin{figure}[htbp]
\centering
\includegraphics[width=1.0\textwidth]{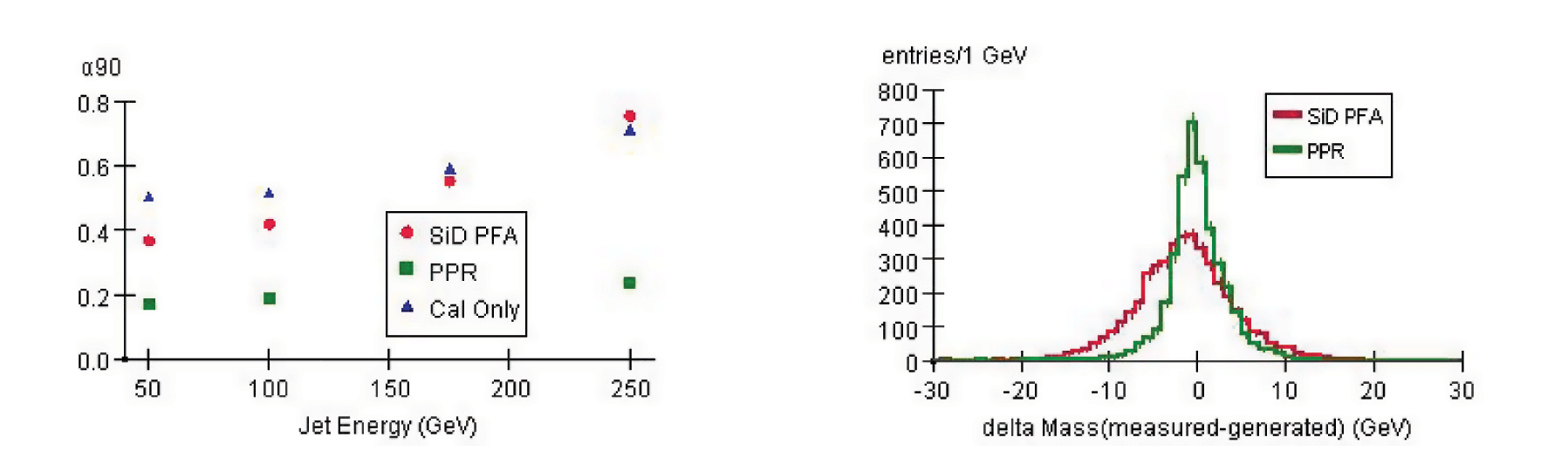}
\caption[Results.] {Left: $\alpha_{\scriptscriptstyle{90}}$ vs jet
energy for the 3 different reconstructions.
 Right: Mass residuals in the ZZ test sample. For PPR rms90 = 2.2 GeV.
 For SiD PFA rms90=4.0GeV}
\label{fig:calperf3}
\end{figure}

Each of these two reconstructions was run on the $qq$ test samples.
Since the samples contain two jets of equal energy, dE$_{\rm
jet}$/$\sqrt{\rm{E}_{\rm jet}}$ = dE$_{\rm
event}$/$\sqrt{\rm{E}_{\rm event}}$. The standard for comparing
results has become rms90, where rms90 is the rms of the 90\% of the
events yielding the smallest rms. Defining
\(\alpha_{\scriptscriptstyle{90}}\) with the expression rms90/E =
$\alpha_{\scriptscriptstyle{90}}$/$\sqrt{\rm E}$,
$\alpha_{\scriptscriptstyle{90}}$ vs E$_{\rm jet}$ is shown in
Fig.~\ref{fig:calperf3} for the reconstructions. Although the PPR
PFA jet energy resolution remains relatively constant with jet
energy, Fig.~\ref{fig:calperf3} shows that the calorimeter-only
resolution degrades somewhat, and the SiD PFA resolution degrades
significantly as the jet energy increases. Future studies must
determine if the SiD PFA performance can be improved at the higher
energies, if it can be complemented with calorimeter-only
measurements, or if a new approach, like that using dual-readout
crystals, described below, is superior.

Finally, the $ZZ$ test sample was reconstructed with the PFA's. The
mass residuals (M$_{\rm measured}$ - M$_{\rm generated}$) are shown
in Fig.~\ref{fig:calperf3}. Using the rms90 of the distributions,
dM/M = 2.5\% for the PPR, and dM/M = 4.5\% for the full PFA. This
demonstrates that the SiD PFA has already approached the desired
mass resolution. Further improvements are expected as work continues
refining the SiD PFA.

\subsubsection{Homogeneous Crystal Dual-readout Option}
\label{sec:LOI_dual_readout_one_page}

%
SiD is studying an alternative to particle flow calorimetry based on
the dual readout approach.
The principal limitations of the hadron energy resolution in traditional
total absorption calorimetry come from two sources:

\begin{itemize}
\item  The sampling nature of the conventional hadron calorimeters

\item  A significant and fluctuating fraction of the incoming hadron energy
is converted into non-observable forms of energy (primarily nuclear binding
energy).
\end{itemize}

\begin{figure}[htbp]
\vspace*{-2mm} \centerline{
\includegraphics[width=6in]{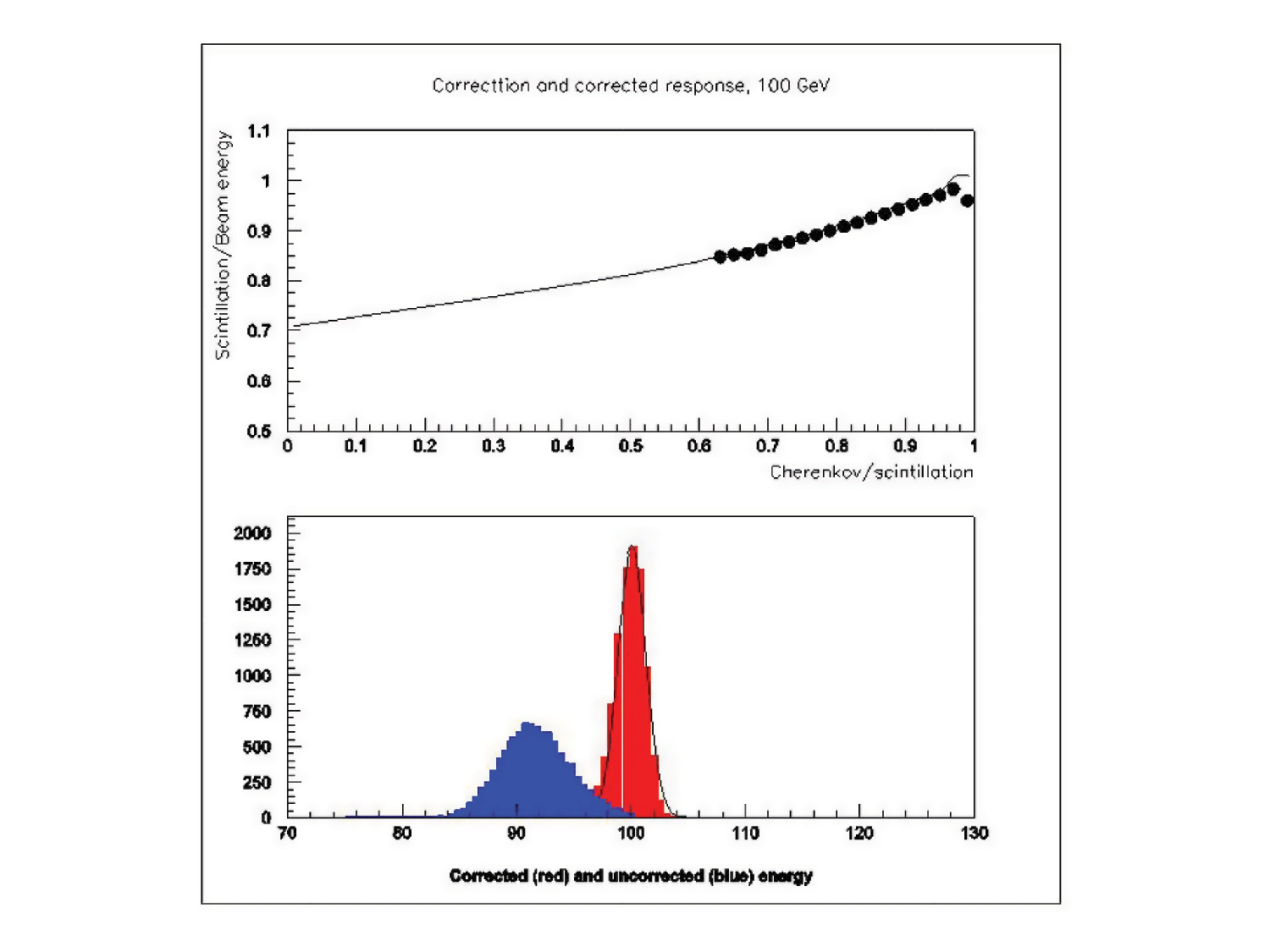}}
\caption[Dual readout response]{Based on a Monte Carlo simulation,
The correlation between the average fraction of the beam energy
detected via scintillation and the ratio of responses measured with
the Cherenkov and the scintillation light (top); A response of the
total absorption calorimeter to 100 GeV pion beam (blue) and the
same response corrected on the event-by-event basis using the
correlation (bottom).} \label{fig:Dimless_correl} \vspace*{-2mm}
\end{figure}

Good jet energy resolution requires a calorimeter where both of these
factors are eliminated or largely reduced. This can
be accomplished with a homogenous, totally active calorimeter with
dual readout, i.e. readout of scintillation and Cherenkov light.
A dimensionless anti-correlation
function may be derived using these components and is
shown in Fig.~\ref{fig:Dimless_correl}, based on a Monte Carlo simulation.
Application of an
event-by-event correction to a sample of hadron induced showers
improves the energy resolution and makes the average hadron response
equal to the beam energy (hence equal to the response to electrons
on neutral pions of the same energy), as shown in
Fig.~\ref{fig:Dimless_correl}.

The construction of a practical hadron calorimeter, especially with the
hermeticity required in the colliding beam environment, is now possible by
two technological breakthroughs:

\begin{itemize}
\item  The development of affordable, high density crystals
\item  The advent of compact, inexpensive silicon-based photodetectors (APD's
and SiPM's) capable of operating in a string magnetic field.
\end{itemize}

A high resolution calorimeter is designed to fit into the space occupied by
the ECAL and HCAL of the baseline design. It is constructed of optical
''crystals'' equipped with two sets of compact silicon photodetectors at the
back. One set, equipped with the low pass optical filter and short
integration gate electronics, is used to detect and measure the Cherenkov
light. The other set, equipped with high pass filter and long integration
gate electronics is used to detect and measure the scintillation component.

Further details of the use of the dual-readout, specifications of a
viable calorimeter, and the required R\&D tasks may be found in
Chapter~\ref{chap:rnd}.


\bibliographystyle{unsrt}



\section{SiD Magnet Subsystem} \label{sec:magnet}

The magnet consists of the following subsystems:

\begin{enumerate}
\item a large 5 T superconducting solenoid with a separated iron plate flux return that is integral with the 
muon tracking system,
\item a power supply, a pressurized water cooled dump resistor, and a conventional mechanical dump switch that 
move with the detector,
\item a helium liquefier that supplies 4.5 K LHe to both the solenoid and to a pair of 2 K cold boxes for each 
of the superconducting focusing quadrupoles,
\item oil flooded helium screw compressors with oil removal and helium gas storage, all located on the surface,
\item a superconducting Detector Integrated ``anti'' Dipole (antiDID), and
\item two superconducting final focus quadrupoles (not discussed here).
\end{enumerate}

\subsection{Superconducting Solenoid}

\begin{figure}[htbp]
\centerline{\includegraphics[width=5in]{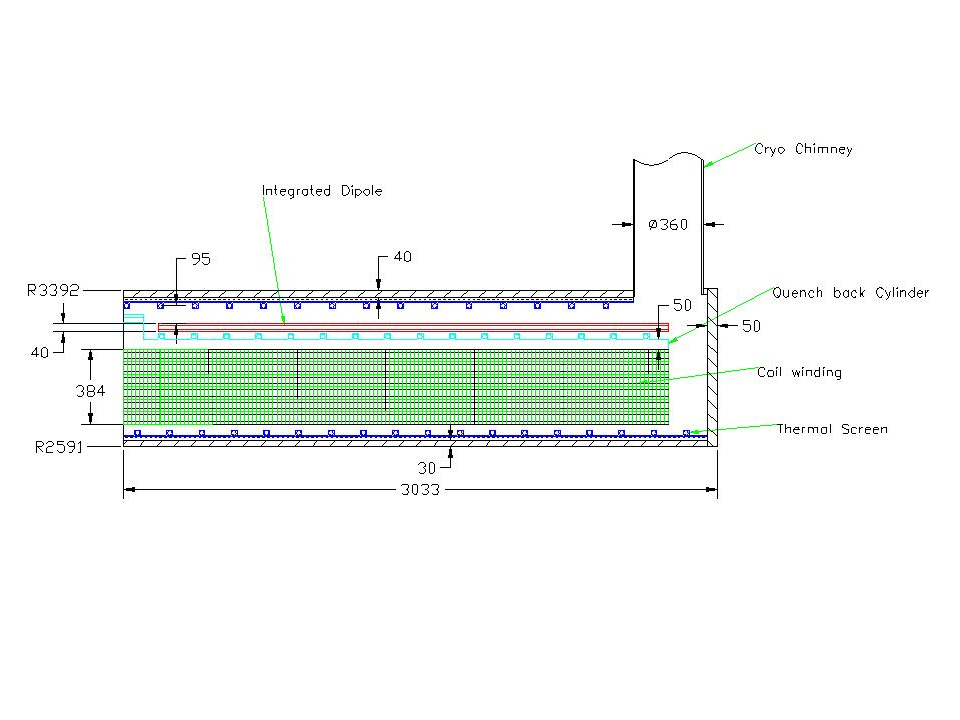}}
\caption{\label{fig:mag1} Longitudinal half-section view of SiD
superconducting solenoid with winding/quench-back cylinder/LHe
cooling tubes, cryostat, thermal shields, and anti-DID.}
\end{figure}

The superconducting solenoid is an expensive and technically
challenging component to procure. Its design is based on the
engineering philosophy, experience and details used for the
successful 4 T CERN CMS superconducting solenoid, the largest in the
world. Thus a direct comparison is warranted, with major parameters
listed Table~\ref{tab:mag1}. The now standard high-purity aluminum
superconductor stabilization, with indirect LHe cooling, technique
will be used. The CMS individual self-supporting winding turn design
philosophy is used for SiD, and is even more important due to the
higher field and the increased radial softness of six layers
versus four layers. Figure~\ref{fig:mag1} shows a cross section of the
SiD magnet.

The SiD solenoid has a stored energy per unit cold mass of 12 kJ/kg, which
is only slightly larger than the CMS value. This value is close to the upper
bound in which this type of large aluminum dominated magnet can be operated in a
fail-safe manner if the quench detection or energy extraction circuit were
to fail. This specific energy density yields an average magnet temperature
of 130 K. Prudent engineering for the SiD solenoid dictates that the total
volume of aluminum stabilizer/structure cannot be reduced significantly relative to the
present baseline design.


\begin{table}[htbp]
\caption[SiD and CMS superconducting solenoid comparison.]
{\label{tab:mag1} SiD and CMS superconducting solenoid comparison. }
\begin{center}
\begin{tabular}{|l|p{79pt}|p{83pt}|l|}
\hline
\textbf{Quantity}&
\textbf{SiD}&
\textbf{CMS} \par &
\textbf{Units} \\
\hline
Central field&
5.0&
4.0&
T \\
\hline
Stored energy&
1.56&
2.69&
GJ \\
\hline
Stored energy per unit cold mass&
12&
11.6&
kJ/kg \\
\hline
Operating current&
17.75&
19.2&
kA \\
\hline
Inductance&
9.9&
14.2&
H \\
\hline
Fast discharge voltage to ground&
300&
300&
V \\
\hline
Number of layers&
6&
4&
 \\
\hline
Total number of turns &
1457&
2168&
 \\
\hline
Peak field on superconductor&
5.75&
4.6&
T \\
\hline
N$^{\rm o}$ CMS superconductor strands&
36&
32&
 \\
\hline {\%} of short sample& 32& 33&
 \\
\hline Temperature stability margin& 1.6 & 1.8 &
K \\
\hline
Total mass of solenoid&
125&
220&
Metric ton \\
\hline
Rmin cryostat&
2.591&
2.97&
m \\
\hline
Rmin coil&
2.731&
3.18&
m \\
\hline
Rmax cryostat&
3.392&
3.78&
m \\
\hline
Rmax coil&
3.112&
3.49&
m \\
\hline
Zmax cryostat&
$\pm 3.033$&
$\pm 6.5$&
m \\
\hline
Zmax coil&
$\pm 2.793$&
$\pm 6.2$&
m \\
\hline
Operating temperature&
4.5&
4.5&
K \\
\hline
Cooling method&
Forced flow&
Thermosiphon&
 \\
\hline
\end{tabular}
\end{center}
\end{table}

\begin{figure}[htbp]
\centerline{\includegraphics[width=5in]{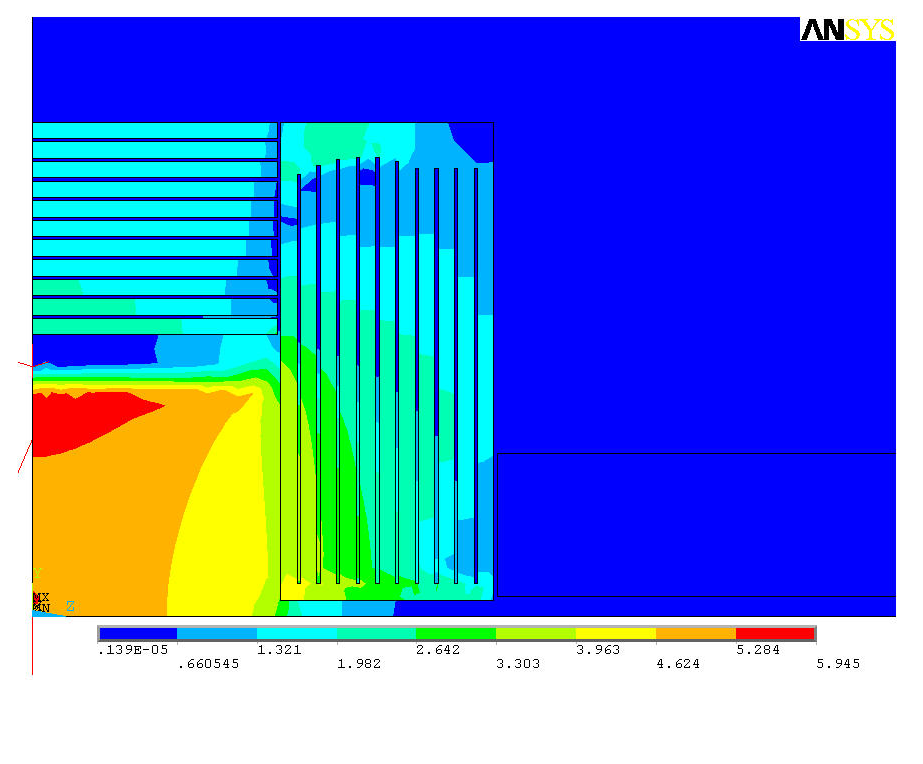}}
\caption{\label{fig:mag2} SiD central field contours in $\vert
$B$\vert $, shown by intensity scale. The iron Pacman is included
and shown as the rectangle in the lower right.}
\end{figure}

The present flux return design has assumed that field uniformity will not require shims or size-graded 
conductors, but this issue will be revisited once forward tracking studies have been conducted with a realistic 
field map.
Fringe field requirements are set at 100 Gauss at
1 m from the iron surface. Two-dimensional and three-dimensional
ANSYS magnetic field calculations of the magnet were performed;
the resulting field shape in the central region of the detector is
shown in Figure~\ref{fig:mag2}. Eleven 20 cm thick iron plates with 4
cm gaps form both the barrel and endcap portions of the flux
return. There is also a 4 cm gap between the barrel and endcaps.
Magnet field optimization studies are underway to improve uniformity while meeting the
fringe field requirements.

\begin{figure}[htbp]
\centerline{\includegraphics[width=4in]{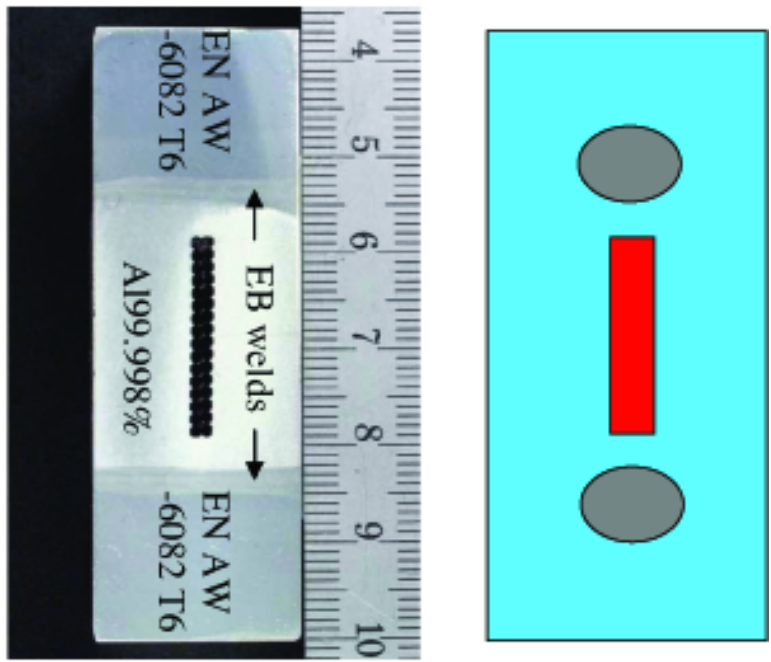}}
\caption{\label{fig:mag3} Photo of CMS conductor and proposed SiD
conductor.}
\end{figure}

The CMS conductor forms a fall-back
option for SiD. All magnet ANSYS finite element stress analysis to
date has been with this conductor. However a more advanced, and most
likely cheaper, conductor is being pursued and is proposed here.
It is based on dilute high-purity aluminum alloys, such as
Al-0.1{\%}Ni, which have been studied more extensively since CMS and
were used in the ATLAS central solenoid. Figure~\ref{fig:mag3} shows a
photo of the CMS conductor and an advanced SiD conductor design
using Al-0.1{\%}Ni and a novel internal high strength stainless
steel reinforcement. An ANSYS coupled transient electromagnetic and
thermal diffusion model is being used to evaluate conductor
stability.
With large-size high-purity
aluminum-stabilized superconductors, the current is slow to
diffuse into the aluminum during a temperature excursion. Thus a
higher resistivity, and therefore higher strength, aluminum
stabilizer can have stability margins equal to a conductor with
lower resistivity aluminum. A stability margin similar to CMS will
be used. ANSYS is also being used to evaluate strength and bending
stiffness for coil winding. Compared to CMS this advanced conductor
has reduced bending stiffness that facilitates winding. Other
conductor stabilizer possibilities are also under consideration and
study.  These include TiB2 grain refinement,
aluminum matrix composites, and cold working via the equal area angle
extrusion process.

The superconductor will be internally wound onto two precise
aluminum 5083-H321 mandrels. All CMS winding procedures such as
conductor milling to size, cleaning, and taping will be employed.
CMS epoxy vacuum impregnation and mandrel joining techniques will be
used. CMS conductor splicing methods shall also be used. CMS
coil-winding experience will significantly reduce SiD time and
expense for setting up and commissioning the coil winding facility, and
vacuum impregnation will take place at the vendor's facility.  The vendor
will probably also perform the initial assembly,
cool down, and low-current checkout.  The magnet will
be shipped as two separate coils and a vacuum shell.


A comparison was made via ANSYS studies of the SiD and CMS solenoid stresses, deflections
and forces. All stresses are evaluated after cool down and
energization. For this comparison the CMS conductor was used in the
SiD analysis. Note that all stress and deflections are very similar
for the two coils.


\begin{table}[htbp]
\caption[Mechanical Comparison of SiD and CMS Solenoids.]
{\label{tab:mag2} Mechanical Comparison of SiD and CMS Solenoids. }
\begin{center}
\begin{tabular}{|l|l|p{72pt}|l|}
\hline
\textbf{Quantity}&
\textbf{SiD}&
\textbf{CMS} & Units \\
\hline
Von Mises stress in high-purity aluminum &
22.4 &
22 & MPa \\
\hline
Von Mises Stress in Structural Aluminum&
165  &
145 & MPa \\
\hline
Von Mises Stress in Rutherford cable &
132  &
128 & MPa \\
\hline
Maximum radial displacement &
5.9 &
$\sim 5$ & mm \\
\hline
Maximum axial displacement &
2.9 &
$\sim 3.5$ & mm \\
\hline
Maximum shear stress in insulation &
22.6 &
21 & MPa \\
\hline
Radial decentering force &
38 &
31 & kN/mm \\
\hline
Axial decentering force &
230 &
85 & kN/mm \\
\hline
\end{tabular}
\end{center}
\end{table}


The power supply, dump resistor and dump switch are attached on the
side of the detector near the top.    These three components are
arranged to be in line with the solenoid current leads, thereby
minimizing the buswork. The power supply and mechanical dump switch are
standard components which will be procured from outside vendors. A novel,
compact, pressurized-water-cooled dump resistor will be used instead
of a very large air-cooled resistor, such as the type used for
CMS and other large superconducting magnets. In the worse case,
in which all 1.56 GJ of stored energy are
deposited as heat in the water of the resistor, an ASME-coated vessel,
holding 3100 l of water, will rise in temperature to a conservative design value
of 150 $\deg$C at 0.48 MPa. Correct
dimensioning of the stainless steel resistor element ensures that
boiling heat transfer is only 1/3 of the peak nucleate boiling flux
at the metal/water interface. A 1.50 m diameter $\times$ 3.5 m tall
cylindrical tank could be used. This design concept eliminates long
runs of 18 kA cable to an air cooled resistor, or long vent pipes to the surface of
boiling water vapor. Internal connections will
provide for both fast dump and normal slow dump modes. A center tap
grounding wire is attached to the electrical center of the resistor.

%
%


Quench protection safety will less stringent than for the CMS coil due to
the lower stored energy and inductance. A conservative 300 V to center-tapped
ground is chosen. Experimental tests and computer simulations show
that the CMS quench propagation velocity around one complete turn is faster
than turn to turn quench propagation through the insulation. Because we have
chosen identical CMS conductor size and insulation thickness, with very
similar conductor electrical and thermal properties, peak temperatures will
be less than the CMS design of 80 K with dump resistor, but equal to the CMS design of 130 K with
dump breaker failure. Both SiD and CMS safety rely on the winding mandrel
serving as a quench-back cylinder for spreading the load on the outer layer
and absorbing some of the stored energy.

\subsection{Cryostat and Magnet Iron}

The cryostat, $\sim $ 60 K thermal shield, current leads, tie rods, and
instrumentation will all be designed using standard cryogenic techniques.
Current leads will be very similar to the CMS design. Vacuum shells
will be built according to the ASME pressure vessel code design rules. Two
separate iron penetrations will be used, a 70 cm $\times$ 40 cm chimney for the
current leads and 36 cm diameter chimney for the cryogenic plumbing. Vacuum
pump-down will take place through both chimneys. Tie rods will be segmented
into three different systems based on direction (axial, vertical and radial),
as in the CMS and BaBar cases. They will be manufactured from Inconel
718 or Ti-5Al2.5Sn ELI. Radial and vertical loads will be carried to the
cryostat outer wall. Axial loads will be carried to the cryostat end plates.
In all cases, the tie rod systems are substantially stiffer than the
magnetic spring constant.

\begin{figure}[htbp]
\centerline{\includegraphics[width=4in]{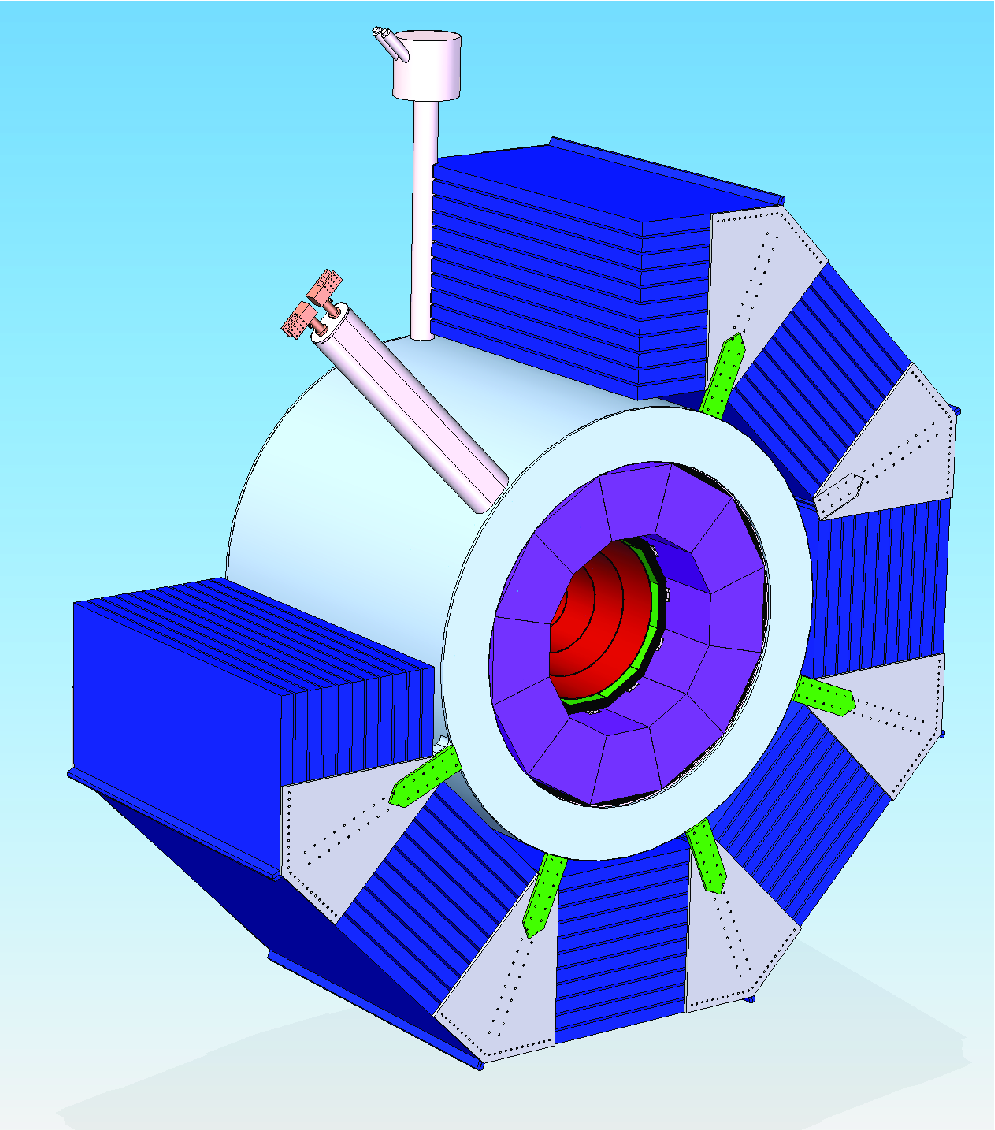}}
\caption{\label{fig:mag4} Solenoid shown with valve box, current
leads and attachments to barrel iron plates.}
\end{figure}


A helium refrigerator/liquefier is required to provide approximately 1 kW of refrigeration
at 4.5 K. Flexible transfer lines connect the liquefier to the detector
valve box to enable push-pull motion. In order to minimize
the length of the transfer line, the liquefier will be mounted
at the same level as the top of the solenoid. The liquefier supplies
forced flows of  4.5 K saturated LHe and 40-80 K helium gas.
The latter flow is  for the thermal shield
and for the thermal intercepts of the support rods.
The liquefier is a custom-built
commercial product whose detailed design and construction will be carried
out by industry as part of the complete cryo plant procurement. A 5000 LHe
storage dewar is stationed next to the refrigerator liquefier and can almost
be considered part of the liquefier. It serves as a pressure buffer for
forced flow operation and as a LHe supply reservoir during liquefier down
times. This technique was used successfully for a decade of running the SC solenoid for the
BaBar detector.

The detector valve box (Figure~\ref{fig:mag4})is used to minimize
flexible connections between the detector and the refrigerator. It also serves as
the distribution point for supply of LHe to the two
final-focus magnet 2 K cold boxes that are fixed on the detector.

It is expected that both ILC detectors will require similar size cryogenic
plants. In order to maintain flexibility, we propose that each detector have
its own complete cryogenic system which includes all the components discussed
above. However, the design should allow connections between these plants at
both the warm compressor level and at the output of the main cold boxes.
These cross connects establish a redundant system for operations and could
provide additional refrigeration capacity when cooling one of the magnets
down from room temperature.


The redundant 1250 kW helium compressors will be oil-flooded screw
compressors. Compressors and their associated oil removal equipment are
commercially available (though typically custom modified). The compressors
will be on the surface to minimize vibration and noise as well as to allow
easier access.
The compressors will be connected to the Cold Box by a 14 -16 bar helium supply
and $\sim $1.05 bar helium return lines. Necessary LN2 and helium gas
buffer/storage tanks will be located near the compressors.

\subsection{Detector Integrated anti-Dipole (anti-DID)}

The 550 kA turn Detector Integrated ``anti'' Dipole (DID) generates a
$\sim$600 G dipole field from each of the coils. These dipole coils are to be
mounted between the end flanges of the outer winding mandrel of the
solenoid. They will be placed just above the solenoid LHe cooling pipes.
Solenoid conductor splices will be secured in the open spaces of the dipole
coils. The most likely method of fabrication is to flat wind the dipole
pancakes in an aluminum case, then bend them to the correct radius and
impregnate the vacuum with a filled epoxy resin. Dozens of bolted connections to
the winding mandrel, with additional thermal straps, will guarantee mechanical
restraint and adequate cooling.

\begin{figure}[htbp]
\centerline{\includegraphics[width=4in]{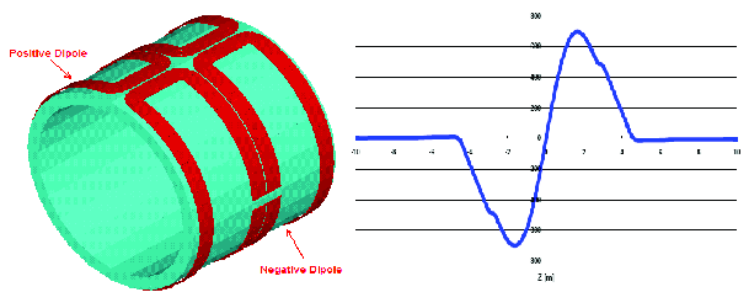}}
\caption{\label{fig:mag5} Anti-DID (left) with magnetic field
profile (right); the dipole field is horizontal.}
\end{figure}

The dipole coils are to be wound with co-extrusion of a high-purity aluminum and a
CMS single superconducting strand.
Four layers of 197
turns of 3 mm $\times$ 4 mm superconductor are proposed, with 0.5 mm of
fiberglass cloth between each turn and each layer.
They operate at
700 A, corresponding to  1/3 of the critical current.
The stored energy for an
independently powered anti-DID is in the range of 150 kJ. Because the
stored energy is so small, the volume fraction of high-purity
aluminum, relative to superconductor, needed for safe energy extraction during
a quench has been reduced from the CMS value (12.4) to 10.3.
Forces on each of the four coils are rather large in sum (4100 kN radial and 7800 kN
axial) but they are spread
uniformly and are manageable. Figure~\ref{fig:mag5} shows the dipole size, location and
magnetic field profile.

On the basis of the CMS experience are no show stoppers for the superconducting solenoid or any other
magnet system component. It remains to
implement the CMS construction techniques, examine other possible conductor
choices, and refine decisions to reduce cost and complexity and improve safety
margins.


\section{Muon System}

\subsection{Overview} \label{subsubsec:mylabel1}
The SiD muon system is designed to identify muons from the
interaction point with high efficiency and to reject almost all
hadrons (primarily pions and kaons). The muon detectors will be
installed in the gaps between steel layers of the solenoid flux
return. The required position and rate capabilities of the detectors
are modest and can be met by several different detector
technologies. The baseline design uses double layers of resistive
plate chambers(RPC). Also under consideration are extruded
scintillator strips read out by silicon photomultipliers (SiPMS).
Cost, reliability and physics capabilities will determine the
preferred choice.

\subsection{Design }
\begin{figure}[htbp]
\centerline{\includegraphics[width=5.00in]{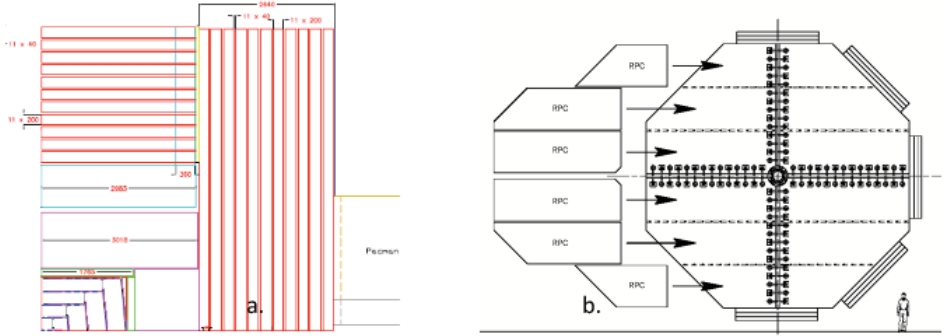}}
\caption{(a) Quarter-section of the SiD flux return. (b)
Sub-division of the endcap into RPC modules.} \label{fig:mu1}
\end{figure}

The muon system will start outside of the highly segmented
electromagnetic and hadronic calorimeters and the 5T solenoid
cryostat at a radius of 3.40 m. In the baseline design shown in
Figure~\ref{fig:mu1}a the flux return is divided into eleven layers
of 20 cm steel in an octagonal barrel geometry. Endcaps of eleven 20
cm thick steel octagons will cap both ends of the barrel. The muon
detectors will be inserted in the 4 cm gaps between the plates.  In
the barrel a detector layer is also inserted between the solenoid
and the first steel plate. The first barrel layer is approximately
2.8 m by 6.0 m and the last layer is $\sim $4.8 m by 6.0 m. The
total detector area needed in the barrel is $\sim $2050 m$^{2}$. As
shown in Figure~\ref{fig:mu1}b, the endcap detectors will be
subdivided into rectangular or trapezoidal modules $\sim $2~m by
6~m. Each endcap has a total detector area of 1220 m$^{2}$.

\subsection{Function}
SiD muon selection will combine information from
the central tracker, calorimeter, and muon detectors to construct 3-dimensional
tracks through the entire detector for each muon candidate. Candidates will be
required to penetrate a number of interaction lengths consistent with the muon
momentum. In addition, the observed number and position of hits along the
fitted track length can be used to further discriminate against hadrons.
The first layers of the muon system may also be useful as a tail-catcher
for the hadronic calorimeter.

\subsection{Requirements}
To date, physics benchmark/detector studies have been carried out
for single muons and pions, Smuon pair production (see
Chapter~\ref{chap:bench}) and b-pair production at 500~GeV. The
single particle studies were done to develop algorithms to use in
determining muon ID efficiency and hadron punch-through probability
vs. momentum\cite{Milstene:1}. The results of these studies show
that muon identification efficiency is greater than 96{\%} above a
momentum of 4~GeV/c. Muons perpendicular to the e+e- beamline reach
the SiD muon system when their momentum exceeds $\sim $3~GeV/c.

\subsection{Detector Design }
Muon systems characteristically cover large areas and have active
layers that are difficult to access or replace.  Reliability and low
cost are major requirements. Over 2.2 meters of steel thickness will
be required for the solenoid flux return, providing 13.2 hadron
interaction lengths to filter hadrons emerging from the hadron
calorimeter and solenoid. Since the central tracker will measure the
muon candidate momentum with high precision, the muon system only
needs sufficient position resolution to unambiguously match
calorimeter tracks with muon tracks. Present studies indicate that a
resolution of $\sim $1-2 cm is adequate.  This can be achieved by
two layers of extruded scintillators or RPC pickup strips of $\sim
$4 cm width.

\begin{figure}[htbp]
\centerline{\includegraphics[width=5.00in,height=3in]{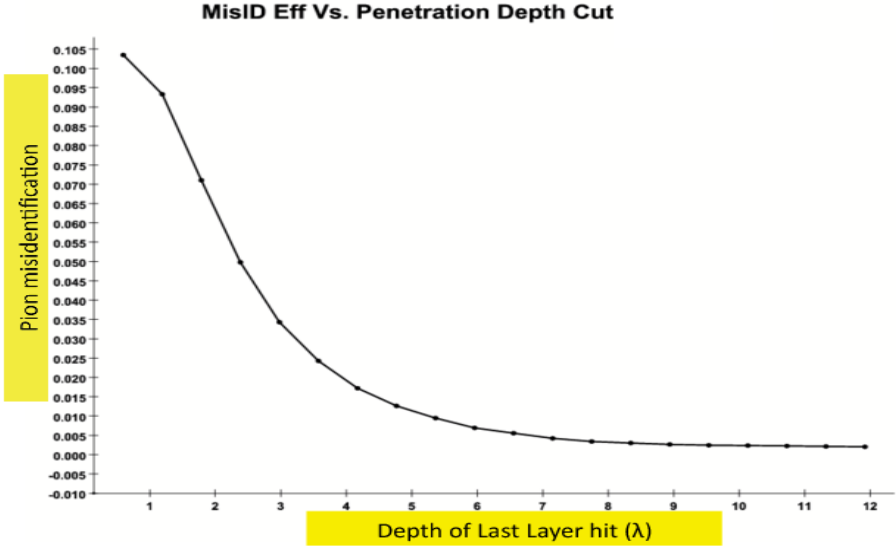}}
\caption{Fraction of pions between 10-50GeV/c that produce a hit as
a function of the number of interaction lengths traversed in the
muon system.} \label{fig:mu2}
\end{figure}

Full optimization of the muon system design has not been completed.
Although the total steel thickness is set by the solenoid requirements,
the optimum number of detector layers is being studied. We have studied
the mis-identification rate of pions between 10-50 GeV/c as a function
of the depth in the flux return\cite{Zhang:2008} to check that the present
design is thick enough. As shown in Figure~\ref{fig:mu2}, requiring that
the track makes hits in some of the outer layers is sufficient to reduce
the pion mis-identification rate to 0.25 {\%}, consistent with the
expected level of pion to muon decays. Pending further R{\&}D it may be
possible to eliminate detector layers without compromising either
performance or reliability of the system.

\subsection{Backgrounds}
Backgrounds in the muon system are expected to come primarily from
beam losses upstream of the detector. The muon system is shielded from
backgrounds generated at the collision point or along the internal beam
lines by the calorimeters, which are greater than 5 absorption lengths thick.
Therefore only penetrating backgrounds, such as high-energy muons or neutrons,
affect the barrel muon detectors. Calculations by N. Mokhov
et al\cite{Mokhov:2005} of the expected background from muons
produced by collimators near the detector hall predict a rate of 0.8
muons/cm$^{2 }$per pulse train ($\sim $1 ms) without muon spoilers,
which is reduced to 3*10$^{-3}$/cm$^{2}$ per pulse train with the
addition of muon spoilers. These background rates are significantly
higher than the typical RPC noise rate of 1 kHz/m$^{2}$(10$^{-4}$/cm$^{2}$
per pulse train). Physics backgrounds from two-photon processes producing
hadrons or muon pairs significantly increase the expected signal rate in
the endcap detectors near the beam-line. At a radius of 30 cm the expected
rate from muons above 2 GeV is $<$ 0.03/cm$^{2}$ per pulse train. The endcap
detectors can in addition be hit by electromagnetic shower debris from local
beam losses and may require additional shielding around the beam-line.

\subsection{Resistive Plate Chambers}
RPCs have often been used as muon detectors (BaBar and BELLE) and
will be used in ATLAS ans CMS. RPCs are inexpensive to build and can
be easily constructed in a variety of shapes and sizes. RPCs use
fluorocarbon gases which may be regulated in the future as they are
greenhouse warming gases and require nontrivial gas delivery
systems. The major concern with RPCs are their aging characteristics
(BaBar was forced to replace its original RPCs and BELLE had startup
problems). However, significant progress has been made in recent
years in understanding aging mechanisms. Many of the aging processes
are proportional to current passing through the gas gap. LHC
detectors will run in avalanche mode which has much lower charge per
track. Aging studies of LHC prototypes have shown good stability
even at the high background rates expected (100 Hz/cm$^{2})$ at the
LHC.The 2$^{nd}$ generation BaBar RPCs and the Belle RPCs have
preformed reliably at low signal rates ($<$0.2 Hz/cm$^{2})$. SiD
RPCs will run in avalanche mode and be subjected to backgrounds of
$<$0.15 Hz/cm$^{2}$ near the beamline and $<$0.015 Hz/cm$^{2}$
elsewhere.

The baseline RPC detector planes will be made by assembling single
gap RPC HV chambers($\sim $1m by 2 m) into modules of the required
size to fill each slot in the octagonal barrel or endcaps. Each
barrel layer will be split into three full length modules to allow
removal even if the gusset plates holding the steel octants are in
place. The supports for the endcap modules will be staggered at
different heights to reduce projective dead spaces. If the single
gap efficiency is 90{\%}, then an average layer efficiency of 93{\%}
can be achieved. Orthogonal pickup strips of a $\sim $3 cm pitch
will cover the sides of each HV module producing two x-y positions
per layer. Muon hits will be digitized by a lower channel
count(64-128) version of the KPIX chip mounted on a transition board
containing circuit protection diodes. The strip area will be limited
to $<$ 300 cm$^{2}$ so less than 1 background hit is expected per
train under the worst background conditions. Use of the KPIX time
information will reduce the expected background occupancy per strip
to $<$ 0.04{\%}. Smaller segmentation maybe necessary near the
beamline to keep the occupancy in this region equally low. One KPIX
chip will digitize a side of a module. Since the total area of
double layer chambers will be $\sim $4500 m$^{2}$ and the average
module is $\sim $2 m$^{2}$ about 4500 RPC modules and 9000 KPIX
chips will be needed. One KPIX concentrator board  will collect all
of the KPIX data from one  layer of each barrel octant or endcap
quadrant.  High voltage, gas, low voltage, and fiberoptic lines will
be routed to the end of the barrel or to the periphery of the
endcaps and connected to gas manifolds or trunk lines.

\subsection{Scintillating Strips with SiPM Readout}

The MINOS experiment has shown that a strip-scintillator
detector works well for identifying muons and for measuring
hadronic energy in neutrino interactions. A MINOS type
scintillator detector design would provide muon identification
and could be used to measure the tails of late developing or
highly energetic hadron showers. However, the MAPMT readout
design utilized by MINOS is a poor match to the geometry of
an ILC detector and would require long runs of optical
fiber to get signals to MAPMTs outside of the detector B field.
The recent development of Silicon PhotoMultipliers (SiPM or Si
avalanche photo-diodes) allows a much simpler design in which
the SiPM is directly coupled to the wave-length shifting fiber
embedded in the scintillating strip. SiD is pursuing a design
with SiPMs as an alternate technology for the muon system detector.
The quantum efficiency of SiPM detectors is about twice that of
vacuum tube photomultipliers, yielding ~ 25 photoelectrons per
minimum-ionizing particle (mip) in recent beam tests. This high
efficiency makes possible the use of long scintillating strip/fibers
with readouts at only one end which reduce the total number of channels.

A possible layout of a forward muon scintillator system divides each
layer into quadrants of two strip planes  rotated by 90$^{o}$ relative
to each other. Each strip plane would contain $\sim $ 142 strips 4.1 cm
wide. There would be 59 strips 5.28m in length and 83 strips that
vary in length from 5.8 m to 2.4 m.  The 20 double planes, ten
forward and ten backward, would result in 22,700 forward/backward
single ended readout strips.

Muon scintillator-strip detectors located in the Fe barrel octant
gaps could be arranged in planes of u-v strips which are oriented at
$\pm $ 45$^{o}$ to the long axis of the barrel or, alternatively,
parallel to the edges in x-y fashion. With the SiPM yield of
photoelectrons per mip it may be possible to reduce the multiplicity of
observed particles in the z-direction axial strips by cutting the
axial strips into two pieces, each 3.03m in length, which results in
16,300 signal sources.  In the perpendicular direction, with an
octagonal barrel geometry it takes eight strips to circumscribe the
beam line. With 10 concentric gaps in the barrel Fe and one inner
layer of detectors there would be 13,000 circumferential strips.
Thus there would be a total of 29,300 barrel single-ended readout
strips.  The total forward/backward and barrel signals with this
geometry would result in 52,000 electronic channels.

\subsection{Muon System R{\&}D}The primary aim of the muon system R{\&}D
is to validate both possible detector choices and to develop
cost-effective read-out designs. The RPC R{\&}D effort is focused on
adapting the KPIX ADC to digitize RPC signals\cite{Band:2009}. Other
studies will measure the aging characteristics of IHEP RPCs and
search for gas mixtures or cathode materials with better aging
properties\cite{Lu:2009}. The groups involved with the scintillating
strip option will evaluate SiPM devices from different manufacturers
and develop mounting, temperature control, and calibration
designs\cite{Karchin:2009}. Both the KPIX and SiPM efforts will be
applicable to the HCAL RPC and scintillator detector options.
Further details of the Muon System R{\&}D plans can be found in
the online appendices and the individual  R{\&}D
proposals~\cite{Band:2009,Lu:2009,Karchin:2009}.

\bibliographystyle{unsrt}



\section{Forward Detector} \label{sec:fcal}

The forward region is defined as polar angles $|\cos\theta| > 0.99$ ($\theta<$ 140mrad)
forward of the SiD Endcap ECAL. The angular coverage is completed by two
detectors, the Luminosity Calorimeter (LumiCal) and the Beam Calorimeter
(BeamCal). The physics missions in this region are:

\begin{itemize}
\item precision measurement of
the integrated luminosity using small-angle Bhabha scattering (LumiCal).
To measure cross section of an event sample $O(10^6)$ expected for
$e^+e^- \rightarrow
W^+W^-$ in 5 years with 500 fb$^{-1}$, the goal is to measure the luminosity
with an accuracy better than $10^{-3}$.

\item precise determination of the luminosity spectrum by measuring
the acolinearity angle of Bhabha scattering (LumiCal).
Due to the beamstrahlung emission the colliding beams loose energies
before collision, and the center-of-mass energy is no longer monochromatic.
The luminosity spectrum affects mass measurements and threshold scan.

\item extend the calorimeter hermeticity into the small angles for physics
searches (LumiCal and BeamCal). An excellent hermeticity is essential
as many new-physics reactions are accompanied with a large missing energy.

\item instantaneous luminosity measurement using beamstrahlung
pairs (BeamCal).

\item two photon veto for new particle searches (BeamCal).
\end{itemize}

The detector challenges are good
energy resolution, radiation hardness, interfacing with the final focus
elements, high occupancy rate requiring special readout, and performing the
physics measurements in the presence of the very high background in the
forward direction.

\subsection{Design criteria}

\subsubsection{LumiCal Physics Requirements}
The lowest order Bhabha cross-section for {\it t} channel one photon
exchange is given by:

$$ \frac{d\sigma}{d\Omega} = \frac{\alpha^2}{8E^2} \left( \frac{1+\rm{cos}^2\theta/2}{\rm{sin}^4\theta/2}\right) $$

We use this simple formula to estimate the polar angle coverage needed,
although the complete electro-weak {\it s} and {\it t} channel cross-section
will later be needed in order to do the physics. The number of Bhabha events
per bunch crossing for a detector with minimum and maximum polar angle coverage
$\theta_{min}$ and $\theta_{max}$ (in mrad) is:

$$ N = 0.5\rm{pb}\frac{L}{R}\int\limits_{\theta_{min}}^{\theta_{max}}\frac{d\rm{cos}\theta}{\rm{sin}^4\theta/2} \sim 8 \left( 
\frac{1}{\theta_{min}^2}-\frac{1}{\theta_{max}^2}\right) $$

\noindent for $\sqrt{s}$=0.5 TeV, L=2$\times10^{34}
\rm{cm}^{-2}\rm{s}^{-1}$, and bunch crossing rate R=$1.4\times10^4
\rm{s}^{-1}$. Our goal is to measure the luminosity normalization
with an accuracy of several $10^{-4}$ for $\sqrt{s}$=0.5 TeV. To do
this one needs $\approx 10^8$ events collected over $\approx10^7$ s,
or about ten events per second. One can then calculate the absolute
luminosity with $\approx10\%$ statistical error every several
minutes during the run. With a bunch crossing rate of $1.4\times10^4
\rm{s}^{-1}$, we need $>10^{-3}$ events per bunch crossing. To
achieve this statistical accuracy, we start the fiducial region for
the precision luminosity measurement well away from the
beamstrahlung pair edge at $\theta_{min}$=20mrad, with a fiducial
region beginning at 46~mrad, which gives $\approx 2 \times 10^{-3}$
events per bunch crossing.

\subsubsection{Luminosity precision and detector alignment}
The integrated luminosity, L is measured using the number of Bhabha
events N and the Bhabha scattering cross section $\sigma$ in the
detector fiducial region as $L = N/\sigma$. Since the Bhabha cross
section is $\sigma \sim 1/\theta^3$, the luminosity precision can be
expressed as

$$\frac{\Delta L}{L} = \frac{2\Delta\theta}{\theta_{min}},$$

\noindent
where $\Delta\theta$ is a systematic error (bias) in polar angle measurement
 and $\theta_{min}$ is the minimum polar angle of the fiducial region.
Because of the steep angular dependence, the precision of the minimum polar
angle measurement determines the luminosity precision.
To reach the luminosity precision goal of $10^{-3}$,
the polar angle must be measured with a precision
$\Delta\theta <$ 0.02 mrad and the radial positions of the sensors
must be controlled within 30 $\mu$m relative to the IP.

\subsubsection{Monitoring the Instantaneous Luminosity with BeamCal}
The colliding electron and positron bunches at the ILC experience
intense electromagnetic fields as they pass each other. These fields
generate large Lorentz forces, which cause radiation of gammas
called beamstrahlung. A small fraction of the beamstrahlung gammas
convert into pairs. Under the ILC Nominal beam parameters at
$\sqrt{s}$ = 0.5 TeV, approximately 75k $e^+/e^-$ are generated. The
BeamCal intercepts $\approx 3\times10^4$ beamstrahlung pairs of
average energy $\approx$0.7 GeV per bunch crossing when the bunches
have maximum overlap. Since the number of pairs is directly
proportional to the beam overlap, the instantaneous luminosity can
be monitored by detecting pairs in the BeamCal. Our goal is at the
ten percent level per beam crossing.

The Detector Integrated Dipole (DID)\cite{did} plays an important role in controlling
the beamstrahlung pairs. The DID is a pair of coils wound on the
detector solenoid, which creates a dipole field whose intensity varies in $z$ like a sine function. In the so-called
Anti-DID configuration, the DID field effectively compensates the crossing
angle for the outgoing beam and directs the low energy pairs into the
extraction aperture, reducing the total pair energy hitting the BeamCal
from 20 TeV to 10 TeV.

\subsubsection{Detector Hermeticity}
By hermeticity we mean complete solid angle coverage down to small polar angles, avoiding cracks and dead areas, which will enable 
accurate measurement of the transverse momentum balance in the event.
This is achieved by good
energy resolution, avoiding cracks, dead areas, and by covering down
to small polar angles. The measurement of $e^+e^- \rightarrow
slepton~pairs$ in the presence of two photon background has been
given as the performance detector design criteria for
hermeticity\cite{stau}. In the stau production and decay,
$$e^+e^- \rightarrow \bar{\tau} \bar{\tau} \rightarrow \tau^+ \chi^0 \tau^- \chi^0$$
the $\chi^0$ is the LSP and escapes the detector. Under the SUSY dark matter
scenario, the mass difference between stau and the LSP becomes less than 10
GeV, the missing $P_T$ is less and the measurements become more difficult.
The main background comes from the two-photon process $e^+e^-\rightarrow e^+e^-X$, where X is
$ee$, $\mu\mu$, or $\tau\tau$ for the slepton search. This background process has
no missing $P_T$; however, if we miss both electrons then the missing $P_T < 2\theta_{min}E_{beam}$. This is the kinematic limit. 
Generally one photon is on-shell, so
usually the missing $P_T<\theta_{min}E_{beam}$. The search region in the missing $P_T$ is
determined by how small angle the two-photon electrons are detected and
thereby vetoed. The electron detection must be made in the BeamCal where the
beamstrahlung pairs deposit 10 TeV of energy.

\subsubsection{Dynamic range and mip sensitivity}
While minimum ionizing particles (MIP) deposit 93 keV in a 320
$\mu$m-thick Si layer, a 250 GeV electron can deposit up to 160 MeV
or 1700 MIP equivalents in a single cell near shower maximum. If we
want a 100\% MIP sensitivity, the S/N ratio for MIP should be
greater than 10, and the dynamic range of electronics need to be at
least 17,000. This dynamic range can be achieved by using a 10-bit ADC
with two gain settings.

\subsubsection{Radiation hardness}
The beamstrahlung pairs will hit the BeamCal, depositing 10 TeV of
energy every bunch crossing. Sensor electronics could be damaged by
the energy deposition, and sensor displacement damage could be
caused by the resulting neutrons. The radiation dose varies significantly with
radius, and a maximum dose of up to 100 MRad/year is expected at
near the beampipe. The main source of neutrons is from secondary
photons in the energy range 5-30 MeV, which excite the giant nuclear
dipole resonance. The neutron flux is
approximately $5\times10^{13}\rm{n}/\rm{cm}^2$ per year.

\subsubsection{Occupancy}
The issue here is how ``deep'' we need to make the readout buffer to
hold one train of events for the LumiCal. The LumiCal occupancy is
studied for the beamstrahlung pairs and two-photon events. The
occupancy from the pairs still dominates. In the outer region of the Lumical,
$R>$ 7.5cm (the `far LumiCal'), just
outside the pair edge, the number of hits will reach 120 hits per
bunch train. The number of hits per bunch train decreases to four at
approximately $R\sim$10 cm (the `near LumiCal').

\subsection{Baseline Design}
\begin{figure}[h]
\center{\includegraphics[height=3in]{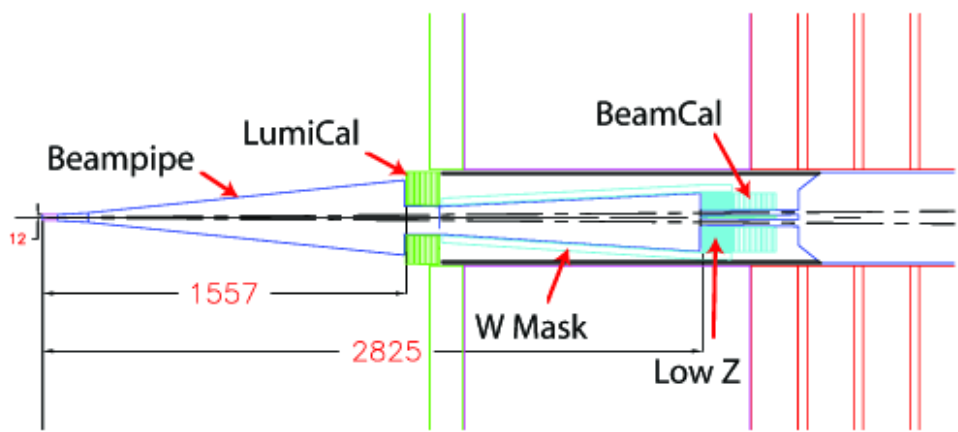}}
\caption{\label{fcal:1} Forward region.}
\end{figure}

The layout of the forward region is illustrated in Figure
\ref{fcal:1}. The LumiCal covers the polar angles from 40 mrad to 90
mrad, and the BeamCal from 3 mrad to 40 mrad. The conical mask made
of 3cm-thick tungsten is located between the LumiCal and BeamCal.
When the beamstrahlung pairs hit the BeamCal, low energy secondary
particles are generated. The mask shields the SiD central detectors
from low energy photons backscattered from the BeamCal. The Low-Z
mask made of 13cm-thick Borated-Polyethylene is located in front of
the BeamCal. The Low-Z mask reduces the low energy electron and
positron albido from the BeamCal by more than an order of magnitude.
Borated-Polyethylene also absorbs low energy neutrons produced in
the BeamCal and directed toward the IP.

The beam pipe at the IP is made of beryllium to minimize multiple scattering.
The inner radius is 1.2 cm, and the thickness is 400 $\mu m$. The conical
section starts at $R$=1.2 cm, $z$=6.25 cm, with a cone angle of 103 mrad, which is outside
the LumiCal fiducial region. The flared beam pipe design allows particles
to exit the beam pipe at normal incidence before entering the LumiCal. The
cylindrical beam pipe inside the LumiCal is 6 cm in radius, and is centered
on the extraction beam line.
The beam pipe does not intercept the beamstrahlung pairs which are confined
within 4 cm radius by the SiD 5 Tesla solenoid.

The entire forward region is supported by the support tube which is
cantilevered from the QD0 cryostat. There is a 1 cm gap between the support
tube and the endcap door to facilitate the door opening. To minimize the
electromagnetic energy loss due to the gap, the LumiCal is 10 cm
closer to the IP relative to the ECAL.

\begin{figure}[h]
\center{\includegraphics[height=3in]{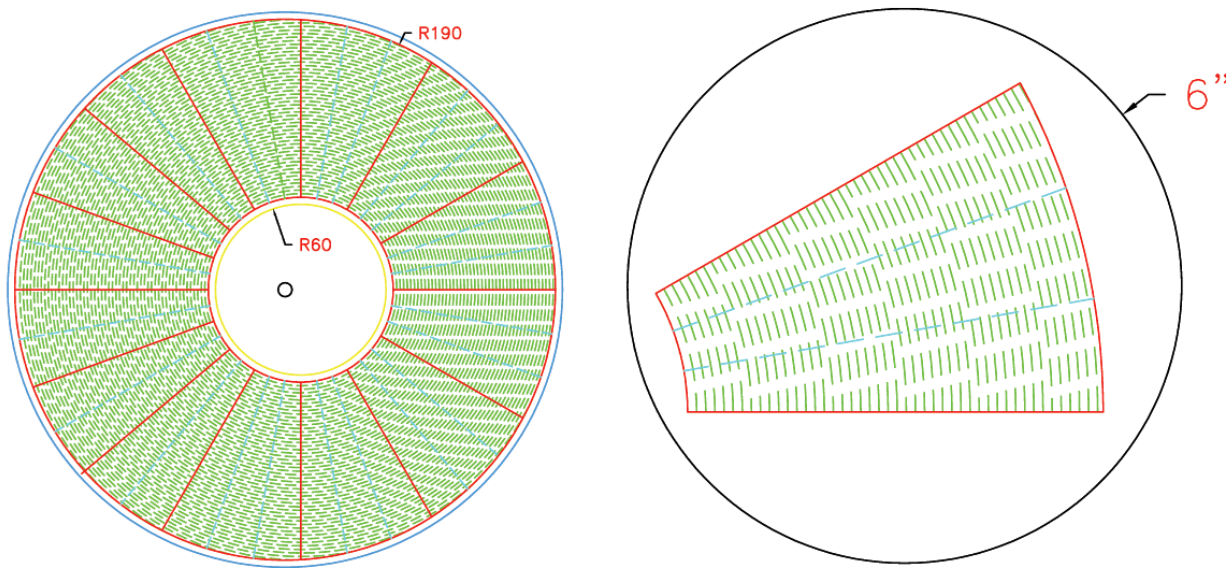}}
\caption{\label{fcal:2} LumiCal sensor and segmentation.}
\end{figure}

The LumiCal consists of two cylindrical C-shaped modules surrounding the beam pipe.
The inner radius is 6 cm centered on the extraction beam line
with a horizontal offset of $\Delta x$ = 1.0 cm (158 cm $\times$ 0.007).
The inner radius is dictated by the requirement that no detector
intercepts the intense beamstrahlung pairs, which are confined within 4 cm
radius by the 5 Tesla solenoid field.
The longitudinal structure follows the ECAL design,
consisting of 30 alternating layers of tungsten and silicon. The first 20
layers of tungsten each have thickness equivalent to 2.5 mm (or 5/7
radiation length) of pure tungsten. The last 10 layers have double this
thickness, making a total depth of about 29 radiation length.
The silicon sensors will be made from 6 inch wafers, and the segmentation is
illustrated in Figure \ref{fcal:2}.
The sensor is segmented in a $R-\phi$ geometry. A fine
radial segmentation with 2.5 mm pitch is used to reach the luminosity
precision goal of $10^{-3}$. The azimuthal division is 36 with each sensor
covering 10 degrees. Table \ref{fcal-table:1} summarizes the LumiCal parameters.

\begin{table}[h]
\begin{center}
\begin{tabular}{|c|c|} \hline
z    & 158 - 173 cm            \\ \hline
Inner radius      & 6 cm            \\ \hline
Outer Radius      & 20 cm     \\ \hline
Fiducial      & 46-86 mrad \\ \hline
Tungsten thickness      & 2.5 mm (20 layers), 5.0 mm (10 layers)  \\ \hline
Silicon sensor thickness      & 320 $\mu$m  \\ \hline
Radial division      & 2.5 mm pitch  \\ \hline
Azimuthal division      & 36  \\ \hline
\end{tabular}
\end{center}
\caption{LumiCal Parameters}
\label{fcal-table:1}
\end{table}

\begin{figure}[h]
\center{\includegraphics[height=3in]{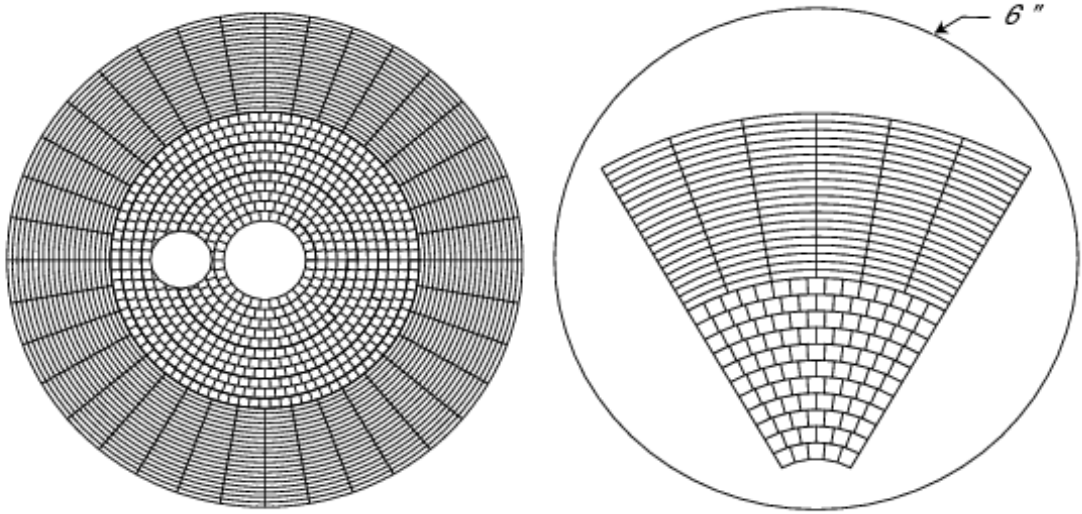}}
\caption{\label{fcal:3}BeamCal sensor and segmentation.}
\end{figure}

The BeamCal consists of two cylindrical C-shaped modules split in half
horizontally to accommodate the incoming beam line.
The inner radius is 2 cm centered on the extraction beam line and the outer
radius is 13.5 cm. The longitudinal structure consists of 50 alternating
layers of tungsten and silicon. The tungsten thickness is 2.5 mm, making a
total depth of 36 radiation lengths. The silicon sensor design based on a 6 inch
wafer is shown in
Figure \ref{fcal:3}. The inner region less than $R<7.5$ cm is the area where the
beamstrahlung pairs would hit. The segmentation in this region is approximately
$5\rm{mm}\times5\rm{mm}$, which is roughly one half of the Moli$\grave{e}$re radius. This segmentation
is optimized so that tell-tale electrons or positrons from two-photon processes can can be detected in the high
beamstrahlung pair background.
The outer region $R>7.5$ cm is the far LumiCal and has the same geometrical
segmentation as shown in Figure \ref{fcal:2}.

Currently two electronic readout chips are being developed. The KPiX chip
with 1024 channels is designed primarily for the ECAL. The chip has four
hits per bunch train to be stored for each channel. The FCAL chip with 64
channels is designed to handle the 100\% occupancy in the BeamCal. The chip
has 2820 buffer space so that a complete bunch train can be stored.
Although the LumiCal
occupancy is not 100\%, the LumiCal region smaller than about 10 cm will
have more than four hits per bunch train. Therefore, the LumiCal will use
the FCAL chip in the inner region and the KPiX chip in the outer region.

\subsection{Expected Performance}
The detector performance has been studied using EGS5 \cite{egs},
FLUKA \cite{fluka},
 and GEANT4 \cite{g4}
simulation packages. The ILC Nominal beam parameters at $\sqrt{s}$=500 GeV are used.
Bhabha scattering is simulated using BHWIDE \cite{bhwide}, and beamstrahlung pairs
are generated using GUINEA-PIG \cite{guinea}.
The magnetic field map for the SiD 5 Tesla solenoid
and the Anti-DID field are used.

\subsubsection{Energy containment and energy resolution in LumiCal}
\begin{figure}[h]
\center{\includegraphics[height=3in]{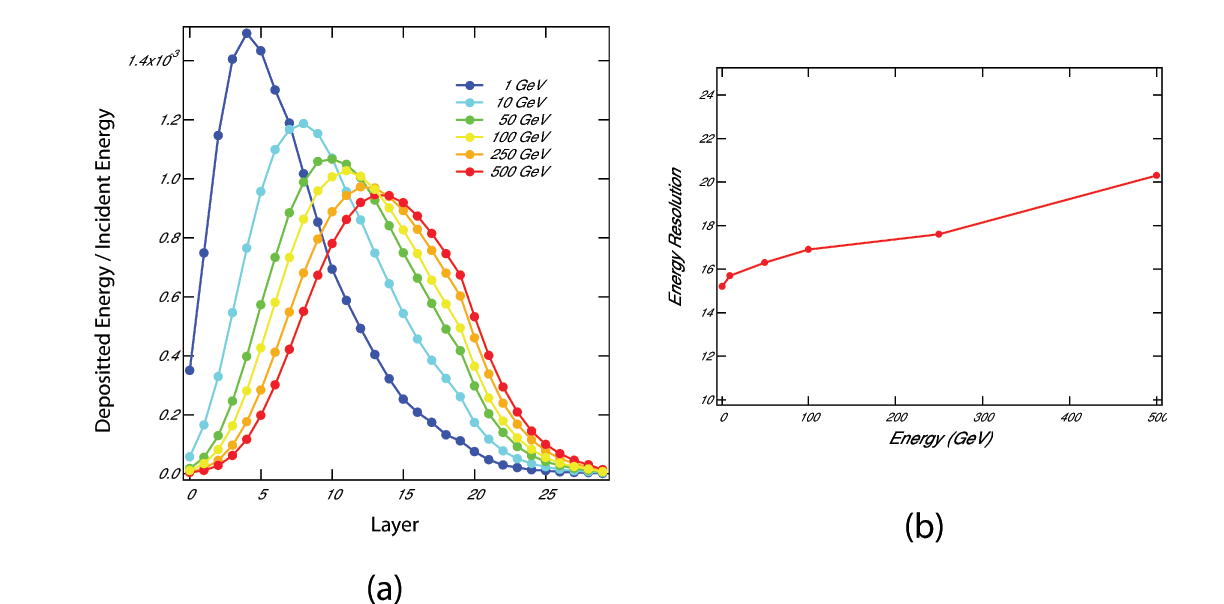}}
\caption{\label{fcal:4} (a) Deposited energy in Si layers (b) Energy
resolution}
\end{figure}

One of the most important requirements for the LumiCal is the energy
containment. If the energy of Bhabha events is not fully contained,
some events will be classified as radiative Bhabhas and fail Bhabha
event selection criteria, introducing a systematic error. Figure
\ref{fcal:4}(a) shows the deposited energy (scaled by the incident
energy) distribution as a function of the silicon layer from 1 GeV
to 500 GeV electrons. At lower energies, the energy is mostly
contained in the front 20 layers. Even at 500 GeV a significant
deposition is in the front 20 layers, and the back 10 layers with
double tungsten thickness accomplish the energy containment. Figure
\ref{fcal:4}(b) shows the energy resolution parameter $\alpha$ in
$\Delta E/E = \alpha / \sqrt{E}$ as a function of energy. The energy
resolution improves at the lower energies, reaching $15\%/\sqrt{E}$
at 1 GeV, and still maintains $20\%/\sqrt{E}$ at 500 GeV.

\subsubsection{Luminosity precision and LumiCal segmentation}
The polar angle bias ($\Delta\theta$) and achievable luminosity
precision have been studied as a function of sensor segmentation.
Each silicon layer is segmented equally in the radial and azimuthal
directions, and energy depositions in individual cells are
calculated. The polar angle is reconstructed by taking a weighted
average of the polar angle $\theta_i$ of individual cells with a
weight $W_i$ as

$$\theta = \frac{\sum_i\theta_i W_i}{\sum_i W_i}.$$

It is well known that the electromagnetic shower development is non-linear and
the weight in linear energy introduces a large bias in coordinate measurements\cite{loge} .
The bias can be reduced by using so-called logarithmic weight,

$$W_i = \rm{max}\{0, C+\ln(E_i/E_{total})\},$$

\noindent
where $E_i$ is energy deposition in each cell, $E_{total}$ total energy deposition,
and $C$ is a constant. The constant provides an effective energy threshold,
and only cells with a large enough energy deposition are used in the polar
angle reconstruction.

The LumiCal is fully simulated using 250 GeV electrons with a $1/\theta^3$
angular distribution. The reconstructed polar angle, $\theta_{rec}$ is compared
with the generated one $\theta_{gen}$. The polar angle bias $\Delta\theta$ is
calculated by an average value of $\theta_{rec} - \theta_{gen}$, and the angular
resolution by the rms value. Table \ref{fcal-table:2} shows the angular
bias, angular resolution
and the luminosity precision as a function of the radial segmentation.
The azimuthal segmentation is fixed to 32. The luminosity precision
improves as the radial segmentation decreases, and the luminosity
precision goal of $10^{-3}$ can be reached when the radial segmentation is smaller
than 3 mm. The dependence on the azimuthal segmentation is studied using a
fixed 2.5 mm radial segmentation. Table \ref{fcal-table:3} shows the result.
The luminosity
precision improves as the azimuthal segmentation is decreased, but the
improvement is small. Finer azimuthal segmentation would be beneficial for
cluster separation.

\begin{table}[h]
\begin{center}
\begin{tabular}{|c|c|c|c|} \hline
$\hspace{0.5cm}\Delta r (mm) \hspace{0.5cm}$ & $\hspace{0.5cm} \Delta\theta
(mrad)\hspace{0.5cm}$ & $\hspace{0.5cm}\sigma(\theta) (mrad)\hspace{0.5cm}$&
$\hspace{1cm}\Delta L/L  \hspace{1cm}$ \\ \hline
2.0    & 0.008 & 0.042 & $3.3\times10^{-4}$ \\ \hline
2.5    & 0.017 & 0.046 & $7.9\times10^{-4}$ \\ \hline
3.0    & 0.023 & 0.050 & $1.0\times10^{-3}$ \\ \hline
4.0    & 0.036 & 0.058 & $1.7\times10^{-3}$  \\ \hline
5.0    & 0.049 & 0.069 & $2.2\times10^{-3}$  \\ \hline
\end{tabular}
\end{center}
\caption{Radial segmentation}
\label{fcal-table:2}
\end{table}

\begin{table}[h]
\begin{center}
\begin{tabular}{|c|c|c|c|} \hline
$\hspace{0.5cm}N\phi \hspace{0.5cm}$ & $\hspace{0.5cm} \Delta\theta
(mrad)\hspace{0.5cm}$ & $\hspace{0.5cm}\sigma(\theta) (mrad)\hspace{0.5cm}$&
$\hspace{1cm}\Delta L/L  \hspace{1cm}$ \\ \hline
16    & 0.017 & 0.046 & $7.7\times10^{-4}$ \\ \hline
32    & 0.017 & 0.046 & $7.9\times10^{-4}$ \\ \hline
48    & 0.017 & 0.045 & $7.6\times10^{-3}$ \\ \hline
64    & 0.014 & 0.045 & $6.6\times10^{-3}$  \\ \hline
\end{tabular}
\end{center}
\caption{Azimuthal segmentation}
\label{fcal-table:3}
\end{table}

\subsubsection{Bhabha scattering}
\begin{figure}[h]
\center{\includegraphics[height=3in]{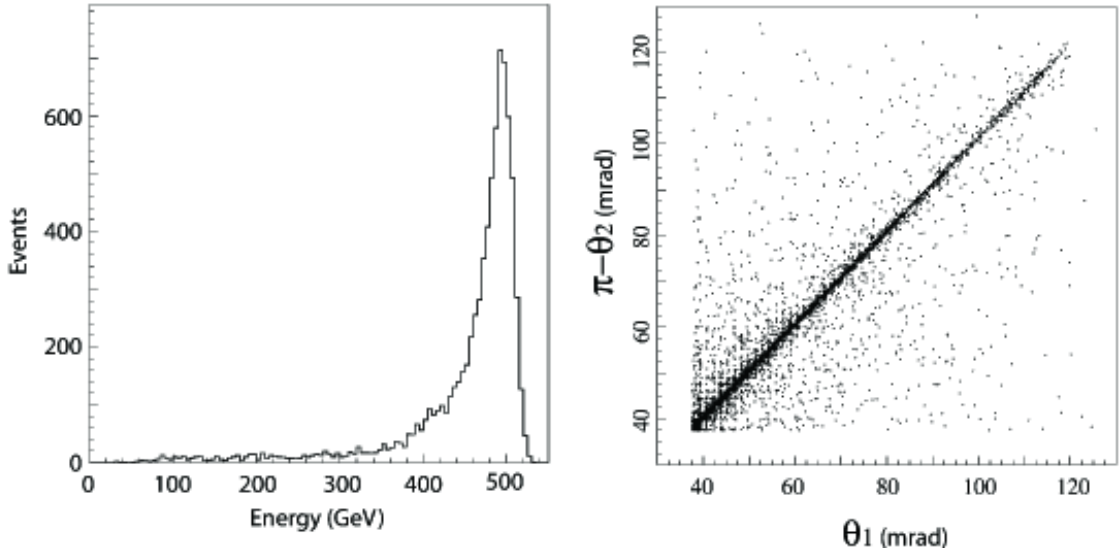}}
\caption{\label{fcal:5} (a) Total energy distribution (b)
Correlation of e+ and e- polar angles}
\end{figure}

The LumiCal performance is studied using Bhabha events at
$\sqrt{s}$=0.5 TeV. A sample of 10,000 Bhabha events is generated in
the CM system by the BHWIDE event generator\cite{bhwide}. Each event
is transferred to the laboratory system with the 14 mrad crossing
angle, and then processed through the LumiCal simulation. After
energy and angle reconstructions, the event is transferred back to
the CM system to understand the characteristics of Bhabha events.
Figure \ref{fcal:5}(a) shows the total energy distribution of the
reconstructed $e^+$ and $e^-$. The distribution has a peak at 500
GeV as expected, and has a long tail of radiative Bhabha events. The
current cluster finder can reconstruct shower energies above 25 GeV. Figure
\ref{fcal:5}(b) shows the correlation between the reconstructed
$e^-$ and $e^+$ polar angles. The 14 mrad crossing angle has been
properly taken care of. The radial segmentation is visible at small
angles.

\subsection{High energy electron detection in BeamCal}
\begin{figure}[h]
\center{\includegraphics[height=3in]{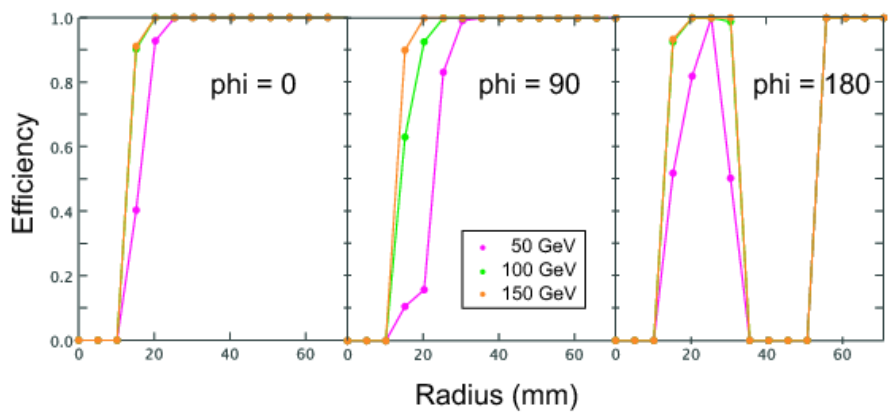}}
\caption{\label{fcal:6} Efficiency to reconstruct high energy
electrons at three azimuthal angles}
\end{figure}

We have studied the ability to identify and reconstruct high energy electron in
the presence of beamstrahlung pair background in the BeamCal\cite{uriel}.
The analysis consists of three parts:
\begin{enumerate}
\item A lookup table which would correlate the energy and position of a
shower on the BeamCal to an incident particle energy. This accounts for the
$\theta$ and $\phi$ dependence of shower energy introduced by energy loss down
the beampipe, and also incorporates the sampling fraction of the detector.
\item A lookup table of average beamstrahlung depositions which would be used
to subtract the expected beamstrahlung energy from the shower of interest.
\item A cluster algorithm which would analyze the total signal on the BeamCal
and attempt to isolate the part of the signal that is due to the high energy
electron.
\end{enumerate}
BeamCal signals are created by overlaying a high energy electron shower and
one bunch crossing of beamstrahlung backgrounds randomly selected from the
10,000 crossings. After subtracting the average beamstrahlung energy, the
cluster algorithm reconstructs the high energy electron if the subtracted
energy is greater than three sigma of the beamstrahlung energy variation.
Since the average energy of individual beamstrahlung pairs hitting
the BeamCal is only 1.7 GeV, the beamstrahlung showers are shallower than
 the showers from high energy electrons. Much better electron detection
efficiency is found possible when only energies deposited at depths of 3cm
 or greater are counted.
As the beamstrahlung energy has a strong radial and azimuthal dependence,
the detection efficiency is calculated as a function of the distance
from the extraction beam axis
at three azimuthal angles (0, 90, and 180 degrees).
Figure \ref{fcal:6} shows the detection efficiency as a function
of radius in the BeamCal for 50, 100, and 150 GeV electrons.
The inefficiency between 30 and 50 mm at phi=$180^\circ$ is due to the incoming
beam hole. Since the beamstrahlung background energy is the highest at phi
$\approx$ $90^\circ$ (and $270^\circ$), the detection efficiency is lower
 at this angular region. The efficiency to detect more than
150 GeV electrons is almost 100\% up to 8 mrad from the beam axis.

\subsubsection{Calorimeter hermeticity}
\begin{figure}[h]
\center{\includegraphics[height=3in]{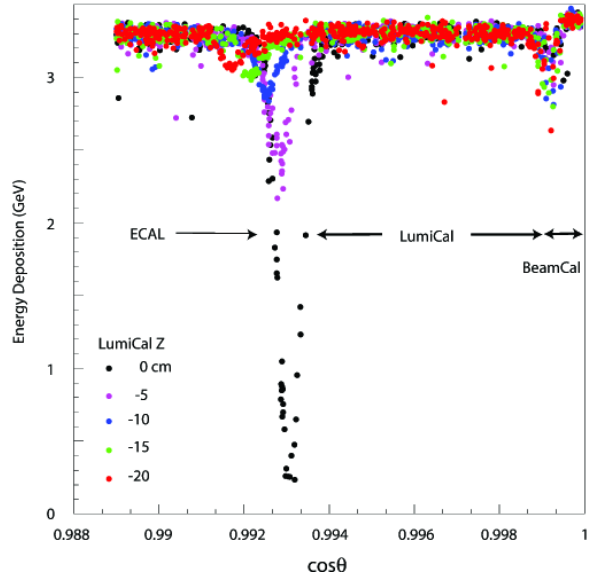}}
\caption{\label{fcal:8} Energy deposition in Si layers as a functon
of $\cos\theta$}
\end{figure}

Calorimeter hermeticity is surveyed using 250 GeV electrons. Figure
\ref{fcal:8} shows the total energy deposition in Silicon layers as
a function of $\cos\theta$. The ECAL, LumiCal and BeamCal angular
coverages are indicated in the figure. Five sets of simulation are
made with different LumiCal Z locations. When the LumiCal is at the
same Z location as the ECAL, a significant energy is lost at
$\cos\theta\sim$0.993. To achieve calorimeter hermeticity, the
LumiCal is moved 10 cm closer to the IP.

\subsubsection{Tungsten mask}
\begin{figure}[h]
\center{\includegraphics[height=3in]{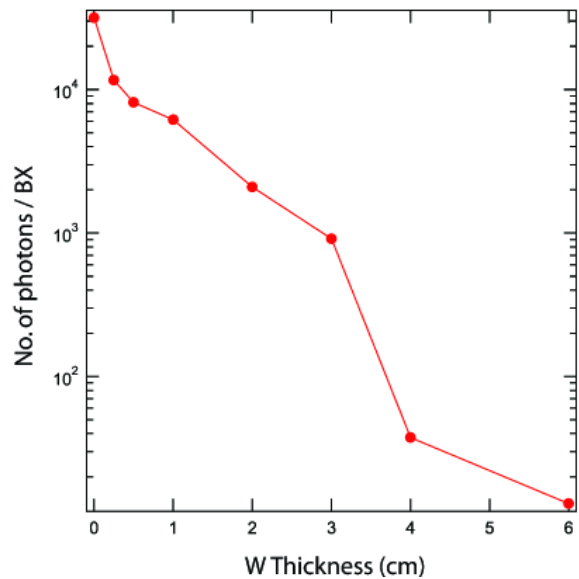}}
\caption{\label{fcal:9} Number of photons penetrating the tungsten
mask as a function of the mask thickness}
\end{figure}

When the beamstrahlung pairs hit the BeamCal, a large number of
secondary photons are produced and about 35k photons per bunch
crossing are back scattered. The tungsten mask is to shield the
central detector from these low energy photons. Figure \ref{fcal:9}
shows the number of photons penetrating the tungsten mask as a
function of the mask thickness. To eliminate the photons, a mask
thicker than 6 cm is necessary. But such a thick mask takes up the
space and significantly increases the weight. The mask thickness of
3 cm is chosen. Although about 1000 photons are penetrating the
mask, they are low energy ($\sim$200 KeV) and are uniformly
distributed over the inner surface of Endcap HCAL.

\subsection{R\&D}
The radiation level in the forward direction implies that one will
need to probably specify specialized Si material. One needs to
select the materials and expose them to radiation levels equivalent
to what will be seen at the ILC forward region over period of 5
years. Two possible radiation hard materials are oxygenated Float
Zone Si wafers and magnetic Czochralski Si wafers.

\bibliographystyle{unsrt}



\section{Data Acquisition and Electronics}  \label{sec:DAQ}

\begin{figure}[htbp]
\centerline{\includegraphics[width=4in]{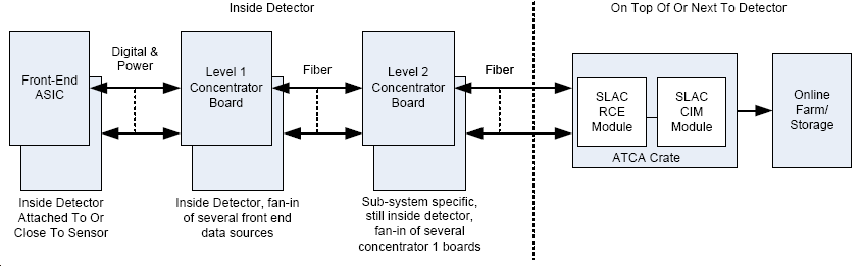}}
\caption{Simplified block-diagram of detector control and readout
electronics using the SLAC ATCA RCE and CIM
modules.}\label{fig:DAQ1}
\end{figure}

SiD has a coherent approach to the electronics architecture that
seems to fit all the baseline subsystems. Figure~\ref{fig:DAQ1}
shows the simplified block diagram for the data-acquisition from the
front-end electronics to the online-farm and storage system. The
subsystems with the exception of the Vertex detector (for which the
sensor technology is not yet selected) and the FCAL (which has
approximately unit occupancy) are read out by variants of KPiX as
the front-end Application-Specific Integrated Circuit (ASIC). KPiX
is a multi-channel system-on-chip, for self triggered detection and
processing of low level charge signals (``KPiX - An Array of Self
Triggered Charge Sensitive Cells Generating Digital Time and
Amplitude Information'' D. Freytag et al, IEEE NSS Dresden 08).

\begin{figure}[htbp]
\centerline{\includegraphics[width=6in]{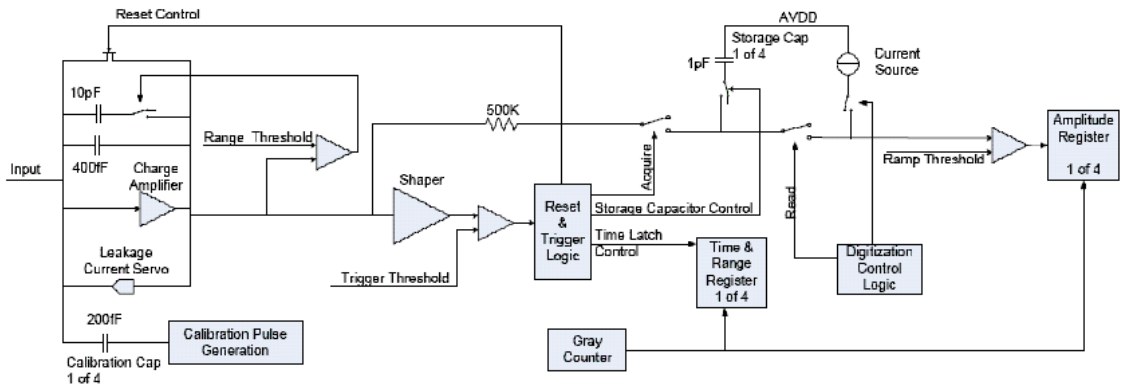}}
\caption{Simplified block-diagram of one channel of the 1024 channel
KPiX Application Specific Integrated Circuit.}\label{fig:DAQ2}
\end{figure}

Figure~\ref{fig:DAQ2} shows a simplified block diagram of the KPiX,
processing signals from 1024 detector channels. The low level charge
signal at the input is processed by the charge amplifier in two
ranges with automatic range switching controlled by the range
threshold discriminator. Calibration is provided covering the full
range. Leakage compensation is available for DC-coupled detectors.
Internal or external trigger options can be selected.

Up to four sets of signals for each channel can be stored in one acquisition
cycle. Time is stored in digital format, the amplitude as a voltage on a
capacitor for subsequent digitization in a Wilkinson-type ADC. At the end of
the acquisition and digitization cycle nine words of digital information are
available for each of the 1024 cells of the KPiX chip. The data is read out
serially from the ASIC before the next acquisition cycle begins.

\begin{table}[htbp]
\begin{center}
\begin{tabular}{|p{133pt}|l|l|}
\hline
Sub-System&
KPiX Count&
Channels/KPiX \\
\hline Tracker&
27464&
1024 \\
\hline ECAL& 99076&
1024 \\
\hline
HCAL&
35412&
1024 \\
\hline
Muons&
8834&
64 \\
\hline
~&
&
~ \\
\hline Total&
170786&
~ \\
\hline
\end{tabular}
\caption{Count of KPiX ASICs for each sub-system.}\label{tab:DAQ1}
\end{center}
\end{table}

Table~\ref{tab:DAQ1} lists the number of KPiX ASICs for each
sub-system. Tracker, ECAL, and HCAL use 1024-channel ASICs while the
Muon sub-system uses a 64-channel version.

As illustrated in Figure~\ref{fig:DAQ1}, several front-end ASICs
(KPiX, FCAL or Vertex ASICs) are connected to a Level-1 Concentrator
(L1C) board using electrical LVDS. The concentrator board main
functions are to fan out upstream signals to the front-end modules,
to fan-in data from the front-end modules for transmission to the
Level 2 Concentrator (L2C) boards, and to perform zero-suppression
and sorting of the event data. Just as an example, for the EMC
Barrel a total of 96 1024-channel KPiX chips would be connected from
8 front-end cables with 12 KPiX's each to one Level-1 Concentrator
board. The number of Level 1 concentrator boards in the detector
depends on the sub-system, e.g for the EMC Barrel there would be 821
L1C boards and 52 L2C boards (80k KPiX, 96 KPiX for each L1C board,
16 L1C boards for each L2C board)

\begin{figure}[htbp]
\centerline{\includegraphics[width=4in]{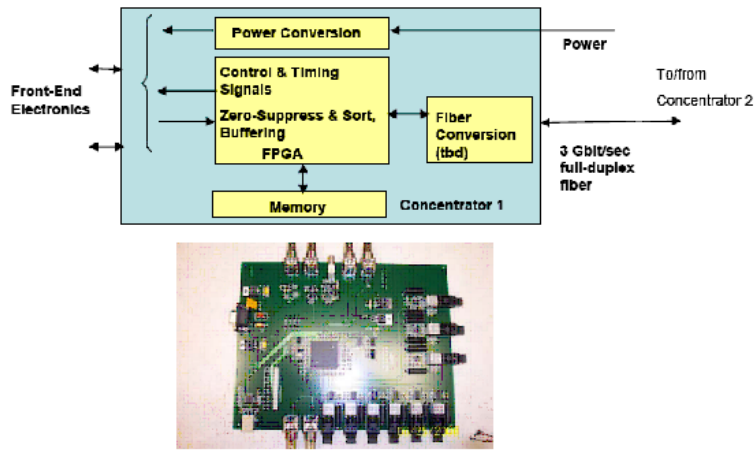}}
\caption{Level 1 Concentrator Board.}\label{fig:DAQ3}
\end{figure}

Figure~\ref{fig:DAQ3} shows a block diagram and a prototype of the
board.

The Level 1 Concentrator boards are in turn connected via 3-Gbit/sec fibers
to the Level-2 Concentrator boards. These are similar to the Level 1
Concentrator boards. They fan out and fan in signals to/from the Level 1
Concentrator boards. In addition the data-streams of sorted event data
received from each Level 1 Concentrator board are merged and sorted before
transmission to the off-detector processor boards. The Level 2 boards are
either located inside the detector or outside, depending on the sub-system.
E.g. for the EMC Barrel there are 36 of such boards inside the detector
volume.

The Level 2 Concentrator boards are connected via fibers to ATCA crates
either on or next to the detector. ATCA stands for ``Advanced
Telecommunications Computing Architecture'' and is the next generation
communication equipment used by the telecommunication industry. It
incorporates the latest trends in high-speed interconnect, processors and
improved Reliability, Availability, and Serviceability (RAS). Essentially
instead of parallel bus back-planes, it uses high-speed serial communication
and advanced switch technology within and between modules, redundant power,
plus monitoring functions. For SiD 10-G Ethernet is used as the serial
protocol.

\begin{figure}[htbp]
\centerline{\includegraphics[width=4in]{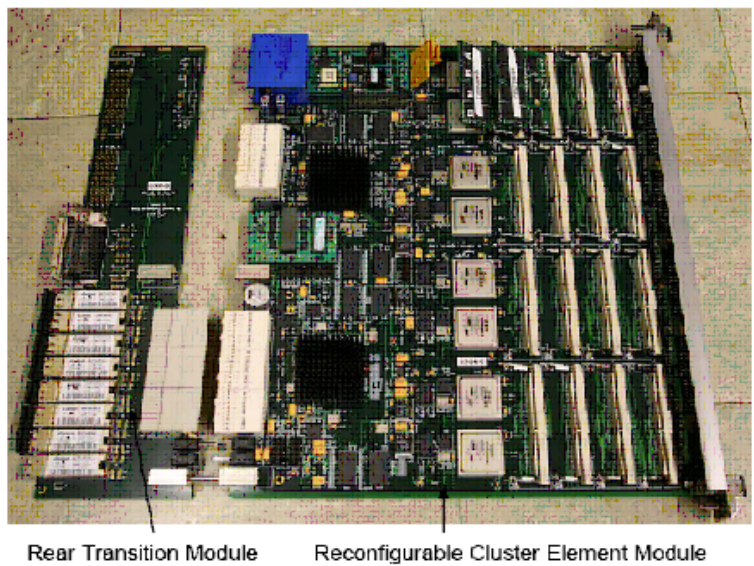}} \caption{SLAC
ATCA Reconfigurable Cluster Element.}\label{fig:DAQ4}
\end{figure}

SLAC designed two custom ATCA boards, the Reconfigurable Cluster
Element (RCE) Module and the Cluster Interconnect Module (CIM) as
shown in Figure~\ref{fig:DAQ1}. Figure~\ref{fig:DAQ4} shows the RCE
which interfaces via eight 3-Gbit/s fiber links to sub-system
detector electronics. It provides timing and configuration commands,
and reads back configuration and event data. It contains 2 sets of
the following: Virtex FPGA with embedded 2 PowerPC processors IP
cores, 512 Mbyte low-latency RLDRAM, important 8 Gbytes/sec cpu-data
memory interface, 10-G Ethernet event data interface, 1-G Ethernet
control interface, RTEMS operating system, up to 512 Gbyte of FLASH
memory or 1 TByte/board. The flash memory does not have to be fully
loaded, it is implemented in SIMMS plug-in modules. One RCE connects
to up to 8 detector cards via 8 3-Gbit/sec fibers connected into the
Read Transition Module (RTM)

\begin{figure}[htbp]
\centerline{\includegraphics[width=3in]{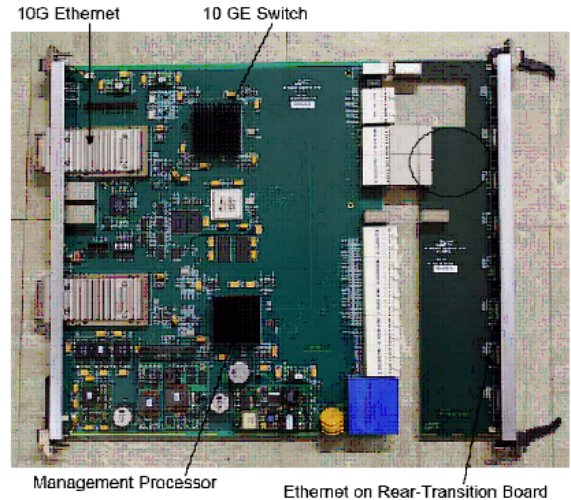}} \caption{SLAC
ATCA Cluster Interconnect Module.}\label{fig:DAQ5}
\end{figure}

Figure~\ref{fig:DAQ5} shows a picture of the SLAC CIM. It is
essentially a switch card with two 24-port 10-G Ethernet Fulcrum
switch ASICs. The switch ASICs are managed via a Virtex FPGA. The
function of the CIM is to provide communication between all the RCE
modules in a crate and to destinations external to the crate at
10Gbit/s data rates.

One of those network cards serves up to 14 in-crate RCE modules. The data
transfer rate is up to 480 Gbit/sec, more than the complete SiD data rate
requirement.

\begin{figure}[htbp]
\centerline{\includegraphics[width=3in]{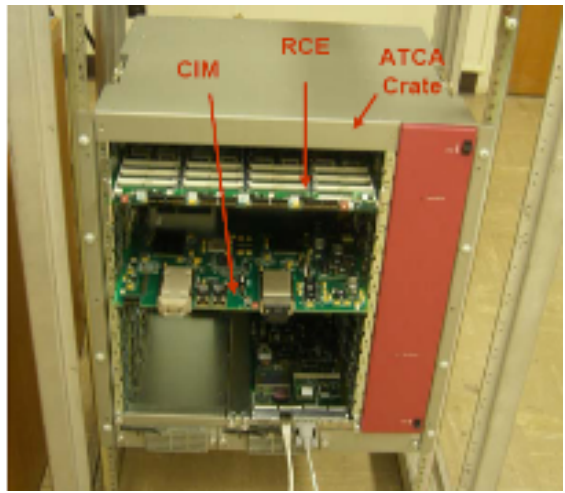}} \caption{ATCA
Crate with RCE and CIM.}\label{fig:DAQ6}
\end{figure}

Figure~\ref{fig:DAQ6} shows an ATCA crate with one RCE and one CIM
inserted. A total of 14 RCE's and 2 CIM's can be inserted into one
crate providing connections to 8 x 14 = 112 3Gbit/sec fiber links
into the detector for a 360 Gbit/sec IO. The estimate for the
complete SiD detector is approximately 320 Gbit/sec including a
factor of 2 margin, so in principle a single ATCA crate could serve
the complete detector. However for partitioning reasons, the ability
to run each of the sub-systems completely independently during
commissioning is desired, and one crate for each sub-system is
planned. The complete SiD DAQ only requires about 4 racks full of
Data Acquisition electronics.

The data is further sorted by event in the ATCA system and sent to the
online processing system for potential data reduction. Whether further data
reduction is required is not determined yet, and the data may directly be
forwarded to the offline system.

Note that the event data is zero-suppressed in the sub-systems
without the need for a global trigger system. All data produced in
the front-ends above a programmable threshold is read out. For
diagnostics and debugging, the DAQ includes the ability to assert
calibration strobe and trigger signals, transmitted to the
front-ends via the Level-2 and Level-1 Concentrator boards using the
fibers shown in Figure~\ref{fig:DAQ1}. The fiber transports encoded
command, clock, and synchronization to all the front-ends.

Power Conversion circuits on the Level 2 and Level 1 Concentrator boards
supply the power to the front-ends, starting with 48V or higher voltages
from off-detector supplies. Alternatively, serial powering architectures are
also under consideration. The power supplies are located in several racks on
or next to the detector.

Environmental and health monitoring circuits are also included on the
concentrator boards. In addition there may be additional monitoring boards
in the detector, connected to RCE fiber interfaces. In addition there are
crates of monitoring modules mounted in several racks on or next to the
detector.

\begin{figure}[htbp]
\centerline{\includegraphics[width=5in]{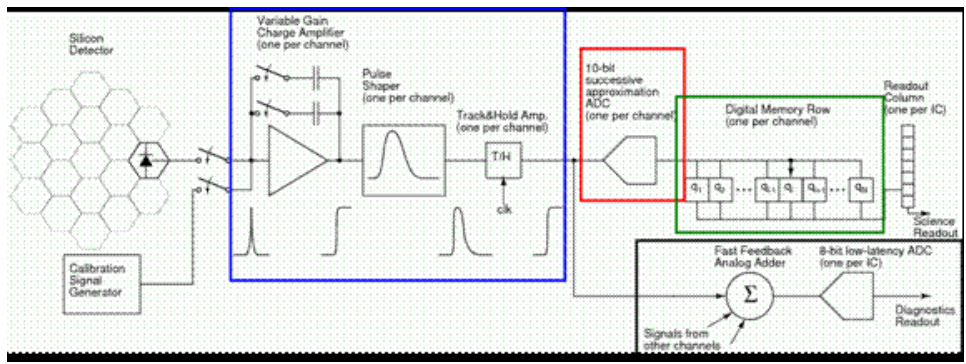}}
\caption{Block diagram of BeamCal Front-End Integrated
Circuit.}\label{fig:DAQ7}
\end{figure}

Most subsystems use the KPiX integrated circuit for reading out the
detector signals. For the FCAL BeamCal subsystem a different custom
integrated circuit has been designed due to its expected high
occupancy (A. Abusleme et al., "Beamcal front-end electronics:
Design and simulation", Proceedings of the Workshop of the
Collaboration on Forward Calorimetry at ILC, 2008;
http://www.vinca.rs/pub/FCAL Belgrade.pdf). Figure~\ref{fig:DAQ7}
shows a simplified block diagram of the integrated circuit.  Input
signals up to 40 pC are amplified, shaped, and digitized with 10-bit
resolution. The amplifier gain can be configured for science (x1)
and calibration (x50) mode. Each of the 32 channels in the device
has its own Analog-to-Digital Converter and 2,820 memory cells to
store the data which is then read out between bunch trains. A
separate low-latency (< 1usec) output digitizing the sum of the 32
channels is provided to allow for fast feedback diagnostics.

\chapter{Machine-Detector Interface and Global Issues}
\label{chap:global}

SiD has contributed to the current definition of the
machine-detector interface (MDI) specifications\footnote{B. Parker
et al, 'Functional Requirements on the Design of the Detectors and
the Interaction Region of an e+e- Linear Collider with a Push-Pull
Arrangement of Detectors'.}. Here we describe our present plans for
assembly of the detector, its alignment, and details of the
push-pull operation.


\section{SiD Assembly} \label{sec:assembly}

SiD expects the VXD, ECAL, and HCAL modules to be built at
collaborating labs and universities and transported to the ILC site.
The iron will be built in sub modules with a mass suitable for
transportation from an industrial site, and bolted together at the
ILC. The solenoid would probably have two sections, and will be
wound industrially.

The detailed strategy of final assembly must await site selection. A shallow
site strategy will likely be different than one for several hundred meters
underground. The optimal mass for steel modules will depend on
transportation modes available between the factory and ILC, and sea, rail,
or road limits will lead to different conclusions. (As a baseline, SiD is
designing for road transportation and approximately 100 tonne maximum unit
weight). The shape of an underground hall and the capacity for underground
bridge cranes may depend on the site geology. In addition, an optimal
strategy will depend on the ILC construction schedule.

\begin{figure}[htbp]
\centerline{\includegraphics[angle=90,height=\textheight]{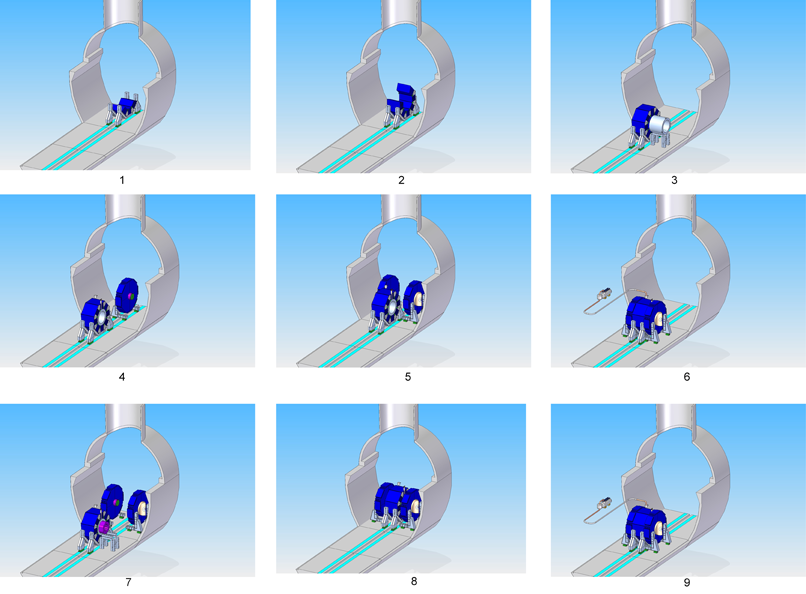}}
\label{fig:assembly}\caption{Schematic of steps in a possible
assembly procedure.}
\end{figure}

A possible assembly procedure, for an underground site, is illustrated in Figure~\ref{fig:assembly}.
The main steps are:

\begin{enumerate}
\item
Underground assembly of Flux Return Barrel.

\item
The Barrel Octants are lowered in pre-assembled pieces of 400t max.

\item
The Solenoid is lowered and inserted in the Flux Return Barrel.

\item
Door are preassembled on surface and lowered in one piece 2500t max.

\item
Flux Return Barrel moves between the doors. Doors close

\item
The magnet (Iron+solenoid) is moved in garage position, hooked up to the cryogenics and commissioned.

\item
Doors move leaving access on the barrel for the insertion of  HCAL (380t), ECAL(60t) and Tracker(2t) (the other 
shaft is used for lowering)

\item
Full assembled SiD barrel moves between the doors to close the detector

\item
Full assembled SiD is moved in garage position, hooked up to the cryogenics and ready for the push pull 
operations.
\end{enumerate}

The complete barrel
assembly weighs approximately 4 Ktonnes, and each door weighs approximately
1.9 Ktonnes. Surface assembly and lowering of each with a substantial gantry crane
is possible, as is assembly on the interaction region floor with a 300 to
400 tonne bridge crane. A possible alternative would be two approximately
200 tonne cranes on the same rails, one nominally for each experiment, with
occasional coordinated use of both cranes for the larger lifts. Dedicated
tooling is assumed for the insertion of the barrel detectors into the
solenoid.

SiD expects to move on hardened steel rails, grouted and locked to the
floor. Rail sets for transverse motion (push pull) and door opening in both
the beamline and garage positions will be needed. SiD expects that locations
for these rails can be found that either can be shared with the other
detector, or not interfere with its system. If ILC is built in a seismic
location, provision will be needed for locking SiD down in both the beamline
and garage positions.

SiD subsystems will have few data and power connections compared to those
for the LHC detectors because of the absence of a trigger and the low
aggregate data rates. The subsystems will be integrated with their
electronics and tested at their assembly points and at staging areas at the
ILC site; final connection, integration and test should be relatively
simple. The utility needs (electricity, water, LHe, compressed air, data
fibers) needs of SiD are modest, and are expected to be via flexible connections
to the wall.


\section{Alignment}

Detector Alignment and positioning is critical to the unprecedented
momentum resolution and vertex detector performance expected of SiD.
The positioning is critical for the push-pull model of ILC.   Well
defined positioning and measuring systems have not yet been
designed but a general concept is described below.

SiD will be moved into the beam position as a single large
unit, carrying the end doors with the barrel.  It is expected that
the detector will move on multi-roller supports, each with an
integrated drive, so that the steel structure is minimally stressed.
The drive will accelerate and decelerate smoothly, and a maximum
speed between 1 to 5 mm/sec is expected. It will be pushed up to
permanent mechanical stops. Operating experience from other large
detectors has shown that this method of positioning achieves position
accuracy of +- 1 mm.    This accuracy is acceptable for the
positioning of the iron structure, muon system, solenoid,
calorimetry and outer tracker.  The QD0s will be moved onto the
beamline defined by the two QFs by remotely-controlled motors. The
QD0s carry the beam pipe and the vertex detector.  The relative
position of the beam pipe/vertex detector to the outer tracker
needs to be known to a few microns.  This relative position
will be measured by a frequency scanning interferometer measuring
system, and will be recorded periodically. It is expected that each
launcher-corner reflector pair will measure with a precision of ~ 1
micron over a range of several meters.

During movement concerns about distortions of the sub-detector
systems need to be addressed so that permanent deformations are not
observed.  This will be achieved by kinetic mounts, progressively
isolating the tracker from any external stress.  The solenoid will
be kinematically mounted to the iron structure.  To be conservative,
kinetic mounts will also be designed between the solenoid and the
calorimetry and between the calorimetry and the outer tracker.


\section{Push Pull}

SiD strongly supports the notion of two detectors at the ILC, and
considers the push-pull interchange of the two detectors on a
frequent schedule necessary for a successful project. Consequently,
SiD has made technical subsystem choices that support a rapid
exchange of position and that will require minimal alignment and
calibration time on beamline. It has also made engineering decisions
to optimize these moves.

SiD assumes an interchange frequency of about a month, with a fixed
(but adjusted) time allocation to each detector. This is long enough
to make the lost data taking time reasonable if the transition time
is approximately a day, and short enough to both avoid the
possibility of significant data disparity opportunities, and to
limit the time for any hardware changes that would require re-commissioning of
the off-beamline detector. We assume that the time a detector takes
to clear the beamline after the beginning of a transition will be
subtracted from its next time allocation, and that a detector may
take as long as it wants to set up on beamline, but that this time
is part of its allocation. We assert that with a properly engineered
detector and adequate experienced staff, the time from beginning
detector exchange to beginning beam-based alignment of the final
quadrupoles should be less than one day. We assume that except for
unusual machine mishaps, both detectors take their chances with the
machine luminosity.

\subsection{Machine Detector Interface Assumptions}

\begin{enumerate}
\item QD0 is located in SiD and moves with it. Supporting it with adequate vibrational stability is a responsibility of SiD. Its 
transverse position within SiD is remotely adjustable to +- 1 mm.
\item QF1 is a fixed part of the machine, and is shared by the two detectors. Its support system includes a longitudinal anchor 
that will be used to stabilize the $z$ position of the QD0 assembly when the SiD door is opened.
\item There is a warm removable spool piece between QD0 and QF1. It has two RF smooth, remotely operated high vacuum valves 
isolating the spool piece, and there is provision for purging, pumping to high vacuum, and leak checking this assembly (presumably 
permanently installed RGAs). There is a bellows system that will accommodate +- 1mm misalignment.
\item SiD will be self shielded. In addition to the SiD calorimeters and iron providing sufficient shielding in the event of an 
accident to the portion of the beamline approximately inboard of the QD0's, the rest of the beamlines are shielded by a system of 
portable iron and concrete (Pacmen). In a conceptual sense, each Pacman consists of a section fixed to the tunnel mouth wall, a 
pair of hinged shields that open to gain access to the beamline connection described above, and a section fixed to the detector 
door. It is expected that the hinged sections and the piece fixed to the tunnel wall will be common with the other detector, and 
that dimensional differences between the two detectors will be made up in the third section.
\item The 2K He cryogenic system for QD0 is carried on a support fixed to SiD and supplied from a 4K He system on SiD. All 
connections to QD0 are undisturbed during a move, and the magnet remains cold.
\item An automatic alignment system with an accuracy of at least +- 1 mm will relate the position of SiD to the beamline 
coordinates (when SiD is near the beamline).
\end{enumerate}

\subsection{Detector Assumptions}

\begin{enumerate}
\item The superconducting solenoid has all of its services and utilities carried with SiD. The He liquefier is mounted on a hall 
wall, and supplies He to SiD by a flexible vacuum insulated line. (It is assumed that He compressors are located remotely for 
vibration isolation.) (It is assumed that each detector has its own liquefier, but that cross connect valves are incorporated in 
the cold boxes to allow mutual assistance.) SiD will have support platforms, attached above, aside, or adjacent to the detector. 
These platforms will support the solenoid power supply, dump resistor, quench protection system, and other auxiliaries.
\item The detector will receive utilities of $\sim $350 KW 480 VAC 3 phase power from a flexible line that remains connected 
during a move. Low Conductivity water, chilled water, and compressed air will also be provided by continuously connected flex 
lines.
\item Data will exit the detector via an optical fiber system continuously connected to the wall. Monitoring and control 
communications will also be via a fiber system. All detector subsystem power will come from supplies that travel with the 
detector. No subsystem signal cable will come off the detector.
\item The detector will roll on hardened steel rails using Hilman rollers or equivalent. SiD will carry its doors when moving in 
$z$, and will have rails in both the beam and garage position to open its doors. The drive mechanism will have minimal energy 
storage, possible a gear drive to a fixed cog rail. It is expected that a velocity between 1 and 5 mm/s is desirable and 
reasonable.
\item The detector transport system will deliver the detector to its appropriate position on the beamline to a precision of +- 
1mm.
\item The detector is designed so that any stresses caused by the force required to move the detector are not transmitted to the 
calorimeter supports, and so the precision calorimeter and tracker systems will not be distorted from the moving process.
\item The detector will have an internal alignment system, probably based on frequency scanning interferometers that will provide 
high precision measurements of the relative positions of the barrel components, endcap systems, and beamline components.
\item The Sid beamline is supported by the inboard ends of the QD0s, and moves with them as they are moved to the ILC beamline 
during beam based alignment. The SiD Vertex Detector (VXD) is mounted to the beamline, and will remain coaxial to the beam during 
the alignment process.
\item The SiD flux return will limit leakage field to be less than 100 Gauss at 1m from the iron.
\end{enumerate}

\subsection{The Push Pull Process}

\begin{figure}[htbp]
\centerline{\includegraphics[width=7.5in]{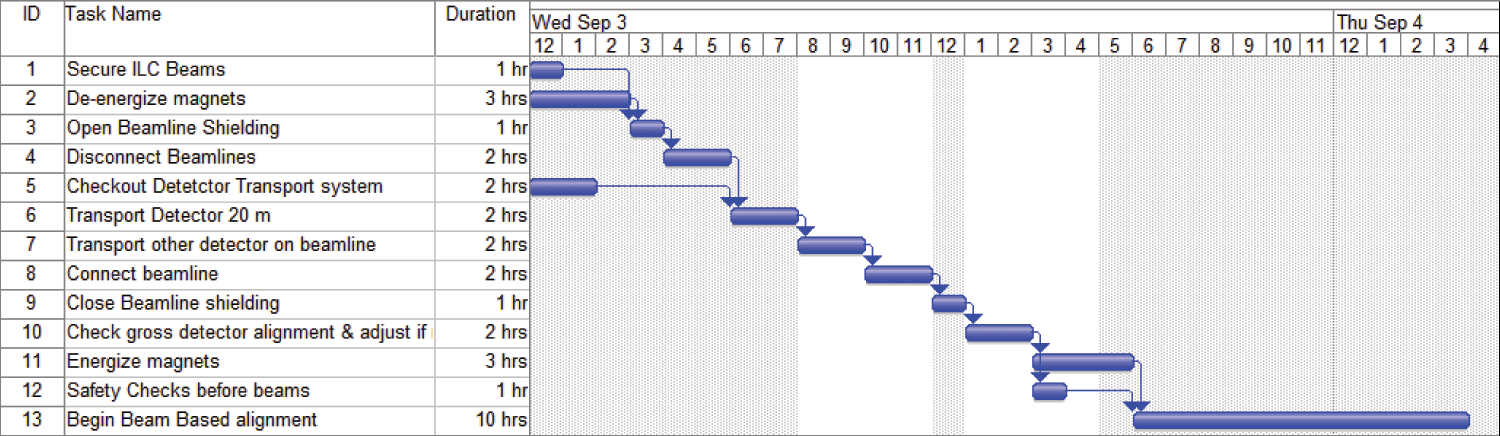}}
\label{fig:push1}\caption{Summary chart of push-pull operational
steps.}
\end{figure}

SiD envisions the process as a series of rehearsed steps which are
described here and in Microsoft Project Gantt chart format in Figure~\ref{fig:push1}.
The time units are hours, and SiD believes
that with suitable experience and engineering, this is a rational
schedule. Some of this belief is based on the experience with the
600 ton SLD doors, in which a door could be opened, the detector
internals accessed for two hours, the door closed, and beam based
alignment begun -- in 8 hours.

\begin{enumerate}
\item Secure ILC Beams: This is the first step of the detector exchange process, and involves ensuring that ILC can not deliver a 
beam to the hall, and releasing the necessary keys (physical or otherwise) to permit opening of the shielding. This step would be 
very similar to that before a detector could open a door. (1 time unit)
\item De-energize magnets: At this time, it is not obvious whether the QD0s and solenoid should be (partly) de-energized. We 
assume here that the QD0s and solenoid will be run down, and that the time will be approximately equal to the charging time for 
the solenoid, which is 3 hours. If the ILC is built in a region with seismic requirements, this time can also be used to release 
seismic restraints.
\item Open Beamline Shielding: It is assumed that the beamline shielding is mechanically locked and requires the previously 
mentioned keys to unlock. The moving parts of the Pacmen are supported by hinges, and are activated by fixed mechanisms so that no 
crane activity is required. It is assumed that all 4 Pacmen are opened together by coordinated but separate crews. (1 time unit).
\item Disconnect Beamlines: This process involves checking that the isolation valves have properly shut, venting the spool piece 
wit an appropriate purge, and disconnecting the flanges. It is assumed that separate crews work on each end. (2 time units).
\item Check Detector Transport System: The detector transport system should be carefully checked to see that it is ready to move 
the detector. It is assumed that the transport system is carefully maintained, and so this is a safety step rather than an 
opportunity for regular maintenance. (2 time units).
\item Transport Detector: The detector is moved in $x$ for 15 to 20 m. The time, including a provision for acceleration and 
deceleration, would under 5 hours for a top velocity of 1 mm/s (15 m travel) and about 1 hour for 5 mm/s. (2 time units)
\item Transport other detector on beamline: It is assumed that the other detector has completed appropriate safety checks, and is 
ready to move when the first detector is fully off beamline. (This may be too conservative.) (2 time units).
\item Connect Beamlines: This process includes making up the spool piece flanges, pumping the spool piece with the in place 
pumping system, and He leak checking with the associated leak checker. Note that the beamline valves are remotely operable and can 
be opened when the spool piece vacuum is satisfactory. (2 time units)
\item Close Beamline Shielding: This step is the reversal of step 3. It is assumed that the Pacmen are engineered to handle +- 1mm 
misalignment with respect to the ideal beamline. ( 1 time unit)
\item Check Gross Detector Alignment: This step is the opportunity to check that the detector is within its tolerance to the 
beamline coordinates, and adjust the position if necessary. (2 time units).
\item Energize magnets: This step is the reversal of step 2. (3 time units)
\item Safety Checks before beams: This step consists of appropriate safety checks and the return of any outstanding ``keys'' to 
the machine. (1 time unit)
\item Begin Beam Based Alignment: This process consists of ILC using beams to move the quadrupoles to their desired positions. It 
is unknown how long this procedure will take, but 10 hours is assumed here, which is pessimistic compared to SLC experience.
\end{enumerate}

The relation and overlap of these steps is indicated in Figure~\ref{fig:push1}.
It seems reasonable to us, that given the engineering assumptions,
the luminosity-luminosity transition could take approximately a day.
SiD expects that its internal alignment system, coupled with
mechanically isolated but robust subsystems, will not need any
calibration following the initial one.

\chapter{Physics Performance and Benchmarking} \label{chap:bench}

\section{Simulation of SiD} \label{sec:simulation} In order to
design a detector which is capable of exploiting the full physics
discovery potential of the ILC, a fairly sophisticated and mature
simulation and reconstruction environment is needed. The Simulation
{\&} Reconstruction Working Group has concentrated its efforts on
assembling a flexible framework to allow different detector designs
to be simulated and multiple reconstruction algorithms to be
implemented and used for physics and detector analyses. The code,
binary executables, and documentation in the form of Application
Programming Interface (API) and tutorials are all available online
at \underline {http://lcsim.org}.

\subsection{Event Generation}

A number of different event samples have been generated for detector
design studies, ranging from single particles to inclusive Standard
Model processes.
The main focus of benchmarking analyses described in this section
were complete sets of unbiased Standard Model processes at 500 and 250
GeV that were generated using the Whizard Monte Carlo program~\cite{Kilian:2007gr}.
All 0, 2, 4 and 6 fermion final states, as well as top quark-dominated 8
fermion processes were generated. PYTHIA~\cite{Sjostrand:2006za} was used for
final state QED and QCD parton showering, fragmentation and decay to
provide final-state observable particles. Included in this sample
are backgrounds arising from interactions between virtual and
beamstrahlung photons.

Samples were generated with electron and positron polarizations of 100{\%}.
Arbitrary polarization samples can be generated by properly
combining events from the four data samples. For the purposes of the
physics benchmark analyses, event samples were created which reflect
the expected ILC baseline parameters of 80\% electron and 30\%
positron polarization. Because of the large size of the full
dataset, only a fraction of these events, about 80M, have been processed through
the full detector simulation, with individual events weighted to
reflect the statistical sampling.

\subsection{Full Detector Simulation}

The simulation of the response of the detector to the generated
events is based on the Geant4 toolkit~\cite{geant4_1}~\cite{geant4_2}, which provides the classes to
describe the geometry of the detector, the transport and the
interactions of particles with materials and fields. A thin layer of
Linear Collider-specific code, SLIC~\cite{slic}, provides access to the Monte Carlo
events, the detector geometry and the output of the detector hits.
The geometries are fully described at runtime, so physicists can
vary detector parameters without having to rebuild the simulation
executable binaries. The output uses the standard LCIO format~\cite{lcio}, so
that detectors modeled using other simulation packages could be
analyzed, and data generated using this system could be analyzed in
other reconstruction and analysis frameworks.

All analyses described in this section use the full detector simulation.

\subsection{Detector Variants}

The XML format allows variations in detector geometries to be easily
set up and studied, e.g. Stainless Steel vs. Tungsten hadronic
calorimeter absorber material, RPC vs. scintillator readout,
calorimeter layering (radii, number of layers, composition,
{\ldots}), readout segmentation (pad size, projective vs. fixed cell
size), tracking detector topologies (``wedding cake'' vs. Barrel +
Cap), magnetic field strength.

In addition to the baseline Silicon Detector (sid02), a number of
variants have been developed in order to study the dependence of the
performance on detector options. The option sid02 used for all
analyses in this section is consistent with the detector described
in the previous sections though some compromises had to be made to
simplify the geometry in specific cases.
A more detailed geometry description is available at
\underline{\textbf{http://lcsim.org/detectors/{\#}sid02}}.

\subsection{Analysis Tools}

The jet finder used in most analyses is a simple y-cut algorithm
usually run with a predetermined number of jets. Additionally the
jet finder also provides $y_{max}$  and $y_{min}$;  these are the
y-cut values that determine the separation between the one and two
jet cases and between the N and N-1 jet cases; where N is the number
of desired jets.
Pattern recognition on hadronic final states is performed using a
PFA algorithm devised for SiD; more details can be found in the
online appendices.

The Marlin Kinematic Fitter (KinFit~\cite{kinfit}) was used in
several analyses to provide the most probable kinematic
configuration of the event topology under specific, user defined
constraints. A wide range of constraints can be selected: total
energy, momentum, mass of selected jets and several other
parameters. The fitter uses the method of Lagrange multipliers to
determine the most probable value for the jet four-momentum given
user specified uncertainties on the energy and on the angles of each
jet.  The algorithm can also select the most probable configuration
of jets out of all possible permutations.

The vertexing package used for several analyses was developed by the
LCFI collaboration~\cite{LCFI}. The main algorithm of the package,
the topological vertex finder ZVTOP, reconstructs vertices in
arbitrary multi-prong topologies. It classifies events on number of
found vertices and combines eight optimized variables for each type
of event in a neural network which is then separately trained on
samples of b-, c- and light quarks. The best discriminating
variables are the corrected vertex mass, the joint probability, the
secondary vertex probability, the impact parameter significance of
the most significant track and number of vertices in the event. The
joint probability is defined as the probability for all tracks in a
jet to be compatible with hypothesis that they  originate at the
primary vertex. Typically nine networks are used with eight inputs,
one hidden layer with 14 sigmoid neurons and one output.

\begin{figure}[htbp]
\begin{center}
\subfigure[]{
\includegraphics[scale=0.45]{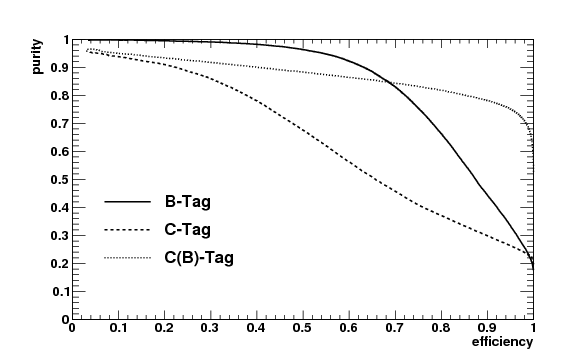}
}
\subfigure[]{
\includegraphics[scale=0.45]{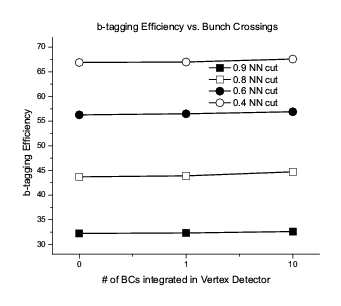}
}
\end{center}
\caption{Flavour tagging with LCFI package optimized for SiD (a) and
dependence of b-tagging efficiency on the amount of beam-beam
background integrated in the vertex detector (b).} \label{fig:btag}
\end{figure}

The performance of the LCFI package optimised for the SiD detector
is shown in Figure~\ref{fig:btag}. The left plot shows dependence of
purity on the efficiency for the di-jet sample at 500 GeV for
b-tagging, c-tagging with b-only background and c-tagging. The SiD
specific optimisation was performed by building a new neural network
which used the same input parameters described above and training it
using di-jet samples which passed through the full SiD simulation
and reconstruction. The right plot of Figure~\ref{fig:btag}
addresses the issue of integration of the beam-beam background in
the vertex detector. It shows the dependence of b-tagging efficiency
on the amount of beam-beam background integrated in the vertex
detector expressed as the number of integrated beam crossing (BC).
The dependence is minimal which confirms the robustness of the
pattern recognition and vertexing in SiD. Bottom quark di-jet events
at 500 GeV with beam-beam background events overlaid at the detector
hit level were used for this study. In all cases the beam-beam
background corresponding to one BC was overlaid with signal events
in the Tracker.

In several analyses the event classification is based on the open
source Fast Artificial Neural Network (FANN)~\cite{fann} package.
The package provides fast and reliable framework written in C
programming language called from within a C++ wrapper. FANN was
partially modified in order to account for event weights during the
Neural Network training.

\section{Benchmark Reactions} \label{sec:reaction}
Physics
performance studies are needed to quantify the performance
of SiD, revisit the performance requirements on the various ILC
detector subsystems, and ultimately optimize the SiD design by
studying how the performance changes as one varies the basic
detector parameters. In a broader context, these studies further the
physics case of the ILC.

A list of physics benchmark reactions~\cite{Battaglia:2006bv} was presented to
the ILC community at Snowmass 2005. This list of about 30 reactions
provides comprehensive coverage of ILC physics topics and detector
challenges, but is too long to be addressed by detector
concept groups at this time. Thus, a reduced list of 6 reactions was
specified by the ILCSC Research Director which is more appropriate for studies on the time
scale of the LOI:

1.  $ e^+e^-\to e^+e^-H\,,\,\; \mu ^+\mu ^-H\,$, $\, \sqrt s $=250 GeV;

2.  $e^+e^-\to ZH\,,\,\; H\to c \bar{c}\, ,\, \; Z\to \nu \bar{\nu}\, , \, q \bar{q}\,$, $\, \sqrt s $=250 GeV;

3.  $e^+e^-\to ZH\,,\,\; H\to \mu^+\mu^-\, ,\, \; Z\to \nu \bar{\nu}\, , \, q \bar{q}\,$, $\, \sqrt s $=250 GeV;

4.  $e^+e^-\to \tau^+ \tau^-\,$, $\, \sqrt s $=500 GeV;

5.  $e^+e^-\to t\bar{t}\, , \,\; t \to bW^+\, , \,\; W^+ \to q \bar{q'}\,$, $\, \sqrt s $=500 GeV;

6.  $e^+e^-\to \tilde{\chi} _1^+ \tilde{\chi} _1^- $/$\tilde{\chi} _2^0 \tilde{\chi} _2^0\,$,  $\, \sqrt s $=500 GeV.

In addition to these compulsory LOI benchmark reactions SiD also investigated the following process:

 7. $e^+e^-\to \tilde{b}\tilde{b}\, ,\,\,\tilde{b}\to b \tilde{\chi}_1^0 , \sqrt s $=500 GeV.

The following sections describe these seven analyses.

\subsection{$ e^+e^-\to e^+e^-H\,,\,\; \mu ^+\mu ^-H\,$, $\, \sqrt s $=250 GeV}

Studies of the Higgs Boson are expected to be center stage at the
ILC. The production of the Higgs through ``Higgs-strahlung'' in
association with a Z, will allow a precision Higgs mass
determination, precision studies of the Higgs branching fractions,
measurement of the production cross section and accompanying tests
of SM couplings, and searches for invisible Higgs decays. When the
associated Z decays leptonically, it is possible to perform a
measurement of the Higgsstrahlung cross section  and to reconstruct
the mass of the Higgs recoiling against the Z with high precision,
independent of how the Higgs boson decays.

A sample of three million Monte Carlo signal events of the type $e^+e^-
\to e^+e^-H,\ \mu^+\mu^-H$ was used along with a
$\sqrt{s}=250$~GeV Standard Model Monte Carlo background event
sample of 47 million events for training and for estimating the
statistical errors for the mass and the cross section.  An independent
signal sample of three million events was used as a testing sample to
verify the statistical errors calculated with the larger training
sample.

\begin{figure}[htbp]
\begin{center}
\subfigure[]{
\includegraphics[scale=0.33]{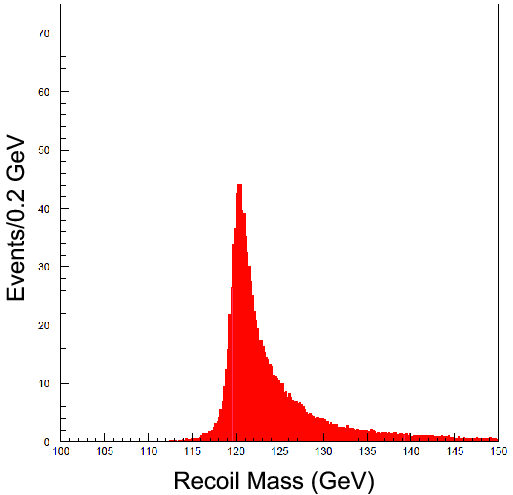}
} \subfigure[]{
\includegraphics[scale=0.33]{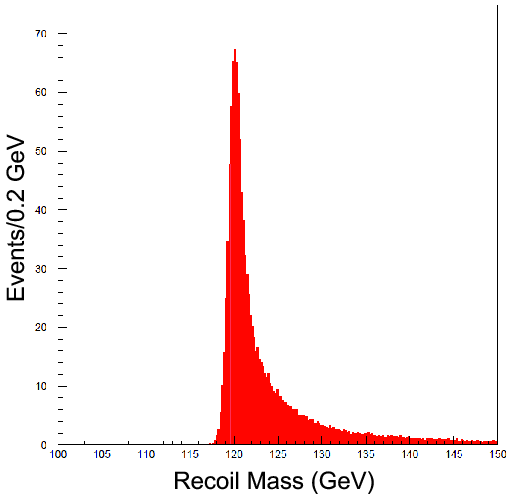}
}
\end{center}
\caption{Recoil mass distributions following selection cuts for
$eeH$ (a) and $\mu\mu H$ (b) assuming 250~fb$^{-1}$ luminosity with
80eR initial state polarization.  Background is not included.}
\label{fig:recoilnobgnd}
\end{figure}

Event  selection begins with the identification of electrons and
muons by the SiD PFA algorithm.  Following lepton identification the
cuts for the the electron and muon channels are identical.   An
event must contain at least two oppositely charged electrons or
muons with $|\cos{\theta_{\rm track}}|<0.99$.  The mass of the
lepton pair must be in the range 87~GeV~$<M_{l^+l^-}<$~95~GeV and
the angle $\theta_{l^+l^-}$ made by the vector sum of the momenta of
the two leptons with the beam axis must satisfy
$|\cos\theta_{l^+l^-}|<0.85$. Finally the angle $\theta_{\rm
missing}$ between the event missing momentum vector and the beam
axis must satisfy $|\cos\theta_{\rm missing}|<0.99$, in order to
reject Bhabha and muon pair events.  The last cut involves objects
other than electron or muon pair;  however we believe the $l^+l^-H$
efficiency for this cut is independent of the Higgs decay mode, and
so this analysis is model independent.  The distributions for the
Higgs signal following cuts is shown in
Figure~\ref{fig:recoilnobgnd} assuming 250~fb$^{-1}$ luminosity with
the initial electron and positron polarization combination $e^-(80\%
R)\ e^+(30\% L)$ and 0~fb$^{-1}$ with the combination $e^-(80\% L)\
e^+(30\%R)$.  These two polarization combinations will be referred
to by their electron polarizations  80eR and 80eL.

\begin{figure}[htbp]
\begin{center}
\subfigure[]{
\includegraphics[scale=0.33]{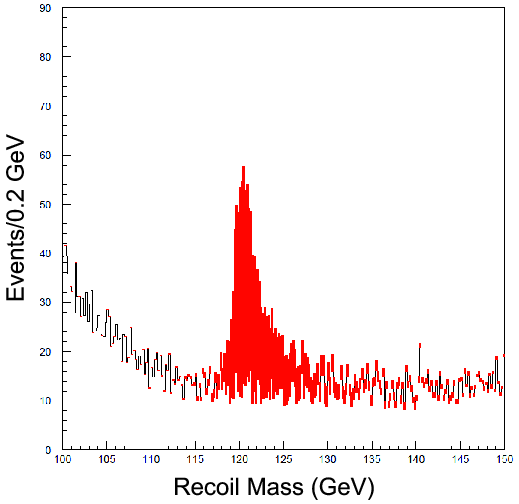}
} \subfigure[]{
\includegraphics[scale=0.33]{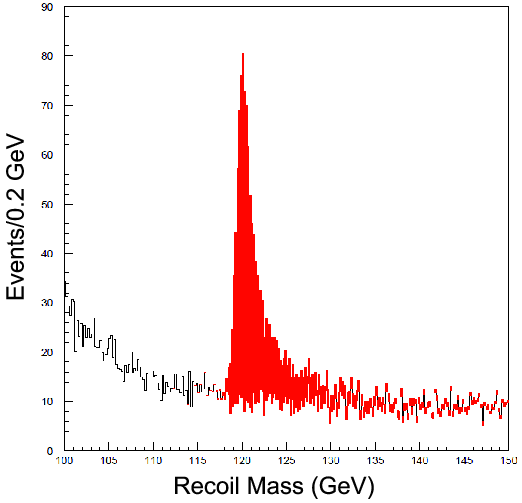}
}
\end{center}
\caption{Recoil mass distributions following selection cuts for
$eeH$ (a) and $\mu\mu H$ (b) assuming 250~fb$^{-1}$ luminosity with
80eR initial state polarization. The signal in red is added to the
background in white; background is assumed constant over 5~GeV
sections.   } \label{fig:recoilsmooth}
\end{figure}

The  major backgrounds following the selection cuts are $e^+e^-\to
\gamma \gamma l^+ l^-$, $e^+e^-\to W^+W^- \to l^+ \nu l^- \bar\nu$,
$e^+e^-\to ZZ^* \to l^+l^-f\bar{f}$, and $e^-\gamma\to e^-e^+e^-$.
The distributions for the
Higgs signal with background included are shown in
Figure~\ref{fig:recoilsmooth}.    The $W^+W^-$ background can be
greatly reduced by running 100\% of the time with the 80eR initial
state polarization combination.  This comes at a price, however, as
the signal cross-section is reduced by 30\% relative to the 80eL
combination.

A linear least squares fit of the Higgs mass $M_H$ and Higgs cross section $\sigma_{ZH}$
near $(M_H,\sigma_{ZH})=(120\ GeV,\sigma^{SM}_H)$ is used to measure the mass and cross section.
All recoil mass bins in the range of 117~GeV~$<M_{\rm recoil}<$~137~GeV are used.
The dependence of the recoil mass bins
on the Higgs mass and cross section is calculated by comparing the reconstructed
recoil mass distributions for the training samples with $M_H=119.7,120.0,120.3$~GeV.
It is assumed that the background cross-section
can be calculated to arbitrary accuracy, that the luminosity
spectrum and polarization can be perfectly measured,  and that there
are no detector systematic errors.     The errors on the Higgs mass and cross section are
summarized in Table~\ref{tab:recoilerr}, assuming a recoil mass bin
size of 0.2 GeV.

\begin{table}[h]
\begin{tabular}{|c|c|c|c|c|}
\hline 80eR lumi   &  80eL lumi    &  Mode                  &
$\Delta M_H$ (GeV)  & $\Delta \sigma_{ZH}/sigma_{ZH}$  \\ \hline
      250 fb$^{-1}$          &         0 fb$^{-1}$       & $e^+e^-H$               &    0.078 &  0.041     \\ \hline
      250 fb$^{-1}$          &         0 fb$^{-1}$       & $\mu^+\mu^- H$          &    0.046 &  0.037     \\ \hline
      250 fb$^{-1}$          &         0 fb$^{-1}$       & $e^+e^-H + \mu^+\mu^- H$ &    0.040    &  0.027     \\ \hline
        0  fb$^{-1}$         &       250 fb$^{-1}$       & $e^+e^-H$               &    0.066   &  0.067      \\ \hline
        0 fb$^{-1}$          &       250 fb$^{-1}$       & $\mu^+\mu^- H$          &    0.037   &  0.057      \\ \hline
        0 fb$^{-1}$          &       250 fb$^{-1}$       & $e^+e^-H + \mu^+\mu^- H$ &    0.032   &  0.043     \\ \hline
\end{tabular}

\caption{Summary of Higgs mass and $ZH$ cross section errors for
different channels and different luminosity assumptions.  The error
includes the measurement statistical error and the systematic error
due to the finite statistics of the Monte Carlo training sample.}
\label{tab:recoilerr}
\end{table}


\subsection{ $e^+e^-\to ZH\,,\,\; H\to c \bar{c}\, ,\, \; Z\to \nu \bar{\nu}\, , \, q \bar{q}\,$, $\, \sqrt s $=250 GeV}

The measurement of the Higgs absolute branching ratios to all
possible species is an important part of the ILC program, giving a
precision test of the Standard Model prediction that the Higgs boson
couples to each particle in proportion to its mass. These
measurements also discriminate between different 'Beyond the SM'
scenarios. The considered decay modes result in two- and four-jet
final states and exercise the tagging of charm quarks which is
particularly sensitive to the vertex detector performance~\cite{ZH}.

At the centre of mass energy $\sqrt{s}=250$~GeV the Higgs boson with
mass of 120 GeV is produced predominantly in the Higgsstrahlung
process, e$^{+}$e$^{-}$ ${\to }$ ZH. The choice of energy maximizes
the the cross-section value for Higgsstrahlung. The expected SM
Higgs boson branching ratio to charm quarks is equal to 3.0\% and
one of the main difficulties of the analysis is to separate this
signal from the background of Higgs decays to b-quarks which has a
substantially larger Br of 68.2\%. For the analysis we considered
the signal and SM samples with total integrated luminosity of
250~fb$^{-1}$ and +80\% electron polarization, -30\% positron
polarization.

\begin{figure}[htbp]
\begin{center}
\subfigure[]{
\includegraphics[scale=0.35]{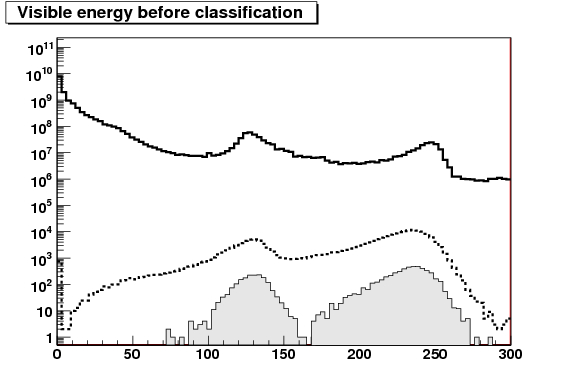}
}
\subfigure[]{
\includegraphics[scale=0.35]{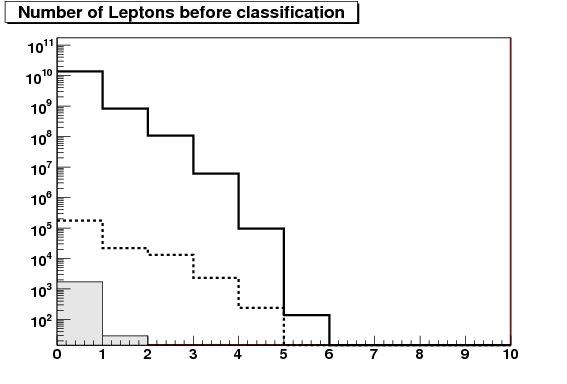}
}
\end{center}
\caption{Visible energy (a) and the number of leptons per event (b)
used for channel classification. Solid curves are SM background,
dashed curves are inclusive Higgs sample and filled histograms are
the signal.} \label{fig:Enlep}
\end{figure}

The analysis signature is dependent on the Z boson decay products
(charged leptons, hadrons or neutrinos). The channels studied in
this analysis are the neutrino mode (Z decaying to neutrinos) and
the hadronic mode (Z decaying to hadrons). The selection of signal
events is performed in three stages. The first step involves the
classification of events into these two channels using the number of
leptons and visible energy in the event. Visible energy is defined
as the sum of energies of all reconstructed particles in the event.
Leptons are defined as reconstructed electrons or muons with minimum
momentum of 15 GeV. Figure~\ref{fig:Enlep} shows the distribution of
the visible energy and the number of leptons for the signal, Higgs
background and SM background before any selections. The neutrino
mode is selected as events with no leptons and with visible energy
in the 90 to 160 GeV interval. The hadronic mode is selected as
events with no leptons and the visible energy above 170 GeV.

The second step is a cut based selection which reduces the
backgrounds in the selected channel followed by a final stage using
two neural network (NN) discriminants trained on signal and on two
types of background samples, the SM and inclusive Higgs ones. The
remaining events after the NN selection are used for the calculation
of the cross sections and branching ratios.

\subsubsection{ The neutrino channel}

\begin{figure}[htbp]
\begin{center}
\subfigure[]{
\includegraphics[scale=0.35]{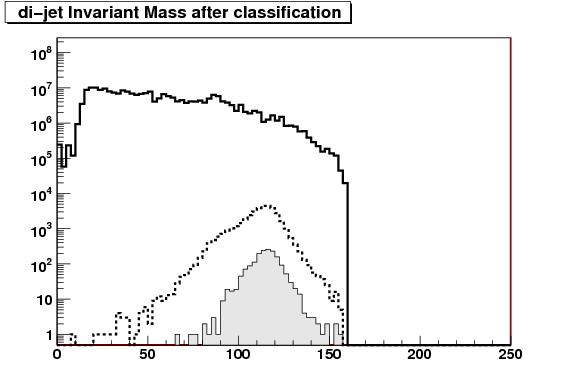}
}
\subfigure[]{
\includegraphics[scale=0.35]{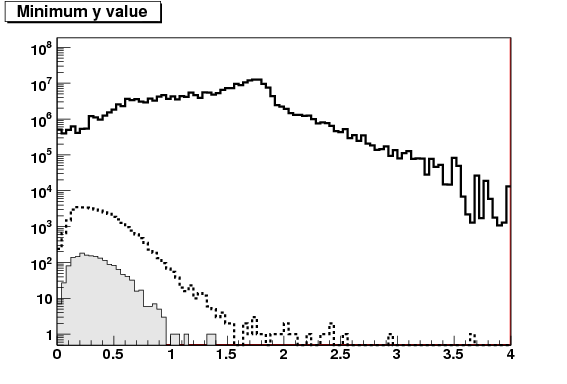}
}

\end{center}
\caption{Distribution of di-jet invariant mass (a) and $y_{min}$
(b). Solid curves are SM background, dashed curves are inclusive
Higgs sample and filled histograms are the signal.} \label{fig:nu}
\end{figure}

The analysis starts with forcing all reconstructed particles to be
clustered into two jets which for the signal are assumed to come
from the Higgs boson recoiling against two neutrinos from the Z
boson decay. The discrimination between the signal and background
with different number of jets is achieved by selection on the
$y_{min}$ parameter. Another powerful discriminant, the
reconstructed invariant mass of two hadronic jets, is expected to be
consistent with the Higgs mass. Figure~\ref{fig:nu} shows
distributions of di-jet invariant mass (a) and $y_{min}$ (b) for the
signal and backgrounds after classification. The background includes
all SM processes but at this stage the most important are 2-fermion
events, ZZ pairs decaying to neutrinos and hadrons and WW pairs
where one W decays hadronically and the other W decays into a
neutrino and a lepton which escapes undetected.

The full list of selections is given in Table~\ref{tab:HZ1}.

\begin{table}[h]
\begin{tabular}{llll}
\hline && selection & value  \\
\hline
20 & $<$ & p$_T$ of jet & $<$ 90 GeV \\
&& number of charged tracks in jet & $>$ 4\\
&& $-\log(y_{min}$) & $<$ 0.8 \\
&& thrust & $< 0.95$ \\
&& $ \cos(\theta_{thrust}) $&$< 0.98$ \\
$100^{\circ}$ &$<$ & angle between jets &$< 170^{\circ}$ \\
100 GeV &$<$ & di-jet invariant mass &$<$ 140 GeV \\
&& energy of isolated photon & $<$ 10 GeV \\
\hline
\end{tabular}
\caption{Selections for the neutrino channel}
\label{tab:HZ1}
\end{table}

\begin{figure}[htbp]
\begin{center}
\subfigure[]{
\includegraphics[scale=0.35]{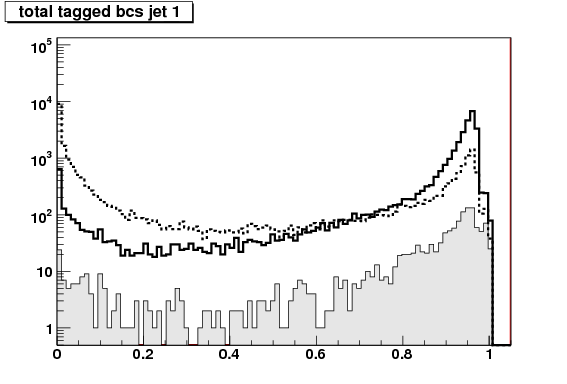}
}
\subfigure[]{
\includegraphics[scale=0.35]{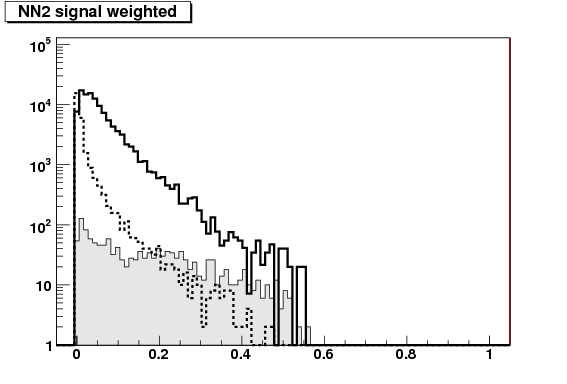}
}

\end{center}
\caption{Neutrino channel: (a) Flavour tag 'c with b only
background' for the first jet; (b) NN$_{Higgs-signal}$ output. Solid
curves are SM background, dashed curves are inclusive Higgs sample
and filled histograms are the signal.} \label{fig:ccNN}
\end{figure}

The remaining events are categorized using a neural network
implemented in FANN [4]. The NN input variables include all the
variables stated above with addition of the LCFI flavour tag outputs
for both jets. Figure~\ref{fig:ccNN} (a) shows the distribution of
one of three possible flavour tags, 'c with b only background' for
the first jet. This particular tag corresponds to the training to
distinguish c-quarks from b-quarks without taking into account light
quarks at all. The NN has been trained two times, the first time to
distinguish the SM background from the inclusive Higgs sample and to
produce the NN$_{SM-Higgs}$ output; and, the second time to
distinguish the signal from the inclusive Higgs sample and to
produce the NN$_{Higgs-signal}$ output, the latter shown in
Figure~\ref{fig:ccNN} (b).

The final signal sample is required to have NN$_{Higgs-signal} >
0.2$ and NN$_{SM-Higgs} > 0.3$. The sample includes 476 signal
events with SM background of 570 events and inclusive Higgs
background of 246 events. The signal efficiency is equal to 28\%.
This corresponds to the measured signal cross section of $6.8 \pm
0.7$ fb. The branching ratio of $3.3 \pm 0.4\%$ was determined by
normalization to the Higgs inclusive cross section as measured in
the Higgs recoil mass analysis. The relative accuracy of 11\% is
dominated by the precision of the $HZ \rightarrow
c\bar{c}\nu\bar{\nu}$ cross section. Here and below the uncertainty
accounts purely for statistical fluctuations of expected number of
events with all other contributions considered negligible.

\subsubsection{The hadronic channel}

\begin{figure}[htbp]
\begin{center}
\subfigure[]{
\includegraphics[scale=0.35]{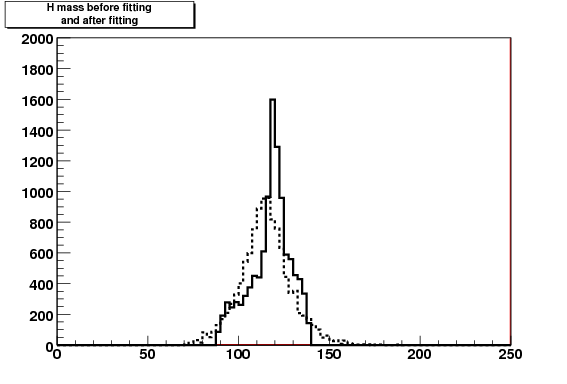}
}
\subfigure[]{
\includegraphics[scale=0.35]{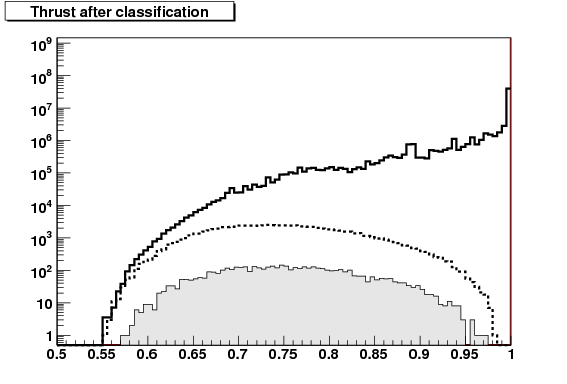}
}

\end{center}
\caption{Higgs invariant mass before and after kinematic fitting (a) and  thrust of the signal and backgrounds after classification (b).
Solid curves are SM background, dashed curves are inclusive Higgs sample and the filled histogram is the signal.}
\label{fig:had}
\end{figure}

In this channel, events are forced to have four reconstructed jets
and as in the two-jet case the $y_{min}$ variable was chosen to
differentiate from backgrounds with different number of jets. For
the signal, two of the jets are assumed to have an invariant mass
consistent with the Higgs boson and the other two having mass
consistent with the Z boson. Kinematic fitting [3] is performed to
improve the assignment of jets to decay products of the Higgs and Z
bosons and therefore to reduce the combinatorial background. The
main backgrounds for this channel are WW and ZZ pairs where the all
the bosons decay to hadrons. Figure~\ref{fig:had} shows the Higgs
invariant mass before and after kinematic fitting (a) and thrust (b)
of the signal and backgrounds after classification.

Selections used for further reduction of the background are presented in Table~\ref{tab:HZ2}.
\begin{table}[h]
\begin{tabular}{llll}
\hline && selection & value  \\
\hline
&& number of charged tracks in jet & $>$ 4\\
&& $-\log(y_{min}$) & $<$ 2.7 \\
&& thrust & $< 0.95$ \\
&& $ \cos(\theta_{thrust})$ & $< 0.96$ \\
$75^{\circ}$ &$<$ & angle between jet 1 and 3 &$< 165^{\circ}$ \\
$50^{\circ}$ &$<$ & angle between jet 2 and 4 &$< 150^{\circ}$ \\
95 GeV &$<$ & invariant mass of Higgs candidate &$<$ 145 GeV \\
45 GeV &$<$ & invariant mass of Z candidate &$<$ 105 GeV \\
&& energy of isolated photon & $<$ 10 GeV \\
\hline
\end{tabular}
\caption{Selections for the four-jet analysis}
\label{tab:HZ2}
\end{table}
For the angular selections the jets were ordered in energy. The
invariant mass selections were employed after the kinematic fitting
which paired the four jets into the Higgs and Z candidates.

\begin{figure}[htbp]
\begin{center}
\subfigure[]{
\includegraphics[scale=0.35]{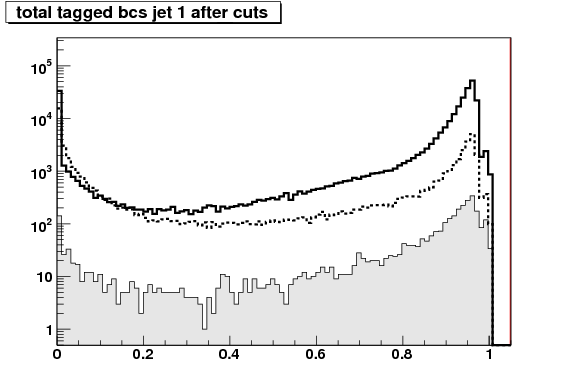}
}
\subfigure[]{
\includegraphics[scale=0.35]{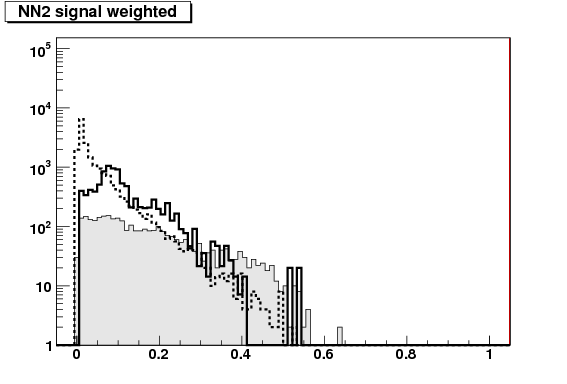}
}

\end{center}
\caption{Four jet channel: (a) Flavour tag 'c with b only background' for the first jet; (b) NN$_{Higgs-signal}$ output.
Solid curves are SM background, dashed curves are inclusive Higgs sample and filled histograms are the signal.}
\label{fig:ccqqNN}
\end{figure}

Similarly to the neutrino channel the variables above and also the
three flavour tag outputs for all jets, are used in a neural network
based selection employing the FANN package. Figure~\ref{fig:ccqqNN}
(a) shows the distribution of one of three possible LCFI flavour
tags, 'c with b only background' for the first jet. The NN has been
trained two times, first to distinguish the SM background from the
inclusive Higgs sample producing the NN$_{SM-Higgs}$ output, and,
second, to distinguish the signal from the inclusive Higgs sample
producing the NN$_{Higgs-signal}$ output, the latter shown in
Figure~\ref{fig:ccqqNN} (b).

The final signal sample is required to have NN$_{Higgs-signal} >
0.2$ and NN$_{SM-Higgs} > 0.3$. The sample includes 814 signal
events with SM background of 569 events and inclusive Higgs
background of 547 events. The signal efficiency is equal to 47\%.
This corresponds to the measured $HZ \rightarrow c\bar{c}q\bar{q}$
cross section of $6.9 \pm 0.4 $ fb. The branching ratio of $3.3 \pm
0.2\%$ was determined by normalization to the Higgs inclusive cross
section as measured in the Higgs recoil mass analysis. The relative
accuracy of 6\% on the latter takes into account contributions from
both cross sections and is dominated by the precision of $HZ
\rightarrow c\bar{c}q\bar{q}$ cross section.

\subsection{ $e^+e^-\to ZH\,,\,\; H\to \mu^+\mu^-\, ,\, \; Z\to \nu \bar{\nu}\, , \, q \bar{q}\,$, $\, \sqrt s $=250 GeV}

The decay of Higgs boson into muons is one of the SM rare decays.
Together with the $c\bar{c}$ final state it gives access to the
Higgs coupling to the second generation fermions. The expected
branching ratios are of the order of 0.01\%. This decay channel will
allow a precise direct determination of the Higgs mass as it only
relies on the performance of the tracking and will also allow to set
an upper limit on the Higgs decay width limited only by the
resolution of tracking system. The challenge of the analysis is to
extract the signal out of an overwhelming background from
two-fermion and four-fermion processes.

For the signal we have a dedicated WHIZARD sample of $H \rightarrow
\mu^+\mu^-$ with $Z \rightarrow f \bar{f}$ (including the fusion
channels) with a Higgs mass of 120 GeV. The branching ratios for the
final states are 67.0 \% for the hadronic final states and 22.9 \%
for the neutrino final state. Contribution due to the WW fusion
process to the neutrino channel, 16.1 \%, complicates this analysis
as this channel does not exhibit a peak in the missing mass. The
integrated luminosity of the sample is 250 fb$^{-1}$.

\begin{figure}[htbp]
\includegraphics[scale=0.4]{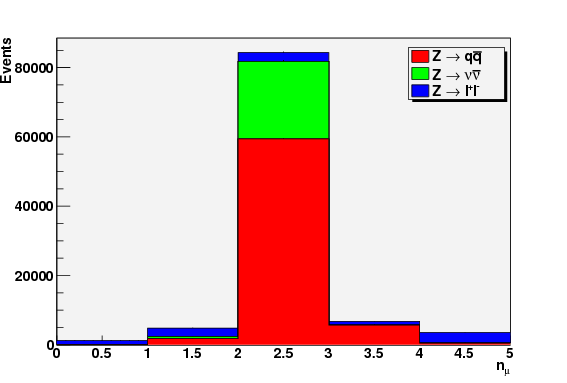}
\includegraphics[scale=0.4]{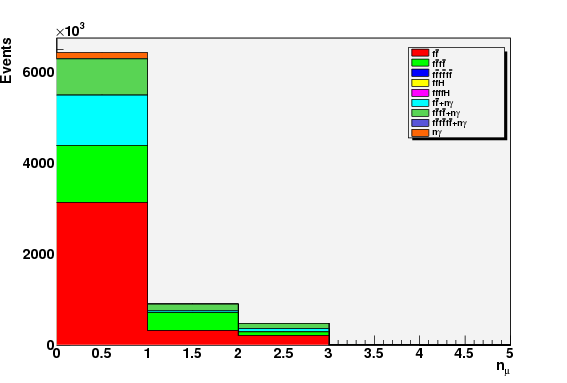}
\caption{The Number of identified muons in the signal sample (left) and the background sample (right).}
\label{hmumu:fig:nmuons}
\end{figure}

\begin{figure}[htbp]
\subfigure[][]{
\includegraphics[scale=0.4]{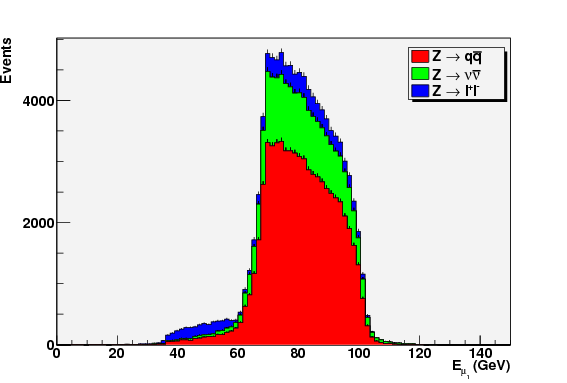}
}
\subfigure[][]{
\includegraphics[scale=0.4]{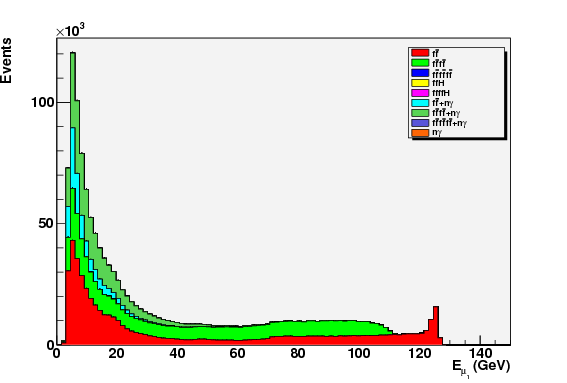}
}
\caption{The energy spectrum of the most energetic muon for the signal (a)
and the SM background (b).}
\label{hmumu:fig:emuons}
\end{figure}

The primary aim of this analysis are the hadronic channel and the
neutrino channel. For both we require in each event a presence of
two muons which are identified by associating a reconstructed track
to a MIP segment in the calorimeter and a stub in the muon system.
For the leading muon we require its energy to be larger than 50 GeV
and for the second muon we require it to be larger than 30 GeV. The
number of identified muons is shown in Figure~\ref{hmumu:fig:nmuons}
for the signal and  background samples.
The energy distribution of the most energetic muon is shown in Figure \ref{hmumu:fig:emuons}.
The signature of the hadronic channel consists of two high-energetic muons from the Higgs decay and two jets from
the hadronically decaying Z boson.
For this channel we force the event into two jets using the Durham algorithm, after excluding the previously identified muons.
We require the $y_{min}$ parameter of the jet clustering to be larger than 0.05.

Preselecting hadronic events compatible with the di-jet di-muon signature we
require the number of charged tracks to be greater than 5 and the visible energy to
be larger than 190 GeV.
We then select events where the leading jet has an energy between 30 and 105 GeV
and second jet has an energy between 10 and 70 GeV. We also require the jet
$P_T$ to be smaller than 90 and 60 GeV respectively.
Finally we require the mass of the di-muon system to be compatible with the
$m_H$=120 GeV $\pm$ 0.9 GeV.

\begin{figure}[htbp]
\includegraphics[scale=0.4]{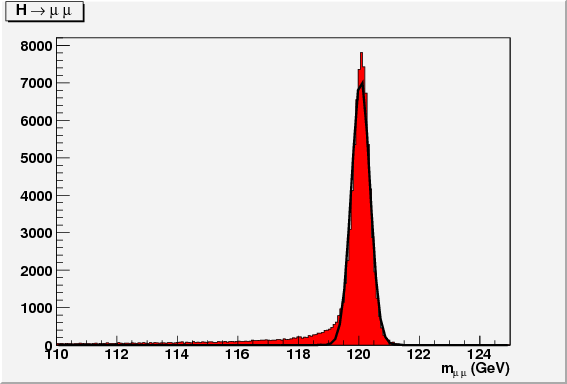}
\includegraphics[scale=0.4]{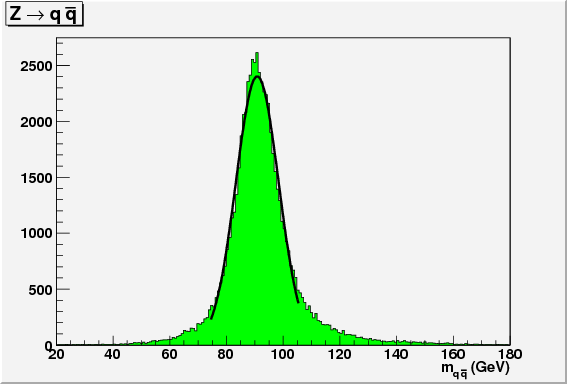}
\caption{The mass resolution for the di-muon system (left) and the di-jet
system(right) for the signal sample $Z \rightarrow q\bar{q};\qquad H\rightarrow
\mu^+\mu^-$.}\label{hmumu:fig:signalres}
\end{figure}

We check the mass resolution for the signal both for the di-muon and
the di-jet system before any kinematic fitting and corrections for energy
losses have been applied. We obtain 120.07 GeV with a width of 304 MeV for
the di-muon system and 90.84 GeV with a width of 7.58 GeV for the di-jet system.
 The results are shown in Figure~\ref{hmumu:fig:signalres}.

After the preselection, the dominant background originates from four-fermion processes,
mainly ZZ production with $e^+e^- \rightarrow ZZ \rightarrow
q\bar{q}\mu^+\mu^-$.
To select ``good'' muons, we cut on the muon opening angle, $\cos(\theta_{\mu\mu}) < -0.5$.
The signal tends to be very back-to-back, and we exploit this fact by cutting on the angle between the two
reconstructed bosons (from the di-jet and di-muon pairs) $\cos(\theta_{BB})<-0.8$.
Furthermore, we require the distance of any muon with respect
to any jet to be at least 0.1 radians. Finally we select events with an acoplanarity
in the reconstructed boson-boson system larger than 2.8 radians.

\begin{figure}[htbp]
\includegraphics[scale=0.4]{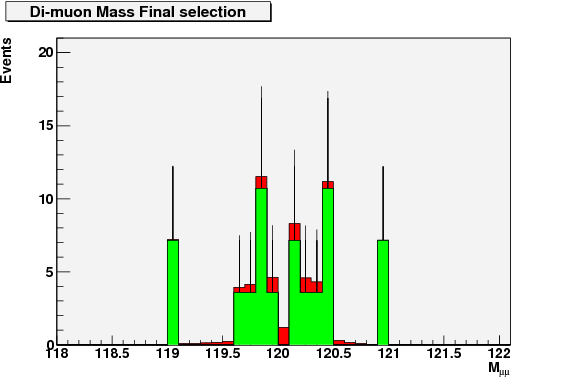}
\includegraphics[scale=0.4]{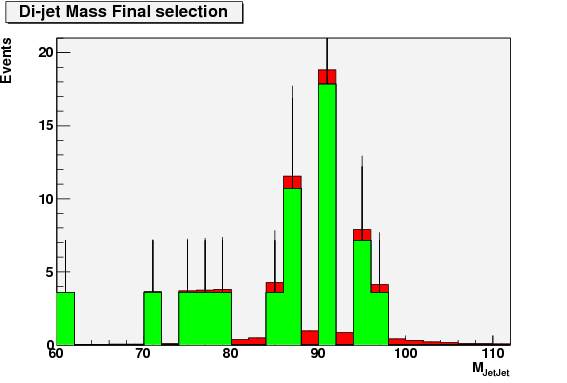}
\caption{The mass of the di-muon pair (a) and the di-jet pair in the final sample.}
\label{hmumu:fig:final_dimuonmass}
\end{figure}

In order to test the compatibility of each event with either the HZ or ZZ hypothesis,
we construct a $\chi^2$ as follows
\begin{eqnarray}
\chi^2_{ZZ}=\left(\frac{m_{JetJet}-m_Z}{\sigma_{m_{JetJet}}}\right)^2+\left(\frac{m_{\mu\mu}-m_Z}{\sigma_{m_{\mu\mu}}}\right)^2
\\
\chi^2_{HZ}=\left(\frac{m_{JetJet}-m_Z}{\sigma_{m_{JetJet}}}\right)^2+\left(\frac{m_{\mu\mu}-m_H}{\sigma_{m_{\mu\mu}}}\right)^2
\end{eqnarray}
For the uncertainties we assume the resolutions derived above, but
for the resolution the di-muon pair in the ZZ case we assume the
natural width of the Z which is considerably larger than the muon
resolution and we assume the Higgs mass to be known as 120 GeV. We
then cut on $80 < \chi^2_{ZZ} < 120$ and $\chi^2_{HZ} < 20$. The
number of events after each cut is summarized in
Table~\ref{hmumu:tab:tablecuts}.
\begin{table}
\begin{center}
\begin{tabular}{|l|r|r|r|r|}\hline
                            &  SM Background
                                &\multicolumn{3}{|c|}{$H\rightarrow
                                \mu^+\mu^-$}\\\cline{2-5}
\raisebox{1.5ex}[-1.5ex]{Cut}   & \multicolumn{1}{|c|}{Events}   & \multicolumn{1}{|c|}{Events} & $\epsilon_{H\rightarrow \mu^+\mu^-}$ &
$\epsilon_{H\rightarrow \mu^+\mu^- Z\rightarrow q\bar{q}}$ \\\hline
           two muons required& 4663389.3 &      17.18  &  0.928  &  0.960 \\\hline
      Charged Tracks hadronic&   26665.9 &      12.03  &  0.650  &  0.959 \\\hline
            Evis hadronic Cut&   19567.1 &      11.98  &  0.647  &  0.957 \\\hline
            Jet Selection Cut&   16683.6 &      11.82  &  0.638  &  0.944 \\\hline
             Muon Mass Window&    1519.3 &      11.49  &  0.621  &  0.918 \\\hline
           MuonMuon Angle Cut&    1297.9 &      11.32  &  0.612  &  0.905 \\\hline
         BosonBoson Angle Cut&    1129.3 &      10.90  &  0.589  &  0.872 \\\hline
      Min Isolation Angle Cut&     870.7 &      10.84  &  0.585  &  0.866 \\\hline
       Boson Acoplanarity Cut&     792.1 &      10.51  &  0.568  &  0.841 \\\hline
       ZZ qqqmm chisquare Cut&     260.7 &      10.04  &  0.543  &  0.804 \\\hline
       HZ qqqmm chisquare Cut&      60.7 &       8.64  &  0.467  &  0.692 \\\hline
                   Z mass cut&      53.6 &       7.74  &  0.418  &  0.622 \\\hline
               Higgs mass cut&      39.3 &       7.66  &  0.414  &  0.616 \\\hline
\end{tabular}
\caption{The event selection for hadronic channel.}\label{hmumu:tab:tablecuts}
\end{center}
\end{table}
As a result of the selection above, we expect about 8 signal events
with the background of 39 events in the final sample. The di-muon
mass distributions after the final selections are shown in
Figure~\ref{hmumu:fig:final_dimuonmass}.

The statistical significance of the abundance of events in the
signal region $S/\sqrt{S+B}$ for this selection is equal to 1.1. The
cross-section of the decay and its uncertainty can be computed
accordingly as $\sigma= 0.074 \pm 0.066$ fb.

For the neutrino channel we select events with a missing mass
compatible with the Z mass, and with two high-energetic muons
consistent with a $H \to \mu \mu$ decay. The analysis selections are
listed in Table~\ref{hmumu:tab:eventSelection}.
\begin{table}
    \begin{tabular}{|l|r|r|r|r|}\hline
                            &  SM Background
                                &\multicolumn{3}{|c|}{$H\rightarrow
                                \mu^+\mu^-$}\\\cline{2-5}
    \raisebox{1.5ex}[-1.5ex]{Cut}   & \multicolumn{1}{|c|}{Events}   & \multicolumn{1}{|c|}{Events} & $\epsilon_{H\rightarrow \mu^+\mu^-}$ &
    $\epsilon_{H\rightarrow \mu^+\mu^- Z\rightarrow \nu\nu}$ \\\hline
           two muons required& 4663389.3 &      17.18  &  0.928  &  0.960\\\hline
             number of Tracks& 4376478.9 &       4.06  &  0.219  &  0.958\\\hline
               visible Energy&  706483.1 &       3.42  &  0.185  &  0.806\\\hline
               charged Energy&  356850.7 &       3.21  &  0.174  &  0.758\\\hline
             missing Momentum&   26937.9 &       3.21  &  0.174  &  0.758\\\hline
      cos theta between muons&   17238.6 &       3.20  &  0.173  &  0.756\\\hline
              muon energy cut&   15952.1 &       3.17  &  0.171  &  0.747\\\hline
                     chi2 cut&    1314.3 &       2.72  &  0.147  &  0.641\\\hline
                 missing Mass&    1280.7 &       2.70  &  0.146  &  0.637\\\hline
                         mass&    1190.7 &       2.70  &  0.146  &  0.637\\\hline
    \end{tabular}
    \caption{Event selection for the missing energy channel.}
    \label{hmumu:tab:eventSelection}
\end{table}

Based the numbers of signal and background events after the final
selections we conclude that in the case of missing energy channel a
cut-and-count analysis is insufficient to obtain a meaningful
result. Multivariate classifiers can aid greatly in the separation
of signal and background. A preliminary analysis employing these
methods showed that a separation between signal and background of
1.8 sigma is achievable in the neutrino channel. Similarly, a
considerable improvement could be expected in the hadronic channel.

\subsection{ $e^+e^-\to \tau^+ \tau^-\,$, $\, \sqrt s $=500 GeV}

The identification  of  250~GeV taus and their decay modes is a
challenge for the tracker and calorimeter. Low multiplicty but
tightly collimated jets must be reconstructed in terms of the
underlying charged hadron and $\pi^0$ constituents.   Particle flow
algorithms designed to measure jet energy may require modification
to provide optimal tau mode identification.

Two sets of observables are used to quantify the ability of the SiD
detector to reconstruct 250~GeV taus.  First, tau decay mode
efficiencies and purities are determined, and then the tau
polarization is measured.  Decay mode efficiencies are determined
for the decays $\tau^-\to e^-\bar{\nu_e}\nu_{\tau}\ ,
\mu^-\bar{\nu_{\mu}}\nu_{\tau}\ , \  \pi^-\nu_{\tau}\ , \
\rho^-\nu_{\tau}\ , \  a_1^-\nu_{\tau}$. With the exception of $
a_1^-\nu_{\tau}$ the same set of decay modes is used for the
measurement of the tau polarization.

\subsubsection{Cross Section and $A_{FB}$ Precision}

\begin{figure}[htbp]
\begin{center}
\includegraphics[scale=0.4]{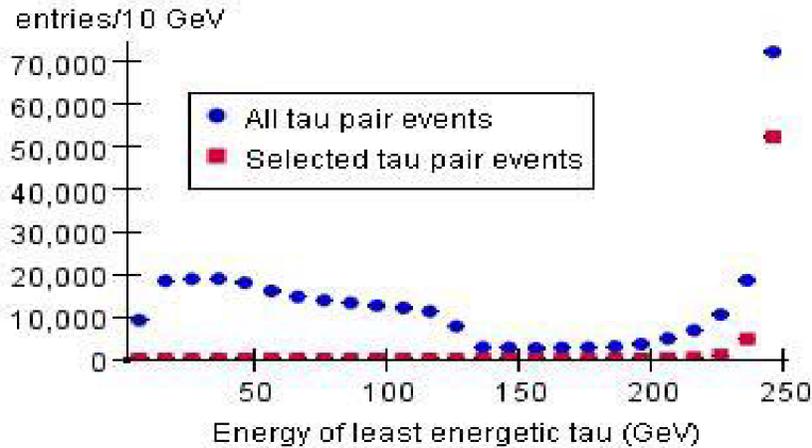}
\end{center}
\caption{Energy of least energetic $\tau$ for full sample (blue) and
following event selection (red).} \label{tau:fig:smalletau}
\end{figure}

Full energy tau pair events were selected by clustering the PFA
reconstructed particles into tau jets. Event cuts were $1 < \#{\rm
tracks} < 7$ and 40 GeV~$< {\rm EVis} <$~450 GeV. Exactly 2 jets
were required, each with $|\cos\theta| < 0.95$. The opening angle
between the jets was required to be greater than 178$^\circ$. Events
with both jets identified as electrons or both jets identified as
muons were also rejected. The results of this selection are shown in
Figure~\ref{tau:fig:smalletau}. The selection efficiency for all tau
pair events is 17.9\%, yielding 128708 events. Using the full SM
data set, 3099 events were selected from non tau pair events, for a
background of  2.4\%.

Using these events the cross section was measured by counting, and
the forward backward asymmmetry $A_{FB}$ was measured by performing
a least squares fit of the $\cos\theta$ distribution of the taus
assuming $d\sigma/d\cos\theta \propto 1+\cos\theta^2+8/3\cdot
A_{FB}\cos\theta$. The precision of the cross section is 0.28\%.
The forward backward asymmetry measurement was $A_{FB}$ = 0.5038
$\pm$ 0.0021 for a 250~fb$^{-1}$ 80eL test sample and $A_{FB}$ =
0.4704 $\pm$ 0.0024 for a 250~fb$^{-1}$ 80eR test sample (80eL and
80eR are defined in the discussion on tau polarization).

\subsubsection{Tau Decay Mode Efficiencies and Purities}

The tau decay modes were identified as follows.  All calorimeter
hits were clustered and assigned to the nearest tau jet. Photon
identification was performed on EM clusters, and photons
reconstructed. All remaining clusters were assigned to tracks. The
total calorimeter energy assigned to the track was required to be no
more than the sum of the track momentum and $n\sigma$, where $\sigma
= 0.65\sqrt{E}$ and $n=$ 2.2 or 2.5 (the EPcut). The selection cuts
are summarized in Table~\ref{tab:taudecselect} where a $\pi^0$ is a
pair of photons satisfying 0.06~GeV~$<M_{2\gamma}<$~0.18~GeV.

\begin{table}[h]
\begin{tabular}{|l|c|c|c|l|}
\hline decay mode & \# $\gamma$   & \# $\pi^0$ & EPcut &
other criteria  \\
\hline $e^-\bar{\nu_e}\nu_{\tau}$ & 0 & 0 & - & HCAL energy~$<4\%$ of track energy. \\
\hline $\mu^-\bar{\nu_{\mu}}\nu_{\tau}$ & 0 & 0 & - & identified as $\mu$ by PFA \\
\hline $\pi^-\nu_{\tau}$ & 0 & 0 &  2.5 & -  \\
\hline $\rho^-\nu_{\tau}\to \pi^-\pi^0\nu_{\tau}$ & 1 & 0 &  2.2 &  0.6 GeV~$<M_\rho<$~0.937~GeV,\ \ \   $E_\gamma > 10$~GeV \\
\hline $\rho^-\nu_{\tau}\to \pi^-\pi^0\nu_{\tau}$ & 2 & 1  &  2.2 &  0.4 GeV~$<M_\rho<$~0.93~GeV  \\
\hline $a^{-}_1\nu_{\tau}\to \pi^-\pi^0\pi^0\nu_{\tau}$ & 3 & 1 &  2.2 &  0.8 GeV~$<M_{a_1}<$~1.5~GeV,\ \ \   $E_\gamma > 10$~GeV   \\
\hline $a^{-}_1\nu_{\tau}\to \pi^-\pi^0\pi^0\nu_{\tau}$ & 4 & 2 &  2.2 &  0.8 GeV~$<M_{a_1}<$~1.5~GeV  \\
\hline $a^{-}_1\nu_{\tau}\to \pi^-\pi^+\pi^-\nu_{\tau}$ & 0 & 0 &  2.5 &  0.8 GeV~$<M_{a_1}<$~1.7~GeV \\
\hline
\end{tabular}
\caption{Decay mode identification criteria.}
\label{tab:taudecselect}
\end{table}

Alternate photon reconstructions were attempted on tau jets that
remained unidentified after this first pass at identification.  The
number of photons was changed by combining the 2 nearest photons,
changing the identification of a photon cluster, or looking for
possible missed photons. After a new set of photons was
reconstructed the criteria in Table~\ref{tab:taudecselect} were
applied.  If the tau jet remained unidentified after this second
pass then further alternate photon reconstructions were attempted.

Using a test sample the purity of the mode identification was
calculated separately for  each of these photon reconstruction
iterations.  The purity was recorded for use in the polarization
analysis. The decay mode purity and efficiency for all identified
taus is shown in Table~\ref{tab:taupurityall}, while the purity and
efficiency for taus identified with a photon reconstruction pass
with purity greater than 85\% is shown in
Table~\ref{tab:taupurityhigh}.  In all semi-leptonic modes, standard
model backgrounds are below 2\%.

\begin{table}[h]
\begin{tabular}{|l|c|c|c|c|c|}
\hline decay mode                          & Correct ID & Wrong ID &
ID eff & ID purity & SM bgnd \\ \hline $e^-\bar{\nu_e}\nu_{\tau}$
& 39602      & 920      &  0.991  &  0.977    & 1703 \\ \hline
$\mu^-\bar{\nu_{\mu}}\nu_{\tau}$               & 39561      & 439
&  0.993  &   0.989   & 1436 \\ \hline $\pi^-\nu_{\tau}$
& 28876      & 2612     &   0.933 &   0.917    & 516  \\ \hline
$\rho^-\nu_{\tau}\to \pi^-\pi^0\nu_{\tau}$      & 55931      & 8094
&   0.790 &   0.874    & 1054 \\ \hline $a^{-}_1\nu_{\tau}\to
\pi^-\pi^0\pi^0\nu_{\tau}$ & 18259      & 11140    &   0.732  &
0.621   &  847 \\ \hline $a^{-}_1\nu_{\tau}\to
\pi^-\pi^+\pi^-\nu_{\tau}$ & 21579      & 2275     &   0.914  &
0.905   & 141  \\ \hline
\end{tabular}
\caption{Tau decay mode reconstruction for all events.}
\label{tab:taupurityall}
\end{table}

\begin{table}[h]
\begin{tabular}{|l|c|c|c|c|c|}
\hline decay mode                          & Correct ID & Wrong ID &
ID eff & ID purity & SM bgnd \\ \hline $e^-\bar{\nu_e}\nu_{\tau}$
& 39602      & 920      &  0.991  &  0.977    & 1703 \\ \hline
$\mu^-\bar{\nu_{\mu}}\nu_{\tau}$               & 39561      & 439
&  0.993  &   0.989   & 1436 \\ \hline $\pi^-\nu_{\tau}$
& 26599      &  491     &   0.859 &   0.982    & 387  \\ \hline
$\rho^-\nu_{\tau}\to \pi^-\pi^0\nu_{\tau}$      & 41013      & 3056
&   0.579 &   0.931    & 674  \\ \hline $a^{-}_1\nu_{\tau}\to
\pi^-\pi^0\pi^0\nu_{\tau}$ & 2720       & 272      &   0.109  &
0.909   &  23  \\ \hline $a^{-}_1\nu_{\tau}\to
\pi^-\pi^+\pi^-\nu_{\tau}$ & 17328      & 1153     &   0.734  &
0.938   & 111  \\ \hline
\end{tabular}
\caption{Tau decay mode reconstruction for the high purity sample of
tau decays.} \label{tab:taupurityhigh}
\end{table}

\subsubsection{Tau Polarization}

A measurement of the tau polarization $P_{\tau}$ for full energy
taus at $\sqrt{s}=500$~GeV would be an important part of a program,
for example, to search for and measure the properties of multi-TeV
$Z'$ resonances at the ILC.  The tau polarization measurement relies
on the quality of the decay mode identification  as well as the
quality of the decay product four-vector reconstruction.  It
therefore serves as an excellent benchmark for detailed particle
reconstruction by ILC detectors.

The optimal observable technique\cite{Davier:1992nw} is used to
measure the mean tau polarization  $<P_{\tau}>$ over all tau
production angles.  The following decay modes are used:
$e^-\bar{\nu_e}\nu_{\tau}$, $\mu^-\bar{\nu_{\mu}}\nu_{\tau}$,
$\pi^-\nu_{\tau}$ , and $\rho^-\nu_{\tau}\to \pi^-\pi^0\nu_{\tau}$.
For lepton and single pion decays the optimal observable $\omega$ is
simply $x$, the ratio of the energy of the lepton or pion and the
beam energy.   For the rho decay $\omega$ is  a complicated
function\cite{Duflot:1993gm} of the angles of the rho in the tau
rest frame and of the charged pion in the rho rest frame.

The average polarization is measured separately for initial electron
and positron polarization combinations of $e^-(80\% L)\ e^+(30\%R)$
and  $e^-(80\% R)\ e^+(30\% L)$.  These two polarization
combinations will be referred to in the following by their electron
polarizations  80eL and 80eR.   In the Standard Model the true
tree-level average tau polarizations at $\sqrt{s}=500$~GeV are
$<P_{\tau}>=$~-0.625 and 0.528 for initial state polarizations of
80eL and 80eR, respectively.

A linear least squares fit of  $<P_{\tau}>$ over all the bins of all
the reconstructed $\omega$ distributions for
$e^-\bar{\nu_e}\nu_{\tau}$, $\mu^-\bar{\nu_{\mu}}\nu_{\tau}$,
$\pi^-\nu_{\tau}$ , and $\rho^-\nu_{\tau}\to \pi^-\pi^0\nu_{\tau}$
is used to measure  $<P_{\tau}>$.  The Monte Carlo tau pair sample
is divided in half with one half used for training and the other for
testing.   The dependence of the reconstructed $\omega$
distributions on  $<P_{\tau}>$ is obtained by comparing the
reconstructed $\omega$ distributions for the training samples for
80eL and 80eR initial state polarizations.  The tau polarization of
the testing sample is then measured.

\begin{figure}[htbp]
\centerline{\includegraphics[scale=0.5]{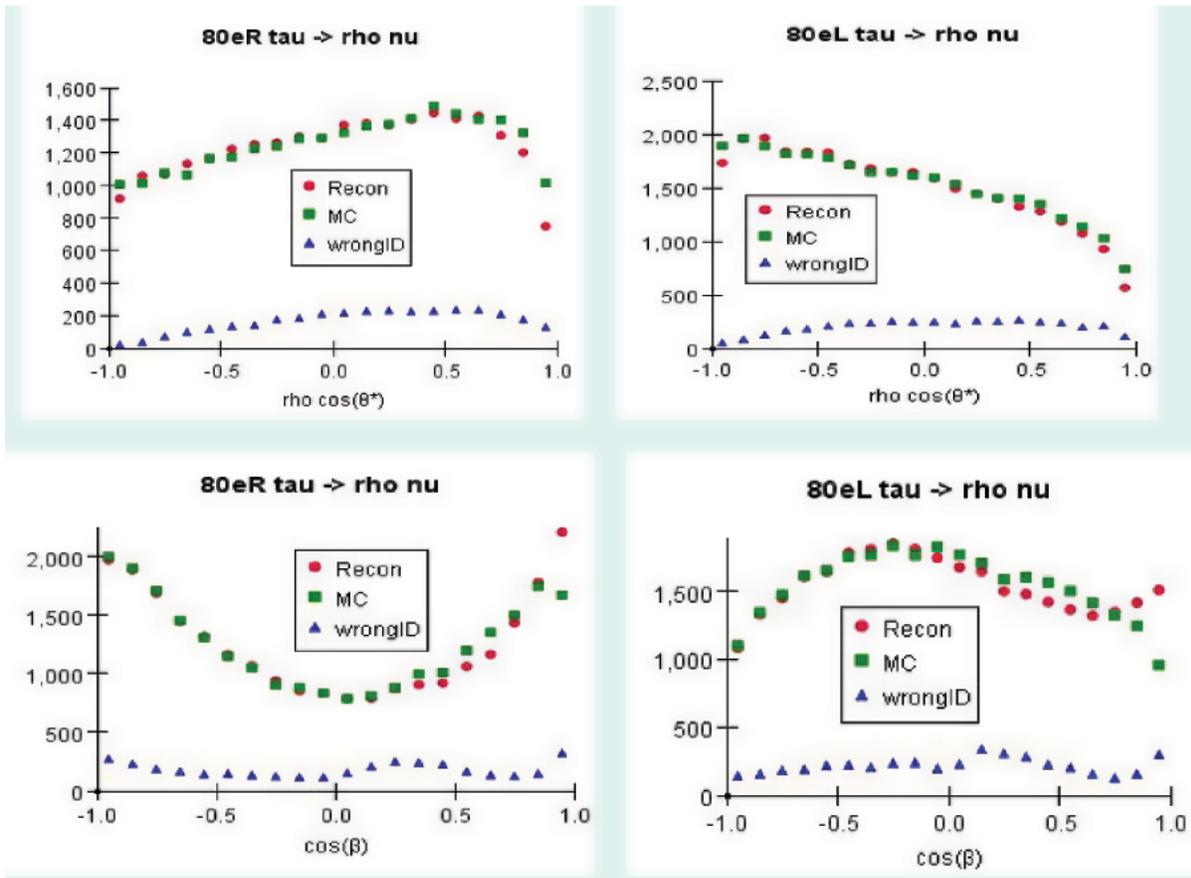}}
\caption{Reconstructed (red) and true (green) kinematic variables
used in the definition of the  optimal observable $\omega$
          for the decay $\rho^-\nu_{\tau}$.   The top two plots are $\cos\theta^*$  and the bottom two plots are $\cos\beta$ (see text).
Also shown are distributions for incorrectly identified taus
(blue).} \label{fig:rhokin}
\end{figure}

\begin{figure}[htbp]
\centerline{\includegraphics[scale=0.5]{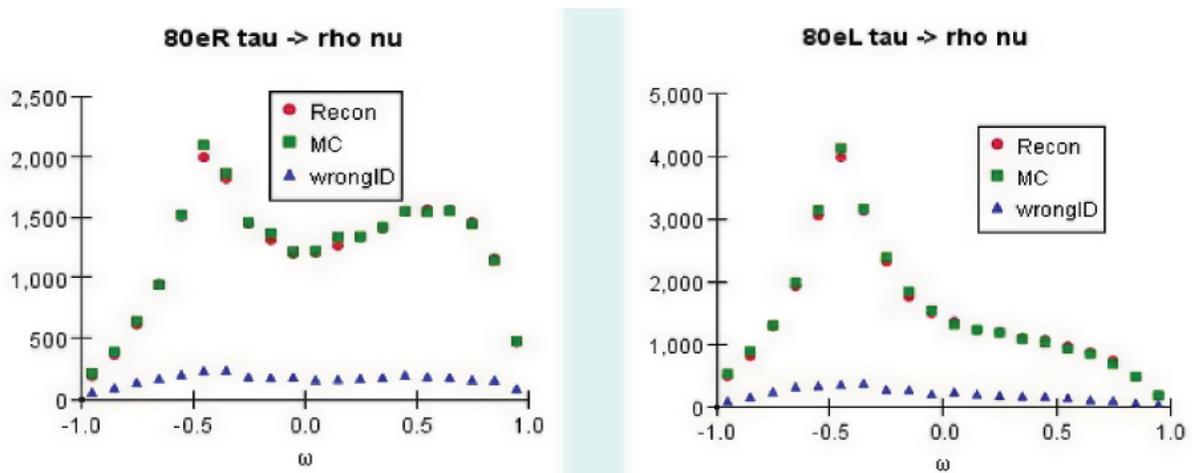}}
\caption{Reconstructed (red) and true (green) distributions for the
optimal observable $\omega$ for the decay $\rho^-\nu_{\tau}$.Also
shown are distributions for incorrectly identified taus (blue).}
\label{fig:omegarho}
\end{figure}

The quality of the tau decay product kinematic variable
reconstruction is exhibited in Figure~\ref{fig:rhokin}.  The
reconstructed and true distributions for the cosine of the angle
$\theta^*$ between the rho and the tau in the tau rest frame, and
the cosine of the angle $\beta$ between the charged pion and rho in
the rho rest frame are shown.   The true and reconstructed
distributions for the optimal observable $\omega$ for the decay
$\rho^-\nu_{\tau}$ are show in Figure~\ref{fig:omegarho}.

The testing tau pair Monte Carlo sample corresponds to a luminosity
of 250~fb$^{-1}$ for each of the intial state polarizations 80eL and
80eR . After performing the least squares fits the mean tau
polarizations are $<P_{\tau}>$=-0.611$\pm$0.009~stat.$\pm$0.005~sys.
for the 80eL sample and
$<P_{\tau}>$=0.501$\pm$0.010~stat.$\pm$0.006~sys. for the 80eR
sample, where the systematic errors are due to the finite training
sample statistics.

\subsection{ $e^+e^-\to t\bar{t}\, , \,\; t \to bW^+\, , \,\; W^+ \to q \bar{q'}\,$, $\, \sqrt s $=500 GeV}

The top quark is substantially more massive than the rest of the
observed quarks~\cite{Amsler:2008zzb}. It is therefore important to search for the physical
mechanism that introduces the mass difference and at the same time to use
the top as a probe into new physics, which often couples to mass. In
order to investigate such mechanisms it is essential to
determine precisely the mass of top and its couplings to the SM bosons. The aim of the analysis is to measure the total top
cross-section of the channel $e^+e^- \rightarrow t\bar{t},
t \rightarrow bW,
W \rightarrow q\bar{q}$, the mass and the
forward backward asymmetry ($A_{FB}$) of the top quark and also of the bottom quark
 from the top quark decay.

The sample used in the analysis is the inclusive SM sample which
includes the top quark production with mass of 174 GeV. Two separate
samples with the top mass 174 and 173.5 GeV have been generated to
evaluate precision of the mass reconstruction.

The analysis considers only the hadronic decay modes with the
largest available statistics. The six jet final state includes two
b-quarks from the top decay. The first step of the analysis is to
identify the hadronically decaying top quarks simultaneously
rejecting the non-top SM background, leptonic top decays and poorly
reconstructed hadronic top decays. The analysis starts with
clustering of reconstructed particles in to six jets using the y-cut
algorithm. All events with jets composed of a single reconstructed
electron or muon are discarded. A list of further selections used in
the analysis is given in Table~\ref{tab:top1}.

\begin{table}[h]
\begin{tabular}{lll}
\hline &selection & value  \\
\hline
& E$_{total}$       & $>$ 400 GeV \\
& log(y$_{min}$) & $>$ -8.5 \\
& number of particles in event & $>$ 80 \\
& number of tracks in event    & $>$ 30 \\
50 GeV $<$ & W mass & $<$ 110 GeV \\
& NN$_{b-tag}$ output for the most b-like jet & $>$ 0.9 \\
& NN$_{b-tag}$ output for the 2$^{nd}$ most b-like jet & $>$ 0.4 \\
& Sum of NN$_{b-tag}$ outputs for all jets & $>$ 1.5 \\
\hline
\end{tabular}
\caption{Selections for the top analysis.}
\label{tab:top1}
\end{table}

\begin{figure}[htbp]
\begin{center}
\subfigure{
\includegraphics[scale=0.35]{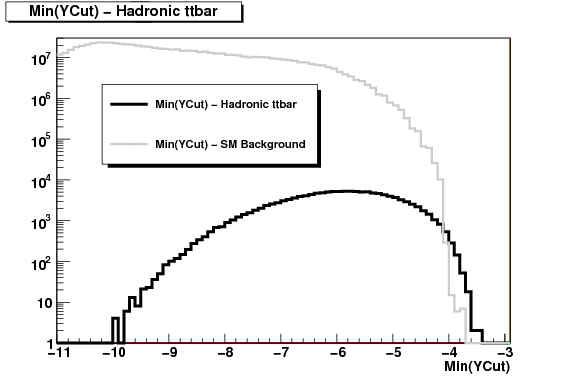}
}
\subfigure{
\includegraphics[scale=0.35]{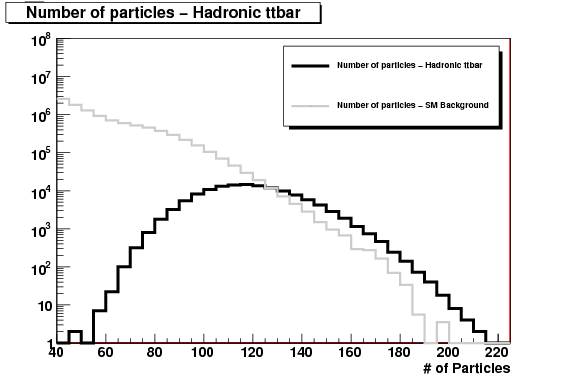}
}
\end{center}
\caption{Distributions of y$_{min}$ (a) and N$_{particles}$ (b) before the selection.}
\label{fig:ycut}
\end{figure}

In particular, the y$_{min}$ variable is a good indicator that an
event is topologically consistent with a six jet hypothesis. For
example, in a case where two jets overlap in the detector, these
jets will be merged by the jet clustering algorithm. However due to
the six jet requirement another jet of the event will have to be
separated into two reconstructed jets. This will result in the
substantially lower y$_{min}$ value and therefore the rejection of
the event.  Events with jets outside of the acceptance will have a
substantial missing energy and will be rejected by the total energy
requirement. Figure \ref{fig:ycut} shows examples of variables,
y$_{min}$ and N$_{particles}$, used in the top selections.

\begin{figure}[htbp]
\begin{center}
\subfigure{
\includegraphics[scale=0.35]{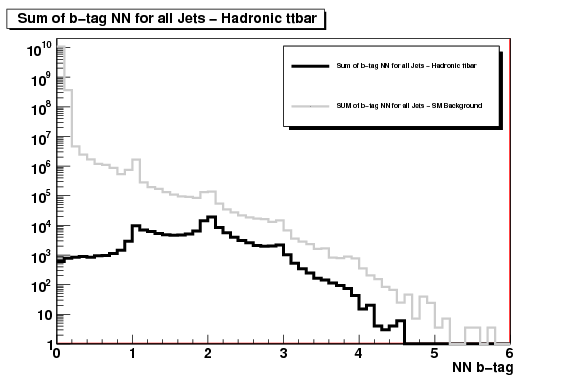}
}
\subfigure{
\includegraphics[scale=0.35]{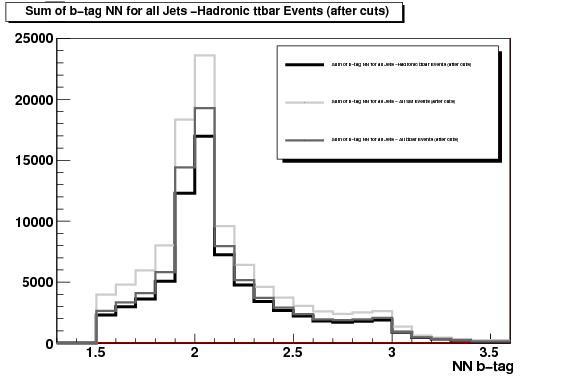}
}
\end{center}
\caption{Sum of b-tagging NN outputs for all jets for top signal and background events before (a) and after (b) selections.}
\label{fig:btagmtop}
\end{figure}

Another powerful selection criteria for the top events is tagging of
two b-quarks necessarily present in the signal. This substantially
reduces the combinatorial background in the top quark events by
decreasing the number of possible jet permutations. Figure
\ref{fig:btagmtop} shows the sum of b-tagging NN outputs for all
jets before (a) and after (b) the selections.

The signal efficiency of the above selections is equal to 51.2\%
which amounts to approximately 73000 signal events in the final
sample. About 99.97\% of di-lepton top decay channels and 93.9\%
single leptonic modes are rejected. The total remaining SM
background comprises about 33000 events which corresponds to the
purity of 68.8\%, where purity is defined as number of events
passing all cuts that originated from two b quarks and four other
quarks at the parton level divided by all events that pass the cuts.

The next step of the analysis is the calculation of the top mass.
A kinematic fitter was employed to reduce the combinatorial background of the six-jet environment and
to take advantage of the existing kinematic constraints of top decays, see Table~\ref{tab:top2} for the full list of constraints
used for this analysis.

\begin{center}
\begin{table}[h]
\begin{tabular}{ll}
\hline
mass(top1) & = mass(top2) \\
mass(W1)   & = 80.4 GeV \\
mass(W2)   & = 80.4 GeV \\
E$_{total }$ & = 500 GeV \\
p$_{x}$; p$_{y}$; p$_{z}$ &= 0 \\
\hline
\end{tabular}
\caption{Kinematic fitting constraints.}
\label{tab:top2}
\end{table}
\end{center}

The two most b-like jets were considered to be originating from
b-quarks and therefore were not used in permutations for
reconstruction of two W bosons. All the events that meet the fitting
constrains with probability less than 10\% were rejected. This
requirement selects only well reconstructed top events and is quite
sensitive to the jet energy and angular resolutions which are the
input parameters for the fitter. Although this selection reduces the
efficiency to 31.3\%, but it  also increases the purity to 85.1\%
and overall improves the precision of mass resolution.

\begin{figure}[htbp]
\begin{center}
\subfigure{
\includegraphics[scale=0.4]{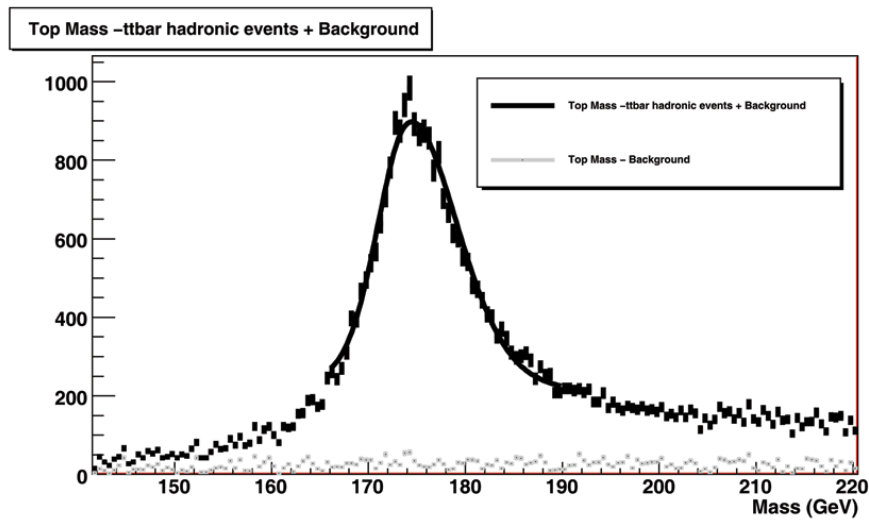}
}
\end{center}
\caption{Distribution of the top invariant mass after the kinematic fitting and probability cut.}
\label{fig:topmass}
\end{figure}

It is now possible to perform a mass fit using an asymmetric double
Gaussian convoluted with a Breit-Wigner as a signal fit function.
The function is fitted on a separate signal sample in the mass range
of 166 GeV - 200 GeV where the parameters of the Gaussians are
determined keeping all other parameters fixed. The Standard Model
background is then described by a second order polynomial function.
The combined function is fitted on the data sample with all the
parameters fixed for the exception of the top mass and the
Breit-Wigner width. This method results in the top mass of $173.918
\pm 0.053$ GeV thus predicting the accuracy of about 50 MeV of the
top mass determination with 500 fb$^{-1}$. The distribution of the
top mass and the corresponding fit is shown in
Figure~\ref{fig:topmass}. It is also possible to estimate the
$t\bar{t}$ cross-section in the hadronic channel as $284.1 \pm 1.4 $
fb.

The determination of top quark mass uncertainty can be calculated
also using a template method. The data sample, without the SM
background, is compared with a template produced at the same mass
and with a template with the top mass shifted by 0.5 GeV. From the
reconstructed top mass distribution the chi-square between the two
can be calculated as:
\begin{equation}
\chi^2_1 = \sum^{Nbins}_{i=0} \frac{(y_{template
1,i}-y_{data,i}+\delta_i)^2}{\sigma^2_{template
1,i}+\sigma^2_{data,i}+\sigma^2_{SM,i}}
\end{equation}
where $y_{...i}$ denotes the content of the $i$th bin. The $\delta$
term is added as a Gaussian smearing of the central value of
$y_{template,i}-y_{data,i}$ to take the SM background into account. The
calculated $\chi^2$ divided by degrees of freedom is close to one and does not depend on the bin width.

Similarly we can obtain $\chi^2_2$ between 'data' and the other
template sample. From $\chi^2_1$ and $\chi^2_2$ we can fit a
$\chi^2$ parabola with respect to $\Delta m_{top}$, assuming
$\chi^2_1$ is the minimum~\cite{lyonsbook}. The half width of the
parabola where $\chi^2 $ equals $ \chi^2_1 + 1$ gives the
uncertainty of top mass. With the method described above it is
possible to measure $m_{top}$ with a resolution of 45 MeV. For this
measurement the top candidate mass was restricted between 150 and
200 GeV. We note that the relation between the value of the measured
top quark mass of this analysis and the mass that comes out of the
measurement of the top quark threshold cross section are both
expected to have a systematic bias.  Both quantities can be related
to m$_t(\bar{MS})$ by precise calculations~\cite{topMS}.

Finally the last aim of the analysis is to determine the top couplings analyzing the forward-backward asymmetry of the top
quark and of the b quark from the top decay. A precise measurement of both will allow to set limits on
the anomalous couplings of Z$\rightarrow t\bar{t}$ and
$t\rightarrow$Wb respectively~\cite{Boos:1999ca}.

\begin{figure}[htbp]
\begin{center}
\subfigure{
\includegraphics[scale=0.35]{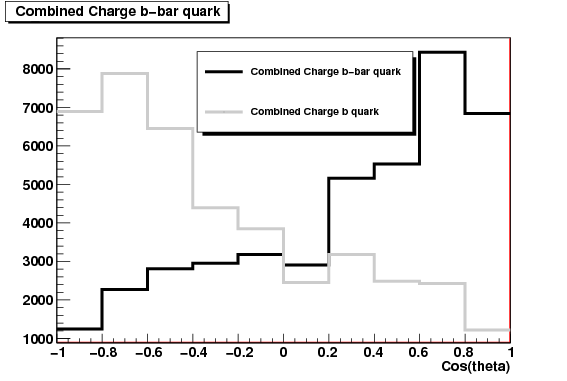}
}
\subfigure{
\includegraphics[scale=0.35]{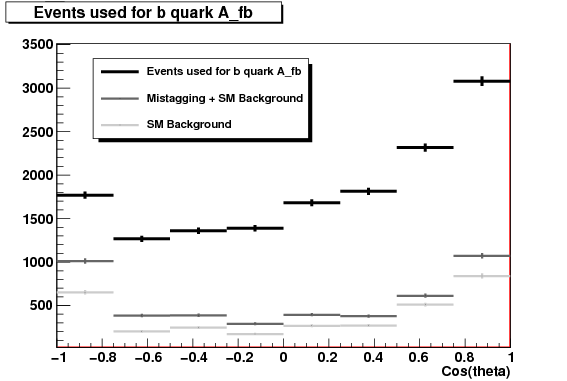}
}
\end{center}
\caption{ Combined quark charge discriminant for b and anti-b quarks
(left).  Number of events as function of $\cos\theta$ used for
calculation of $A_{FB}$ for the b-quark (right). The histograms are
inclusive corresponding to the top signal with correct and incorrect
charge assignment, and SM background.} \label{fig:chrgafb}
\end{figure}

    For this purpose, one must be able to discriminate between b quarks and
b anti-quarks. The conceptually simplest way to do this is by
measuring the vertex charge, defined as the sum of the charges of all
particles associated to a secondary vertex.  However, there are other
variables that are correlated with the primary quark charge, and these can
be combined to make a more powerful discriminant.  For this analysis, we
use a discriminant built from the momentum-weighted vertex charge
and the momentum-weighted jet charge~\cite{Abazov:2006qp}. The performance of this
discriminant is shown in Figure~\ref{fig:chrgafb} (a).

Figure~\ref{fig:chrgafb}(b) shows the reconstructed number of top
candidates as function of $\cos\theta$ of the b-quark direction with
respect to the direction of the electron beam. This information is
used as input for calculation of the forward-backward asymmetry of
b-quark in top decays which was determined using the combined quark
charge method described above. By using this method the resulting
b-quark forward-backward asymmetry in 500 fb$^{-1}$ is determined to
be $A^b_{FB} = 0.272 \pm 0.015$, consistent with the generated one.

Calculation of the top quark asymmetry represents another
complication as the reconstructed b-quark must be correctly
associated with its W boson. This pairing is done using the output
of the kinematic fitter. We note that the flavour of the top quark
is determined by the flavour of the b-quark so the charge
determination does not change; however the flight direction of the
quark may change. The resulting top quark forward-backward asymmetry
is calculated to be equal to $A^t_{FB} = 0.342 \pm 0.015$.

The selection efficiency for events where the quark charge can be
determined with sufficiently high purity is substantially lower,
7.1\%, than the one previously presented for the top mass. This is
due to the additional 155-195 GeV mass window selection and the fact
that the selection  on the combined charge effectively rejects a
large sample of the B$^{0}$ mesons which have low purity of the
quark charge determination and also rejects events in which a
secondary vertex has not been reconstructed in both b-tagged jets.
The selection on the top charge was done by requiring the product of
combined charges of two b-quarks in the event to be smaller than
-0.3.

\subsection{ $e^+e^-\to \tilde{\chi} _1^+ \tilde{\chi} _1^- $/$\tilde{\chi} _2^0 \tilde{\chi} _2^0\,$,  $\, \sqrt s $=500 GeV}

\newcommand{\charginop} {\tilde{\chi} _1^+}
\newcommand{\charginom} {\tilde{\chi} _1^-}
\newcommand{\charginopm} {\tilde{\chi}_1^{\pm}}
\newcommand{\neutralinoone} {\tilde{\chi} _1^0}
\newcommand{\neutralinotwo} {\tilde{\chi} _2^0}

The masses of charginos and neutralinos are important parameters in
supersymmetry which can be measured at ILC with high precision. In this analysis the
following processes were used for the measurement:
\begin{eqnarray*}
e^+ e^-\;&\rightarrow&\;\charginop \charginom \;\rightarrow\;
\neutralinoone \neutralinoone W^+ W^- \\
e^+ e^-\;&\rightarrow&\;\neutralinotwo \neutralinotwo
\;\rightarrow\; \neutralinoone \neutralinoone Z^0 Z^0
\end{eqnarray*}

Here we only discuss the SUSY point at which the $\charginopm$ and
$\neutralinotwo$ decay by emission of on-shell W and Z
bosons~\cite{tsukamoto:1995}. If we consider the all-hadronic decay
of the gauge bosons in the final states, the above processes will
both have a signature of four jets with large missing energy. Once
the W or Z bosons are successfully reconstructed the shapes of their energy
distributions will provide information on the chargino or neutralino
masses. The uncertainty of the mass measurement is determined by the
template fitting.

\begin{figure}[htbp]
\begin{center}
\subfigure[]{
\includegraphics[scale=0.35]{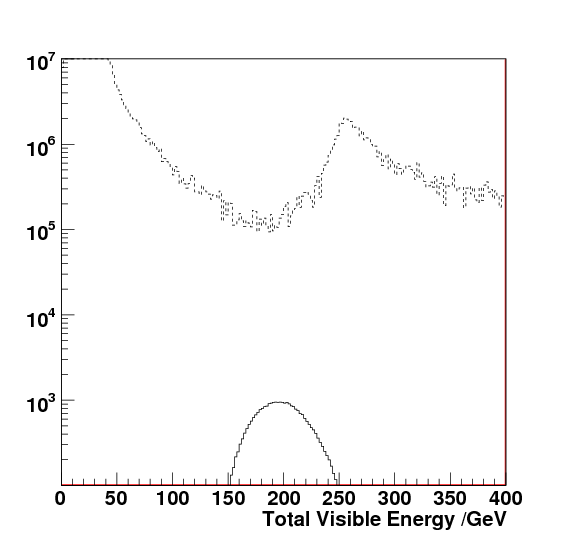}
} \subfigure[]{
\includegraphics[scale=0.35]{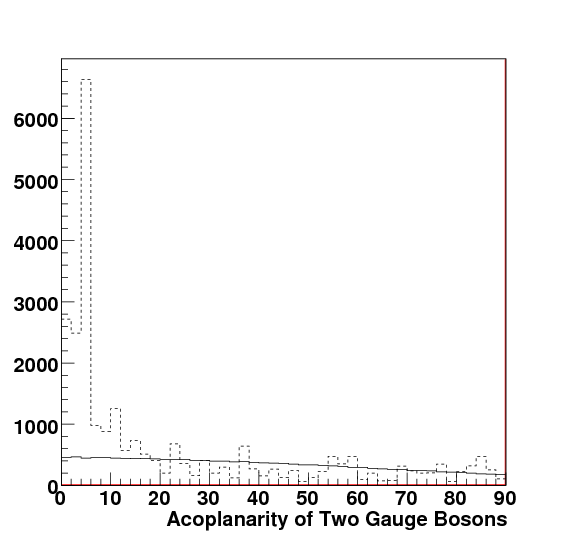}
} \caption{ Examples of selections for chargino
analysis. Solid lines are chargino signal and dashed lines are
SM background. (a)Total visible energy; (b)Acoplanarity
of two reconstructed W bosons. }
\label{fig:cuts}
\end{center}
\end{figure}

A Monte-Carlo SUSY sample with $\sqrt{s}$ of 500 GeV used for this
study included the chargino/neutralino signals and SUSY backgrounds
such as $e^+e^-\rightarrow \neutralinoone\neutralinotwo$ and slepton
pair production. Apart from this sample, six more samples with the
same statistics were generated in order to estimate the uncertainty of
these mass measurements, each with a corresponding SUSY mass shifted by either
$\pm 0.5$~GeV,
see Table~\ref{tab:susysample}. All the samples have one
polarization, namely 80\% electrons left-handed and 30\% positions
right-handed and correspond to the integrated luminosity of 500
fb$^{-1}$. The production cross-section of chargino events with all
hadronic final state is $57$ fb which is considerably larger than
that of the neutralino, $11$ fb.
\begin{center}
\begin{table}[h]
\begin{tabular}{cccc}
\hline sample & $m_{\neutralinoone}$ (GeV) & $m_{\charginopm}$
(GeV)& $m_{\neutralinotwo}$ (GeV)  \\
\hline Reference & 115.7 & 216.7 & 216.5\\
  $m_{\neutralinoone}$ + 0.5 & 116.2 & 216.7 & 216.5 \\
  $m_{\charginopm}$ + 0.5 & 115.7 & 217.2 & 216.5 \\
  $m_{\neutralinotwo}$ + 0.5 & 115.7 & 216.7 & 217.0 \\
  $m_{\neutralinoone}$ - 0.5 & 115.2 & 216.7 & 216.5 \\
  $m_{\charginopm}$ - 0.5 & 115.7 & 216.2 & 216.5 \\
  $m_{\neutralinotwo}$ - 0.5 & 115.7 & 216.7 & 216.0 \\
\hline
\end{tabular}
\caption{Parameters of the MC SUSY samples.} \label{tab:susysample}
\end{table}
\end{center}
To identify events of  the all-hadronic channel all reconstructed
particles are clustered into four jets and selections described in
Table~\ref{tab:cuts} are applied.
\begin{center}
\begin{table}[h]
\begin{tabular}{ll}
\hline cut & value  \\
\hline
$E_{jet}$ & $>$ 10 GeV \\
Fraction of EM energy in each jet & $<$ 80\% \\
Number of tracks & $>$ 20\\
Total visible energy & $<$ 250 GeV \\
Thrust & $<$ 0.85\\
$\cos\theta_{thrust}$ & $<$ 0.9\\
$\theta$(1, 2) & $>$ 60$^{\circ}$\\
$\theta$(1, 3), $\theta$(1, 4) , $\theta$(1, 3)& $>$ 40$^{\circ}$\\
$\theta$(2, 4), $\theta$(3, 4) & $>$ 20$^{\circ}$\\
Acoplanarity of two reconstructed gauge bosons & $>$
$10^{\circ}$\\
\hline
\end{tabular}
\caption{Pre-selection cuts. $\theta (i, j)$ defines the angle
between the $i$th and $j$th jet. The jets are ordered in energy,
e.g. jet 1 is the most energetic jet.}
\label{tab:cuts}
\end{table}
\end{center}
These selections efficiently suppress the Standard Model background
as illustrated in Figure~\ref{fig:cuts} which shows distributions of
total visible energy and acoplanarity before the selections.

\begin{figure}[htbp]
\centerline{\includegraphics[scale=0.7]{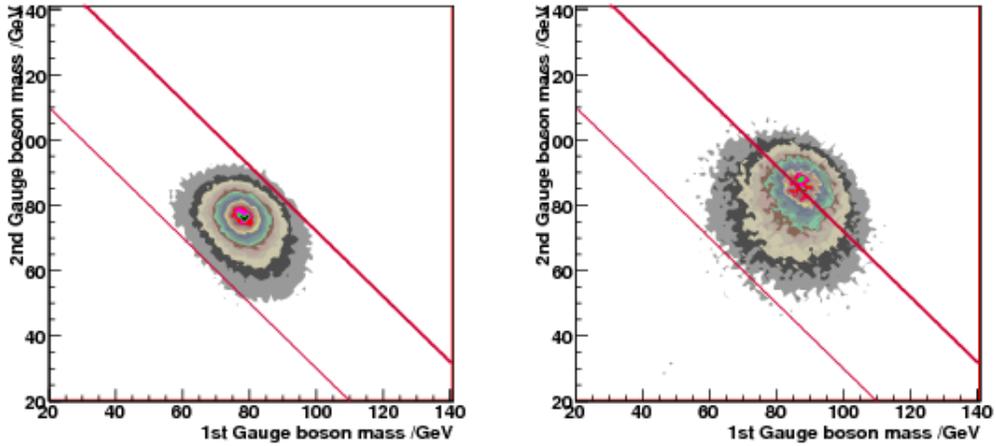}}
\caption{The reconstructed boson masses from the four jets,
selecting chargino events. Left: pure chargino signal; right: pure
neutralino signal. The region between the two straight lines shows
the allowed chargino selection window.} \label{fig:mass2dCh}
\end{figure}

To reconstruct the two gauge bosons from four jets we have to
determine the correct jet pairing.
It is done by minimizing the quantity of $(m(j_1,
j_2)-m_{W/Z})^2+(m(j_3, j_4)-m_{W/Z})^2$, where $m(j_1, j_2)$ is the
invariant mass of the jets $j_1$ and $j_2$ and all possible jet
permutations are considered. W mass is used when selecting
$\charginop \charginom$ events, and Z mass for $\neutralinotwo
\neutralinotwo$ events. The correlation of the two reconstructed
boson masses is shown in Figure~\ref{fig:mass2dCh}. The
chargino and neutralino events will populate different regions in the
mass plane and therefore can be distinguished by a selection indicated by a thick line in
Figure~\ref{fig:mass2dCh}. Events above the line are
classified as $\charginop \charginom$ events and below the line as
$\neutralinotwo \neutralinotwo$ events. Another selection is applied for the chargino events
to remove the events on the lower
left corner of the mass plane, which are due to the
remaining leptonic decays of the bosons.

\begin{figure}[htbp]
\begin{center}
\subfigure[]{
\includegraphics[scale=0.35]{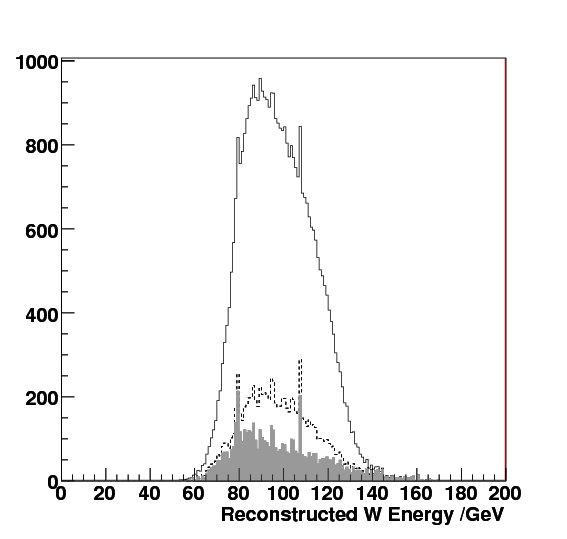}
} \subfigure[]{
\includegraphics[scale=0.35]{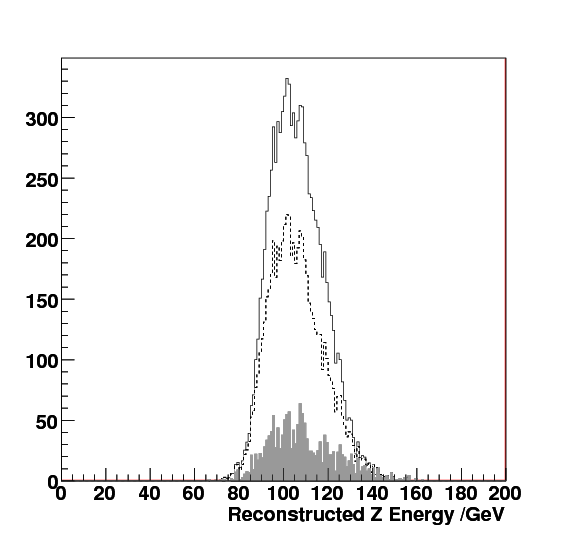}
} \caption{ (a) The stack histogram of reconstructed W energy for the
chargino selection. The solid line shows the combined chargino
signal and all backgrounds including neutralino, the dashed line is neutralino events and SM background, the solid grey line is the
SM background only. (b) Reconstructed Z energy for neutralino
selection, the solid line is the neutralino signal and all backgrounds including chargino, dashed line is the
chargino events and SM background, the solid grey line is the
SM background only. The spikes
in (a) are due to the large weight of a few events in the SM background
sample. }\label{fig:selChNeu}
\end{center}
\end{figure}

After these selections the reconstructed boson energy distributions
are shown in Figure~\ref{fig:selChNeu}. The chargino sample contains
15362 signal events with 1997 neutralino pair background and 2980 SM background events. Similarly
the neutralino sample contains 1659 signal events with 2395
chargino and 865 SM background events. Contribution of other SUSY backgrounds are negligible in both cases.
The uncertainty on the cross section measurement for chargino and
neutralino pair production with all hadronic final state is 0.9\%
and 4.2\% respectively neglecting the uncertainty on efficiency.

\begin{figure}[htbp]
\begin{center}
\includegraphics[scale=0.4]{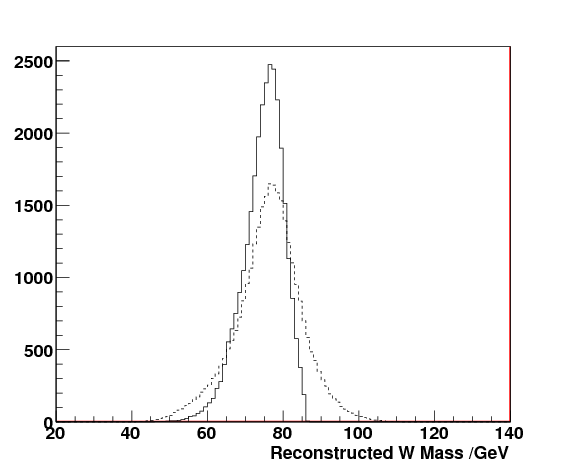}
\caption{Mass distribution of reconstructed bosons before (dashed
line) and after (solid line) kinematic fitting. }\label{fig:kinfit}
\end{center}
\end{figure}

In the rest frame of $\charginopm$/$\neutralinotwo$, W$^{\pm}$/Z
bosons are monochromatic and their energy is determined by
$m_{\charginopm}/m_{\neutralinotwo}$, $m_{W^{\pm}}/m_Z$ and
$m_{\neutralinoone}$. For a boosted parent
$\charginopm$/$\neutralinotwo$, the boson energy distribution
depends on the SUSY particle masses which will determine the edges
of the energy spectrum.  The masses can be measured
comparing the energy spectrum in the 'data' to reference MC samples as in
Table~\ref{tab:susysample}. Thus, the best possible energy
resolution is required for a high precision mass measurement. Using
the fact that the two bosons reconstructed from the four jets have
the same mass in both $\charginop\charginom$ and
$\neutralinotwo\neutralinotwo$ cases, kinematic fitting with one
constraint ($m_{boson1}=m_{boson2}$) can be used for improving the
energy distribution assuming that the detector energy and angular resolution for jets is known.
Kinfit in Marlinreco package is used for the fitting. The comparison
of reconstructed W mass with and without fitting is shown in
Figure~\ref{fig:kinfit}.

The determination of SUSY particle mass uncertainty is explained
below on the example of  $m_{\charginopm}$. The reference
$\charginop\charginom$ sample (without SM background) is split into
two unequal sub-samples, 'data' and 'template 1', with the latter one
having much higher statistics. From the normalized reconstructed W energy
distribution the chi-square between the two can be calculated as:
\begin{equation}
\chi^2_1 = \sum^{Nbins}_{i=0} \frac{(y_{template
1,i}-y_{data,i}+\delta_i)^2}{\sigma^2_{template
1,i}+\sigma^2_{data,i}+\sigma^2_{SM,i}}
\end{equation}
where $y_{...i}$ denotes the content of the $i$th bin. The $\delta$
term is added as a Gaussian smearing of the central value of
$y_{template,i}-y_{data,i}$ to account for fluctuations of the SM background. The
resulting $\chi^2/NDF$ is close to one and does not depend on the binning.

Similarly we can obtain $\chi^2_2$ and $\chi^2_3$ using template samples
with $m_{\charginopm}$ shifted by $+0.5$ and $-0.5$~GeV, respectively.
We fit a $\chi^2$ parabola with respect to $\Delta
m_{\charginopm}$, assuming that $\chi^2_1$ is the minimum and using
$\chi^2_2$ and $\chi^2_2$ as second and third points. The half width
of the parabola where $\chi^2 $ equals $ \chi^2_1 + 1$ gives the
uncertainty on the chargino mass~\cite{lyonsbook}.

With the method described above for the $\charginop\charginom$
process we can measure $m_{\charginopm}$ and $m_{\neutralinoone}$
with uncertainties of 450 MeV and of 160 MeV respectively. The
$\neutralinotwo\neutralinotwo$ process gives uncertainties of 490
MeV for $m_{\neutralinotwo}$ and 280 MeV for $m_{\neutralinoone}$.
The larger uncertainties for the latter are expected because of the
smaller cross-section with respect to the $\charginop\charginom$
process, which yields lower statistics as well as inferior purity of
the sample. A small improvement of the precision can be achieved by
a simultaneous fit to the both chargino and neutralino samples.

\subsection{$e^+e^-\to \tilde{b}\tilde{b}\, ,\,\,\tilde{b}\to b \tilde{\chi}_1^0 $}

\begin{figure}[htbp]
\centerline{
\includegraphics[width=9cm,height=7cm]{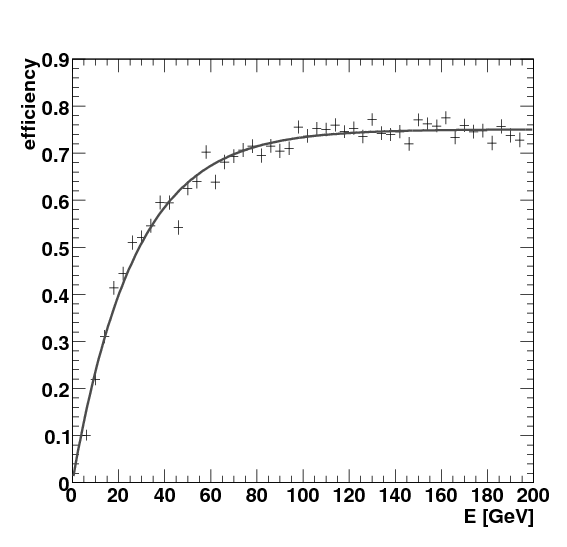}
\includegraphics[height=7cm]{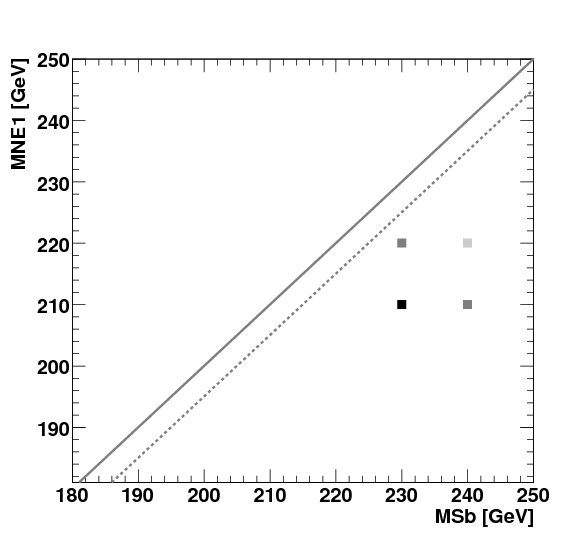}
} \caption{Jet b-tagging efficiency as a function of the jet energy
for b-jets (left). Signal mass points selected for this analysis
(right). The dashed line corresponds to the kinematic limit due to a
small difference between $m_{\tilde{b}}$ and
$m_{\tilde{\chi}_1^0}$.} \label{fig:btag_efficiency}
\end{figure}

Recent measurements of the neutralino relic density suggest a small
mass splitting between the neutralino and the next to lightest SUSY
particle (NLSP) assuming that neutralino accounts for the measured
dark matter content of the universe. The ILC has a potential to
cover a large fraction of the parameter space motivated by this SUSY
scenario. In this analysis we assume that the supersymmetric partner
of b-quark, sbottom, is the NLSP \cite{sbottom}. Such a measurement
is challenging due to very low energy of jets, below 20-30~GeV,
which pushes the jet clustering and tagging algorithms to their
limits. Figure~\ref{fig:btag_efficiency} shows that while the
b-tagging efficiency is about 75\% above 60~GeV, it is falling
steeply for lower jet energies.

In this study the CalcHEP event generator~\cite{calchep} was
employed to generate SUSY signal events. The events were converted
to Les Houches format, passed to Pythia for their fragmentation and
particle decays and consequently to the usual SiD full simulation
and reconstruction chain. The event classification is based on a
Neural Network using the FANN package~\cite{fann}. Several points in
the MSSM parameter space were chosen close to kinematic limits in
order to investigate the discovery potential. These points
correspond to various masses of the sbottom quark and neutralino:
(230, 210); (240, 210); (230,220); (240,220), all in GeV, see the
right plot in Figure~\ref{fig:btag_efficiency}. For each signal
point 200k events were generated, also accounting for ISR, FSR and
beamstrahlung.

The mass of sbottom quark $m_{\tilde{b}}$ essentially determines the
cross section of the process, $1.3$ fb for  $m_{\tilde{b}} =
230\,$GeV but the cross section is 3.5 times smaller for
$m_{\tilde{b}} = 240\,$GeV. The mass difference between
$m_{\tilde{b}}$ and $m_{\tilde{\chi}_1^0}$, on the other hand,
determines the energy of b-jets.

Jets were reconstructed by employing the Durham $k_T$ algorithm with
$k_T^{min} = 10\,$GeV and limiting the maximum number of jets to
two. Events with a single reconstructed jet were later rejected as
well. The majority of the SM background is suppressed by the
$E_{visible} < 80\,$GeV selection. Signal events are misbalanced in
energy due to neutralinos while the dominant two-photon and $e^+e^-
\to q\bar{q}$ backgrounds produce jets in back-to-back topology
motivating the $\Delta R_{\eta\phi} < 3.0$ selection. To further
suppress the background a selection $max(|\eta_1|,|\eta_2|) < 2.0$
is required, where $\eta_1$ and $\eta_2$ are jet pseudorapidities.
Finally, the total number of particles in an event is required to be
within $10 \leq N_{particles} \leq 60$ for the mass point (230,210)
and it is adjusted for the other mass points individually.

\begin{figure}[htbp]
\centerline{
\includegraphics[width=7cm]{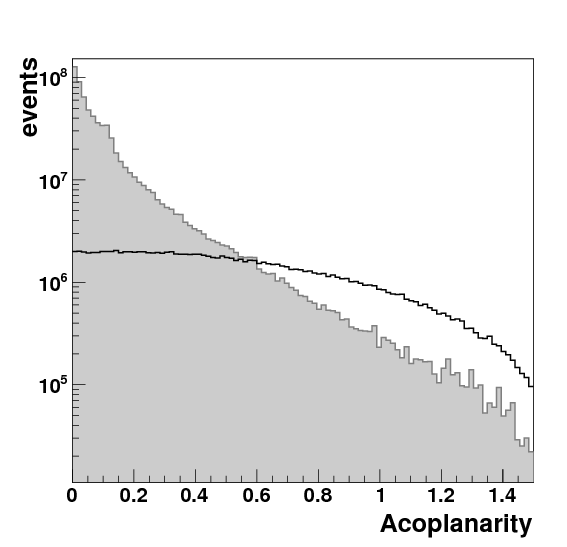}
\includegraphics[width=7cm]{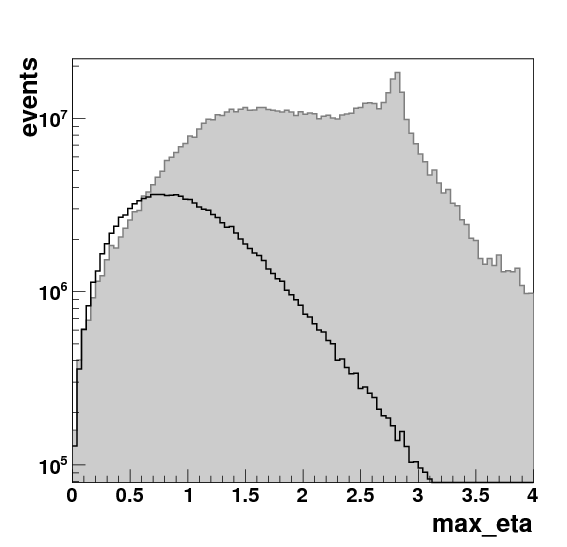}
} \caption{Examples of selection variables: acoplanarity (left) and
maximum pseudorapidity (right) of jets. Signal with $m_{\tilde{b}} =
230\,$GeV and $m_{\tilde{\chi}_1^0} = 210\,$GeV (black) is shown
together with the inclusive SM background (filled histogram). The
signal was scaled up by a factor of 100000.} \label{fig:variables}
\end{figure}

Figure~\ref{fig:variables} shows distributions of acoplanarity and
maximum pseudorapidity for the signal with $m_{\tilde{b}} =
230\,$GeV and $m_{\tilde{\chi}_1^0} = 210\,$GeV and inclusive SM
background. The signal was scaled up by a factor of 100000 for the
presentation purposes.

An important ingredient in this analysis is an electromagnetic veto
using the forward detector at very low polar angles, above $\theta =
10\,$mrad. This is used to suppress the two-photon background as
well as other backgrounds with a large ISR contribution. The forward
detector acceptance was estimated using a simple geometrical model.
An event is rejected if a photon or electron with $E>300\,$MeV is
detected within the acceptance.

\begin{figure}[htbp]
\centerline{
\includegraphics[width=7cm]{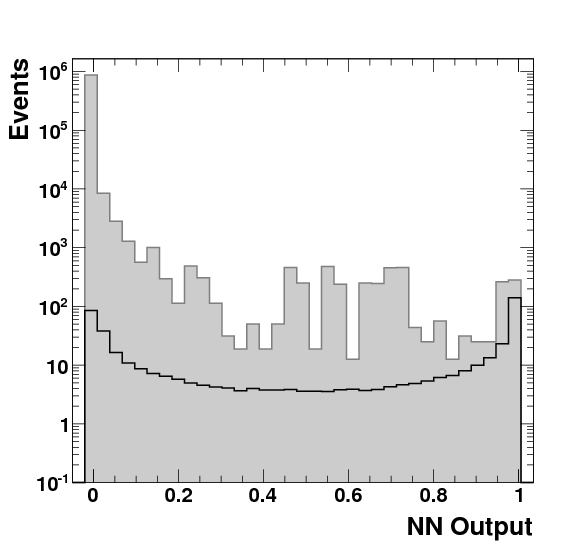}
\includegraphics[width=7cm,height=7cm]{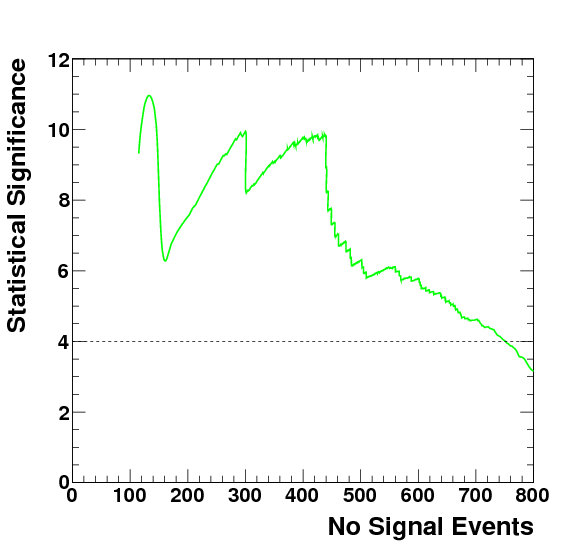}
} \caption{Neural Network output (left) is shown for events which
passed the basic selections for signal with $m_{\tilde{b}} =
230\,$GeV and $m_{\tilde{\chi}_1^0} = 210\,$GeV (black) and
inclusive SM background (filled histogram). Statistical significance
(right) is shown as a function of number of selected signal events.
Dashed line shows a limit of four standard deviations. Big
variations are caused by SM events with large weights.}
\label{fig:confidence_level}
\end{figure}

For the final event selection a Neural Network based on the above
variables was defined and trained on independent signal and
background samples. The Neural Net inputs were completed by adding
the flavour tag information for both jets, momentum isotropy,
acoplanarity and the total number of charged particles. The
resulting NN output is shown in the left plot of
Figure~\ref{fig:confidence_level}. The measurement is interpreted in
terms of signal significance calculated as $S/\sqrt{S+B}$, where $S$
is the number of selected signal events and $B$ is the number of
selected background events. The event numbers are normalised to the
total luminosity of $1000~$fb$^{-1}$.
Figure~\ref{fig:confidence_level} shows a distribution of
$S/\sqrt{S+B}$ as function of the number of selected signal events
with each bin corresponding to a particular cut on the NN classifier
output. Big variations are caused by the SM events with large
weights. Based on the above the signal cross section statistical
uncertainty was calculated to be equal to 15\% for the (230,210)
mass point. The mass points (240, 210); (230,220); (240,220) all can
be excluded at the 95\% CL. We conclude that the exclusion region
can be extended very close to the ILC kinematic limits and, in case
of discovery, the sbottom production cross section can be measured
with a resonable statistical precision. This measurement relies on
good performance of the forward detector used to veto two-photon
background events and soft di-jet events with a dominant ISR
contribution.

\bibliographystyle{unsrt}



\chapter{Cost Estimate}
\label{chap:cost}

The SiD cost estimate is a construction cost estimate; it does not include
R{\&}D, commissioning, operating costs, or physicist salaries.

The SiD design incorporates a parametric cost model, which has been essential for ongoing detector optimization. Detector 
parameters (e.g. dimensions or masses) have been transferred to a Work Breakdown Structure (WBS) where a subsystem can be 
described to an arbitrary level of detail. Here we describe the LOI version of SiD as such a point design.

The parametric model of the detector comprises a large set of Excel
spreadsheets that encapsulate a self-consistent model of SiD. It is
straightforward to vary parameters ranging from the most basic, such
as the tracker radius and aspect ratio, to parameters such as the
number of tracking layers, the number and thickness of HCAL layers,
and calorimeter radiator material. The tracking layers and disks are
adjusted to fit the allocated space. The calorimeters are adjusted
to nest properly with the tracker. The solenoid model is adjusted
for its radius and field, and the flux return is adjusted to roughly
contain the return flux. For each system, the cost driving component
count, such as tungsten plates, silicon detectors, and readout chips
for the ECAL, are calculated. The model has tables for material
costs and estimates both materials and services (M{\&}S) and labor
costs that are associated with the actual scale of SiD. Costs that
are approximately fixed, for example engineering, fixturing, or
solenoid He plants, are imported from separate Work Breakdown
Structure programs. A set of macros are used to calculate the costs
of SiD as parameters are varied.

\begin{figure}[htbp]
\centerline{\includegraphics[angle=90,height=9in]{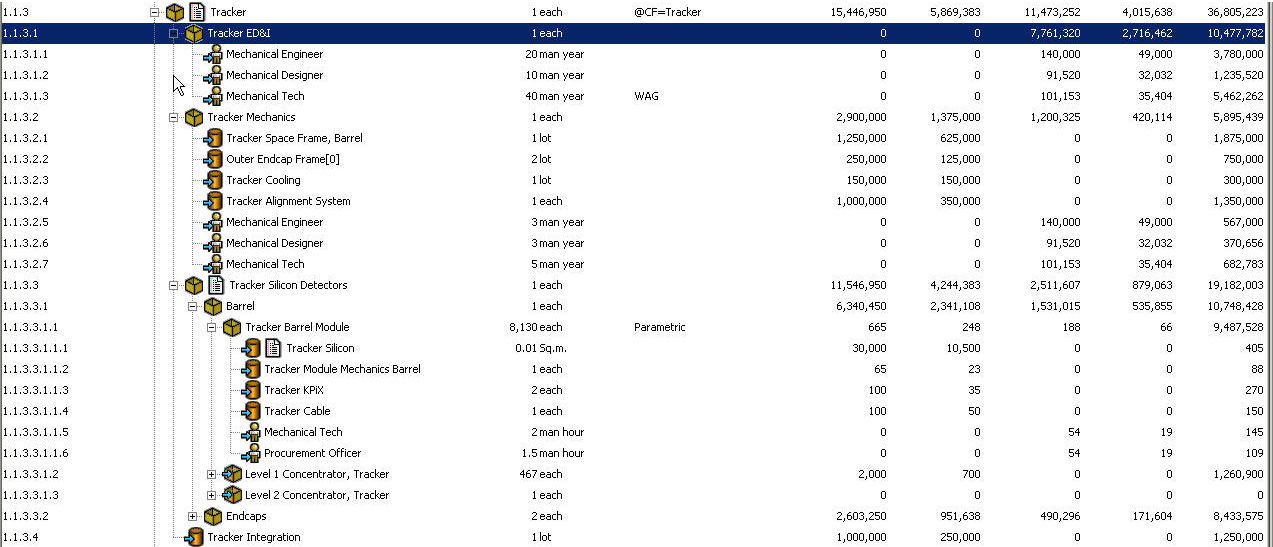}}
\caption{Example of WBS.}\label{fig:cost8}
\end{figure}

The WBS facilitates the description of the costs as a hierarchical
breakdown with increasing levels of detail. Separate tables describe cost
estimates for purchased M{\&}S and labor. These tables include contingencies
for each item, and these contingencies are propagated through the WBS.
The cost tables have been reviewed by the subsystem groups. The
quality of the estimate is considered roughly appropriate for a
Letter of Intent; it would not be defensible for a construction
readiness review. For illustration, sample of the WBS, for the tracker, is shown in
Figure~\ref{fig:cost8}.

The M{\&}S costs are estimated in 2008 US {\$}. Labor is estimated in man-hours or
man-years as convenient. For this estimate labor types are
condensed to engineering, technical, and clerical. The
statement of base M{\&}S and labor in man-years by the three categories
results in a cost which we believe is comparable to that used by the ILC
GDE, and is referred to here as the ILC cost.\footnote{ To expand this
to U.S. DOE style costs, the following steps are made:\par Contingencies are
assigned to M{\&}S and Labor. The purpose of the contingency is primarily to
make funding available to hold a schedule in the face of unforeseen
problems; to fund items that were forgotten in the estimate; and to provide
some relief from underestimates.\par Labor is transformed to a dollar value
by using SLAC salaries that include benefits but not overhead.\par Indirects
are computed as a fraction of the M{\&}S and Labor. SLAC large project
values are used. \par Escalation is computed assuming a start date, a 6 year
construction cycle, and an inflation rate. For this study, a start date of
2016 and an inflation rate of 3.5{\%}/year is assumed. The escalation is
both quite substantial and uncertain.\par \par \par }

While contingency is not used in
the ILC value system, it does give some idea of the uncertainties in the
costs of the detector components. Items which are close to commodities, such
as iron, have seen wild cost swings over the last few years. It
is also difficult to estimate silicon detector costs for such large
quantities.

There are a substantial set of interfaces in the IR hall.
For the purpose of this estimate, the following has been assumed:

\begin{itemize}
\item The hall itself, with finished surfaces, lighting, and HVAC are provided.
\item Utilities, including 480 VAC power, LCW, compressed air, and internet connections are provided on the hall wall.
\item An external He compressor system with piping to the hall is provided. The refrigeration and associated piping is an SiD 
cost.
\item Any surface buildings, gantry cranes, and hall cranes are provided.
\item Data storage systems and offline computing are provided.
\item Detector motion rails, for push-pull and opening in the beamline and garage positions, are installed by SiD in suitably 
prepared areas in the concrete floor.
\item The QD0s and their cryogenic systems are provided. The beampipe is an SiD cost.
\end{itemize}

\begin{figure}[htbp]
\centerline{\includegraphics[width=4in]{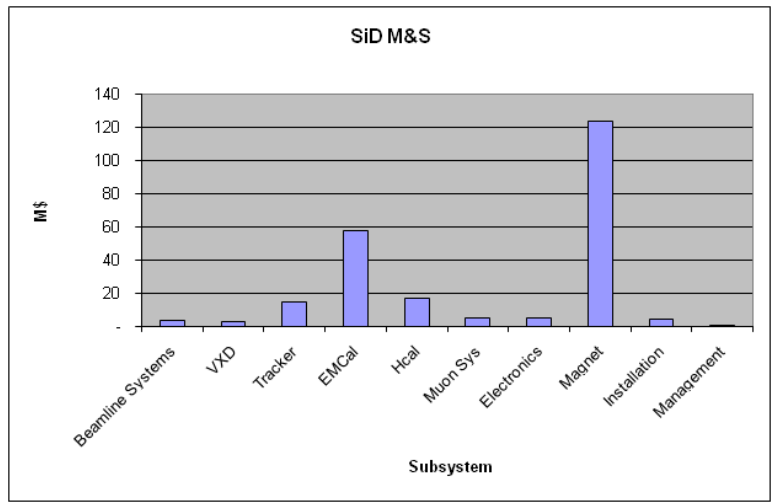}}
\caption{Subsystem M\&S Costs.}\label{fig:cost1}
\end{figure}

\begin{figure}[htbp]
\centerline{\includegraphics[width=4in]{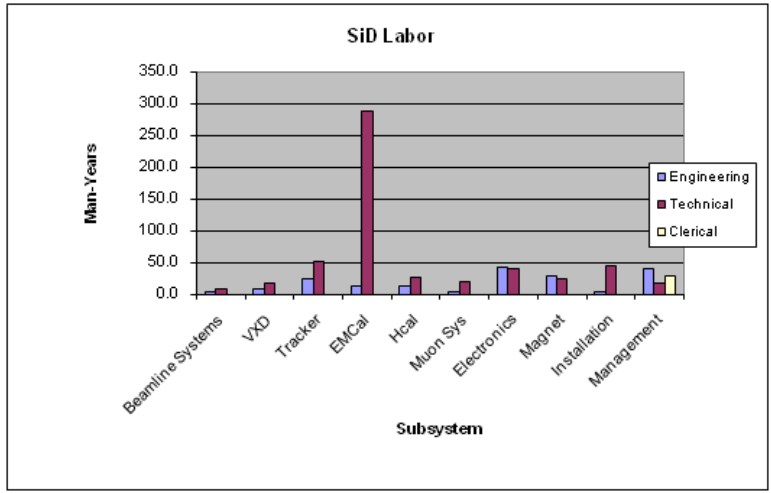}}
\caption{Subsystem Labor.}\label{fig:cost2}
\end{figure}

The subsystem level summary is shown in Table~\ref{tab:cost1}, the
M{\&}S costs are plotted in Figure~\ref{fig:cost1}, and the labor
costs are shown in Figure~\ref{fig:cost2}. The costs are dominated
by the Magnet and the ECAL. The magnet has roughly equal costs for
the superconducting coil and the iron. The ECAL is dominated by the
silicon detectors. The ECAL manpower is dominated by the operations associated with handling and testing the Si detectors and 
their associated KPiX and cable. The ECAL has relatively complex cables related to the thin gap between radiators. At this time, 
it is uncertain whether significant fractions of the handling and testing can be automated, so the estimate is conservative. Also, 
the tungsten plate assembly is complex, with a large number of spacer columns. This is again conservatively estimated.

\begin{table}[htbp]
\begin{center}
\begin{tabular}{|c|c|c|c|c|c|c|}
  & M\&S  & M\&S  & Engineering & Technical & Administrative \\
  & Base  & Contingency  &  &  &  \\
  &(M\$)       &(M\$)              & (MY)        & (MY) & (MY) \\ \hline
  LumCal and BeamCal & 3.68  & 1.42  & 4.0 & 10.0 & 0.0 \\
  VXD & 2.80  & 2.04  & 8.0 & 17.7 & 0.0 \\
  Tracker & 14.45  & 5.71  & 24.0 & 53.2 & 0.0 \\
  ECAL & 57.74  & 23.02  & 13.0 & 287.8 & 0.0 \\
  HCAL & 16.72  & 6.15  & 13.0 & 28.2 & 0.0 \\
  Muon System & 5.35  & 1.65  & 5.0 & 20.1 & 0.0 \\
  Electronics & 4.90  & 1.65  & 44.1 & 41.7 & 0.0 \\
  Magnet & 123.74  & 42.58  & 29.2 & 25.0 & 0.0 \\
  Installation & 4.10  & 1.08  & 4.5 & 46.0 & 0.0 \\
  Management & 0.92  & 0.17  & 42.0 & 18.0 & 30.0 \\ \hline
    Totals &   234  & 85  & 187  & 548  & 30  \\
\end{tabular}
\caption[Subsystem summary of costs.]{Subsystem summary of costs.
Labor contingency is not shown.}\label{tab:cost1}
\end{center}
\end{table}

\begin{figure}[htbp]
\centerline{\includegraphics[width=4in]{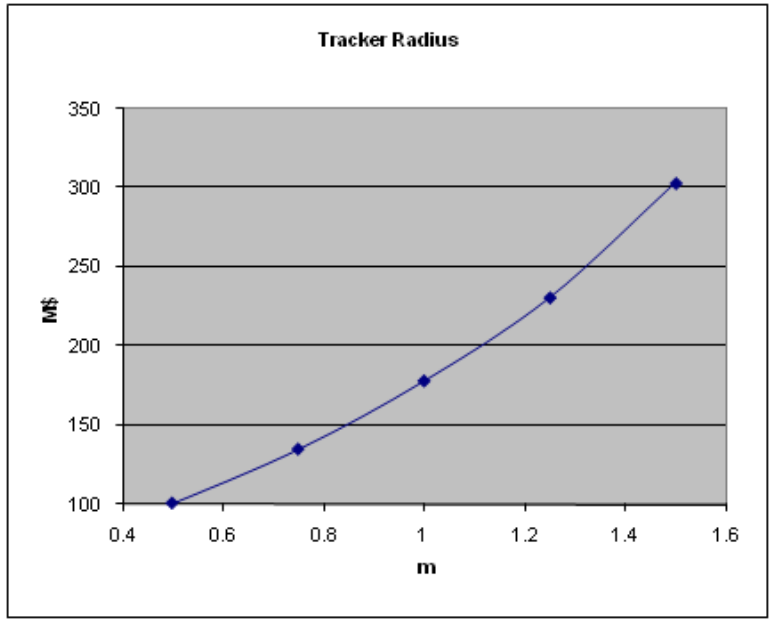}}
\caption{SiD Base M\&S Cost vs Tracker Radius.}\label{fig:cost3}
\end{figure}

\begin{figure}[htbp]
\centerline{\includegraphics[width=4in]{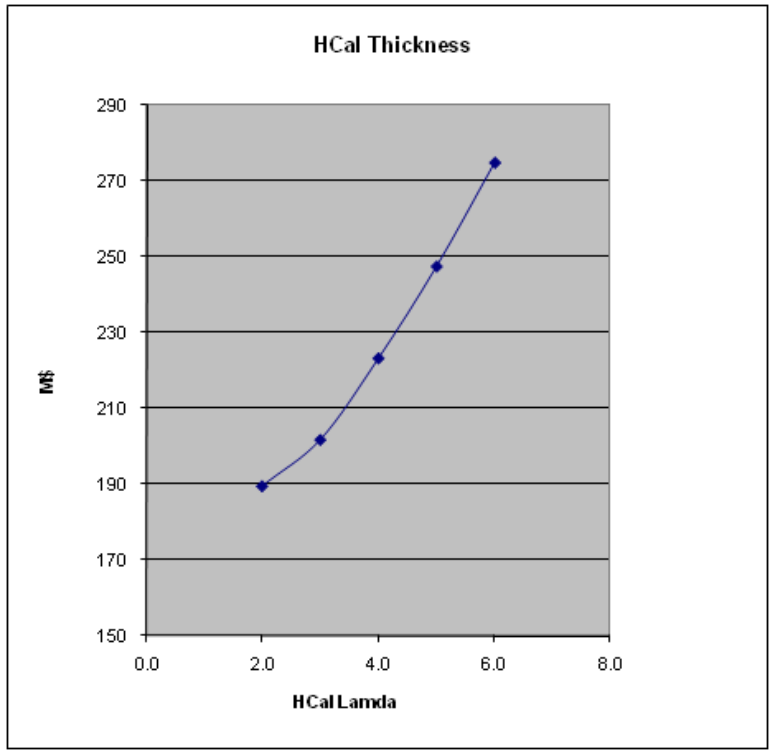}}
\caption{SiD Base M\&S Cost vs Hadronic Calorimeter
Thickness.}\label{fig:cost4}
\end{figure}

\begin{figure}[htbp]
\centerline{\includegraphics[width=4in]{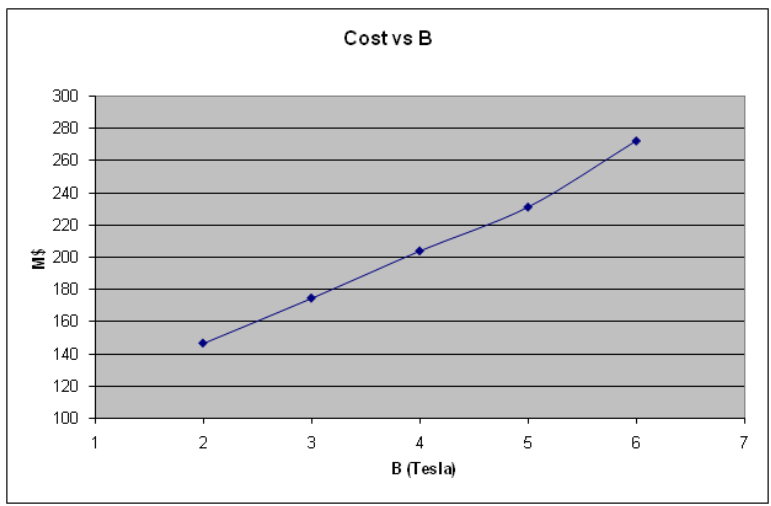}}
\caption{SiD Base Cost vs B.}\label{fig:cost5}
\end{figure}

\begin{figure}[htbp]
\centerline{\includegraphics[width=4in]{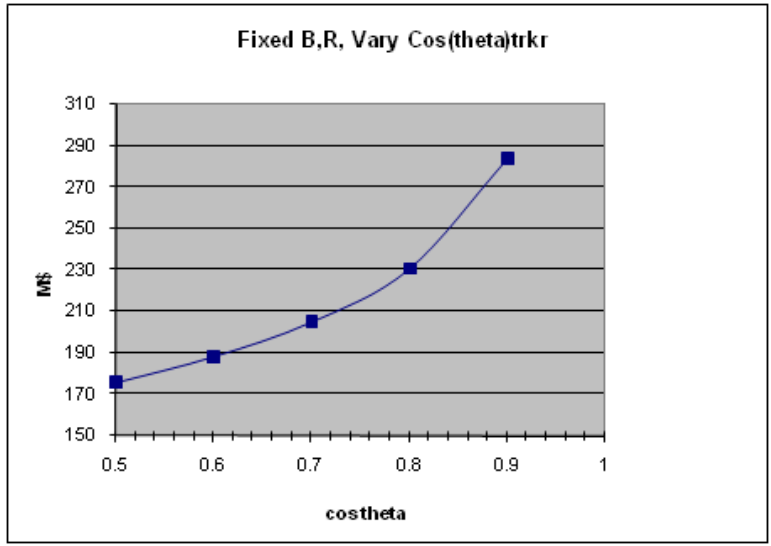}}
\caption{SiD Base Cost vs Tracker Aspect Ratio.}\label{fig:cost6}
\end{figure}

Figure~\ref{fig:cost3} through Figure~\ref{fig:cost6} show simple
examples of the parametric costs of SiD. In all cases, the non-varied parameters are the SiD baseline: tracker radius = 1.25 m, 
tracker polar-angle coverage, $|$cos$|\theta$ = 0.8, B= 5 T, and
HCAL thickness = 4.5 $\lambda $.
Finally,
physics performance versus PFA resolution is used to study physics
performance versus SiD cost. This is discussed in more detail in chapter~\ref{chap:introduction}.

The cost estimate has several important ``commodity'' items whose
costs have recently been fluctuating significantly. These
include most metals and processed Si detectors.
Table~\ref{tab:cost2} illustrated the cost sensitivity to these
prices by indicating the unit cost used in the estimate and the
effect on the SiD cost of doubling the unit cost.

\begin{table}[htbp]
\begin{center}
\begin{tabular}{|p{147pt}|l|p{185pt}|}
\hline
Item&
Nominal Unit Cost&
$\Delta $SiD Base M{\&}S Cost (M{\$}) \\
\hline
Magnet Iron (finished and delivered)&
{\$}7/Kg&
56 \\
\hline
Tungsten (powder alloy) plate&
{\$}88/Kg&
7 \\
\hline
Si Detector&
{\$}3/cm$^{2}$&
39 \\
\hline
HCAL Detector&
{\$}2000/m$^{2}$&
7 \\
\hline
\end{tabular}
\caption{Cost Sensitivity to selected unit costs.}
 \label{tab:cost2}
\end{center}
\end{table}

\begin{figure}[htbp]
\centerline{\includegraphics[width=4in]{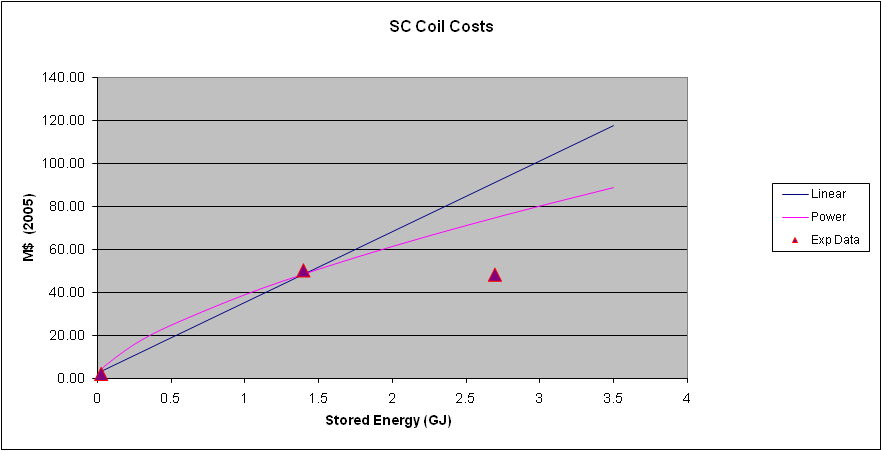}}
\caption[Superconducting coil costs.]{Superconducting coil costs.
The triangles are, from the left, the estimated cost of the BaBar
coil, an industrial estimate for an earlier version of SiD, and our
understanding of the CMS coil cost.}\label{fig:cost7}
\end{figure}

The superconducting coil cost is difficult to estimate, because
there is little data and experience with coils of this size and
field. An attempt was made to extract the CMS coil cost, and it is
believed to be $\sim ${\$}48M for the cold mass and vacuum tank. An
industrial estimate for the SiD coil was obtained, and it was
approximately the same as CMS, but for a coil with roughly half the
stored energy. Cost functions linear in the stored energy and with a
0.66 exponential dependence have been studied, and are shown in
Figure~\ref{fig:cost7}. SiD has taken a conservative approach and
for the parametric study has used a linear model fit to the Babar
coil at the low end and the industrial estimate at the high end. The
result for the current SiD design is {\$}55M, higher than the CMS
cost, but inflation and currency exchange variations have been
ignored. SiD is doing R{\&}D on advanced conductor design, and there
is some reason to expect the coil cost estimate to decrease.

The SiD cost in ILC value units is {\$}238M for M{\&}S, 206 MY
engineering, 613 MY technical, and 25 MY administrative labor. The estimated
M{\&}S contingency, reflecting uncertainty in unit costs and some estimate
of the maturity of this study, is {\$}88M. The cost in US accounting,
assuming a construction start in 2016 and 3.5{\%} per year inflation and US
National Laboratory labor rates, is {\$}700M. For the SiD Detector Outline
Document in 2006, the base M{\&}S cost estimate was {\$}203M, to be compared
with the {\$}238M present value. About {\$}20M of the difference is in the
approach used for estimating the superconducting solenoid.


\begin{thebibliography}{99}

\bibitem{role-pol}
G.~Moortgat-Pick et~al.
\newblock The role of polarized positrons and electrons in revealing
  fundamental interactions at the linear collider.
\newblock {\em Phys. Rept.}, 460:131--243, 2005.

\bibitem{rdr}
N.~Phinney, N.~Toge and N.~Walker Editors, {\it International Linear
Collider
  Reference Design Report - Volume 3: Accelerator},
  http://www.linearcollider.org/cms/?pid=1000437 (2007).

\bibitem{pole-technote}
S. Boogert et al., {\it Polarimeters and Energy Spectrometers for
the ILC Beam
  Delivery System}, ILC-NOTE-2009-049.

\bibitem{Aurand}
B. Aurand et al., {\it Executive Summary of the Workshop on
Polarisation and
  Beam Energy Measurement at the ILC}, DESY-08-099, ILC-NOTE-2008-047,
  SLAC-PUB-13296 (2008); Workshop website is
  https://indico.desy.de/conferenceDisplay.py?confId=585.

\bibitem{Thomson:2008zz}
Mark Thomson.
\newblock {Particle flow calorimetry}.
\newblock {\em J. Phys. Conf. Ser.}, 110:092032, 2008.

\bibitem{Thomson:2007xb}
Mark~A. Thomson.
\newblock {Progress with Particle Flow Calorimetry}.
\newblock 2007.

\bibitem{Schulte:1997tk}
D.~Schulte.
\newblock {Study of electromagnetic and hadronic background in the interaction
  region of the TESLA Collider}.
\newblock DESY-TESLA-97-08.

\bibitem{Chen:1993ye}
P.~Chen, T.~L. Barklow, and Michael~Edward Peskin.
\newblock {Beamstrahlung and minijet background from gamma gamma collisions in
  linear colliders}.
\newblock Prepared for 2nd International Workshop on Physics and Experiments
  with Linear e+ e- Colliders, Waikoloa, Hawaii, 26-30 Apr 1993.

\end{thebibliography}

\begin{thebibliography}{99}
\bibitem{ref:Boogert_2002jr}
  S.~T.~Boogert and D.~J.~Miller,
  arXiv:hep-ex/0211021.

\bibitem{ref:ECAL_intro_pi0GW}G.W. Wilson, "pi0 Reconstruction in Jets", Jet and
Photon Energy Measurements session, ALCPG 2007, Fermilab,
{\verb$http://ilcagenda.linearcollider.org/conferenceDisplay.py?confId=1556$}

\bibitem{ref:ECAL_intro_GEANT4}
  S.~Agostinelli {\it et al.}  [GEANT4 Collaboration],
  Nucl.\ Instrum.\ Meth.\  A {\bf 506} (2003) 250.

\bibitem{ref:Ray_Calor02} R. Frey, {\it et al.},
{\it Proc. 10th International Conference on Calorimetry in High
Energy Physics, pg. 54}, World Scientific, 2002.

\bibitem{ref:WWS_review} World-wide Study review of Calorimeter R\&D program,
talks by D.~Strom, agenda URL:
{\verb$http://ilcagenda.linearcollider.org/sessionDisplay.py?sessionId=$}\\
{\verb$38&slotId=22&confId=1296$}

\bibitem{ref:ANL_review} DoE/NSF review of the U.S. ILC R\&D program,
talk by R.~Frey, agenda URL:\\
{\verb$http://ilcagenda.linearcollider.org/conferenceDisplay.py?confId=1640$}

\bibitem{ref:ECAL_MAPS_basics} N.~K.~Watson {\it et al.},
  ``A MAPS-based readout of an electromagnetic calorimeter for the ILC,''
  J.\ Phys.\ Conf.\ Ser.\  {\bf 110} (2008) 092035.

\bibitem{ref:ECAL_MAPS_details} J.~A.~Ballin {\it et al.}, ``Monolithic
Active Pixel Sensors (MAPS) in a quadruple well technology for  nearly 100
factor and full CMOS pixels'', Sensors 2007 Journal, arXiv:0807.2920

\bibitem{ref:ECAL_MAPS_LCWS08} J.~A.~Ballin {\it et al.},``A digital
ECAL based on MAPS,'' arXiv:0901.4457 [physics.ins-det].




\bibitem{ref:HCAL_RPC_designs} G.Drake et al., Nucl. Instrum.Meth.,A578, 88-97(2007).

\bibitem{ref:HCAL_RPC_calice} https://twiki.cern.ch/twiki/bin/view/CALICE/WebHome

\bibitem{ref:HCAL_RPC_groups} http://www.hep.anl.gov/repond/DHCAL\_US.html

\bibitem{ref:HCAL_RPC_muons} B.Bilki et al., JINST {\bf{3}} P05001 (2008).

\bibitem{ref:HCAL_RPC_positrons} B.Bilki et al., ``Measurement of Positron Showers
with a Digital Hadron Calorimeter", ArXiV 0902.1699, submitted to
JINST.

\bibitem{ref:HCAL_RPC_pions} B.Bilki et al., ``Hadron Showers in a Digital Hadron
Calorimeter'', to be submitted for publication.

\bibitem{ref:HCAL_RPC_rate} B.Bilki et al., ``Measurement of the Rate Capability of
Resistive Plate Chambers'', ArXiV 0901.4371, submitted to JINST.

\bibitem{ref:HCAL_RPC_environ} See talk by Q. Zhang at the 2008 Linear Collider
Workshop LCWS2008, Chicago, IL.

\bibitem{ref:HCAL_GEM_DHCAL}For an early discussion of the GEM DHCAL concept please
see:
{\verb$http://www-hep.uta.edu/~white/SiD/LOI/GEM_DHCAL_FNAL_092205.ppt$}

\bibitem{ref:HCAL_GEM_KPiX}KPiX, An Array of Self Triggered Charge Sensitive Cells
Generating Digital Time and Amplitude Information. D. Freytag et al.
SLAC-PUB-13462, Dec 11, 2008. 4pp. Presented at 2008 IEEE Nuclear
Science Symposium (NSS) and Medical Imaging Conference (MIC) and
16th International Workshop on Room-Temperature Semiconductor X-Ray
and Gamma-Ray Detectors (RTSD), Dresden, Germany, 18-25 Oct 2008.


\bibitem{ref:HCAL_GEM_TGEM}
   R. Chechik, A. Breskin, C. Shalem, D. Mormann,
   ``Thick GEM-like hole multipliers: Properties and possible applications,''
   Nucl.Instrum.Meth.{\bf A535}, 303 (2004).


\bibitem{ref:HCAL_GEM_RETGEM}V Peskov et al. 2007 Vienna Conference on
Instrumentation, {\verb$http://vci.oeaw.ac.at/2007/$}.



\bibitem{micro1}
Y, Giomataris, Ph. Rebourgeard, J.P Robert and G. Charpak,
"MICROMEGAS: A High granularity position sensitive gaseous detector
for high particle flux environments", NIM A376, 1996, pp 29-35

\bibitem{micro2}
D. Roy, "CENTAURE Acquisition Program",
http://www-subatech.in2p3.fr/~electro/infoaq/CENTAURE/main\_centaure.html

\bibitem{micro3}
S. Callier, F. Dulucq, Ch. de La Taille, G. Martin-Chassard, N.
Seguin-Moreau, R. Gaglione, I. Laktineh, H. Mathez, V. Boudry, J-C.
Brient, C. Jauffret, "HARDROC1, readout chip of the Digital HAdronic
CALorimeter of ILC", IEEE-NSS Conference Record, vol 3, 2007, pp
1851-1856

\bibitem{micro4}
R. Gaglione, "DIRAC: Digital Readout ASIC for hAdronic Calorimeter",
IEEE-NSS/MIC 2008

\bibitem{micro5}
B. Hommels, "Data Acquisition Systems for Future Calorimetry at the
International Linear Collider", IEEE-NSS/MIC 2008

\bibitem{ref:HCAL_scint_sipm}B. Dolgoshein et. al, NIM A504:48-52, 2003.

\bibitem{ref:HCAL_scint_ham}Hamamatsu Photonics K. K.,
Solid-state Division, 1126-1 Ichino-cho, Hamamatsu City, 435-8558,
Japan.

\bibitem{ref:HCAL_scint_analnotes}https://twiki.cern.ch/twiki/bin/view/CALICE/CaliceAnalysisNotes

\bibitem{ref:HCAL_scint_minical}``A high granularity scintillator
hadronic-calorimeter with SiPM readout for a linear collider
detector.'' V. Andreev et al, NIM A540, 2005.

\bibitem{ref:slic}
  http://www.lcsim.org/software/slic/doxygen/html/.

\end{thebibliography}

\begin{thebibliography}{7}
\bibitem{Milstene:1} Milstene, C. , H.E. Fisk, A. Para (Fermilab),
``Tests of the Charged Particle Stepper with Muons'',
\textbf{FERMILAB-TM-2274-E}, Oct 2004. 17pp.
\bibitem{Zhang:2008} Q. Zhang, ``Primary Muon ID Results''
\newline http://silicondetector.org/display/SiD/Muon+Talks (unpublished) 2008.
Band, H. R. and H. E. Fisk,``SiD Muon System,''
\newline http://silicondetector.org/download/attachments/33358377
\newline SiD{\_}Muon{\_}Oxford{\_}hrb.ppt,(unpublished) Apr 2008.
\bibitem{Mokhov:2005} Mokhov, N.~V., A.~I.~Drozhdin, M.~A.~Kostin,
``\emph{Beam Collimation and Machine Detector Interface at the
International Linear Collider}, '' PAC 2005 Paper
\textbf{Fermilab-Conf-05-154-AD} May 2005.

\bibitem{Band:2009} H.~R.~Band,``\emph{RPC/KPiX Studies for Use in Linear Collider
Detectors},''\newline
{http://silicondetector.org/download/attachments/37323762/RPC{\_}Wisc.pdf}
\bibitem{Lu:2009} C.~Lu \emph{et al}. ``\emph{Aging Study for SiD Hcal and
Muon System RPCs},''\newline
http://silicondetector.org/download/attachments/37323762/\newline
${RPC{\_}Princeton{\_}k.pdf}$
\bibitem{Karchin:2009} P. Karchin \emph{et al.},``\emph{Scintillator
Based Muon System {\&} Tail-CatcherR{\&}D}'',\newline
http://silicondetector.org/download/attachments/37323762/
${ILC{\_}Muon{\_}2009{\_}proposal{\_}v2.pdf}$



\end{thebibliography}

\begin{thebibliography}{99}
\bibitem{stau} ``Experimental implications for a linear collider of the
SUSY dark matter scenario,'' P. Bambade et al.,
arXiv:hep-ph/0406010v1, june 2004.

\bibitem{did} ``IR Optimization, DID and Anti-DID,'' A. Seryi, T. Maruyama and
B. Parker, SLAC-PUB-11662, Jan. 2006.

\bibitem{loge} ``A simple method of shower localization and identification in
laterally segmented calorimeters,'' Nucl. Instrum. Methods A, vol.
311, 130 (1992).
\bibitem{egs} ``The EGS5 code system,'' H. Hirayama, Y. Namito, A.F. Bielajew,
S.J. Wilderman and W.R. Nelson, SLAC-R-730 (2005)

\bibitem{fluka} ``FLUKA: a multi-particle transport code,'' CERN-2005-10 (2005), INFN/TC-05/11, SLAC-R-773.

\bibitem{g4} ``GEANT4 developments and applications,'' IEEE Transactions on
Nuclear Science, 53, 270 (2006)

\bibitem{bhwide} ``$O(\alpha)$ YFS exponentiated monte carlo for Bhabha scattering at wide angles for LEP/SLC and LEP2,'' Phys. 
Lett. B390, 298 (1997)

\bibitem{guinea} ``Study of Electromagnetic and Hadronic Background in the
Interaction Region of the TESLA Collider,'' D. Schulte, Ph.D Thesis,
Hamburg University 1996.

\bibitem{uriel} ``Electron Identification in the BeamCal of the Silicon Detector at ILC,'' G. Oleinik, J. Gill and U. Nauenburg, 
ILC-SID-TechNote-003. For details, see
http://hep-www.colorado.edu/\~{}uriel/Beamstrahl\_TwoPhoton-Process/grp\_results.html
.

\end{thebibliography}

\begin{thebibliography}{99}

\bibitem{Kilian:2007gr}
  W.~Kilian, T.~Ohl and J.~Reuter,
  ``WHIZARD: Simulating Multi-Particle Processes at LHC and ILC,''
  arXiv:0708.4233 [hep-ph].


\bibitem{Sjostrand:2006za}
  T.~Sjostrand, S.~Mrenna and P.~Skands,
  ``PYTHIA 6.4 physics and manual,''
  JHEP {\bf 0605}, 026 (2006)
  [arXiv:hep-ph/0603175].

\bibitem{Davier:1992nw}
  M.~Davier, L.~Duflot, F.~Le Diberder and A.~Rouge,
  Phys.\ Lett.\  B {\bf 306}, 411 (1993).

\bibitem{Duflot:1993gm}
  L.~Duflot,
  ``New method of measuring tau polarization. Application to the tau $\to$ a1
  tau-neutrino channel in the ALEPH experiment'', LAL-93-09.

\bibitem{geant4_1}
  S. ~Agostinelli et al, ``GEANT4�a simulation toolkit'', {\it Nuclear Instruments and
  Methods in Physics Research  A 506 (2003) 250-303.}

\bibitem{geant4_2}
  J.~Allison et al, ``Geant4 developments and applications'', {\it IEEE Transactions on
  Nuclear Science  53 No. 1 (2006) 270-278.}

\bibitem{slic}
  http://www.lcsim.org/software/slic/doxygen/html/ .

\bibitem{lcio}
  F.~Gaede, T.~Behnke, N.~Graf and T.~Johnson, LC-Note LC-TOOL-2003-053.

\bibitem{kinfit}
  B.~List, J.~List, ``MarlinKinfit: An Object--Oriented Kinematic
Fitting Package'', LC-TOOL-2009-001.

\bibitem{fann} http://leenissen.dk/fann/ [leenissen.dk]


\bibitem{LCFI}
http://hepwww.rl.ac.uk/LCFI/; LCFI Collaboration, "LCFIVertex
package: vertexing, flavour tagging and vertex charge reconstruction
for the design  of an ILC vertex detector", NIM paper in
preparation.

\bibitem{Battaglia:2006bv}
  M.~Battaglia, T.~Barklow, M.~E.~Peskin, Y.~Okada, S.~Yamashita and P.~M.~Zerwas,
  ``Physics benchmarks for the ILC detectors,''
{\it In the Proceedings of 2005 International Linear Collider
Workshop (LCWS 2005), Stanford, California, 18-22 Mar 2005, pp 1602}
  [arXiv:hep-ex/0603010].

\bibitem{tsukamoto:1995}
  T.~Tsukamoto et al. ``Precision study of supersymmetry at future linear $e^+ e^-$ colliders'',
  {\it Phys. Rev. D 51, 3153 (1995).}

\bibitem{lyonsbook}
  L.~Lyons, {\it ``Statistics for nuclear and particle physicists'' Cambridge University Press, 1986.}

\bibitem{ZH} T. Kuhl, K. Desch, {\it ``Simulation of the measurement of the hadronic branching
ratios for a light Higgs boson at the ILC''}, LC-PHSM-2007-001.


\bibitem{Amsler:2008zzb}
  C.~Amsler {\it et al.}  [Particle Data Group],
  Phys.\ Lett.\  B {\bf 667}, 1 (2008).

\bibitem{Boos:1999ca}
  E.~Boos, M.~Dubinin, M.~Sachwitz and H.~J.~Schreiber,
  Eur.\ Phys.\ J.\  C {\bf 16}, 269 (2000)
  [arXiv:hep-ph/0001048].


\bibitem{Abazov:2006qp}
  V.~M.~Abazov {\it et al.}  [D0 Collaboration],
  Phys.\ Rev.\  D {\bf 74}, 112002 (2006)
  [arXiv:hep-ex/0609034].

\bibitem{calchep} A. Pukhov, arXiv:hep-ph/0412191v1.

\bibitem{topMS} S. Fleming {\it et al.}, PR D77, 114003 (2008).

\bibitem{sbottom} S. Profumo, Phys.Rev. D68 (2003) 015006.

\end{thebibliography}


\chapter{SiD R\&D}
\label{chap:rnd}

The SiD R\&D Plan is designed to provide a technical basis for an
optimized SiD detector in 2012.

\begin{center}
\begin{large}\textbf{Critical R\&D for the SiD Detector Concept}\end{large}
\end{center}

Members of the SiD detector concept have been engaged in R\&D for
detector components and software systems at the International Linear
Collider (ILC) for a number of years. In this section we summarize
the critical R\&D tasks for each area of SiD.

\textbf{1) General.} For the overall performance of the SiD
detector, we need to demonstrate that the detector can adequately
address the full spectrum of the physics at a 500GeV ILC, with
extension to 1 TeV, including studies of benchmark reactions. This
requires a full and realistic simulation of the detector, track
reconstruction code, and development of a fully functional Particle
Flow Algorithm (PFA). While we have working versions of the
simulation, reconstruction, and the PFA, we anticipate significant
further developments, which will provide the critical tools to
optimize and finalize the detector design.

\textbf{2) Vertex Detector.} No ILC ready vertex detector sensor yet
exists. The main needs are to develop one or more solutions for the
sensors, a demonstrably stable and low mass mechanical support, and
pulsed power/cooling solutions. Sensor technologies are being
developed, as well as mechanical support concepts, materials, pulsed
power, and cooling.

\textbf{3) Tracking Detector.} The priorities for tracking are
testing a multi-sensor prototype with readout in the absence of a
magnetic field and at 5T, refining the track finding and fitting
performance, understanding the optimal forward sensor configuration,
developing more detailed understanding of the mechanical stability
and power distribution, and designing the alignment system. Work is
underway in all of these areas, some by members of the SiLC
Collaboration.

\textbf{4) Electromagnetic Calorimetry. } For the baseline
silicon-tungsten ECAL design, the operability of a fully integrated
active layer inside the projected 1.25mm gap between absorber plates
must be demonstrated. Sufficient S/N, successful signal extraction,
powering, and adequate cooling must be shown as well. Mechanical
prototypes with steel rather than tungsten will first be built,
followed by a full depth tower appropriate for beam tests. Sensor
operation at 5T will be demonstrated. For the alternative MAPS
technology, the physics advantage must be clarified, large sensors
produced with sufficient yield, and power requirements understood.

\textbf{5) Hadronic Calorimetry.} The priority for hadronic
calorimetry is to demonstrate the feasibility of assembling a fully
integrated, full-size active layer within an 8 mm gap between
absorber plates. Several technologies are being investigated, which
need technical demonstration: RPC's, GEM's, Micromegas, and
scintillating tiles/SiPM's. All of this work is being carried out in
conjunction with the CALICE Collaboration, and the results will form
a critical component of SiD's future technology selection. An
alternative approach, using homogeneous crystal calorimetry with
dual readout, is also being studied. This effort needs to
demonstrate good hadronic energy linearity and resolution in a test
beam, to develop suitable crystals, to produce a realistic
conceptual design, and to simulate physics performance.

\textbf{6) Electronics. } One critical item in electronics is the
demonstration of the operation of the 1024 channel version of the
KPiX chip and full documentation of its performance. Another is to
develop power distribution schemes with DC-DC conversion or serial
powering. A third is development  of a readout chip for the Beamcal
that can handle unit occupancy and a high radiation environment.

\textbf{7) Magnet.} For the superconducting solenoid, it is required
to demonstrate that a 5T field can be achieved with acceptable
forces, reliability and cost. To address cost reduction, a new
conductor is being studied. R\&D for the Detector Integrated
anti-Dipole coils is also required. The field uniformity required
for pattern recognition in the tracker must be understood, and
accommodated with refined designs for the flux return steel.

\textbf{8) Engineering Issues. } A credible scheme for push-pull
operation is required that achieves acceptable repositioning of the
detector, preserving internal alignment, in an acceptably short
cycle time. Equally important is achieving the required mechanical
stability of the quadrupole focusing lenses.

\textbf{9) Forward Calorimetry.  } A sensor that can survive the
radiation environment in the beamcal is required. Some of this work
is collaborative with the FCAL Collaboration.

\textbf{10) Muon system.  } Emphasis is placed on development of
reliable, and robust RPCs. SiPMs for scintillator strips are an
alternative technology of interest, also under development.

\begin{center}
\begin{large}\textbf{Prioritization of SiD R\&D}\end{large}
\end{center}

The previous section enumerated the critical R\&D areas for the SiD
detector concept. Prioritization of R\&D is a complicated
undertaking, and we understand that not every aspect of the R\&D can
have the same emphasis. Many considerations need to be balanced in
setting priorities: importance of the R\&D to viability of the
detector concept; expected return on investment; growth of the
collaboration; and proper exploitation of developments world-wide.

Of critical importance for any high energy collider detector is
achieving superb jet energy resolution. Although significant
progress has been made establishing the viability of the Particle
Flow concept, it has not been demonstrated experimentally. Proving
that SiD's choice of hadron calorimeter can do the physics and is
cost effective is our top priority. Calorimetry is thus seen as
critical to the SiD detector concept. A very fine-grained
Silicon-Tungsten electromagnetic calorimeter and an RPC based
hadronic calorimeter form the baseline for the detector, with GEM,
micromegas, and scintillator as interesting alternative
technologies. It is proposed that both baseline technologies be
readied for large scale prototype efforts as quickly as possible,
together with the development of the associated particle flow based
software. It is strategically important to also start R\&D on a
crystal based total absorption calorimeter, which is complementary
to particle flow calorimetry. The SiD tracker employs a hybrid-less
readout scheme. The sensors have been procured in previous funding
cycles. To establish the hybrid-less concept, the development of an
array of sensors mounted on a support structure, with full readout
in actual detector configuration in a test beam is given high
priority. Readout through the baseline architecture, employing the
KPiX chip, is given priority. Because the deployment of two
detectors will be achieved through the push-pull mechanism, it is
deemed critical to engineer structures that will preserve the
alignment of the tracker  during this operation. The 3D vertical
integrated silicon technology and the further development of the
Chronopixel are given priority in the area of the vertex detector,
since no viable technology for an ILC vertex detector exists yet.

\begin{center}
\begin{large}\textbf{R\&D Efforts}\end{large}
\end{center}

Ongoing R\&D efforts closely linked to SiD are described below along
with the expected evolution of the R\&D in 2010-12 (see Tables
\ref{tab:RnD1} and \ref{tab:RnD2}). Other efforts world-wide may
also contribute to SiD's R\&D goals.


\underline{\textbf{Vertex Detector}
}

\textit{High efficiency identification of heavy flavor quarks, with the ability to distinguish quarks from anti-quarks, will be of crucial importance for the anticipated spectrum of physics signals at the ILC. While it is foreseen that the needed performance can be developed, to date no vertex detector technology meets the ILC requirements. This is reflected in the fact that vertex detector technologies are an SiD priority. Many different technologies are being pursued, such as DEPFET detectors, monolithic active pixel sensors (MAPS), in-situ storage imaging sensors (ISIS), and charge-coupled detectors (CCD), such as fine pixel, column parallel and short column CCDs. Because it is not a priori clear which technology will give the best overall performance, SiD has not adopted a baseline technology for the vertex detector. But SiD is engaged in developing technologies which can time stamp individual bunch crossings, which it considers highly desirable. }

\textbf{Future year vertex detector developments}

The technology for an ILC vertex detector has not been established.
Efforts are proceeding globally on a number of different solutions
(ISIS, DEPFET, CCDs, 3D Vertical Integrated Silicon, and MAPS, for
example). On the other hand, there is time to find a solution. Once
technologies are identified as holding promise for an ILC
environment, the R\&D will move towards full-scale sensors with the
associated readout. Given the expense of silicon technology, we
expect this level of R\&D to require funding levels beyond the
current level. There should be a technology selection by the time of
ILC construction approval. The selected technology will then be
applied to the design and construction of a small prototype. R\&D in
parallel will include understanding the number of background pulses
that can be integrated without destroying tracking capability,
developing mechanics, powering schemes, and understanding the
realistic limits on material reduction.

\begin{table}
\begin{tabular}[]{| l | l | l | l | l |}
\hline
Project Descriptor &Institution(s)&  Past & Current & Future    \\
   &   & Funding & Funding & Funding\\ \hline
\hline
\textbf{ENGINEERING}&  &&   &\\ \hline
SC Solenoid & Fermilab, SLAC &DOEL &DOEL &DOEL \\ \hline
Flux Return & Fermilab, SLAC, Wisc. &DOEL &DOEL &DOEL\\ \hline
MDI/Push-pull & Fermilab, SLAC &DOEL &DOEL &DOEL\\ \hline
General &Fermilab, SLAC, etc. & DOEL &DOEL &DOEL\\ \hline
\hline
\textbf{VERTEX}&  &&&   \\ \hline
Chronopix  &Yale, Oregon, SLAC &     LCDRD &sLCDRD, & ULCDP,    \\
   & & &DOEL & DOEL \\ \hline
CAPS      &Hawaii & LCDRD &sLCDRD & ULCDP   \\ \hline
ISIS &Bristol, Oxford, RAL & UK,   &internal & new fund.  \\
        &                                     & term. 2008 &     & \\ \hline
MAPS &Strasbourg &France &France &France \\ \hline
Sensor Powering &Fermilab &DOEL &DOEL&DOEL \\ \hline
Sensor Thinning &Fermilab, Cornell &DOEL &DOEL&DOEL \\ \hline
Vertex Mechanics  &Washington,  & LCDRD, &DOEL & ULCDP,     \\
   &Fermilab & DOEL & & DOEL\\  \hline
3D Sensor &Fermilab &DOEL &DOEL&DOEL \\ \hline
3D Sensor Sim.    &Cornell &    &DOEL& ULCDP    \\ \hline
\hline
\textbf{TRACKING}&  &&& \\ \hline
Alignment     &Michigan, Spain, &   LCDRD,  &Spain & ULCDP, \\
      &Spain &Spain  & &Spain   \\ \hline
DC-DC converters  &Yale, SLAC &LCDRD,   &DOEL   & ULCDP,    \\
  &  & DOEL & & DOEL\\ \hline
DS/SS Si Det. &Kyungpook N Univ. & Korea & Korea & Korea \\
\& calor (eg. lum)& & & & \\ \hline
Forward Tracking &Spain &Spain
&Spain
&Spain \\ \hline Sensor QA, cables  &New Mexico &    &   & ULCDP \\
\hline Sensor QA  &UC Santa Cruz & LCDRD &sLCDRD & ULCDP   \\ \hline
Tracking Sim. &Santa Cruz, Oregon,  &DOEL &DOEL&DOEL \\
 & SLAC, Fermilab & & &ULCDP\\ \hline
\end{tabular}
\caption{SiD R\&D Projects, part 1. Definitions: LCDRD is 3 year US
DOE/NSF Linear Collider Detector R\&D grant beginning 2005; sLCDRD
is 2008-9 LCDRD supplement; DOEL is US DOE laboratory funding; ULCDP
is SiD US University LC Detector Proposal to DOE/NSF on February 18,
2009.}\label{tab:RnD1}
\end{table}


\underline{
\textbf{Tracker}}

\textit{The ILC experiments demand tracking systems unlike any previously envisioned. In addition to efficient and robust track-finding over the full solid angle, the momentum resolution required to enable precision physics at ILC energies must improve significantly beyond that of previous trackers. The design must minimize material in front of the calorimeter that might compromise particle-flow jet reconstruction. Establishing and maintaining the alignment for the tracker is critical. Even with the largest feasible magnetic field, the tracking volume is quite large, demanding optimized tracker components which facilitate mass production. Finally, the tracker must be robust against beam-related accidents and aging. All these requirements must be maintained within a "push-pull" scenario. }

\textbf{Future year tracking detector developments}

The emphasis in the area of the tracking detector is currently on
the development of the double-metal sensor with the associated
readout. These sensors need to work in a 5 T magnetic field and
remain stable and aligned  during power pulsing. The forward tracker
design must be optimized. A small scale system consisting of a few
sensors with full readout will be tested in a test beam under these
operating conditions. Only then can issues associated with the
Lorentz forces and mechanical stability be tested. Tracking R\&D
will require an increased level of support to address these
questions.


\underline{\textbf{Calorimeters}}

\textit{
To measure hadronic jets of particles produced in high-energy collisions of electrons and positrons, with sufficient precision, it is widely accepted that a new approach is necessary. The most promising method, utilizing Particle Flow Algorithms (PFAs), uses the tracker to measure the energy of charged particles and the calorimeter to measure the energy of neutrals. The baseline SiD detector concept accepts that a PFA is necessary, and is designed to optimize the PFA performance with the goal of obtaining jet energy resolutions of the order of 3-4\% of the jet energy.}

\textit{
SiD has adopted initial baseline technologies for both the electromagnetic and hadron calorimeters, and is also pursuing alternative technologies, since it is not, a priori, clear which combination(s) of hardware and PFA software will produce the most cost-effective performance.}

\textit{
We have defined performance criteria for the calorimeter systems, both in terms of basic design requirements for use with a PFA, and detailed hardware performance for efficient and reliable operation. These criteria will be used in the final selection of electromagnetic and hadron calorimeter technologies once all the required R\&D has been completed.}

\begin{table}
\begin{tabular}[] {| l | l | l | l | l |}
\hline
Project Descriptor &Institution(s)&  Past & Current & Future    \\
   &   & Funding & Funding & Funding\\ \hline
\hline
\textbf{CALORIMETRY}&   &&  &\\ \hline
Beamcal, Readout/Sim      &Colorado, SLAC & LCDRD, &DOEL & ULCDP,   \\
  &  &DOEL  &  &DOEL\\ \hline
Dual Readout  &Caltech, Fermilab & DOEL &DOEL & DOEL        \\
& & & & ULCDP \\ \hline
ECAL Mechanics &Annecy, SLAC &France, &France, &France,  \\
    &  & DOEL & DOEL & DOEL \\ \hline
GEM HCAL      &UTA, MIT &   LCDRD & & ULCDP \\ \hline Micromegas
HCAL &Annecy & France &France&France \\ \hline
MAPS ECAL &B'ham, Bristol, & UK &SPiDeR& SPiDeR \\
  &Imperial, Oxford, RAL & && \\ \hline
PFA   &Iowa, MIT,  &    LCDRD,  &DOEL & ULCDP,  \\
  & ANL, SLAC &DOEL &&DOEL \\ \hline
Radhard Bmcal Sens. & UCSC, SLAC & &DOEL & ULCDP,\\
    &  &  &  &DOEL\\ \hline
RPC HCAL      &BU, Iowa, ANL, FNAL &  LCDRD, &sLCDRD, & ULCDP,    \\
  &  & DOEL &DOEL& DOEL\\ \hline
Scint HCAL  &NIU    & LCDRD & & ULCDP       \\ \hline SiPM Readout
&Fermilab, SLAC & DOEL &DOEL&DOEL \\ \hline
Si-W ECAL \& Readout      &Oregon, Davis,  &LCDRD, &sLCDRD, &ULCDP, \\
& SLAC, BNL &    DOEL &DOEL & DOEL  \\ \hline \hline \textbf{MUON} &
&&&   \\ \hline RPC Muon/HCAL  &Princeton & LCDRD & & ULCDP \\
\hline RPC Muon System   &Wisconsin    &LCDRD & & ULCDP    \\ \hline
SiPM Muon System  &WS, IU, ND, NIU, &LCDRD, &sLCDRD, &ULCDP, \\
 & Fermilab&DOEL &DOEL & DOEL       \\ \hline
\hline
\textbf{BEAMLINE MEAS.} & &&& \\ \hline
Energy Spectrometer & Notre Dame, Oregon & LCDRD &sLCDRD & ILC accel? \\ \hline
Polarimeter & Iowa & LCDRD & & ILC accel? \\ \hline

\end{tabular}
\caption{SiD R\&D Projects, part 2. See table \ref{tab:RnD1} for
definitions.}\label{tab:RnD2}
\end{table}


\textbf{Future year calorimeter developments}

For the Si-W ECAL, we anticipate that the test-beam related data
taking and analysis will continue beyond August 2010. First the
module will be tested in an electron beam (possibly at SLAC),
followed later by a beam test with hadrons. Completion of this R\&D
is expected by 2012. For the MAPS ECAL the goals are to assess the
physics potential for the device with simulation studies and to make
a second generation chip which is sufficiently large to make an ECAL
stack to study digital eletromagnetic calorimetry in detail.

The RPC option for the HCAL will continue with testing the CALICE $1
m^{3}$ stack beyond the first year. The calorimeter will be exposed
to muons and pions and positrons of various energies. The response
and energy resolution will be measured together with characteristics
of hadronic showers, such as the lateral and longitudinal shower
shapes, which are relevant for Particle Flow Algorithms. The GEM
option will test its $1 m^{2}$ layers as part of the CALICE hadron
calorimeter prototype, and will design and build a complete,
integrated layer with minimal thickness and full services. Thick GEM
prototypes will also be assembled and tested as large sections of
thick GEMs become available. Gas studies for thick GEMs will also
continue. The micromegas option will see the continued testing and
analysis of results from the $1 m^{3}$ stack. Finally, the major
activity for the scintillator/SiPM option will be the insertion of
the integrated readout layer planes fabricated with the CALICE/EUDET
electronics into the CALICE absorber stack. This installation will
be followed by the commissioning and exposure of this prototype to a
test beam.

The homogeneous dual-readout calorimetry will continue development
of suitable crystals, photodetectors, and associated readout
electronics, all in preparation for a demonstration of linearity and
energy resolution for hadrons in a test beam, while developing a
conceptual design for inclusion of this technology into SiD, and
evaluating physics performance.

Continued effort will also be directed towards improving and
refining the SiD Particle Flow algorithm and simulation of the
dual-readout concept. This will involve several SiD groups and focus
on improvements for higher jet energies.

Progress on all calorimetry options will be monitored closely, with
a technology selection for SiD during 2011.


\underline{\textbf{Muons}}

\textit{The SiD muon system is designed to identify muons from the interaction point with high efficiency and to reject almost all hadrons (primarily pions and kaons). The muon detectors will be installed in gaps between steel layers of the solenoid flux return. Since the central tracker will measure the muon candidate momentum with high precision, the muon system only needs sufficient position resolution to unambiguously match calorimeter tracks with muon tracks. Present studies indicate that a resolution of 1-2 cm is adequate. These position resolutions and the required rate capabilities can be met by more than one detector technology. The baseline design uses double layers of resistive plate chambers(RPC).
Also under consideration are extruded scintillator strips read out by silicon photomultipliers (SiPMS). Cost, reliability and physics capabilities should determine the preferred choice between RPCs and scintillators.  }

\textbf{Future Year Muon Developments}

The muon detector effort will work toward a technology choice during
2011. RPC efforts include cost-effective readouts and aging studies.
The scintillator SiPM work is expected to include evaluation of SiPM
manufacturers, mounting, temperature control, and monitoring, as
well as development or acquisition of an SiPM compatible ASIC,
possibly a modified KPiX. Planning for construction of a
larger-scale prototype of the selected technology would begin, with
simulation studies of overall detector performance for muons. The
simulation model of the muon system would be derived from
measurements with prototypes.


\underline{\textbf{Forward Calorimetry}}

\textit{Two photon processes present a major background to searches for certain supersymmetric particles.  These processes can be identified by detecting a high energy electron or positron in the beamcal, above the beamstrahlung background. Developing highly efficient techniques for tagging electrons and suppressing background is being done with full GEANT simulation.}

\textit{
Through its participation in the CERN-based RD50 R\&D collaboration, SiD collaborators are working on the development of Czochralski-process silicon diode sensors for application to high-radiation environments. Due to the relatively high oxygen content, it is expected that Czochralski sensors will be much more resistant to damage than standard sensors produced from float-zone process silicon. Studies are investigating the suitability of Czochralski sensors for forward calorimetry.}

\textbf{Future Year Forward Calorimeter Developments}

The outcomes of these studies of two photon backgrounds will lead to
refined designs of the beamcal. Radiation hard detectors and readout
electronics for the beamcal must be developed.


\underline{\textbf{Beamline Instrumentation}}

SiD is interested in beamline instrumentation, particularly work on
energy spectrometry and polarimetry. This area of needed R\&D can be followed
jointly with other detector groups, as well as machine groups.

\begin{center}
\begin{large}\textbf{Schedule and Milestones}\end{large}
\end{center}


{\parskip=0pt\parindent=0in

\hspace{.5 in}\underline{\textbf{2009}}

\hangindent=.2in
{\bf $\bullet$ Simulation/Reconstruction:} PFA improvements; tracking
simulation and reconstruction improvements and background
studies; simulation of dual readout concept; optimization of SiD design.

{\bf $\bullet$ Electronics:} Full KPiX chip; develop beamcal readout.

{\bf $\bullet$ Tracker:}
Sensor test; sensor with readout test; develop alignment concept.

{\bf $\bullet$ ECAL:} Sensor test; sensor with readout test.

{\bf $\bullet$ Beamcal:} Evaluation of beamcal sensor technologies.

\hangindent=.2in {\bf $\bullet$ HCAL:} RPC and GEM  with readout
tests; GEM slice test; Micromegas slice test; RPC construct 1~$m^3$;
engineering design of HCAL module; dual readout crystal candidate
selection and photon detection studies.

{\bf $\bullet$ Vertex:}
Develop sensors; continue mechanical and power distribution designs.

{\bf $\bullet$ Muon:}
Test RPCs, scintillating fiber, and RPC longevity.


\vspace{.2in}

\hspace{.5 in}\underline{\textbf{2010}}

\hangindent=.2in
{\bf $\bullet$ Simulation/Reconstruction:}
Update physics studies; dual readout full simulation and physics performance.

{\bf $\bullet$ Electronics:}
Test beamcal sensor readout; develop SiPM readout (if needed).

\hangindent=.2in
{\bf $\bullet$ Tracker:}
Test alignment concept; beam test sensors with readout and support
system in B field; test Lorentz
forces and mechanical stability with pulsed power.

{\bf $\bullet$ ECAL:} Build and test ECAL tower; build mechanical
prototype for ECAL module.

\hangindent=.2in {\bf $\bullet$ HCAL:} Produce engineering design of
module with integrated readout; dual readout beam test of concept;
continue beam tests and analysis; ready 1 $m^2$ modules of GEM and
Micromegas; continue development of suitable crystals.

{\bf $\bullet$ Vertex:}
Develop sensors; continue mechanical and power distribution designs.

{\bf $\bullet$ Muon:}
Prototype muon chambers; longevity test; study costs.

{\bf $\bullet$ MDI:}
Develop push pull designs; vibration studies; study alignment issues.

{\bf $\bullet$ Magnet:}
Develop new conductor jointly with others.

{\bf $\bullet$  Beamcal:}
Design sensors.


\vspace{.2in}

\hspace{.5 in}\underline{\textbf{2011}}

{\bf $\bullet$ Technology Selections: } ECAL, HCAL, Muon.

\hangindent=.2in {\bf $\bullet$ Engineering: } Complete engineering
designs for ECAL, HCAL and Muons for chosen technologies and forward
systems; plan preproduction and detailed design phase.

\hangindent=.2in
{\bf $\bullet$ Simulation/Reconstruction:}
Complete detector optimization; realistic GEANT4 detector description
based on technology choices; generate MC data.

\hangindent=.2in {\bf $\bullet$ Complete beam testing: } SiW ECAL,
RPC, GEM, Micromegas, Scint \& SiPM HCAL; proof of principle
development of suitable crystals  and photodetectors.

{\bf $\bullet$ Tracker:}
Test large scale system.

{\bf $\bullet$ Vertex:}
Test sensors; continue mechanical and power distribution designs.

{\bf $\bullet$ Benchmarking: }
Studies with final detector choices and optimized design.

{\bf $\bullet$ Magnet:}
Continue new conductor development.


\vspace{.2in}

\hspace{.5 in}\underline{\textbf{2012}}

{$\bullet$ Complete optimized SiD detector design.}

{$\bullet$ Begin tests of magnet material.}

{$\bullet$ Begin full scale prototyping.}

{$\bullet$ Write SiD proposal.}

}

\begin{center}
\begin{large}\textbf{Conclusion}\end{large}
\end{center}

The SiD R\&D Plan has been evolving for several years, and the
definition of SiD's critical needs has sharpened through the LoI
process.  The SiD R\&D Plan is designed to deliver the results
needed to provide a solid technical basis of an optimized SiD
detector design in 2012.

\end{document}